\documentclass[12pt]{article}
\usepackage{amsmath}
\usepackage{graphicx}
\usepackage{enumerate}
\usepackage{natbib}
\usepackage{url} 
\usepackage{binomexp}
\usepackage{mathtools}
\usepackage{amsfonts}
\usepackage{bbm}
\usepackage{dsfont}
\usepackage{longtable}
\usepackage{threeparttable}
\usepackage{yhmath}
\usepackage{bm}
\usepackage{amssymb}
\usepackage[nodisplayskipstretch]{setspace}
\usepackage{tikz}
\usepackage{float}
\usepackage{color}
\usepackage{stmaryrd}
\usepackage{paralist}
\usepackage{multirow}
\usepackage{algorithm,algpseudocode}
\makeatletter
\newcommand{\algmargin}{\the\ALG@thistlm}
\makeatother
\newlength{\whilewidth}
\algdef{SE}[parWHILE]{parWhile}{EndparWhile}[1]
  {\parbox[t]{\dimexpr\linewidth-\algmargin}{%
     \hangindent\whilewidth\strut\algorithmicwhile\ #1\ \algorithmicdo\strut}}{\algorithmicend\ \algorithmicwhile}%
\algnewcommand{\parState}[1]{\State%
  \parbox[t]{\dimexpr\linewidth-\algmargin}{\strut #1\strut}}

\newcommand{\bcx}{{\bm X}}

\newcommand{\bco}{{\bm O}}

\newcommand{\bx}{{\bm x}}

\newcommand{\bbE}{\mathbb{E}}

\newcommand{\E}{\mathbb{E}}
\newcommand{\Prob}{\mathbb{P}}
	
\newcommand{\indep}{\perp \!\!\! \perp}
\newcommand{\differential}{\text{d}}

\usepackage[colorlinks=true,linkcolor=red,citecolor=black]{hyperref}

\newtheorem{theorem}{Theorem}
\newtheorem{assumption}{Assumption}
\newtheorem{proposition}{Proposition}
\newtheorem{lemma}{Lemma}
\newtheorem{remark}{Remark}
\newtheorem{example}{Example}

\newcommand{\blind}{1}

\allowdisplaybreaks

\usetikzlibrary{shapes,decorations,arrows,calc,arrows.meta,fit,positioning}
\tikzset{
     -Latex,auto,node distance =1 cm and 1 cm,semithick,
     state/.style ={ellipse, draw, minimum width = 0.7 cm},
     point/.style = {circle, draw, inner sep=0.04cm,fill,node contents={}},
    bidirected/.style={Latex-Latex,dashed},
     el/.style = {inner sep=2pt, align=left, sloped}
 }

\begin{document}

\def\spacingset#1{\renewcommand{\baselinestretch}%
{#1}\small\normalsize} \spacingset{1}


\if1\blind
{
  \title{\Large \bf Identification and multiply robust estimation in causal mediation analysis across principal strata}
  \author{Chao Cheng$^{1,2}$ and Fan Li$^{1,2}$\vspace{0.2cm}\\
    $^1$Department of Biostatistics, Yale School of Public Health\\
    $^2$Center for Methods in Implementation and Prevention Science,\\ Yale School of Public Health}
  \maketitle
} \fi

\if0\blind
{
  \bigskip
  \bigskip
  \bigskip
  \begin{center}
    {\Large \bf Identification and multiply robust estimation in causal mediation analysis across principal strata
    }
\end{center}
  \medskip
} \fi

\bigskip
\begin{abstract}
We consider assessing causal mediation in the presence of a post-treatment event (examples include noncompliance, a clinical event, or death). We identify natural mediation effects for the entire study population and for each principal stratum characterized by the joint potential values of the post-treatment event. We derive the efficient influence function for each mediation estimand, which motivates a set of multiply robust estimators for inference. The multiply robust estimators are consistent under four types of misspecifications and are efficient when all nuisance models are correctly specified. {We also develop a nonparametric efficient estimator that leverages data-adaptive machine learners to achieve efficient inference and discuss sensitivity methods to address key identification assumptions.} We illustrate our methods via simulations and two real data examples.
\end{abstract}

\noindent%


\spacingset{1.75} 

\section{Introduction}

\subsection{Background and motivation}

Causal mediation analysis (\citealp{imai2010identification}) is widely used to investigate the role of a mediator ($M$) in explaining the causal mechanism from a treatment ($Z$) to an outcome ($Y$). 
Under the potential outcomes framework, a primary step in causal mediation analysis is to decompose the \textit{total treatment effect} into an \textit{indirect effect} that works through $M$ and a \textit{direct effect} that works around $M$. While alternative definitions exist, the natural indirect and direct effects are the most relevant for studying causal mechanisms (\citealp{nguyen2021clarifying}). 
The natural indirect effect compares potential outcomes by switching $M$ from the value it would have taken under the control condition to that under the treated condition, while fixing $Z$ to the treated condition. The natural direct effect compares potential outcomes by switching $Z$ from the control to the treated condition, while fixing $M$ to the value it would have taken under the control condition. Parametric regressions (e.g., \citealp{cheng2021estimating,cheng2023product}), semiparametric methods (e.g., \citealp{tchetgen2012semiparametric}), and nonparametric methods (e.g., \citealp{kim2017framework}) have been proposed for estimating natural mediation effects, typically by assuming that $M$ is the only variable sitting on the causal chain connecting treatment and outcome.

Increasingly for experimental and observational studies, a post-treatment event ($D$) may occur prior to the measurement of the mediator. This event may be a post-treatment action or decision regarding uptake (e.g., noncompliance or treatment discontinuation), a clinical event (e.g., worsening of disease, adverse medication effect), or a terminal event precluding the observation of any data afterward (e.g., death). In each context, the post-treatment event provides important information in defining partially observed subgroups of the study population. {An emerging interest lies in learning the treatment effect within each subgroup, and possibly the effect heterogeneity across subgroups.} Under the principal stratification framework, one can characterize each subgroup with the joint potential values of $D$ under alternative conditions, referred to as principal strata more generally \citep{frangakis2002principal}, or as endogenous subgroups in social sciences \citep{page2015principal}. {Beyond understanding the total effect within each subgroup, here we are further interested in evaluating the natural mediation effects via $M$ within each subgroup characterized by a post-treatment event $D$.} Below, we give two examples that motivate such an objective (an additional example on mediation analysis with death-truncated mediator and outcome is provided in the Supplementary Material).

\begin{example}\label{Example1}
(Mediation analysis with noncompliance) Noncompliance occurs if the actual treatment received ($D$) differs from the treatment assignment ($Z$). {By \citet{angrist1996identification}, the study population is partitioned into four subgroups, including (i) \textit{always-takers} who take the treatment regardless of assignment; (ii) \textit{never-takers} who take the control regardless of assignment; (iii) \textit{compliers} who comply with the assignment; (iv) \textit{defiers} who take the opposite assignment. Typically, the compliers are of central interest because this is the only group for whom the average causal effect due to assignment reflects the average causal effect due to actual treatment received.} As noncompliance is a post-randomization action or decision, a relevant research question in the presence of a mediator, measured prior to the outcome, is whether the treatment works through $M$ among the compliers (i.e., the complier natural mediation effect). A follow-up question is whether there is heterogeneity in the treatment effect mechanisms among the subgroups formed by noncompliance patterns.
\end{example}

\begin{example}\label{Example2}
(Mediation analysis with an intercurrent event) In health research, disease progression, adverse reaction, or other early outcome may occur due to treatment, which are collectively referred to as an intercurrent event by the ICH E9 Estimands Framework \citep{kahan2023eliminating}. Section \ref{sec:WHO-LARES} studies the role of perceived control ($M$) in mediating the effect of residence in a damp and moldy dwelling ($Z$) on depression ($Y$), but dampness/mold related disease ($D$) occurred among some study units. It is then of interest to study the indirect effect due to $M$ among those who would always develop dampness/mold related disease regardless of their living condition (i.e., the doomed stratum), those who never develop dampness/mold related disease regardless of their living condition (i.e., the immune stratum), as well as those who would develop dampness/mold related disease only if living in such condition but otherwise not (i.e., the harmed stratum). A follow-up question is whether there is variation in the natural indirect effects among these subgroups.
\end{example}

In both examples, the \textit{principal causal effect} (PCE)---the total treatment effect within a principal stratum---can be decomposed into a \textit{principal natural indirect effect} (PNIE) through the mediator and a \textit{principal natural direct effect} (PNDE) around the mediator. Addressing the PNIEs and their variation allows us to unpack the overall natural indirect effect to understand for whom and under what circumstances $M$ plays a crucial role in explaining the underlying mechanism. Such analyses could also help determine whether the overall natural indirect effect is driven by one particular subgroup, in which case future interventions might be restructured to better serve an intended subpopulation. {To this end, estimating principal natural mediation effects is informative in itself, but comparing them across strata may also provide additional insights. In fact, studying variations in principal natural mediation effects is related to moderated mediation analysis given measured baseline covariates \citep{qin2024causal}. However, the difference is that we focus on endogenous subgroups defined by a post-treatment event rather than those defined by covariates. Consequently, the scientific question addressed by principal natural mediation effects generally cannot be answered by merely exploring the conditional mediation effects given observed covariates alone (an empirical comparison is provided in Section \ref{sec:WHO-LARES}).} 

\subsection{Prior work and our contribution}

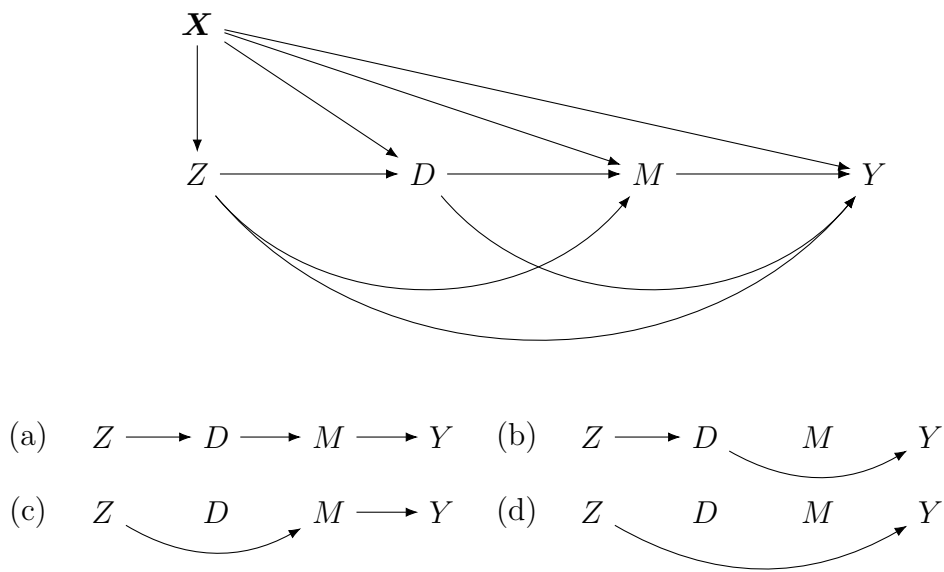
\begin{figure}[htbp]
\centering
\begin{tikzpicture}
    \node(z) at (-3,0) {$Z$};
    \node(s) at (0,0) {$D$};
    \node(m) at (3,0) {$M$};
    \node(y) at (6,0) {$Y$};
    \node(x) at (-3,2) {$\bcx$};

    \path (x) edge (z);
    \path (x) edge (s);
    \path (x) edge (m);
    \path (x) edge (y);

    \path (z) edge (s);
    \path (z) edge[bend left=-50] (m);
    \path (z) edge[bend left=-50] (y);
    \path (s) edge[bend left=-50] (y);
    \path (s) edge (m);
    \path (m) edge (y);


    \node (ind1) at (-5.25,-3.5) {(a)};
    \node (z1) at (-4.25,-3.5) {$Z$};
    \node (s1) at (-2.75,-3.5) {$D$};
    \node (m1) at (-1.25,-3.5) {$M$};
    \node (y1) at (0.25,-3.5) {$Y$};
    \path (z1) edge (s1);
    \path (s1) edge (m1);
    \path (m1) edge (y1);

    \node (ind2) at (1.25,-3.5) {(b)};
    \node (z2) at (2.25,-3.5) {$Z$};
    \node (s2) at (3.75,-3.5) {$D$};
    \node (m2) at (5.25,-3.5) {$M$};
    \node (y2) at (6.75,-3.5) {$Y$};
    \path (z2) edge (s2);
    \path (s2) edge[bend left=-30] (y2);

    \node (ind3) at (-5.25,-4.5) {(c)};
    \node (z3) at (-4.25,-4.5) {$Z$};
    \node (s3) at (-2.75,-4.5) {$D$};
    \node (m3) at (-1.25,-4.5) {$M$};
    \node (y3) at (0.25,-4.5) {$Y$};
    \path (z3) edge[bend left=-30] (m3);
    \path (m3) edge (y3);

    \node (ind4) at (1.25,-4.5) {(d)};
    \node (z4) at (2.25,-4.5) {$Z$};
    \node (s4) at (3.75,-4.5) {$D$};
    \node (m4) at (5.25,-4.5) {$M$};
    \node (y4) at (6.75,-4.5) {$Y$};
    \path (z4) edge[bend left=-30] (y4);
    
\end{tikzpicture}
\caption{A directed acyclic diagram for mediation analysis across principal strata. Notation: $Z$ is the treatment, $D$ is the post-treatment event, $M$ is the mediator, $Y$ is the outcome, and $\bcx$ is a vector of pre-treatment covariates. There are four possible causal pathways from the treatment to outcome, indexed by (a)--(d). The exclusion restriction assumption 
excludes pathways (c) and (d).\label{fig:dag}}
\end{figure}

The post-treatment event poses unique challenges on identification of the principal natural mediation effects, because (i) $D$ is a treatment-induced variable confounding the mediator-outcome relationship and (ii) the principal stratum membership is only partially observed. To tackle these challenges, a prevalent approach, usually developed under the noncompliance setting (Example \ref{Example1}), is to view the treatment assignment as an instrument variable for $D$ and use exclusion restriction to identify the mediation effects among the compliers. Exclusion restriction requires that all causal pathways from the treatment assignment to the mediator and outcome are only through $D$ (see Figure \ref{fig:dag} for illustration), and therefore no treatment or mediation effects exist among principal strata where $Z$ does not affect $D$. For example,  \citet{yamamoto2013identification}, \citet{park2018causal,park2020two} considered a combination of exclusion restriction, monotonicity of treatment assignment on the treatment receipt, and mediator ignorability to nonparametrically identify the complier natural mediation effects. Instead of assuming mediator ignorability, \cite{frolich2017direct} identified the complier natural mediation effect by drawing a second instrumental variable for the mediator. Under monotonicity and exclusion restriction, \cite{rudolph2024using} further proposed semiparametrically efficient estimators of (i)  complier natural/interventional mediation effects with a single instrumental variable for the treatment and (ii) double complier interventional mediation effects with two instrumental variables for the treatment and mediator.

A critical assumption in the prior work is the exclusion restriction, which may not always hold in open-label studies where the assignment can exert a direct \emph{psychological effect} on the mediator and the outcome not through the treatment receipt. {In Example \ref{Example2}, the PCE and PNIE among the doomed and immune strata are of interest and may be non-zero; in this instance, the exclusion restriction does not apply either.} 
Relaxing exclusion restriction, \cite{park2020sensitivity} 
developed a maximum likelihood approach with full distributional assumptions to empirically identify principal natural mediation effects; however, the consistency of their estimators requires all parametric models to be correctly specified and bias may arise under misspecification. 

Our primary interest is to identify the principal natural mediation effects in endogenous subgroups defined by a post-treatment event. We assume monotonicity and ignorability conditions for nonparametric identification, but require neither the exclusion restriction nor fully parametric modeling assumptions. In a similar context, \citet{tchetgen2014identification} studied the marginal natural mediation effect of $M$ in the presence of a post-treatment confounder $D$ based on a nonparametric structural equation model with independent errors and monotonicity, but did not provide identification results for the finer causal mediation estimands within the principal stratum. As we explain in Section \ref{sec:connection}, our identification assumptions are weaker than those considered by \cite{tchetgen2014identification} in a technical sense, but remain sufficient to unpack the stratum-specific mediation effects that contribute to the marginal natural mediation effect. Leveraging the semiparametric efficiency theory \citep[][see \cite{hines2022demystifying} for an overview]{bickel1993efficient}, we characterize the efficient influence function for each estimand under the nonparametric model to motivate semiparametric estimators. Our estimators are consistent under four types of working model misspecification, and are quadruply robust. As a further improvement, a nonparametric extension is also provided to incorporate data-adaptive machine learners for efficient inference \citep{chernozhukov2018double}. Finally, we develop strategies for sensitivity analyses under violations of the key ignorability assumptions in cases when insufficient baseline covariates are collected, or when there is unmeasured treatment-induced confounding of the mediator-outcome relationship.

\section{Notation, causal estimands and identification}
\label{sec:notation}

Suppose that we observe $n$ independent  copies of the quintuple $\bm O = \{\bcx,Z,D,M,Y\}$, where $Z\in\{0,1\}$ represents treatment assignment with 1 indicating the treated condition and 0 indicating the control condition, $D\in\{0,1\}$ is the occurrence of the post-treatment event, $M$ is a mediator measured after $D$, $Y$ is the final outcome of interest, and $\bcx$ is a vector of pre-treatment covariates. A directed acyclic graph (DAG)
summarizing the relationships among variables is in Figure \ref{fig:dag}, where $Z$ is allowed to affect $Y$ either directly or through the intermediate variables $D$ and $M$. For a generic variable $W$, we use $\Prob_W(w)$ to denote its distribution function, $f_W(w)$ to denote the probability mass/density function, $\E_W[W]$ to denote its expectation.  Whenever applicable, we abbreviate $\Prob_W(w)$, $\E_W[W]$, and $f_W(w)$ as $\Prob(w)$, $\E[W]$, and $f(w)$ without ambiguity. Moreover, we define  $\Prob_n[W]=\frac{1}{n}\sum_{i=1}^n W_i$ as the empirical average operator, $\mathbb{I}(\cdot)$ as the indicator function, $|\cdot |$ as the absolute value, $\|\cdot\|$ as the $L_2(\Prob)$-norm such that $\|g\|^2=\int g^2 d\Prob$.

We pursue the potential outcomes framework to define causal mediation estimands (\citealp{imai2010identification}). Let $D_z$ be the potential value of the post-treatment event under treatment $z$, $M_{zd}$ the potential value of the mediator when the treatment is $z$ and  $D$ is set to $d$, $Y_{zdm}$ the potential outcome when the treatment is set to $z$, the post-treatment event is set to $d$, and the mediator is set to $m$. Furthermore, we write $M_z=M_{zD_z}$ such that the potential value of the mediator under treatment $z$ is identical to that under treatment $z$ when $D$ takes its natural value under treatment $z$. Similarly, we have $Y_z=Y_{zD_zM_z}$ and $Y_{zm}=Y_{zD_zm}$. The equalities $M_z = M_{zD_z}$, $Y_z=Y_{zD_zM_z}$, and $Y_{zm}=Y_{zD_zm}$ are collectively referred to as the composition of potential values (\citealp{vanderweele2009conceptual}).

To proceed, we adopt the  principal stratification framework (\citealp{frangakis2002principal}) and use the joint potential values of the post-treatment event $U=(D_1,D_0)$ to partition the study population into four subgroups, $\{(1,0),(1,1),(0,0),(0,1)\}$. To contextualize the development, we take the noncompliance as a running example throughout, in which case these four strata are named as compliers, always-takers, never-takers, and defiers. For notational convenience, we re-express the joint potential values $(D_1,D_0)$ as $D_1D_0$ so that $U\in\{10,11,00,01\}$. A central property is that $U$ is unaffected by the treatment and can be treated as a baseline covariate; therefore causal comparisons conditional on $U$ are well-defined subgroup causal effects. We define $e_{d_1d_0}(\bcx)=f_{U|\bcx}(d_1d_0|\bcx)$ and $e_{d_1d_0}=f_U(d_1d_0)$ as the proportion of principal stratum $U$ conditional on and marginalized over covariates $\bcx$, where $e_{d_1d_0}(\bcx)$ is referred to as the principal score (\citealp{ding2017principal}). Since the stratum membership $U=D_1D_0$ is only partially observed, the principal score $e_{d_1d_0}(\bcx)$ and its marginal counterpart $e_{d_1d_0}$ cannot be estimated without further assumptions.

The PCE is defined as the effect of treatment assignment in each principal stratum (\citealp{jo2009use,ding2017principal}) and is written as:
$$
\text{PCE}_{d_1d_0} = \E\left[Y_{1}-Y_{0}|U=d_1d_0\right] \quad (d_1d_0=10,11,00,01)
$$
which equals $\E\left[Y_{1M_1}-Y_{0M_0}|U=d_1d_0\right]$ by composition of the potential outcome. To assess mediation, we decompose $\text{PCE}_{d_1d_0}$ into a \textit{principal natural indirect effect} ($\text{PNIE}_{d_1d_0}$) and a \textit{principal natural direct effect} ($\text{PNDE}_{d_1d_0}$):
\begin{align}
 \underbrace{\E\left[Y_{1M_1}-Y_{0M_0}|U=d_1d_0\right]}_{\text{PCE}_{d_1d_0}} = \underbrace{\E\left[Y_{1M_1}-Y_{1M_0}|U=d_1d_0\right]}_{\text{PNIE}_{d_1d_0}} + \underbrace{\E\left[Y_{1M_0}-Y_{0M_0}|U=d_1d_0\right]}_{\text{PNDE}_{d_1d_0}}.\label{e:principal_effects}
\end{align}
Intuitively, $\text{PNDE}_{d_1d_0}$ captures the effect of treatment assignment on outcome among units in stratum $d_1d_0$ when the mediator is fixed to its natural value without treatment. The $\text{PNIE}_{d_1d_0}$, on the other hand, captures the mean difference of the potential outcomes among units in stratum $d_1d_0$, when the assignment $Z=1$, but the mediator changes from its natural value under treatment to its counterfactual value under control. Therefore, $\text{PNIE}_{d_1d_0}$ measures the extent to which the causal effect of treatment assignment is mediated through $M$ among the subpopulation in stratum $d_1d_0$. Similarly, the \textit{intention-to-treat effect} (ITT), defined as $\E[Y_1-Y_0]$, can be decomposed in the usual fashion into an \textit{intention-to-treat natural indirect effect} (ITT-NIE) and a \textit{intention-to-treat natural direct effect} (ITT-NDE):
\begin{align}
 \underbrace{\E\left[Y_{1M_1}-Y_{0M_0}\right]}_{\text{ITT}} = \underbrace{\E\left[Y_{1M_1}-Y_{1M_0}\right]}_{\text{ITT-NIE}} + \underbrace{\E\left[Y_{1M_0}-Y_{0M_0}\right]}_{\text{ITT-NDE}}.\label{e:itt}
\end{align}
One can verify that ITT is a weighted average of the PCEs such that
$
\text{ITT} = \displaystyle\sum_{d_1d_0\in \mathcal U_{\rm{all}}} e_{d_1d_0} \times \text{PCE}_{d_1d_0}
$
, where $\mathcal U_{\rm{all}}=\{10,11,00,01\}$. Similarly,  $\text{ITT-NIE} = \displaystyle\sum_{d_1d_0\in \mathcal U_{\rm{all}}} e_{d_1d_0} \times \text{PNIE}_{d_1d_0}$ and $\text{ITT-NDE} = \displaystyle\sum_{d_1d_0\in \mathcal U_{\rm{all}}} e_{d_1d_0} \times \text{PNDE}_{d_1d_0}$.

In what follows, we focus on identification of $\theta_{d_1d_0}^{(zz')}=\E\left[Y_{zM_{z'}}|U=d_1d_0\right]$ for any $z$ and $z'\in\{0,1\}$, based on which all PCEs and their effect decompositions in \eqref{e:principal_effects} can be obtained. Notice that ITT, ITT-NIE, and ITT-NDE can be also obtained as they are weighted averages of $\theta_{d_1d_0}^{(zz')}$. Identification of $\theta_{d_1d_0}^{(zz')}$ requires the following structural assumptions.

\begin{assumption}\label{assum:consistency}
(Consistency) For any $z$, $d$, and $m$, we have $D_z=D$ if $Z=z$, $M_{zd}=M$ if $Z=z$ and $D=d$, and $Y_{zdm}=Y$ if $Z=z$, $D=d$ and $M=m$. 
\end{assumption}

\begin{assumption}\label{assum:ignora_z}
(Ignorability of the treatment assignment) $\{D_z,M_{z'd'},Y_{z^*d'm^*}\} \indep Z|\bcx$ $\forall$ $z$, $z'$, $z^*$, $d'$,  and $m^*$, where ``$\indep$" stands for independence.
\end{assumption}

Assumption \ref{assum:consistency} is commonly invoked to exclude unit-level interference and enables us to connect the observed variables with their potential values. Assumption \ref{assum:ignora_z} is the no unmeasured confounding condition for treatment assignment that is often required to identify the ITT estimand in the absence of randomization. It is considered plausible when sufficient baseline covariates $\bcx$ are collected such that no hidden confounders would give rise to systematic differences between the post-randomization variables in the treated and the control groups. A stronger statement of Assumption \ref{assum:ignora_z}, $\{D_z,M_{z'd'},Y_{z^*d'm^*},\bcx\} \indep Z$, is often satisfied in randomized experiments. 

We additionally require the monotonicity of treatment on the post-treatment event to identify the distribution of the principal stratum membership $U$. We consider two types of monotonicity---standard monotonicity (Assumptions {\color{red}\ref{assum:mono}a}) and strong monotonicity (Assumptions {\color{red}\ref{assum:mono}b}). The standard version requires that treatment has a non-negative impact on the post-treatment event, whereas the stronger version further assumes $D_0=0$. Strong monotonicity is satisfied under one-sided noncompliance where all control units had no access to treatment (\citealp{frolich2013identification}). 

\begin{assumption}\label{assum:mono}
(Monotonicity) (a) Under standard monotonicity, $D_1\geq D_0$ for all units; (b) under strong monotonicity, $D_0= 0$ for all units. 
\end{assumption}

Assumption {\color{red}\ref{assum:mono}a} rules out defiers ($U=01$), and enables the identification of the principal scores $e_{d_1d_0}(\bcx)$ for the remaining three principal strata ($U\in\{10,11,00\}$) \citep{ding2017principal}. Defining $p_{zd}(\bcx)=f_{D|Z,\bcx}(d|z,\bcx)$ and $p_{zd}=\E[p_{zd}(\bcx)]$ for $z,d\in\{0,1\}$, because the observed data with $(Z=0,D=1)$ includes only always-takers, we have $e_{11}(\bcx)=p_{01}(\bcx)$. Similarly, $e_{00}(\bcx)=p_{10}(\bcx)$ and $e_{10}(\bcx)=p_{11}(\bcx)-p_{01}(\bcx)$ since $\displaystyle\sum_{d_1d_0 \in \mathcal U_{\text{all}}} \!\!\! e_{d_1d_0}(\bcx)=1$, and the strata proportions are
$e_{10}=p_{11}-p_{01}$, $e_{00}=p_{10}$, and $e_{11}=p_{01}$. Assumption {\color{red}\ref{assum:mono}b} rules out both always-takers and defiers ($U=11$ and 01). Under strong monotonicity, the principal scores are given by $e_{10}(\bcx)=p_{11}(\bcx)$ and $e_{00}(\bcx)=p_{10}(\bcx)$, and the strata proportions are $e_{10}=p_{11}$ and $e_{00}=p_{10}$. To unify the presentation under standard and strong monotonicity, we re-express $e_{d_1d_0}(\bcx)$ and $e_{d_1d_0}$ as:
\begin{align}
e_{d_1d_0}(\bcx)=p_{z^*d^*}(\bcx)-kp_{01}(\bcx) \text{, \quad } e_{d_1d_0}=p_{z^*d^*}-kp_{01} \label{e:principal_scores}
\end{align}
for $d_1d_0 \in \{10,00,11,01\}$, where $k=|d_1-d_0|$, $z^*d^*=11$, 10, 01, and 01 if $d_1d_0=10$, 00, 11, and 01, respectively. Note that $p_{01}(\bcx)\equiv p_{01} \equiv 0$ under strong monotonicity. Because $\{e_{d_1d_0}(\bcx),e_{d_1d_0}\}$ is equivalent to $\{p_{z^*d^*}(\bcx)-kp_{01}(\bcx),p_{z^*d^*}-kp_{01}\}$, we will use them interchangeably.

We next introduce two additional ignorability assumptions for mediation analysis within principal strata.

\begin{assumption}\label{assum:ignora_p}
(Generalized principal ignorability) $\{M_{zd},Y_{z'dm'}\} \indep U|\bcx$ $\forall$ $z$, $z'$, $d$, $m'$.
\end{assumption}

Principal ignorability has been previously introduced to identify the PCEs (\citealp{jo2009use,ding2017principal,forastiere2018principal}). Assumption \ref{assum:ignora_p} generalizes the usual assumption to accommodate the mediator as an additional intermediate outcome. This assumption requires sufficient pre-treatment covariates to remove the confounding between $U$ and $M$ and that between $U$ and $Y$; in other words, no systematic differences exist in the distribution of the potential mediator and outcome across principal strata, given covariates. 
Next, we require ignorability of the mediator (\citealp{yamamoto2013identification,park2020two}):

\begin{assumption}\label{assum:ignora_m}
(Ignorability of the mediator) $M_{zd} \indep Y_{z'd'm'} | \{Z, U, \bcx$\} $\forall$ $z$, $z'$, $d$, $d'$, $m'$.
\end{assumption}

{ Assumption \ref{assum:ignora_m} assumes that the potential mediator is independent of the potential outcome, given the observed covariates $\bcx$, within each assignment group and principal stratum. This assumption rules out unmeasured baseline confounders in the mediator-outcome relationship and requires that, apart from $D$, there are no other treatment-induced confounders affecting the mediator-outcome relationship.} Assumption \ref{assum:ignora_m}, coupled with Assumptions \ref{assum:ignora_z} and \ref{assum:ignora_p}, generalizes the standard \textit{sequential ignorability} assumption for causal mediation analysis (\citealp{imai2010identification}) to address a post-randomization event $D$. In addition, when Assumptions \ref{assum:ignora_z}--\ref{assum:ignora_p} hold, Assumption \ref{assum:ignora_m} is equivalent to $M_{zd} \indep Y_{z'd'm'} | \bcx$ without the need to condition on treatment assignment and principal stratum (see Lemma S6 in the Supplementary Material). 
Lastly, we state the following positivity assumption. 
\begin{assumption}\label{assum:positivity}
(Positivity)  Assume that $f_{Z|\bcx}(z|\bx)>0$, $f_{D|Z,\bcx}(d|1,\bx)>0$, $f_{D|Z,\bcx}(0|0,\bx)>0$, $f_{M|Z,D,\bcx}(m|1,d,\bx)>0$, and $f_{M|Z,D,\bcx}(m|0,0,\bx)>0$ for any $z$, $d$, $m$, and $\bm x$. Additionally assume {$p_{11}-p_{01}>0$ with} $f_{D|Z,\bcx}(1|0,\bx)>0$ and $f_{M|Z,D,\bcx}(m|0,1,\bx)>0$ under standard monotonicity. 
\end{assumption}

Let $\mathcal U_{\text{a}}=\{10,00,11\}$ be the three strata under standard monotonicity and let $\mathcal U_{\text{b}}=\{10,00\}$ be the two strata under strong monotonicity. Theorem \ref{thm:identification} below shows that $\theta_{d_1d_0}^{(zz')}$ is nonparametrically identified under the aforementioned assumptions.

\begin{theorem}\label{thm:identification}
(Nonparametric identification) Suppose that Assumptions \ref{assum:consistency}--\ref{assum:positivity} hold. For any $z,z'\in\{0,1\}$, $d_1d_0\in \mathcal U_{\text{a}}$ under standard monotonicity, and $d_1d_0\in \mathcal U_{\text{b}}$ under strong monotonicity, $\theta_{d_1d_0}^{(zz')}$ is identified as follows:
$$
\theta_{d_1d_0}^{(zz')} = \int_{\bx}\frac{e_{d_1d_0}(\bx)}{e_{d_1d_0}}\int_{m} \E_{Y|Z,D,M,\bcx}\left[Y|z,d_z,m,\bx\right]d\Prob_{M|Z,D,\bcx}\left(m|z',d_{z'},\bx\right)d\Prob_{\bcx}\left(\bx\right),
$$
where $d_z=\mathbb{I}(z=1)d_1 + \mathbb{I}(z=0)d_0$ and $d_{z'}=\mathbb{I}(z'=1)d_1 + \mathbb{I}(z'=0)d_0$. Here, $e_{d_1d_0}(\bx)=p_{z^*d^*}(\bx)-kp_{01}(\bx)$ and $e_{d_1d_0}=p_{z^*d^*}-kp_{01}$ are identified in \eqref{e:principal_scores}.
\end{theorem}

By Theorem \ref{thm:identification}, we have $\text{PNDE}_{d_1d_0}=\theta^{(10)}_{d_1d_0}-\theta^{(00)}_{d_1d_0}$ and $\text{PNIE}_{d_1d_0}=\theta^{(11)}_{d_1d_0}-\theta^{(10)}_{d_1d_0}$, and the decomposition of the ITT effect can also be identified by 
\begin{equation}\label{e:nme_formula}
\text{ITT-NDE}=\sum_{d_1d_0 \in \mathcal U}e_{d_1d_0}\times\left(\theta^{(10)}_{d_1d_0}-\theta^{(00)}_{d_1d_0}\right), \quad \text{ITT-NIE}=\sum_{d_1d_0 \in \mathcal U}e_{d_1d_0}\times\left(\theta^{(11)}_{d_1d_0}-\theta^{(10)}_{d_1d_0}\right),
\end{equation}
where $\mathcal U=\mathcal U_{\text{a}}$ under standard monotonicity and $\mathcal U=\mathcal U_{\text{b}}$ under strong monotonicity.

\begin{remark}\label{remark:rudolph}
\cite{rudolph2024using} considered instrumental variables to identify the interventional and natural mediation effects among compliers. In a comparable scenario, they showed that under (i) exclusion restriction, (ii) standard monotonicity, and (iii) sequential randomization (Assumption \ref{assum:ignora_z} plus $Y_{zm}\indep M|Z,D,\bcx$), $\text{PNIE}_{10}$ is identified by $\text{ITT-NIE}/e_{10}$. A similar identification formula follows for $\text{PNDE}_{10}$. Due to exclusion restriction, the mediation effects among other strata are automatically zero, and thus only the compliers stratum contributes information to the ITT natural mediation effect. Replacing exclusion restriction with principal ignorability, Theorem \ref{thm:identification} allows additional strata to contribute information to the ITT natural mediation effect and enables point identification of each stratum-specific natural mediation effect.
\end{remark}

\section{Connections to mediation analysis with a treatment-induced confounder or two mediators}\label{sec:connection}
 
Although we consider $D$ as a primary source to sub-classify the study population, there exist two complementary perspectives for the role of $D$ in causal mediation analysis; that is, $D$ can be viewed as a binary post-treatment confounder or another mediator sitting in the causal pathway between $Z$ and $M$. We discuss the connections between the present work and existing mediation methods for addressing treatment-induced confounding (\citealp{robins2010alternative,tchetgen2014identification,vanderweele2014effect,miles2020semiparametric,diaz2021nonparametric,xia2021identification}) or two causally-ordered mediators (\citealp{albert2011generalized,daniel2015causal,steen2017flexible,zhou2022semiparametric}). When $D$ is considered as another mediator, methods have been proposed for identification of the path-specific effects through the four causal pathways given by Figure \ref{fig:dag}(a)--(d) (e.g., \citealp{daniel2015causal,zhou2022semiparametric}).  If $D$ is treated as a post-treatment confounder, methods have been developed for identifying different versions of mediation effects through $M$, including the interventional mediation effects (\citealp{vanderweele2014effect,diaz2021nonparametric}), the natural mediation effects (\citealp{robins2010alternative,tchetgen2014identification,xia2021identification}), and the path-specific effect on the causal pathway $Z \shortrightarrow M \shortrightarrow Y$ in Figure \ref{fig:dag}(c) (\citealp{vanderweele2014effect,miles2017quantifying,miles2020semiparametric}).

In contrast to methods that view $D$ as a post-treatment confounder, our work addresses a different scientific question. Both our approach and methods for two causally-ordered mediators aim to disentangle the roles of $M$ and $D$ in jointly explaining the causal mechanism, whereas mediation methods with a post-treatment confounder focus on the primary role of $M$ in explaining the causal mechanism. For example, \cite{tchetgen2014identification} and \cite{xia2021identification} studied identification of the natural mediation effects defined in \eqref{e:itt}, which summarize the causal sequence through $M$ on the outcome marginalized over different levels of $D$. Similarly, the interventional mediation effects in \cite{vanderweele2014effect} and \cite{diaz2021nonparametric} only considered the causal sequence through mediator $M$ to define the estimand of interest. Finally, \cite{miles2017quantifying} and \cite{miles2020semiparametric} mainly considered the path-specific effects on the causal pathway $Z \shortrightarrow M \shortrightarrow Y$ in Figure \ref{fig:dag}(c), where other causal pathways passing through $D$ were assumed of less interest. 

Comparing the present work to methods for identifying path-specific effects, a notable difference lies in the causal estimands of interest. We specifically focus on decomposing the causal effects within endogenous subgroups characterized by the joint potential values of the $D$, whereas the path-specific effects are defined for the entire study population. A further difference lies in the identification assumptions. Identifying path-specific effects requires certain ignorability assumptions regarding the observed post-treatment variable $D$ directly. For example, \cite{daniel2015causal} requires that $M_{zd}\indep D | \{Z,\bcx\}$ for any $d$ and $z$. On the other hand, our identification assumptions require the use of the potential values of $D$ to define the principal stratum ($U$) and then invoke ignorability assumptions across the principal stratum $U=D_1D_0$. 

Despite the aforementioned differences, there exist mathematical connections across the requisite identification conditions. We provide two further remarks about such connections, in particular to the work by \cite{tchetgen2014identification} (on treatment-induced confounding) and \cite{zhou2022semiparametric} (on two causally-ordered mediators). All proofs are provided in the Supplementary Material. 

\begin{remark}\label{remark:relationship1}
\cite{tchetgen2014identification} used a nonparametric structural equations model with independent errors (NPSEM-IE) for the DAG in Figure \ref{fig:dag}, coupled with monotonicity, to identify the natural mediation effects \eqref{e:itt} with a binary post-treatment confounder. Suppose that the consistency (Assumption \ref{assum:consistency}) and the monotonicity (either Assumption {\color{red}\ref{assum:mono}a} or {\color{red}\ref{assum:mono}b}) hold,  if further the NPSEM-IE for the DAG in Figure \ref{fig:dag} (i.e., Assumption S1 in Supplementary Material) hold, then Assumptions \ref{assum:ignora_z}, \ref{assum:ignora_p}, and \ref{assum:ignora_m} also hold, but not vice versa. 
\end{remark}
By Remark \ref{remark:relationship1}, under the consistency and monotonicity, the ignorability assumptions (Assumptions \ref{assum:ignora_z}, \ref{assum:ignora_p}, and \ref{assum:ignora_m}) are directly implied from a NPSEM-IE corresponding to the DAG in Figure \ref{fig:dag}. Remark \ref{remark:relationship1} further implies that the identification formulas for the natural mediation effects \eqref{e:nme_formula} are equivalent to the identification formulas in \cite{tchetgen2014identification}, except that the present work invokes technically weaker assumptions.

\begin{remark}\label{remark:relationship2}
\cite{zhou2022semiparametric} considered a set of generalized sequential ignorability assumptions to identify path-specific effects with multiple mediators, and they are comparable to the present work in the special case when $D$ is binary. Suppose that the consistency (Assumption \ref{assum:consistency}) holds, then, under monotonicity (either Assumption {\color{red}\ref{assum:mono}a} or {\color{red}\ref{assum:mono}b}), the set of generalized sequential ignorability assumptions in \cite{zhou2022semiparametric} (i.e., Assumption S2 in Supplementary Material) are equivalent to Assumptions \ref{assum:ignora_z}, \ref{assum:ignora_p}, and \ref{assum:ignora_m}.
\end{remark}

By Remark \ref{remark:relationship2}, the assumptions in the present work are stronger than those in \cite{zhou2022semiparametric}, since the latter does not require monotonicity. This is expected as stronger assumptions are necessary to identify our finer-grained estimands that provide insights into the pathways within each subpopulation. Finally, if the monotonicity is plausible by the treatment $Z$ on the first mediator $D$, our assumptions are equivalent to the set of generalized sequential ignorability assumptions in \citet{zhou2022semiparametric}.

\section{Estimation of natural mediation effects}\label{sec:estimation}

\subsection{Nuisance functions and parametric working models}\label{sec:working_models}

We first define several nuisance functions of the observed-data distributions. Let $\pi_z(\bx) = \Prob_{Z|\bcx}(z|\bx)$ be the probability of treatment conditional on $\bcx$, where $\pi_1(\bx)$ is the propensity score. 
Note that $\pi_z(\bx)$ degenerates to a constant value $\pi_z$ in randomized experiments. Let $r_{zd}(m,\bx)=f_{M|Z,D,\bcx}(m|z,d,\bx)$ be the probability of the mediator conditional on $Z$, $D$, and $\bcx$. Let $\mu_{zd}(m,\bx)=\E_{Y|Z,D,M,\bcx}[Y|z,d,m,\bx]$ be the conditional expectation of $Y$ given $Z$, $D$, $M$, and $\bcx$. Let $h_{nuisance}=\{\pi_z(\bx),p_{zd}(\bx),r_{zd}(m,\bx),\mu_{zd}(m,\bx)\}$ contain all nuisance functions, where $p_{zd}(\bx)=f_{D|Z,\bcx}(d|z,\bx)$ is defined in Section \ref{sec:notation}. It should be noted that, within our definitions of the nuisance functions, the two variables $(Z, D)$--which directly relate to the principal strata--are presented as the subscript, while all other variables are presented as arguments.

One can specify parametric working models $h_{nuisance}^{\text{par}}=\{\pi_z^{\text{par}}(\bx),  \allowbreak p_{zd}^{\text{par}}(\bx), \allowbreak r_{zd}^{\text{par}}(m,\bx), \allowbreak \mu_{zd}^{\text{par}}(m,\bx)\}$ for $h_{nuisance}$. 
Specification of the parametric working models can be flexible.  For example, logistic regressions can be used for $\pi_z^{\text{par}}(\bx)$ and  $p_{zd}^{\text{par}}(\bx)$. When the mediator is continuous or binary, a linear regression or a logistic regression can be employed for $r_{zd}^{\text{par}}(m,\bx)$. Similarly, a generalized linear model can be used for $\mu_{zd}^{\text{par}}(m,\bx)$. Detailed model examples are provided in the Supplementary Material. Hereafter, we use $\mathcal M_{\pi}$ to denote the submodel of the nonparametric model $\mathcal M_{np}$ with a correctly specified $\pi_z^{\text{par}}(\bx)$ for $\pi_z(\bx)$ and unspecified other components. Analogously, define $\mathcal M_{e}$, $\mathcal M_{m}$, and $\mathcal M_{o}$ as the submodel of $\mathcal M_{np}$ with a correctly specified $p_{zd}^{\text{par}}(\bx)$, $r_{zd}^{\text{par}}(m,\bx)$, and $\mu_{zd}^{\text{par}}(m,\bx)$, respectively. In addition, we use ``$\cup$" and ``$\cap$" to denote union and intersection of submodels such that $\mathcal M_{\pi}\cap \mathcal M_{e}$ denotes the correct specification of both $\pi_z^{\text{par}}(\bx)$ and $p_{zd}^{\text{par}}(\bx)$, and $\mathcal M_{\pi}\cup \mathcal M_{e}$ denotes the correct specification of either $\pi_z^{\text{par}}(\bx)$ or $p_{zd}^{\text{par}}(\bx)$.

Suggested in Theorem \ref{thm:identification}, one also needs to estimate $e_{d_1d_0}$, or equivalently $p_{zd}\equiv \E[p_{zd}(\bcx)]$, in order to estimate $\theta_{d_1d_0}^{(zz')}$. There are multiple ways to estimate $p_{zd}$, as one can simply use the plug-in estimator $\Prob_n[\widehat p_{zd}^{\text{par}}(\bcx)]$ and the inverse probability weighting estimator $\Prob_n\left[{\mathbb{I}(Z=z,D=d)}/{\widehat{\pi}_z^{\text{par}}(\bcx)}\right]$. In randomized experiments, because $\pi_z(\bcx)=\pi_z$, one can also estimate $p_{zd}$ by ${\Prob_n[\mathbb{I}(Z=z,D=d)]}/{\Prob_n[\mathbb{I}(Z=z)]}$. In this article, we consider the doubly robust estimator developed in \citet{jiang2022multiply},
\begin{equation}\label{e:p_dr}
\widehat p_{zd}^{\text{dr}}=\Prob_n\left[\frac{\mathbb{I}(Z=z)\left\{\mathbb{I}(D=d)-\widehat p_{zd}^{\text{par}}(\bcx)\right\}}{\widehat{\pi}_z^{\text{par}}(\bcx)}+\widehat p_{zd}^{\text{par}}(\bcx)\right],
\end{equation}
which is consistent to $p_{zd}$ under $\mathcal M_{\pi} \cup \mathcal M_{e}$ and is locally efficient under $\mathcal M_{\pi} \cap \mathcal M_{e}$.

\subsection{Moment-type estimators}\label{sec:moment_type}

We provide four distinct identification expressions of $\theta_{d_1d_0}^{(zz')}$; each expression uses only part, but not all, of the nuisance functions $h_{nusiance}$ and the principal stratum proportion $p_{zd}$.

\begin{theorem}\label{thm:single_estimator}
For $z,z'\in\{0,1\}$, $d_1d_0\in \mathcal U_{\text{a}}$ or $d_1d_0\in \mathcal U_{\text{b}}$ under standard or strong monotonicity, 
we have $\theta_{d_1d_0}^{(zz')}=\theta^{(zz'),\textrm{a}}_{d_1d_0}=\theta^{(zz'),\textrm{b}}_{d_1d_0}=\theta^{(zz'),\textrm{c}}_{d_1d_0}=\theta^{(zz'),\textrm{d}}_{d_1d_0}$, where
\begin{align*}
\theta^{(zz'),\textrm{a}}_{d_1d_0} & = \E\left[\frac{p_{z^*d^*}(\bcx)-kp_{01}(\bcx)}{p_{z^*d^*}-kp_{01}}\frac{\mathbb{I}(D=d_z,Z=z)}{p_{zd_z}(\bcx)\pi_{z}(\bcx)}\frac{r_{z'd_{z'}}(M,\bcx)}{r_{zd_z}(M,\bcx)}Y\right], \\
\theta^{(zz'),\textrm{b}}_{d_1d_0} & = \E\left[\left\{\frac{\mathbb{I}(Z=z^*,D=d^*)}{\pi_{z^*}(\bcx)} - k \frac{(1-Z)D}{\pi_0(\bcx)}\right\}\frac{\eta_{zz'}(\bcx)}{p_{z^*d^*}-kp_{01}}\right], \\
\theta^{(zz'),\textrm{c}}_{d_1d_0} & = \E\left[\frac{p_{z^*d^*}(\bcx)-kp_{01}(\bcx)}{p_{z^*d^*}-kp_{01}}\frac{\mathbb{I}(D=d_{z'},Z=z')}{p_{z'd_{z'}}(\bcx)\pi_{z'}(\bcx)}\mu_{zd_z}(M,\bcx)\right], \\
\theta^{(zz'),\textrm{d}}_{d_1d_0} & = \E\left[\frac{p_{z^*d^*}(\bcx)-kp_{01}(\bcx)}{p_{z^*d^*}-kp_{01}}\eta_{zz'}(\bcx)\right].
\end{align*}
with $\eta_{zz'}(\bcx)=\int_m \mu_{zd_z}(m,\bcx)r_{z'd_{z'}}(m,\bcx) dm$, 
$k=|d_1-d_0|$, $z^*d^*=$11, 10, 01 if $d_1d_0=$10, 00, and 11, respectively. 
\end{theorem}

The first expression is an average of outcome by the product of four different weights, where the first weight, $\frac{p_{z^*d^*}(\bcx)-kp_{01}(\bcx)}{p_{z^*d^*}-kp_{01}} ={e_{d_1d_0}(\bcx)}/{e_{d_1d_0}}$, is the principal score weight for creating a pseudo-population within stratum $U=d_1d_0$ (\citealp{jiang2022multiply}). The remaining three weights in $\theta^{(zz'),\textrm{a}}_{d_1d_0}$---the inverse probability of treatment weight, the inverse probability of the post-treatment event weight, and the mediator density ratio weight---correct for selection bias associated with the treatment, post-treatment event, and the observed mediator value, within the pseudo population created by the principal score weight. The second expression is a product of two components, where the first component, $\left\{\frac{\mathbb{I}(Z=z^*,D=d^*)}{\pi_{z^*}(\bcx)} - k \frac{(1-Z)D}{\pi_0(\bcx)}\right\}\Big/(p_{z^*d^*}-kp_{01})$, plays a similar role to the principal score weight to create a pseudo-population of stratum $U=d_1d_0$, and the second component, $\eta_{zz'}(\bcx)=\E[Y_{zM_{z'}}|U=d_1d_0,\bcx]$ is a conditional version of $\theta_{d_1d_0}^{(zz')}$ given fixed values of baseline covariates and within stratum $U=d_1d_0$. Construction of the third expression bears some resemblance to the first expression, both of which use the principal score weight, except that the third expression uses a slightly different weighting scheme coupled with the conditional expectation of outcome $\mu_{zd_z}(M,\bcx)$ instead of weighting directly on the observed outcome. The fourth expression shares a similar form to the second expression, but now involves the product between the principal score weight and $\eta_{zz'}(\bcx)$.

According to Theorem \ref{thm:single_estimator}, we can obtain the four moment-type estimators, $\{\widehat\theta^{(zz'),\rm{a}}_{d_1d_0},\allowbreak \widehat\theta^{(zz'),\rm{b}}_{d_1d_0},\allowbreak \widehat\theta^{(zz'),\rm{c}}_{d_1d_0},\allowbreak \widehat\theta^{(zz'),\rm{d}}_{d_1d_0}\}$, by replacing the unknown nuisance functions with their estimates from parametric working models and substituting the outer expectation operator $\E$ by the empirical average operator $\Prob_n$. As an example,  $\widehat\theta^{(zz'),\rm{d}}_{d_1d_0}$ is given by $$\Prob_n\left[\frac{\widehat p_{z^*d^*}^{\text{par}}(\bcx)-k\widehat p_{01}^{\text{par}}(\bcx)}{\widehat p_{z^*d^*}^{\text{dr}}-k\widehat p_{01}^{\text{dr}}}\widehat \eta_{zz'}^{\text{par}}(\bcx)\right],$$ 
where $\widehat\eta_{zz'}^{\text{par}}(\bcx)=\int_m \widehat{\mu}_{zd_z}^{\text{par}}(m,\bcx)\widehat r_{z'd_{z'}}^{\text{par}}(m,\bcx) dm$. 
Here, the integral in $\widehat\eta_{zz'}^{\text{par}}(\bcx)$ becomes simple summations when the mediator is categorical and, if the mediator is continuous, numerical integration can be used for an approximate calculation. We summarize the asymptotic properties of the four moment-type estimators below.
\begin{proposition}\label{prop:moment_estimator}
Suppose that the regularity conditions outlined in the Supplementary Material hold. Then, $\widehat\theta^{(zz'),\rm{a}}_{d_1d_0}$, $\widehat\theta^{(zz'),\rm{b}}_{d_1d_0}$, $\widehat\theta^{(zz'),\rm{c}}_{d_1d_0}$, and $\widehat\theta^{(zz'),\rm{d}}_{d_1d_0}$ are consistent and asymptotic normal under $\mathcal M_{\pi} \cap \mathcal M_{e} \cap \mathcal M_{m}$, $\mathcal M_{\pi} \cap \mathcal M_{m} \cap \mathcal M_{o}$, $\mathcal M_{\pi} \cap \mathcal M_{e} \cap \mathcal M_{o}$, and $\mathcal M_{e} \cap \mathcal M_{m} \cap \mathcal M_{o}$, respectively.
\end{proposition}

\subsection{From efficient influence function to multiply robust estimator}

Denote $\mathcal M_{np}$ as the nonparametric model over the observed data density function $f_{\bm O}$. 
The efficient influence function (EIF) of $\theta_{d_1d_0}^{(zz')}$ under $\mathcal M_{np}$ is derived in Theorem \ref{thm:eif} based on the semiparametric estimation theory (\citealp{bickel1993efficient}), which also implies the semiparametric efficiency bound, i.e., the lower bound of the asymptotic variance among all regular and asymptotic linear estimators of $\theta_{d_1d_0}^{(zz')}$ under the nonparametric model $\mathcal M_{np}$. 

\begin{theorem}\label{thm:eif}
The EIF of  $\theta_{d_1d_0}^{(zz')}$ over $\mathcal M_{np}$ is
$$
\mathcal D^{(zz')}_{d_1d_0}(\bm O) = \frac{\psi_{d_1d_0}^{(zz')}(\bm O)-\theta_{d_1d_0}^{(zz')}\delta_{d_1d_0}(\bm O)}{p_{z^*d^*}-kp_{01}},
$$
 where 
\begin{align*}
\psi_{d_1d_0}^{(zz')}(\bco) = & \left(\frac{\mathbb{I}(Z=z^*)\left\{\mathbb{I}(D=d^*)-p_{z^*d^*}(\bcx)\right\}}{\pi_{z^*}(\bcx)} - k\frac{(1-Z)\left\{D-p_{01}(\bcx)\right\}}{\pi_{0}(\bcx)}\right)\eta_{zz'}(\bcx) \\
& + \left\{p_{z^*d^*}(\bcx)-kp_{01}(\bcx)\right\}\frac{\mathbb{I}(D=d_z,Z=z)}{p_{zd_z}(\bcx)\pi_{z}(\bcx)}\frac{r_{z'd_{z'}}(M,\bcx)}{r_{zd_z}(M,\bcx)}\left\{Y-\mu_{zd_z}(M,\bcx)\right\} \\
& + \left\{p_{z^*d^*}(\bcx)-kp_{01}(\bcx)\right\}\frac{\mathbb{I}(D=d_{z'},Z=z')}{p_{z'd_{z'}}(\bcx)\pi_{z'}(\bcx)}\left\{\mu_{zd_z}(M,\bcx)-\eta_{zz'}(\bcx)\right\} \\
& + \left\{p_{z^*d^*}(\bcx)-kp_{01}(\bcx)\right\}\eta_{zz'}(\bcx),\\
\delta_{d_1d_0}(\bco) =& \frac{\mathbb{I}(Z=z^*)\left\{\mathbb{I}(D=d^*)-p_{z^*d^*}(\bcx)\right\}}{\pi_{z^*}(\bcx)}- k\frac{(1-Z)\left\{D-p_{01}(\bcx)\right\}}{\pi_{0}(\bcx)} + p_{z^*d^*}(\bcx) - k p_{01}(\bcx), 
\end{align*}
$k=|d_1-d_0|$, $z^*d^*=$11, 10, 01 if $d_1d_0=$10, 00, and 11, respectively.
Therefore, the semiparametric efficiency bound for estimation of $\theta_{d_1d_0}^{(zz')}$ is $\E\left[\left\{\mathcal D^{(zz')}_{d_1d_0}(\bm O)\right\}^2\right]$.
\end{theorem}

Theorem \ref{thm:eif} inspires a new estimator of $\theta_{d_1d_0}^{(zz')}$ by solving the following EIF-induced estimating equation 
$$
\Prob_n\left[\frac{\psi_{d_1d_0}^{(zz')}(\bm O)-\theta_{d_1d_0}^{(zz')}\delta_{d_1d_0}(\bm O)}{p_{z^*d^*}-kp_{01}}\right] =  0
,$$ 
where $\psi_{d_1d_0}^{(zz')}(\bm O)$ and $\delta_{d_1d_0}(\bm O)$  depend on nuisance functions $h_{nuisance}$ and the denominator $p_{z^*d^*}-kp_{01}$ is a constant that does not affect the solution of the estimating equation. Therefore, the new estimator, which we hereafter refer to as the multiply robust estimator, can be constructed as 
$$
\widehat\theta^{(zz'),\text{mr}}_{d_1d_0} = \frac{\Prob_n\left[\widehat\psi_{d_1d_0}^{(zz'),\text{par}}(\bm O)\right]}{\Prob_n\left[\widehat\delta_{d_1d_0}^{\text{par}}(\bm O)\right]}.
$$ 
Theorem \ref{thm:mr} summarizes the asymptotic properties of the multiply robust estimator.
\begin{theorem}\label{thm:mr}
    Suppose that the regularity conditions outlined in Supplementary Material hold. Under either $\mathcal M_{\pi} \cap \mathcal M_{e} \cap \mathcal M_{m}$, $\mathcal M_{\pi} \cap \mathcal M_{m} \cap \mathcal M_{o}$, $\mathcal M_{\pi} \cap \mathcal M_{e} \cap \mathcal M_{o}$, or $\mathcal M_{e} \cap \mathcal M_{m} \cap \mathcal M_{o}$, the multiply robust estimator $\widehat\theta^{(zz'),\text{mr}}_{d_1d_0}$ is consistent and asymptotically normal such that $\sqrt{n}\left(\widehat\theta^{(zz'),\text{mr}}_{d_1d_0}-\theta_{d_1d_0}^{(zz')}\right)$ converges to a zero-mean normal distribution with finite variance $V_{\text{mr}}$. Moreover, $V_{\text{mr}}$ achieves the semiparametric efficiency bound under $\mathcal M_{\pi} \cap \mathcal M_{e} \cap \mathcal M_{m} \cap \mathcal M_{o}$.
\end{theorem}

An attractive property of $\widehat\theta^{(zz'),\text{mr}}_{d_1d_0}$ is that it offers four types of protection against misspecification of the parametric working models. 
Notice that the four moment-type estimators provided in Section \ref{sec:moment_type} are only single robust; for example, $\widehat\theta^{(zz'),\text{a}}_{d_1d_0}$ is only consistent under $\mathcal M_{\pi} \cap \mathcal M_{e} \cap \mathcal M_{m}$. By contrast, $\widehat\theta^{(zz'),\text{mr}}_{d_1d_0}$ is quadruply robust such that it is consistent for $\theta_{d_1d_0}^{(zz')}$ even if one of the four working models, $\mathcal M_{\pi}$, $\mathcal M_{e}$, $\mathcal M_{m}$, and $\mathcal M_{o}$, is misspecified. 
In addition, $\widehat\theta^{(zz'),\text{mr}}_{d_1d_0}$ is also locally efficient when all of the four working models are correctly specified.  A proof of the quadruple robustness property is given in Supplementary Material. {As a caveat, the quadruple robustness is more stringent than the double robustness property, as the former requires three out of four working models to be correct whereas the latter only require one out of two working models to be correct.} In practice, one can use nonparametric bootstrap to construct the standard error and confidence interval of $\widehat\theta^{(zz'),\text{mr}}_{d_1d_0}$.

\begin{remark}
{The monotonicity assumption can place a restriction on the observed data density $f_{\bco}$; that is, the standard monotonicity indicates that $f_{D|Z,\bcx}(1|1,\bx)\geq f_{D|Z,\bcx}(1|0,\bx)$ and the strong monotonicity further constrains $f_{D|Z,\bcx}(1|0,\bx)\equiv 0$. Following previous efforts in obtaining efficient causal estimators under a principal stratification framework \citep{rudolph2024using,jiang2022multiply}, the EIF in Theorem \ref{thm:eif} is derived under the nonparametric model $\mathcal M_{np}$, which does not leverage the monotonicity restriction on $f_{\bco}$ to potentially sharpen the efficiency bound. Therefore, $\widehat\theta^{(zz'),\text{mr}}_{d_1d_0}$ is only locally efficient under $\mathcal M_{np}$, rather than under a more restrictive model space assuming monotonicity.}
\end{remark}

\subsection{Nonparametric efficient estimation}
\label{sec:np}

We extend the proposed multiply robust estimator by  estimating the nuisance functions, $h_{\text{nuisance}}$, via flexible nonparametric methods or modern data-adaptive machine learning methods. We denote the new estimator as $\widehat{\theta}^{(zz'),\text{np}}_{d_1d_0}$ with the superscript ``np" to indicate using nonparametric algorithms. The cross-fitting procedure (\citealp{chernozhukov2018double}) is employed to circumvent the  bias due to overfitting of nonparametric estimation on the nuisance functions. Specifically, we randomly partition the dataset into $V$ groups with approximately equal size such that the group size difference is at most 1. For each $v$, let $\mathcal O_v$ be the data in $v$-th group and $\mathcal O_{-v}=\displaystyle \cup_{v'\in\{1,\dots,V\}\setminus v} \mathcal O_{v'}$ be the data excluding the $v$-th group. For $v=1,\dots,V$, we calculate the nuisance function estimates on data $\mathcal O_{v}$, denoted by $\widehat h_{nuisance}^{\text{np},v}=\{\widehat\pi_{z}^{\text{np},v}(\bx),\widehat p_{zd}^{\text{np},v}(\bx),\widehat r_{zd}^{\text{np},v}(m,\bx),\widehat \mu_{zd}^{\text{np},v}(m,\bx)\}$, based on machine learning or nonparametric methods trained on data $\mathcal O_{-v}$. The nuisance function estimates evaluated over the entire dataset, $\widehat h_{nuisance}^{\text{np}}$, is therefore a combination of $\widehat h_{nuisance}^{\text{np},1}$, $\widehat h_{nuisance}^{\text{np},2}$, $\dots$, $\widehat h_{nuisance}^{\text{np},V}$. Finally,  $\widehat{\theta}^{(zz'),\text{np}}_{d_1d_0}$ is given by the solution to $$\Prob_n\left[\frac{\widehat\psi_{d_1d_0}^{(zz'),\text{np}}(\bm O)-\theta_{d_1d_0}^{(zz')}\widehat\delta_{d_1d_0}^{\text{np}}(\bm O)}{p_{z^*d^*}-kp_{01}}\right] =  0$$ 
so that $\widehat{\theta}^{(zz'),\text{np}}_{d_1d_0}=\Prob_n[\widehat\psi_{d_1d_0}^{(zz'),\text{np}}(\bm O)]/\Prob_n[\widehat\delta_{d_1d_0}^{\text{np}}(\bm O)]$, where $\widehat\psi_{d_1d_0}^{(zz'),\text{np}}(\bm O)$ and $\widehat\delta_{d_1d_0}^{\text{np}}(\bm O)$ are $\psi_{d_1d_0}^{(zz')}(\bm O)$ and $\delta_{d_1d_0}(\bm O)$ evaluated based on $\widehat h_{nuisance}^{\text{np}}$.

\begin{theorem}\label{thm:nonpar}
Under Assumptions \ref{assum:consistency}--\ref{assum:positivity}, $\widehat{\theta}^{(zz'),\text{np}}_{d_1d_0}$ is consistent if any three of the four nuisance functions in $\widehat h_{nuisance}^{\text{np}}$ are consistently estimated in the $L_2(\Prob)$-norm.  Furthermore, if all elements in $\widehat h_{nuisance}^{\text{np}}$ are consistent in the $L_2(\Prob)$-norm and $\|\widehat l^{\text{np}} - l\| \times \|\widehat g^{\text{np}} - g\|=o_p(n^{-1/2})$ for any $l\neq g \in \{\pi_z(\bx),p_{zd}(\bx),r_{zd}(m,\bx),\mu_{zd}(m,\bx)\}$, then $\widehat{\theta}^{(zz'),\text{np}}_{d_1d_0}$ is asymptotically normal and its asymptotic variance achieves the efficiency lower bound.
\end{theorem}

Theorem \ref{thm:nonpar} indicates that $\widehat{\theta}^{(zz'),\text{np}}_{d_1d_0}$ is consistent, asymptotically normal, and also achieves semiparametric efficiency lower bound, if all nuisance functions can be consistently estimated with a $o_p(n^{-1/4})$ rate, which can be achieved by several machining learning algorithms (e.g., the boosting approach by \citet{luo2016high}, and the random forest by \citet{wager2015adaptive}, and the neural networks by \citet{chen1999improved}). When nuisance functions are estimated via data-adaptive methods, 
we use the empirical variance of the estimated EIF to construct the variance estimator for $\widehat{\theta}^{(zz'),\text{np}}_{d_1d_0}$; that is
$$\widehat{\text{Var}}(\widehat{\theta}^{(zz'),\text{np}}_{d_1d_0}) = \frac{1}{n}\Prob_n\left[\left\{\frac{\widehat\psi_{d_1d_0}^{(zz'),\text{np}}(\bm O)-\widehat\theta_{d_1d_0}^{(zz'),\text{np}}\widehat\delta_{d_1d_0}^{\text{np}}(\bm O)}{\widehat p_{z^*d^*}^{\text{np}}-k\widehat p_{01}^{\text{np}}}\right\}^2\right],$$ 
where $\widehat p_{zd}^{\text{np}}$ is constructed analogous to $\widehat p_{zd}^{\text{dr}}$ in \eqref{e:p_dr} but evaluated using $\widehat h_{nuisance}^{\text{np}}$.

\subsection{Estimation of natural mediation effects}\label{sec:gmf_to_me}

Once we obtain $\widehat{\theta}_{d_1d_0}^{(zz')}$, estimators of $\text{PNIE}_{d_1d_0}$ and $\text{PNDE}_{d_1d_0}$ can be constructed based on \eqref{e:principal_effects}. For example, we can construct $\widehat{\text{PNIE}}_{d_1d_0}^{\text{s}} = \widehat\theta^{(11),\text{s}}_{d_1d_0}-\widehat\theta^{(10),\text{s}}_{d_1d_0}$ and $\widehat{\text{PNDE}}_{d_1d_0}^{\text{s}} = \widehat\theta^{(10),\text{s}}_{d_1d_0}-\widehat\theta^{(00),\text{s}}_{d_1d_0}$, if either the moment-type method (s=a, b, c or d), multiply robust estimator (s=mr), or nonparametric efficient estimator (s=np) is used for $\theta^{(zz')}_{d_1d_0}$. Analogously, ITT-NIE and ITT-NDE can estimated via using \eqref{e:nme_formula} by replacing $e_{d_1d_0}=p_{z^*d^*}-kp_{01}$ and $\theta_{d_1d_0}^{(zz')}$ with their corresponding estimators. Specifically, we can construct estimators of ITT-NIE and ITT-NDE as 
$\widehat{\text{ITT-NDE}}^{\text{s}}=\displaystyle\sum_{d_1d_0 \in \mathcal U}\widehat e_{d_1d_0}^{\text{dr}}\times\left(\widehat \theta^{(10),\text{s}}_{d_1d_0}-\widehat \theta^{(00),\text{s}}_{d_1d_0}\right)$ and $\widehat{\text{ITT-NIE}}^{\text{s}}=\displaystyle\sum_{d_1d_0 \in \mathcal U}\widehat e_{d_1d_0}^{\text{dr}}\times\left(\widehat\theta^{(11),\text{s}}_{d_1d_0}-\theta^{(10),\text{s}}_{d_1d_0}\right)$ if either the moment-type method (s=a, b, c or d) or the multiply robust estimator (s=mr) is used for estimating $\theta_{d_1d_0}^{(zz')}$, where  $\widehat e_{d_1d_0}^{\text{dr}}=\widehat p_{z^*d^*}^{\text{dr}}-k\widehat p_{01}^{\text{dr}}$ and $\mathcal U=\mathcal U_{\text{a}}$ and $\mathcal U_{\text{b}}$ under standard and strong monotonicity assumptions, respectively. In particular, the multiply robust estimators have the following explicit expressions
\begin{align}
\widehat{\text{ITT-NDE}}^{\text{mr}}=&\Prob_n\left[\sum_{d_1d_0 \in \mathcal U}\left\{\widehat\psi_{d_1d_0}^{(10),\text{par}}-\widehat\psi_{d_1d_0}^{(00),\text{par}}\right\}\right],\label{e:nme_estimate_1} \\
\widehat{\text{ITT-NIE}}^{\text{mr}}=&\Prob_n\left[\sum_{d_1d_0 \in \mathcal U}\left\{\widehat\psi_{d_1d_0}^{(11),\text{par}}-\widehat\psi_{d_1d_0}^{(10),\text{par}}\right\}\right].\label{e:nme_estimate_2}
\end{align}
Similarly, the nonparametric estimators $\widehat{\text{ITT-NDE}}^{\text{np}}$ and $\widehat{\text{ITT-NIE}}^{\text{np}}$ can be obtained by replacing all $\widehat\psi_{d_1d_0}^{(zz'),\text{par}}$ in \eqref{e:nme_estimate_1} and \eqref{e:nme_estimate_2} with $\widehat\psi_{d_1d_0}^{(zz'),\text{np}}$. In the Supplementary Material, we show that, for all $\tau \in \{\text{PNIE}_{d_1d_0},\text{PNDE}_{d_1d_0},\text{ITT-NIE}, \text{ITT-NDE}\}$,  $\widehat\tau^{\text{np}}$ is consistent and semiparametrically efficient if conditions in Theorem \ref{thm:nonpar} are satisfied and $\widehat\tau^{\text{mr}}$  is still quadruply robust and locally efficient when all working models in $h_{nuisance}$ are correctly specified. Details on inference are given in the Supplementary Material.

Although we primarily discuss mediation effects on a mean difference scale, all effects can be defined on other scales as needed. For example, with a binary outcome one can consider a risk ratio scale and use $\text{PNIE}_{d_1d_0}^{\text{RR}}=\theta_{d_1d_0}^{(11)}/\theta_{d_1d_0}^{(10)}$ and $\text{PNDE}_{d_1d_0}^{\text{RR}}=\theta_{d_1d_0}^{(10)}/\theta_{d_1d_0}^{(00)}$ to quantify the natural indirect and direct effects within principal stratum $U=d_1d_0$. Similarly, one can use $\text{ITT-NIE}^{\text{RR}} = \E[Y_{1M_1}]/\E[Y_{1M_0}]$ and $\text{ITT-NDE}^{\text{RR}} = \E[Y_{1M_0}]/\E[Y_{0M_0}]$ to measure the natural indirect and direct effects among the entire study population. Estimation of ratio mediation effects is straightforward based on $\widehat{\theta}_{d_1d_0}^{(zz')}$, and is omitted for brevity.

\section{A simulation study}\label{sec:sim}

We investigate the finite-sample performance of the proposed methods via simulation studies. We consider the following data generation process modified from that in \cite{kang2007demystifying}, in which the positivity assumptions are practically violated under model misspecification. Specifically, we generate 1000 Monte Carlo samples with $n=1000$ by the following process. We draw baseline covariates $\bcx  = [X_1,X_2,X_3,X_4] \sim N(\bm{0}_{4\times 1},\bm{I}_{4\times 4})$, and
\begin{align*}
& Z|\bcx \sim \text{Bernoulli}\left(\text{expit}([-1,0.5,-0.25,-0.1]^T\bcx)\right),\\
& D|Z,\bcx  \sim \text{Bernoulli}\left(\text{expit}(-1+2Z+[1,-0.8,0.6,-1]^T\bcx)\right), \\
& M|D,Z,\bcx \sim \text{Bernoulli}\left(\text{expit}(-1.8+2Z+1.5D+[1,-0.5,0.9,-1]^T\bcx)\right),\\
& Y|M,D,Z,\bcx \sim N\left(210+1.5Z-D+M+[27.4,13.7,13.7,13.7]^T\bcx,1\right).
\end{align*}
In correctly specified parametric working models, we directly include the true baseline covariates $\bcx$ into each working model, where specifications of the working models are given in Supplementary Material. Otherwise, we include a set of transformed covariates, $\widetilde{\bcx}=[\widetilde{X}_1,\widetilde{X}_2,\widetilde{X}_3,\widetilde{X}_4]$, into misspecified working models, where $\widetilde{X}_1=\exp(0.5X_1)$, $\widetilde{X}_2=\frac{X_2}{1+X_1}$, $\widetilde{X}_2=\left(\frac{X_2X_3}{25}+0.6\right)^3$, and $\widetilde{X}_4=\left(X_2+X_4+20\right)^2$. We evaluate each of the proposed moment-type estimators and multiply robust estimators under 6 different scenarios: (\romannumeral1) all components in $h_{nuisance}^{\text{par}}$ are correctly specified; (\romannumeral2) only $\pi_z^{\text{par}}(\bx)$ is misspecified; (\romannumeral3) only $p_{zd}^{\text{par}}(\bx)$ is misspecified; (\romannumeral4) only $r_{zd}^{\text{par}}(m,\bx)$ is misspecified; (\romannumeral5) only $\mu_{zd}^{\text{par}}(m,\bx)$ is misspecified; (\romannumeral6) all components in $h_{nuisance}^{\text{par}}$ are  misspecified.


For the nonparametric estimator, we consider a five-fold cross-fitting with the nuisance functions estimated by the Super Learner (\citealp{van2007super}) with a combination of random forest and generalized linear models libraries. Although the Super Learner is more flexible than parametric working models, its performance still depends on the quality of the input feature matrix. In each of Scenarios (\romannumeral1)--(\romannumeral6), we use the true covariates $\bcx$ as the feature matrix under the correctly specified nuisance scenario and the transformed covariates $\widetilde{\bcx}$ as the feature matrix under the misspecified nuisance scenario. 

\begin{figure}[t]
\begin{center}
\includegraphics[width=0.8\textwidth]{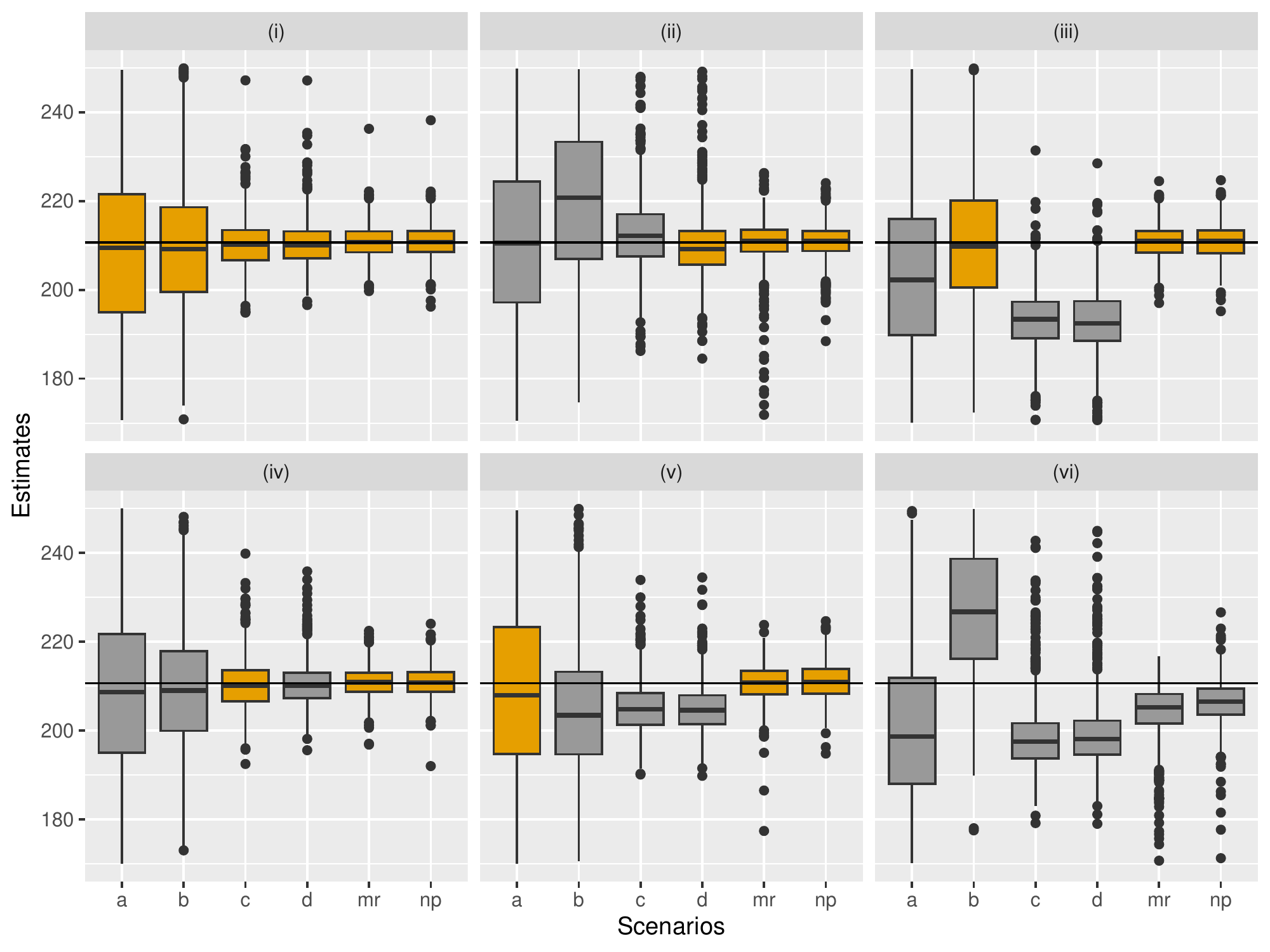}
\end{center}
\caption{Simulation results for estimators of $\theta_{10}^{(10)}$ among 6 different scenarios with sample size $n=1,000$. Scenarios (\romannumeral1)--(\romannumeral6) are described in Section \ref{sec:sim}. The horizontal line in each panel is the true value of $\theta_{10}^{(10)}$. The yellow highlighted boxplots indicate that the corresponding estimators are expected to be consistent by large-sample theory.}
\label{fig:sim}
\end{figure} 

Figure \ref{fig:sim} presents the boxplots of different estimators of $\theta_{10}^{(10)}=\E[Y_{1M_0}|U=10]$ over 1000 Monte Carlo simulations, with each panel corresponding to a specific simulation scenario. As expected, the moment-type estimators are centered around the true value if the required parametric working models are all correctly specified but may diverge from the true value otherwise. The multiply robust estimator exhibits minimum bias in Scenarios (\romannumeral1)--(\romannumeral5), confirming the quadruply robust property; however, it exhibits bias when all of the working models are misspecified as demonstrated in scenario (\romannumeral6).  The nonparametric efficient estimator performs fairly well with minimal bias in scenarios (\romannumeral1)--(\romannumeral5), and its bias in scenario (\romannumeral6) is also smaller than that of the multiply robust estimator with parametric working models. For each scenario, we also investigate the 95\% Wald-type confidence interval coverage rate in Table \ref{tab:ci_sim}, where the variance is estimated by bootstrap in moment-type and multiply robust estimators and by the empirical variance of the EIF in the nonparametric method. {We observe that both $\widehat\theta_{10}^{(10),\text{mr}}$ and $\widehat\theta_{10}^{(10),\text{np}}$ present close to nominal coverage in scenarios (\romannumeral1)--(\romannumeral5), but their coverage rates are attenuated in scenario (\romannumeral6) due to misspecification. Moreover, the moment estimator $\widehat\theta_{10}^{(10),\text{a}}$ appears to have nominal coverage except for Scenario (iii), likely due to the over-estimation of the true sampling variance for this weighting estimator under misspecified weights.} We also evaluate estimators of $\theta_{11}^{(10)}$ and $\theta_{00}^{(10)}$ and results are qualitatively similar. The detailed additional simulation results are provided in the Supplementary Material Figures S1--S2.

\begin{table}[t]
\centering
\caption{Simulation results of the 95\% confidence interval coverage rate among different estimators of $\theta_{10}^{(10)}$. Scenarios (\romannumeral1)--(\romannumeral6) are described in Section 4.6. The numbers displayed in bold signify that the corresponding estimators are expected to be consistent by large-sample theory. \label{tab:ci_sim}}
\begin{tabular}{ccccccc}
  \hline
 Scenario & $\widehat{\theta}_{10}^{(10),\text{a}}$ & $\widehat{\theta}_{10}^{(10),\text{b}}$ & $\widehat{\theta}_{10}^{(10),\text{c}}$ & $\widehat{\theta}_{10}^{(10),\text{d}}$ & $\widehat{\theta}_{10}^{(10),\text{mr}}$ & $\widehat{\theta}_{10}^{(10),\text{np}}$ \\ 
  \hline
  (i) All nuisance correctly specified & \textbf{0.951} & \textbf{0.910}  & \textbf{0.924} & \textbf{0.930} & \textbf{0.929} & \textbf{0.969} \\ 
  (ii) misspecified $\pi_z^{\text{par}}(\bx)$  & 0.955 & 0.874  & 0.941 & \textbf{0.921} & \textbf{0.932} & \textbf{0.953} \\ 
  (iii) misspecified $p_{zd}^{\text{par}}(\bx)$ & 0.871 & \textbf{0.926}  & 0.738 & 0.717 & \textbf{0.918} & 
  \textbf{0.951} \\ 
  (iv) misspecified $r_{zd}^{\text{par}}(m,\bx)$ & 0.938 & 0.907  & \textbf{0.921} & 0.932 & \textbf{0.943} & \textbf{0.953} \\ 
  (v) misspecified $\mu_{zd}^{\text{par}}(m,\bx)$ & \textbf{0.954} & 0.873  & 0.799 & 0.748 & \textbf{0.946} & \textbf{0.958} \\ 
  (vi) All nuisance misspecified & 0.953 & 0.657  & 0.919 & 0.870 & 0.732 & 0.812 \\ 
   \hline
\end{tabular}
\end{table}

\section{Two real-data applications}

\subsection{A job training program with noncompliance}\label{sec:jobs_ii}

JOBS II is a randomized field experiment among 1,801 unemployed workers to examine the effect of a job training workshop to promote mental health and high-quality reemployment (\citealp{price1992impact}). Participants in the treatment group ($Z=1$) were assigned to a job skills workshop, but 45\% of the participants did not show up and dropped into control group. Let $D$ be the indicator of whether the individual attends the workshop, where $D=0$ among all participants in the control group because they had no access to the workshops. As strong monotonicity holds by design, we have two principal strata: compliers and never-takers. { In the JOBS II study, it is of interest to assess the effect of attending job skills workshop ($Z$) on depression ($Y$) among the compliers, which quantifies the efficacy of the training program to mental health \citep{vanderweele2011principal}. Previous efforts (e.g., \citealp{park2020two}) have also investigated the role of sense of mastery ($M$) in mediating the effect from the job skills workshop ($Z$) on depression ($Y$) among the complier stratum, typically under the exclusion restriction assumption.} Because JOBS II is not double-blinded, the exclusion restriction may not hold due to psychological effects \citep{park2020two,stuart2015assessing}. For example, \cite{stuart2015assessing} pointed out that participants assigned to the workshop may feel more optimistic about their reemployment opportunity, suggesting direct pathways from the assignment to depression not via their actual attendance status. 

We assess causal mediation under our proposed assumptions, which permit the exploration of causal mechanism among never-takers. The mediator is sense of mastery at 6 weeks after randomization, with $M=1$ indicating a higher sense of mastery. The outcome is a continuous measure of depression at 6 months after randomization, which ranges from 1 to 4 with a higher value indicating worse depression. Baseline covariates ($\bm X$) include age, gender, race, marital status, education, assertiveness, level of economic hardship, level of depression, and motivation. For the moment-type and multiply robust estimators, we used the parametric working models described in the Supplementary Material for the nuisance functions. Of note, the propensity score is known by randomization, and therefore the working logistic regression of $\pi_z^{\text{par}}(\bcx)$ is not subject to misspecification; we still include all baseline covariates into this working logistic regression to adjust for chance imbalance. For the nonparametric efficient estimator, we used Super Learner with the random forest and generalized linear model libraries for estimating the nuisance functions. We only present the results from the multiply robust estimator and nonparametric efficient estimator; complete results from other estimators are in Supplementary Material Tables S1--S2. 

\begin{table}[t]
\centering
\caption{Estimated causal effects and 95\% confidence intervals in the JOBS II study}\label{tab:jobs2}
\scalebox{0.84}{
\begin{threeparttable}
\begin{tabular}{ccccc}
\hline
\multirow{2}{*}{Population} & \multirow{2}{*}{Estimand} & \multicolumn{3}{c}{Method} \\ \cline{3-5} 
                            &                           & mr  & np  & Rudolph et al.$^\mathsection$   \\ 
\hline
\multirow{3}{*}{Overall} & ITT-NIE & $-$0.015 ($-$0.032, $-$0.003) & $-$0.017 ($-$0.031, $-$0.004) & $-$0.017 ($-$0.031, $-$0.004) \\ 
& ITT-NDE & $-$0.072 ($-$0.151, $-$0.001) & $-$0.067 ($-$0.146, 0.013) & $-$0.067 ($-$0.146, 0.013) \\ 
& ITT & $-$0.088 ($-$0.167, $-$0.016) & $-$0.084 ($-$0.162, $-$0.005) & $-$0.084 ($-$0.162, $-$0.005) \\ 
\hline 
\multirow{3}{*}{Compliers} & PNIE & $-$0.026 ($-$0.055, 0.006) & $-$0.029 ($-$0.052, $-$0.006) & $-$0.030 ($-$0.053, $-$0.006) \\ 
& PNDE & $-$0.083 ($-$0.170, 0.002) & $-$0.066 ($-$0.156, 0.023)  & $-$0.115 ($-$0.251, 0.022) \\ 
& PCE & $-$0.109 ($-$0.191, $-$0.027) & $-$0.096 ($-$0.182, $-$0.009) & $-$0.145 ($-$0.280, $-$0.008) \\ 
\hline
\multirow{3}{*}{Never-takers} & PNIE & 0.000 ($-$0.006, 0.006) & $-$0.001 ($-$0.004, 0.003) & -- \\ 
& PNDE & $-$0.058 ($-$0.160, 0.031) & $-$0.066 ($-$0.163, 0.032) & -- \\ 
& PCE & $-$0.058 ($-$0.160, 0.030) & $-$0.067 ($-$0.163, 0.032) & -- \\ 
\hline
\end{tabular}
\begin{tablenotes}
      \item[$\mathsection$] `Rudolph et al.' is the nonparametric efficient estimator in \cite{rudolph2024using}; see Remark \ref{remark:rudolph} for more details of this approach. Their identification formulas of the ITT natural mediation effects are identical to the present work; this explains the numerical equivalence between `np' and `Rudolph et al.' for the ITT analysis. All effects among never-takers given by \cite{rudolph2024using} are zero due to exclusion restriction.  
    \end{tablenotes}
\end{threeparttable}}
\end{table}

Table \ref{tab:jobs2} (upper panel) presents the estimated ITT effect and its indirect and direct effect decomposition. Both multiply robust and nonparametric efficient estimators present similar results, indicating that the job skills workshop corresponds to negative ITT and ITT-NIE estimates, confirming that sense of mastery is a mediator of the total effect on depression. However, the ITT estimands do not provide resolution to the potential heterogeneity of mediation effects between compliers and never-takers. The estimated proportions of compliers and never-takers are 55\% and 45\%, respectively (under the nonparametric efficient estimator).  We present the stratum-specific mean and standard deviation of the baseline characteristics in Supplementary Material Table S3. Compared to the never-takers, the compliers are older with higher education; a larger fraction of compliers are female, white, married, and are more motivated to participate in the study, but less assertive. To offer a complete picture of the mediation mechanism for compliers and never-takers, Table \ref{tab:jobs2} (middle and bottom panel) additionally presents the estimated PCEs, together with their mediation effect decomposition. For the compliers stratum, both multiply robust and nonparametric efficient estimators suggest that the JOBS II intervention exerts a statistically significant effect on reducing depression, and approximately one quarter of the $\text{PCE}_{10}$ can be explained by the improvement of sense of mastery. For the never-takers, we observe a smaller but still beneficial effect of the intervention on reducing depression; the 95\% confidence interval for $\text{PCE}_{00}$ crosses zero. {For both the multiply robust and nonparametric efficient estimators, the indirect effect among the never-takers is estimated to be almost zero (e.g., $\widehat{\text{PNIE}}_{00}^{\text{np}}=-0.001$ with 95\% confidence interval $[-0.004,0.003]$).}

We also compare our results to estimates under exclusion restriction. Table \ref{tab:jobs2} (right column) shows the estimates using the approach developed by \cite{rudolph2024using} (see Remark \ref{remark:rudolph} for more details of this approach). Due to exclusion restriction, all mediation effects among never-takers are assumed zero. We observe that the point and interval estimates of $\text{PNIE}_{10}$ under exclusion restriction are close to their counterparts under principal ignorability, which is anticipated as the PNIE estimate among never-takers is minimal under principal ignorability. However, the point estimates of $\text{PNDE}_{10}$ and $\text{PCE}_{10}$ under exclusion restriction are larger than those under principal ignorability.

\subsection{An epidemiological study with an intercurrent event}\label{sec:WHO-LARES}

We re-analyze the World Health Organization’s Large Analysis and Review of European Housing and Health Status (WHO-LARES) study with 5,882 individuals for the effect of living in damp or moldy conditions ($Z=1$ if yes and 0 if no) on depression ($Y=1$ if yes and 0 if no), where perceived control on one's home ($M$) is the mediator of interest \citep{vanderweele2010odds}. However, some individuals developed dampness or mold related diseases ($D=1$ if yes and 0 otherwise). \cite{steen2017flexible} viewed $D$ as another mediator prior to $M$ and assesses the path-specific effects through $D$ and/or $M$. {To provide a complementary perspective, we view $D$ as an intercurrent event and partition the population into four strata: the doomed stratum ($U=11$) including those who would always be diseased regardless of living conditions, the immune stratum ($U=00$) including those who would never be diseased, the harmed stratum ($U=10$) including those who would be diseased only if living in damp or moldy conditions, and the benefiters stratum ($U=01$) including those who would only be diseased if not living in damp or moldy conditions. Among them, the doomed and harmed strata are two subpopulations of typical interest because their physical health is more sensitive to living conditions. That is, studying their treatment effects can help understand the impact of living in damp or moldy conditions on the mental health among the more physically vulnerable subgroups. As a further step, addressing the principal natural mediation effects can uncover the extent to which this impact is attributed to the perceived control on one's home.}

We consider the standard monotonicity assumption to rule out benefiters, which is plausible because living in damp and moldy conditions would generally only make an individual more likely to develop dampness or mold related diseases. The exclusion restriction is unlikely to hold because living in damp or moldy conditions can still directly affect mental health even in the absence of dampness or mold related diseases. We adjust for the following confounders: gender, age, marital status, education, employment, smoking, home ownership, home size, crowding (number of residents per room), heating, and natural light. The proportions of doomed, harmed, and immune strata based on the nonparametric efficient estimator are 51\%, 8\%, and 41\%, respectively.  We summarize the stratum-specific mean and standard deviation of the baseline characteristics in Supplementary Material Table S4. The doomed stratum includes more females, followed by harmed stratum, whereas the immune stratum includes the fewest females. As compared to the doomed and harmed strata, members in the immune stratum are more likely to be married and employed; they are also more satisfied with the heating system and natural light condition in their dwellings.

\begin{table}[t]
\centering
\caption{Estimated causal effects and 95\% confidence intervals in the WHO-LARES study.}\label{tab:who_lares}
\begin{threeparttable}
\scalebox{0.9}{\begin{tabular}{cccc}
\hline
\multirow{2}{*}{Population} & \multirow{2}{*}{Estimand} & \multicolumn{2}{c}{Method} \\ \cline{3-4} 
                            &                           & mr  & np$^\mathsection$    \\ 
\hline
\multirow{3}{*}{Overall} & ITT-NIE$^{\text{RR}}$ & 1.021 (1.003, 1.043)  & 1.031 (1.010, 1.053) \\ 
& ITT-NDE$^{\text{RR}}$  & 1.248 (1.114, 1.405) & 1.219 (1.078, 1.361) \\ 
& ITT$^{\text{RR}}$ &  1.274 (1.137, 1.438) & 1.257 (1.114, 1.400)  \\ 
\hline 
\multirow{3}{*}{Doomed} & PNIE$^{\text{RR}}_{11}$  & 1.017 (0.999, 1.034) &  1.025 (1.001, 1.050) \\ 
& PNDE$^{\text{RR}}_{11}$  & 1.223 (1.044, 1.403) &  1.181 (1.014, 1.348) \\ 
& PCE$^{\text{RR}}_{11}$  & 1.244 (1.070, 1.433) &  1.212 (1.044, 1.379) \\ 
\hline
\multirow{3}{*}{Harmed} & PNIE$^{\text{RR}}_{10}$  & 1.029 (1.001, 1.059) & 1.046 (1.013, 1.079) \\ 
  & PNDE$^{\text{RR}}_{10}$  & 2.296 (1.841, 2.982) & 2.142 (1.724, 2.560) \\ 
& PCE$^{\text{RR}}_{10}$  & 2.363 (1.897, 3.045) & 2.241 (1.817, 2.666) \\ 
\hline
\multirow{3}{*}{Immune} & PNIE$^{\text{RR}}_{00}$  & 1.025 (0.983, 1.074) &  1.035 (0.995, 1.075) \\ 
& PNDE$^{\text{RR}}_{00}$  & 1.102 (0.911, 1.328) & 1.111 (0.881, 1.340) \\ 
& PCE$^{\text{RR}}_{00}$  & 1.129 (0.945, 1.336) & 1.150 (0.916, 1.385) \\ 
\hline
\end{tabular}}
\vspace{0.05in}
\end{threeparttable}
\begin{minipage}{\linewidth}\scriptsize
$\mathsection$ Based on the nonparametric efficient method, the contrasts (and 95\% confidence intervals) between the PNDE in different principal strata pairs are $\log(\widehat{\text{PNDE}}_{10}^{\text{RR}})-\log(\widehat{\text{PNDE}}_{00}^{\text{RR}}) =0.657$ $(0.344, 0.967)$, 
$\log(\widehat{\text{PNDE}}_{10}^{\text{RR}})-\log(\widehat{\text{PNDE}}_{11}^{\text{RR}}) =0.595$ $(0.315, 0.874)$, and 
$\log(\widehat{\text{PNDE}}_{11}^{\text{RR}})-\log(\widehat{\text{PNDE}}_{00}^{\text{RR}}) =0.061$ $(-0.189, 0.311)$, respectively. The contrasts (and 95\% confidence intervals) between the PNIE in different principal strata pairs are $\log(\widehat{\text{PNIE}}_{10}^{\text{RR}})-\log(\widehat{\text{PNIE}}_{00}^{\text{RR}}) =0.011$ $(-0.039,0.061)$, 
$\log(\widehat{\text{PNIE}}_{10}^{\text{RR}})-\log(\widehat{\text{PNIE}}_{11}^{\text{RR}}) =0.020$ $(-0.033, 0.072)$, 
$\log(\widehat{\text{PNIE}}_{11}^{\text{RR}})-\log(\widehat{\text{PNIE}}_{00}^{\text{RR}}) =-0.009$ $(-0.055, 0.036)$, respectively.
\end{minipage}
\end{table}

Table \ref{tab:who_lares} presents the results based on the multiply robust and nonparametric efficient estimators. With a binary outcome, we define all causal estimands on the risk ratio scale. Results based on moment-type estimators are given in Supplementary Material Tables S5--S6, exhibiting similar patterns. Table \ref{tab:who_lares} (upper panel) presents the ITT natural mediation effects, and {suggests that living in damp or moldy conditions has a causal effect on elevating the risk of depression; the 95\% confidence intervals for ITT, ITT-NIE and ITT-NDE estimands all exclude the null. Table \ref{tab:who_lares} (lower panels) presents the principal natural mediation effects. For the harmed stratum who are most sensitive to living conditions, we observe a large PNDE (risk ratio $>2$), but PNDEs are smaller in the other strata (risk ratio $<1.3$). 
We further obtain the difference in log PNDEs across the three subgroups using the nonparametric efficient estimator, and confirm that the PNDE within the harmed stratum is substantially different from that within the other two strata. For example, $\log(\widehat{\text{PNDE}}_{10}^{\text{RR}})-\log(\widehat{\text{PNDE}}_{00}^{\text{RR}})=0.657$ with 95\% confidence interval (0.344, 0.967) and $\log(\widehat{\text{PNDE}}_{10}^{\text{RR}})-\log(\widehat{\text{PNDE}}_{11}^{\text{RR}})=0.595$ with 95\% confidence interval (0.315, 0.874). 
On the other hand, the PNIEs are rather comparable in magnitude across strata. Based on the nonparametric efficient estimator, although only the 95\% confidence intervals of the PNIE in the doomed and harmed strata exclude null, the 95\% confidence interval for each pairwise difference in log PINE includes null. 
Finally, the proportion mediated varies across principal strata. That is, the perceived control on one's home, as a mediator, explains the largest fraction of PCE$^{\text{RR}}_{00}$ among the immune strata ($\log(\text{PNIE}^{\text{RR}}_{00})$/$\log(\text{PCE}^{\text{RR}}_{00})\approx 25\%$), and explains the smallest fraction of PCE$^{\text{RR}}_{10}$ among the harmed strata ($\log(\text{PNIE}^{\text{RR}}_{10})$/$\log(\text{PCE}^{\text{RR}}_{10})\approx 6\%$).} 

{As an additional exploratory comparison, we also carry out moderated mediation analysis with respect to baseline covariates. We evaluate the conditional natural indirect and direct effects on a risk ratio scale given each covariate, using the R package \texttt{moderate.mediation}  \citep{qin2024causal}. For each covariate considered, we partition the covariates vector into the moderator of interest and all remaining covariates as confounding adjustment variables. Next, we fit logistic models for the mediator and outcome to assess conditional mediation based on the identification formulas given in \cite{qin2024causal}. 
The conditional natural (in)direct effects are summarized in Supplementary Material Figure S3 and Table S7. 
The results suggest that the mediation effect heterogeneity across different covariate levels is milder as compared to the results under the principal stratification mediation analysis. An important distinction of the moderated mediation analysis from our proposed methods is that the former fails to address $D$ as a potential post-treatment confounder, and may be biased even for quantifying the conditional mediation effect estimands.}

\section{A framework for sensitivity analysis}\label{sec:sensitivity}

The principal ignorability (Assumption \ref{assum:ignora_p}) and ignorability of the mediator (Assumption \ref{assum:ignora_m}) are two crucial assumptions for identification of $\theta_{d_1d_0}^{(zz')}$. These two assumptions, however, cannot be empirically verified. Sensitivity analysis is therefore a useful tool to assess causal effects under assumed violations of these assumptions. In the Supplementary Material, we develop a semiparametric sensitivity analysis framework to assess the impact of violation of Assumption \ref{assum:ignora_p} and Assumption \ref{assum:ignora_m} on inference about $\theta_{d_1d_0}^{(zz')}$ and mediation effects. The proposed sensitivity analysis strategy relies on the confounding function approach \citep{tchetgen2012semiparametric,ding2017principal}. Once the confounding functions are developed, we further provide a multiply robust estimator for $\theta_{d_1d_0}^{(zz')}$ and natural mediation effects and prove its large-sample properties, assuming a known confounding function. In practice, the confounding function is unknown and users can specify a working sensitivity function with interpretable sensitivity parameters and then report the causal estimates under a range of values of sensitivity parameters, in order to identify tipping points that might reverse the causal conclusions. In the Supplementary Material, we also illustrate the proposed sensitivity analysis methods in the context of the JOBS II study. 

\section{Discussion}\label{sec:discussion}

In this article, we consider a set of new identification assumptions for studying the natural mediation effects across several principal strata. This provides an important complementary perspective to existing methods that view $D$ as either a post-treatment confounder or another mediator, and enables the investigation of mediation mechanisms within subpopulations.  
We then derive the EIF for the principal natural mediation effects and further propose a quadruply robust estimator. Finally, a nonparametric extension has been developed to alleviate parametric model misspecification bias and to achieve efficient estimation. 

{
While each principal stratum often represents a scientifically relevant subpopulation, the stratum membership is not always fully observed for each individual in a particular study, leading to potential barriers in optimizing future interventions to target subpopulations. Although there has been no consensus in mitigating such barriers, we offer three considerations that may improve the policy relevance for addressing mediation across principal strata. First, as a routine practice, we recommend summarizing the baseline characteristics for each stratum to help distinguish partially observed subpopulations in measurable dimensions. Given that the baseline summary is widely available from published social science and biomedical studies, summary statistics such those in Web Tables 3 and 4 (applications in Sections \ref{sec:jobs_ii} and \ref{sec:WHO-LARES}) can facilitate a direct comparison to existing study populations and determine the relevance of the current results to alternative populations. Importantly, these summary statistics can be readily obtained once the principal score is estimated, and an example case study can also be found in Section 5.2 of \citet{cheng2023multiply}. 
Second, special study design features may enable an explicit characterization of the endogenous subgroups. As a concrete example, randomization plus strong monotonicity---design features of the JOB II study in Section \ref{sec:jobs_ii}---ensure that individuals attending the workshop and those not attending the workshop in the treatment group are unbiased representations of the compliers and never-takers in the entire study, and evidence about their principal natural mediation effects serves to improve interventions that can at least target individuals in the treatment group. Third, 
for individuals with unobserved stratum membership, the estimated principal score model is useful for membership prediction. In the noncompliance scenario, \citet{kennedy2020sharp} discussed a two-stage treatment policy, where in the first stage one predicts the compliance status (for example, based on estimated principal scores), and in second stage, one recommends the optimal intervention in each predicted stratum. Predicting principal stratum membership was also an intermediate step in \citet{chen2024bayesian}, who have quantified conditional average treatment effects among the partially observed always-survivors in the truncation-by-death setting. Although the optimal methods for predicting membership and the best practice for operationalizing a multi-stage treatment policy remain important topics for future research, we believe this perspective continues to endorse the value of the principal natural mediation effect estimates for informing improved interventions to target partially observed subpopulations.}

{This article addresses a univariate mediator. When $M$ is multi-dimensional with several continuous components, it may be cumbersome to leverage the EIF in Theorem \ref{thm:eif} to assess mediation, because one needs to estimate a multi-dimensional density $r_{zd}(m,\bcx)$ and to further calculate a multi-dimensional integration $\eta_{zz'}(\bcx)=\int_m \mu_{zd_z}(m,\bcx)r_{z'd_{z'}}(m,\bcx) dm$. These challenges may be mitigated by reparametrizing the nuisance functions in EIF \citep{diaz2021nonparametric,zhou2022semiparametric}. For example, one can retain the current parameterization of $\delta_{d_1d_0}(\bco)$ but re-express $\psi_{d_1d_0}^{(zz')}(\bco)$ to $\psi_{d_1d_0}^{(zz'),\dagger}(\bco)$ as
\begin{align*}
 & \left(\frac{\mathbb{I}(Z=z^*)\left\{\mathbb{I}(D=d^*)-p_{z^*d^*}(\bcx)\right\}}{\pi_{z^*}(\bcx)} - k\frac{(1-Z)\left\{D-p_{01}(\bcx)\right\}}{\pi_{0}(\bcx)}\right)\eta_{zz'}^{\dagger}(\bcx) \\
+ &  \left\{p_{z^*d^*}(\bcx)-kp_{01}(\bcx)\right\}\frac{\mathbb{I}(D=d_z,Z=z)}{p_{z'd_{z'}}(\bcx)\pi_{z'}(\bcx)}\frac{g_{z'd_{z'}}(M,\bcx)\kappa_{d_{z'}}(M,\bcx)}{g_{zd_z}(M,\bcx)\kappa_{d_z}(M,\bcx)}\left\{Y-\mu_{zd_z}(M,\bcx)\right\} \\
+ &  \left\{p_{z^*d^*}(\bcx)-kp_{01}(\bcx)\right\}\frac{\mathbb{I}(D=d_{z'},Z=z')}{p_{z'd_{z'}}(\bcx)\pi_{z'}(\bcx)}\left\{\mu_{zd_z}(M,\bcx)-\eta_{zz'}^{\dagger}(\bcx)\right\} \\
+ &  \left\{p_{z^*d^*}(\bcx)-kp_{01}(\bcx)\right\}\eta_{zz'}^{\dagger}(\bcx).
\end{align*}
Here, $\psi_{d_1d_0}^{(zz'),\dagger}(\bco)$ depends on a set of nuisance functions $h_{\text{nuisance}}^{\dagger}=\{\pi_z(\bcx),\allowbreak p_{zd}(\bcx),\allowbreak \mu_{zd}(M,\bcx),\allowbreak \kappa_d(M,\bcx), \allowbreak g_{zd}(M,\bcx),\allowbreak \eta_{zz'}^{\dagger}(\bcx)\}$, where the first three are identical to these from $h_{\text{nuisance}}$, $\kappa_d(M,\bcx)=f_{D|M,\bcx}(d|M,\bcx)$ and $g_{zd}(M,\bcx)=f_{Z|D,M,\bcx}(z|d,M,\bcx)$ are two conditional probabilities, and $\eta_{zz'}^{\dagger}(\bcx) = \E[\mu_{zd_z}(M,\bcx)|Z=z',D=d_{z'},\bcx]$ is a nested expectation that can be estimated by regressing $\widehat \mu_{zd_z}(M,\bcx)$ on $\bcx$ within strata $\{Z=z',D=d_{z'}\}$. Notice that this alternative set of nuisance functions only involves one-dimensional conditional expectations or probabilities regardless of the dimensionality of $M$, and has potential to simplify the modeling process. 
The semiparametric efficient estimator based on the reparameterized EIF can now be defined as $\widehat \theta_{d_1d_0}^{(zz'),\dagger} = \Prob_n\left[\widehat\psi_{d_1d_0}^{(zz'),\dagger}(\bm O)\right]\Big/\Prob_n\left[\widehat\delta_{d_1d_0}(\bm O)\right]$, whose asymptotic properties and finite-sample performance require future work.}

To support the implementation of the proposed methodology, we have developed the \texttt{psmediate} R package along with a brief vignette, which can be accessed at  \url{https://github.com/chaochengstat/psmediate} and \url{https://rpubs.com/chaocheng/psmediate}.

\section*{Acknowledgement}

This work is partially supported by the Patient-Centered Outcomes Research Institute\textsuperscript{\textregistered} (PCORI\textsuperscript{\textregistered} Award ME-2023C1-31350). 
We thank the World Health Organization’s European Centre for Environment and Health, Bonn office, for providing the WHO-LARES data. We thank Johan Steen for connecting us with the Bonn office to apply for data access. The statements in this article are solely the responsibility of the authors and do not necessarily represent the views of PCORI\textsuperscript{\textregistered} or World Health Organization.

\singlespacing

\bibliographystyle{jasa3}
\bibliography{PImediate}

\clearpage

\section*{\centering Supplementary Material to ``Identification and multiply robust estimation in mediation analysis across principal strata"}

\renewcommand{\thelemma}{S\arabic{lemma}}
\renewcommand{\theassumption}{S\arabic{assumption}}
\renewcommand{\theproposition}{S\arabic{proposition}}

\renewcommand\thesection{\Alph{section}}
\renewcommand{\theequation}{s\arabic{equation}}
\renewcommand{\thefigure}{S\arabic{figure}}
\renewcommand{\thetable}{S\arabic{table}}
\newenvironment{proof}{\paragraph{\textit{Proof.}}}{\hfill$\square$}

\setcounter{section}{0}
\setcounter{figure}{0}
\setcounter{table}{0}
\setcounter{assumption}{0}
\setcounter{proposition}{0}

\spacingset{1} 

Section \ref{sec:example} provides an additional motivating example. Section \ref{sec:par} provides practical strategies on specification of the parametric working models.  In Section \ref{sec:a}, we provide a semiparametric sensitivity analysis framework for the principal ignorability assumption and ignorability of the mediator assumption. In Section \ref{sec:b}, we provide the proofs for all theorems, propositions, and remarks in the main manuscript. In Section \ref{sec:c}, we present Supplementary Material Tables and Figures.

\section{An additional motivating example}\label{sec:example}

\begin{example}\label{Example3}
(Mediation analysis with death-truncated mediator and outcome) Consider a case when the mediator and outcome are truncated due to a terminal event before measurements of the mediator, but no other terminal event occurs between the mediator and outcome. One concrete example is the Obstetrics and Periodontal Therapy (OPT) trial \citep{michalowicz2006treatment}, where one may address the role of gestational age ($M$) in mediating the effect of a periodontal treatment during pregnancy ($Z$) on birthweight ($Y$), but $M$ and $Y$ are only measured if infants born alive. Here, the survival status of the infants ($D$) serves as a terminal event, where $M$ and $Y$ are not well defined for dead units ($D=0$) \citep{merchant2021quantile}. {In this scenario, it is of interest to estimate the average treatment effect and mediation effect among the subset of infants who would always survive regardless of the treatment (i.e., the always-survivor stratum). Specifically, assessing the average treatment effect among always-survivors (or referred to as the survivor average causal effect) can help answer the central research question in the OPT trial on whether the periodontal therapy has adverse/positive effect on newborn infant's birthweight \citep{michalowicz2006treatment}, without the complications due to death as a terminal event. As a next step, investigating the principal natural mediation effect within always-survivors, one can clarify the role of gestational age in explaining the survivor average causal effect.} 
\end{example}

\section{Specification of the parametric working models}\label{sec:par}

We can specify working models $h_{nuisance}^{\text{par}}=\{\pi_z^{\text{par}}(\bx),  p_{zd}^{\text{par}}(\bx),  r_{zd}^{\text{par}}(m,\bx), \mu_{zd}^{\text{par}}(m,\bx)\}$ for $h_{nuisance}$.  Specification of $h_{nuisance}^{\text{par}}$ can be flexible, and we provide one example below. This specification strategy is also used in our simulation and application studies.  

For $\pi_z^{\text{par}}(\bx)$, one can consider $f_{Z|\bcx}(1|\bcx) = \text{expit}\left(\bm\alpha^T[1,\bcx^T]\right)$ as a logistic regression with coefficients $\bm\alpha$ such that $\pi_{z}^{\text{par}}(\bx)=\left\{\text{expit}\left(\bm\alpha^T[1,\bx^T]\right)\right\}^{z}\left\{1-\text{expit}\left(\bm\alpha^T[1,\bx^T]\right)\right\}^{1-z}$, where $\text{expit}(x)=\frac{1}{1+\exp(-x)}$ is the logistic function. Specification of $p_{zd}^{\text{par}}(\bx)$ differs between the one-sided and two-sided noncompliance scenarios. Under two-sided noncompliance, one can consider $f_{D|Z,\bcx}(1|Z,\bcx)=\text{expit}\left(\bm\beta^T[1,Z,\bcx^T]\right)$ as a logistic regression with coefficients $\bm\beta$, leading to $p_{zd}^{\text{par}}(\bx) = \left\{\text{expit}\left(\bm\beta^T[1,z,\bx^T]\right)\right\}^{d}\left\{1-\text{expit}\left(\bm\beta^T[1,z,\bx^T]\right)\right\}^{1-d}$. Under one-sided noncompliance, we already know $p_{0d}(\bx) \equiv 1-d$ by the strong monotonicity and therefore we can fix $p_{0d}^{\text{par}}(\bx) = 1-d$ and only specify a working model for $p_{1d}^{\text{par}}(\bx)$; for example, one can consider $f_{D|Z,\bcx}(1|1,\bcx)=\text{expit}\left(\bm\beta^T[1,\bcx^T]\right)$ such that $p_{1d}^{\text{par}}(\bx) = \left\{\text{expit}\left(\bm\beta^T[1,\bx^T]\right)\right\}^{d} \left\{1-\text{expit}\left(\bm\beta^T[1,\bx^T]\right)\right\}^{1-d}$. If $M$ is binary, we can further consider $f_{M|Z,D,\bcx}(1|Z,D,\bcx)=\text{expit} \left(\bm\gamma^T[1,Z,D,\bcx^T]\right)$ as a logistic regression with coefficients $\bm\gamma$ such that $r_{zd}^{\text{par}}(m,\bx) = \left\{\text{expit}\left(\bm\gamma^T[1,z,d,\bx^T]\right)\right\}^{m}\left\{1-\text{expit}\left(\bm\gamma^T[1,z,d,\bx^T]\right)\right\}^{1-m}$. For a continuous $M$, a feasible working model is $M|Z,D,\bcx\sim N(\bm\gamma^T[1,Z,D,\bcx^T],\sigma_\gamma^2)$, which implies that $r_{zd}^{\text{par}}(m,\bx)=\phi(\bm\gamma^T[1,z,d,\bx^T],\sigma_\gamma^2)$, where $\phi(\mu,\sigma^2)$ is the density function of $N(\mu,\sigma^2)$. When $Y$ is a continuous or binary, one can specify $\E[Y|Z,D,M,\bcx]=\text{expit}\left(\bm\kappa^T[1,Z,D,M,\bcx^T]\right)$ or $\bm\kappa^T[1,Z,D,M,\bcx^T]$ with coefficients $\bm\kappa$, leading to $\mu_{zd}^{\text{par}}(m,\bx;\bm\kappa) = \text{expit}\left(\bm\kappa^T[1,z,d,m,\bx^T]\right)$ or $\bm\kappa^T[1,z,d,m,\bx^T]$. Estimators of the parameters in the parametric working models, $\{\widehat{\bm\alpha},\widehat{\bm\beta},\widehat {\bm\gamma},\widehat{\sigma}_\gamma^2,\widehat{\bm\kappa}\}$, can proceed by maximum likelihood. Estimators of nuisance functions are therefore $\widehat h_{nuisance}^{\text{par}}=\left\{\widehat\pi_z^{\text{par}}(\bx), \widehat p_{zd}^{\text{par}}(\bx), \widehat r_{zd}^{\text{par}}(m,\bx), \widehat \mu_{zd}^{\text{par}}(m,\bx)\right\}$, which is $h_{nuisance}^{\text{par}}$ evaluated at $\{\widehat{\bm\alpha},\widehat{\bm\beta},\widehat {\bm\gamma},\widehat{\sigma}_\gamma^2,\widehat{\bm\kappa}\}$.

\section{Sensitivity analysis}\label{sec:a}

The principal ignorability (Assumption 4) and ignorability of the mediator (Assumption 5) are required for the identification of $\theta_{d_1d_0}^{(zz')}$ in Theorem 1. However, these two assumptions are not empirically verifiable based on the observed data and may not hold in randomized experiments. We propose sensitivity analysis strategies to assess the impact of violation of these two assumptions on inference about  $\theta_{d_1d_0}^{(zz')}$. When evaluating the sensitivity to violation of one specific assumption, we shall assume all other structural assumptions hold. To fix ideas, we consider that the mediator $M$ is a multi-valued variable with finite support $M\in\{0,1,\dots,m_{\max}\}$, and the methodology can be generalized to a continuous mediator. 

\subsection{Sensitivity analysis for the principal ignorability assumption}\label{sec:a1}

We first focus on the scenario under standard monotonicity, and methods under strong monotonicity are discussed at the end of this section. To begin with, we notice that Theorem 1 holds under a weaker version of Assumption 4 which consists of two statements:
\begin{itemize}
\item[(\romannumeral1)] $\E_{Y_{1m}|U,\!\bcx}[Y_{1m}|10,\!\bcx]\!=\!\E_{Y_{1m}|U,\bcx}[Y_{1m}|11,\!\bcx]$ and $\E_{Y_{0m}|U,\!\bcx}[Y_{0m}|10,\!\bcx]\!=\!\E_{Y_{0m}|U,\!\bcx}[Y_{0m}|00,\!\bcx]$ for any $m\in\{0,\dots,m_{\max}\}$.

\item[(\romannumeral2)] $f_{M_1|U,\bcx}(m|10,\bcx)=f_{M_1|U,\bcx}(m|11,\bcx)$ and $f_{M_0|U,\bcx}(m|10,\bcx)=f_{M_0|U,\bcx}(m|00,\bcx)$ for any $m\in\{1,\dots,m_{\max}\}$.

\end{itemize}
Statement (\romannumeral1) requires that the expectation of $Y_{1m}$ is same between the complier and always-takers strata and the expectation of $Y_{0m}$ is same between the complier and never-takers strata, conditional on all observed covariates. Statement (\romannumeral2)  implicitly suggests that $f_{M_1|U,\bcx}(0|10,\bcx)=f_{M_1|U,\bcx}(0|11,\bcx)$ and $f_{M_0|U,\bcx}(0|10,\bcx)=f_{M_0|U,\bcx}(0|00,\bcx)$. Therefore, Statement (\romannumeral2) requires that the distribution of $M_1$ is same between the complier and always-takers strata and the distribution of $M_0$ is same between the complier and never-takers strata, conditional on all observed covariates. Our sensitivity analysis is based on the following confounding functions 
measuring departure from the weaker version of principal ignorability:
\begin{align*}
&\xi_{Y}^{(1)}(m,\bx) = \frac{\E_{Y_{1m}|U,\bcx}[Y_{1m}|10,\bx]}{\E_{Y_{1m}|U,\bcx}[Y_{1m}|11,\bx]}, \quad \xi_{Y}^{(0)}(m,\bx) =\frac{\E_{Y_{0m}|U,\bcx}[Y_{0m}|10,\bx]}{\E_{Y_{0m}|U,\bcx}[Y_{0m}|00,\bx]} \quad \text{for $m=0,\dots,m_{\max}$}\\
&\xi_{M}^{(1)}(m,\bx) = \frac{f_{M_1|U,\bcx}(m|10,\bx)}{f_{M_1|U,\bcx}(m|11,\bx)}, \quad \xi_{M}^{(0)}(m,\bx) = \frac{f_{M_0|U,\bcx}(m|10,\bx)}{f_{M_0|U,\bcx}(m|00,\bx)} \quad \text{for $m=1,\dots,m_{\max}$}.
\end{align*}
The first two confounding functions measure deviation of principal ignorability in the outcome variable, where $\xi_{Y}^{(1)}(m,\bx)$ measures the ratio of the mean of $Y_{1m}$ among compliers versus always-takers and $\xi_{Y}^{(0)}(m,\bx)$ measures the ratio of the mean of $Y_{0m}$ among compliers versus never-takers, conditional on $\bcx=\bx$. On the other hand, the last two confounding functions measure deviation of principal ignorability in the mediator variable, where $\xi_{M}^{(1)}(m,\bx)$ measures the relative risk of compliers against always-takers on the treated potential mediator at level $m$ and $\xi_{M}^{(0)}(m,\bx)$ measures the relative risk of compliers against never-takers on the control potential mediator at level $m$, conditional on $\bcx=\bx$. Notice that $\xi_{M}^{(1)}(m,\bx)$ and $\xi_{M}^{(0)}(m,\bx)$ are only defined for $m\geq 1$, which will determine the values of $\xi_{M}^{(1)}(0,\bx):=\frac{f_{M_1|U,\bcx}(0|10,\bx)}{f_{M_1|U,\bcx}(0|11,\bx)}$ and $\xi_{M}^{(0)}(0,\bx):=\frac{f_{M_0|U,\bcx}(0|10,\bx)}{f_{M_0|U,\bcx}(0|00,\bx)}$ as shown in Section \ref{sec:b8}, where we have provided the following explicit expressions in terms of $\{\xi_{M}^{(1)}(m,\bx),\xi_{M}^{(0)}(m,\bx) \text{ for } m=1,\dots,m_{\max}\}$:
\begin{align*}
\xi_{M}^{(1)}(0,\bx)  &  = \frac{1-\displaystyle\sum_{j=1}^{m_{\max}}\frac{\xi_{M}^{(1)}(j,\bx)p_{11}(\bx)}{\xi_{M}^{(1)}(j,\bx)(p_{11}(\bx)-p_{01}(\bx))+p_{01}(\bx)}r_{11}(j,\bx)}{1-\displaystyle\sum_{j=1}^{m_{\max}} \frac{p_{11}(\bx)}{\xi_{M}^{(1)}(j,\bx)(p_{11}(\bx)-p_{01}(\bx))+p_{01}(\bx)}r_{11}(j,\bx)}, \\
\xi_{M}^{(0)}(0,\bx) &  = \frac{1-\displaystyle\sum_{j=1}^{m_{\max}}\frac{\xi_{M}^{(0)}(j,\bx)p_{00}(\bx)}{\xi_{M}^{(0)}(j,\bx)(p_{11}(\bx)-p_{01}(\bx))+p_{10}(\bx)}r_{00}(j,\bx)}{1-\displaystyle\sum_{j=1}^{m_{\max}} \frac{p_{00}(\bx)}{\xi_{M}^{(0)}(j,\bx)(p_{11}(\bx)-p_{01}(\bx))+p_{10}(\bx)}r_{00}(j,\bx)}. 
\end{align*}

Theorem 1 holds if all sensitivity functions in $
\xi = \Big\{\left(\xi_{M}^{(1)}(m,\bx), \xi_{M}^{(0)}(m,\bx)\right) \text{ for }m=1,\dots,m_{\max}\text{ and }\left(\xi_{Y}^{(1)}(m,\bx),\xi_{Y}^{(0)}(m,\bx)\right) \text{ for } m=0,\dots,m_{\max}\Big\}
$
are equal to 1. The following proposition generalizes Theorem 1 to the scenario when at least one confounding function has a value different from 1.
\begin{proposition}\label{proposition:sen_pi}
Suppose that Assumptions 1, 2, 3a, 5, and 6 hold with known values of the confounding functions ($\xi$), we can identify $\theta_{d_1d_0}^{(zz')}$ by
$$
\theta_{d_1d_0}^{(zz')} = \int_{\bx}\frac{e_{d_1d_0}(\bx)}{e_{d_1d_0}}\left\{\sum_{m=0}^{m_{\max}}w_{d_1d_0}^{(zz')}(m,\bx)\mu_{zd_z}(m,\bx) r_{z'd_{z'}}(m,\bx)\right\} \differential\Prob_{\bcx}(\bx),
$$
for any $d_1d_0\in \mathcal U_{\text{a}}$. Here  $w_{d_1d_0}^{(zz')}(m,\bx)$ is a sensitivity weight defined in Section \ref{sec:b8}, which depends on the confounding functions $\xi$ and the observed-data nuisance functions $p_{zd}(\bx)$ and $r_{zd}(m,\bx)$. As an example,\begingroup\makeatletter\def\f@size{8.5}\check@mathfonts 
\begin{align*}
w_{10}^{(10)}(m,\bx) =
\begin{cases}
    \displaystyle\frac{\xi_{M}^{(0)}(m,\bx)p_{00}(\bx)}{\xi_{M}^{(0)}(m,\bx)(p_{11}(\bx)-p_{01}(\bx))+p_{10}(\bx)}\displaystyle\frac{\xi_{M}^{(1)}(m,\bx)(p_{11}(\bx)-p_{01}(\bx))+p_{01}(\bx)}{p_{01}(\bx)/\xi_Y^{(1)}(m,\bx)+\xi_{M}^{(1)}(m,\bx)(p_{11}(\bx)-p_{01}(\bx))},& \text{if } m\geq 1,\\
    \left\{\displaystyle\frac{1}{r_{00}(0,\bx)}-\displaystyle\sum_{j=1}^{m_{\max}} \frac{\xi_{M}^{(0)}(j,\bx)p_{00}(\bx)r_{00}(j,\bx)/r_{00}(0,\bx)}{\xi_{M}^{(0)}(j,\bx)(p_{11}(\bx)-p_{01}(\bx))+p_{10}(\bx)}\right\}\displaystyle\frac{\xi_{M}^{(1)}(0,\bcx)(p_{11}(\bx)-p_{01}(\bx))+p_{01}(\bx)}{p_{01}(\bx)/\xi_Y^{(1)}(0,\bx)+\xi_{M}^{(1)}(0,\bx)(p_{11}(\bx)-p_{01}(\bx))},              & \text{if } m = 0.
\end{cases}
\end{align*}\endgroup
\end{proposition}


If $\xi$ is known, we can construct a new estimator of $\theta_{d_1d_0}^{(zz')}$ by carefully re-weighting each term in the original multiply robust estimator $\widehat\theta^{(zz'),\text{mr}}_{d_1d_0}$ by the sensitivity weight $w_{d_1d_0}^{(zz')}(m,\bx)$. Specifically, the new estimator, $\widehat\theta^{(zz'),\text{mr}}_{d_1d_0}(\xi)$, takes the following form:
\begin{align}
\widehat\theta^{(zz'),\text{mr}}_{d_1d_0}(\xi) = & \Prob_n\Big\{\left(\frac{\mathbb{I}(Z=z^*)\left\{\mathbb{I}(D=d^*)-\widehat p_{z^*d^*}^{\text{par}}(\bcx)\right\}}{\widehat\pi_{z^*}^{\text{par}}(\bcx)} - k\frac{(1-Z)\left\{D-\widehat p_{01}^{\text{par}}(\bcx)\right\}}{\widehat\pi_{0}^{\text{par}}(\bcx)}\right) \frac{\widehat\eta_{zz'}^{w,\text{par}}(\bcx)}{\widehat p_{z^*d^*}^{\text{dr}} - k\widehat p_{01}^{\text{dr}}} \nonumber \\
& + \frac{\widehat p_{z^*d^*}^{\text{par}}(\bcx)-k\widehat p_{01}^{\text{par}}(\bcx)}{\widehat p_{z^*d^*}^{\text{dr}} - k\widehat p_{01}^{\text{dr}}}\frac{\mathbb{I}(D=d_z,Z=z)}{\widehat p_{zd_z}^{\text{par}}(\bcx)\widehat\pi_{z}^{\text{par}}(\bcx)}\frac{\widehat r_{z'd_{z'}}^{\text{par}}(M,\bcx)}{\widehat r_{zd_z}^{\text{par}}(M,\bcx)}\widehat w_{d_1d_0}^{(zz')}(M,\bcx)\left\{Y-\widehat \mu_{zd_z}^{\text{par}}(M,\bcx)\right\} \nonumber \\
& + \frac{\widehat p_{z^*d^*}^{\text{par}}(\bcx)-k\widehat p_{01}^{\text{par}}(\bcx)}{\widehat p_{z^*d^*}^{\text{dr}} - k\widehat p_{01}^{\text{dr}}}\frac{\mathbb{I}(D=d_{z'},Z=z')}{\widehat p_{z'd_{z'}}^{\text{par}}(\bcx)\widehat \pi_{z'}^{\text{par}}(\bcx)}\left\{\widehat w_{d_1d_0}^{(zz')}(M,\bcx)\widehat \mu_{zd_z}^{\text{par}}(M,\bcx)-\widehat \eta_{zz'}^{w,\text{par}}(\bcx)\right\} \nonumber \\
& + \frac{\widehat p_{z^*d^*}^{\text{par}}(\bcx)-k\widehat p_{01}^{\text{par}}(\bcx)}{\widehat p_{z^*d^*}^{\text{dr}} - k\widehat p_{01}^{\text{dr}}}\widehat \eta_{zz'}^{w,\text{par}}(\bcx) \Big\},\label{e:theta_pi}
\end{align}
where $\widehat \eta_{zz'}^{w,\text{par}}(\bx)=\displaystyle\sum_{m=0}^{m_{\max}}\widehat w_{d_1d_0}^{(zz')}(m,\bx)\widehat \mu_{zd_z}^{\text{par}}(m,\bx) \widehat r_{z'd_{z'}}^{\text{par}}(m,\bx)$ and $\widehat w_{d_1d_0}^{(zz')}(m,\bx)$ is $w_{d_1d_0}^{(zz')}(m,\bx)$ evaluated at $\left\{\widehat p_{zd}^{\text{par}}(\bx),\widehat r_{zd}^{\text{par}}(m,\bx)\right\}$. The following proposition shows that $\widehat\theta^{(zz'),\text{mr}}_{d_1d_0}(\xi)$ is a doubly robust estimator under $\mathcal M_{\pi} \cap \mathcal M_{e} \cap \mathcal M_{m}$ or $\mathcal M_{e} \cap \mathcal M_{m} \cap \mathcal M_{o}$. 
\begin{proposition}\label{proposition:mr_xi}
   Suppose that Assumptions 1, 2, 3a, 5, and 6 hold. Then, the estimator $\widehat\theta^{(zz'),\text{mr}}_{d_1d_0}(\xi)$ is consistent and asymptotically normal for any $d_1d_0\in \mathcal U_{\text{a}}$ under $\mathcal M_{\pi} \cap \mathcal M_{e} \cap \mathcal M_{m}$ or $\mathcal M_{e} \cap \mathcal M_{m} \cap \mathcal M_{o}$.
\end{proposition}
In practice, the confounding functions in $\xi$ are unknown. To conduct the sensitivity analysis, one can specify a parametric form of $\xi$ indexed by a finite-dimensional parameter $\bm\lambda$, say $\xi_{\bm\lambda}$.  Then, one can report $\widehat\theta^{(zz'),\text{mr}}_{d_1d_0}(\xi_{\bm\lambda})$ and its confidence intervals over a sequence of values of $\bm\lambda$, which summarizes how sensitive the inference is affected under assumed departure from the principal ignorability assumption.

The above sensitivity analysis strategy can be easily extended to the scenario under strong monotonicity. Because there are no always-takers under strong monotonicity, we only need to quantify the departure of the principal ignorability between the never-takers and compliers strata, that is, only $\xi_M^{(0)}(m,\bx)$ and $\xi_Y^{(0)}(m,\bx)$ are needed for sensitivity analysis. Similar to the construction of \eqref{e:theta_pi}, we can develop an estimator, $\widehat\theta^{(zz'),\text{mr}}_{d_1d_0}(\kappa)$, based on a set of confounding functions, $\kappa = \{\xi_M^{(0)}(m,\bx) \text{ for } m=1,\dots,m_{\max},~\xi_Y^{(0)}(m,\bx) \text{ for } m=0,\allowbreak\dots,\allowbreak m_{\max}\}$, and this estimator is consistent to $\theta^{(zz')}_{d_1d_0}$ under $\mathcal M_{\pi} \cap \mathcal M_{e} \cap \mathcal M_{m}$ or $\mathcal M_{e} \cap \mathcal M_{m} \cap \mathcal M_{o}$ for any $d_1d_0\in \mathcal U_{\text{b}}$. Details of $\widehat\theta^{(zz'),\text{mr}}_{d_1d_0}(\kappa)$ are given in Section \ref{sec:b9}. Analogously, one can report $\widehat\theta^{(zz'),\text{mr}}_{d_1d_0}(\kappa_{\bm\lambda})$ over a set of choices of $\bm\lambda$ to quantify the values of $\theta_{d_1d_0}^{(zz')}$ under assumed departure from principal ignorability, where $\kappa_{\bm\lambda}$ is user-specified parametric functions of $\kappa$. 

\subsection{Sensitivity analysis for the ignorability of the mediator assumption} 
\label{sec:a2}


We develop a sensitivity analysis framework to assess the extent to which the violation of Assumption 5 might affect the inference of $\theta_{d_1d_0}^{(10)}$; identification of $\theta_{d_1d_0}^{(11)}$ and $\theta_{d_1d_0}^{(00)}$, however, does not depend on Assumption 5. In Section \ref{sec:b10}, we show that the expression of $\theta_{d_1d_0}^{(10)}$ in Theorem 1 holds under a weaker version of Assumption 5 such that $\E_{Y_{zm}|Z,M,U,\bcx}[Y_{zm}|z,m,d_1d_0,\bcx] = \E_{Y_{zm}|Z,M,U,\bcx}[Y_{zm}|z,0,d_1d_0,\bcx]$, for all $m> 0$, $z\in\{0,1\}$, $d_1d_0 \in \mathcal U_{a}$ under standard monotonicity, and $d_1d_0 \in \mathcal U_{b}$ under strong monotonicity. This weaker assumption only requires mean independence between the potential outcome and the mediator conditional on the treatment assignment, principal strata, and baseline covariates. Recognizing the sufficiency of this weaker assumption, we propose the following sensitivity function to assess violations of the weaker version of Assumption 5:
$$
t(z,m,d_1d_0,\bx) = \frac{\E_{Y_{1m}|Z,M,U,\bcx}[Y_{1m}|z,m,d_1d_0,\bx]}{\E_{Y_{1m}|Z,M,U,\bcx}[Y_{1m}|z,0,d_1d_0,\bx]},
$$
for $m \in \{0, 1,\dots,m_{\max}\}$, where $t(z,0,d_1d_0,\bx) \equiv 1$ by definition. If $t(z,m,d_1d_0,\bx)$ differs from 1, then the identification formula of $\theta^{(10)}_{d_1d_0}$ in Theorem 1 no longer holds. The following proposition generalizes Theorem 1 to the scenario for a known $t(z,m,d_1d_0,\bx)$.
\begin{proposition}\label{proposition:sen_m}
Suppose that Assumptions 1--4 and 6 hold. Based on the confounding function $t(z,m,d_1d_0,\bx)$, we can identify $\theta^{(10)}_{d_1d_0}$ as
$$
\theta^{(10)}_{d_1d_0} = \int_{\bx}\frac{e_{d_1d_0}(\bx)}{e_{d_1d_0}}\left\{\sum_{m=0}^{m_{\max}} \rho_{d_1d_0}^{(10)}(m,\bx)\mu_{1d_1}(m,\bx) r_{0d_0}(m,\bx)\right\} \differential\Prob_{\bcx}(\bx),
$$
for any $d_1d_0\in \mathcal U_{\text{a}}$ under standard monotonicity and any $d_1d_0\in \mathcal U_{\text{b}}$ under strong monotonicity, 
where 
$$
\rho_{d_1d_0}^{(10)}(m,\bx) = \left\{\sum_{j=0}^{m_{\max}} \frac{t(1,j,d_1d_0,\bx)}{t(1,m,d_1d_0,\bx)}r_{1d_1}(j,\bx)\right\}\Big/\left\{\sum_{j=0}^{m_{\max}} \frac{t(0,j,d_1d_0,\bx)}{t(0,m,d_1d_0,\bx)}r_{0d_0}(j,\bx)\right\}
$$ 
is the sensitivity weight which depends on the confounding function $t(z,m,d_1d_0,\bx)$ and the observed-data nuisance function $r_{zd}(m,\bx)$.
\end{proposition}

If the sensitivity function $t = t(z,m,d_1d_0,\bx)$ is known, we show in Section \ref{sec:b10} that a consistent estimator of $\theta^{(10)}_{d_1d_0}$ can be obtained by re-weighting each term in the multiply robust estimator by the sensitivity weight $\rho_{d_1d_0}^{(10)}(m,\bcx)$, and takes the following form:
\begin{align*}
\widehat\theta^{(10),\text{mr}}_{d_1d_0}(t) = & \Prob_n\Big\{\left(\frac{\mathbb{I}(Z=z^*)\left\{\mathbb{I}(D=d^*)-\widehat p_{z^*d^*}^{\text{par}}(\bcx)\right\}}{\widehat\pi_{z^*}^{\text{par}}(\bcx)} - k\frac{(1-Z)\left\{D-\widehat p_{01}^{\text{par}}(\bcx)\right\}}{\widehat\pi_{0}^{\text{par}}(\bcx)}\right) \frac{\widehat\eta_{10}^{\rho,\text{par}}(\bcx)}{\widehat p_{z^*d^*}^{\text{dr}} - k\widehat p_{01}^{\text{dr}}} \nonumber \\
& + \frac{\widehat p_{z^*d^*}^{\text{par}}(\bcx)-k\widehat p_{01}^{\text{par}}(\bcx)}{\widehat p_{z^*d^*}^{\text{dr}} - k\widehat p_{01}^{\text{dr}}}\frac{\mathbb{I}(D=d_1,Z=1)}{\widehat p_{1d_1}^{\text{par}}(\bcx)\widehat\pi_{1}^{\text{par}}(\bcx)}\frac{\widehat r_{0d_0}^{\text{par}}(M,\bcx)}{\widehat r_{1d_1}^{\text{par}}(M,\bcx)}\widehat{\rho}_{d_1d_0}^{(10)}(M,\bcx)\left\{Y-\widehat \mu_{1d_1}^{\text{par}}(M,\bcx)\right\} \nonumber \\
& + \frac{\widehat p_{z^*d^*}^{\text{par}}(\bcx)-k\widehat p_{01}^{\text{par}}(\bcx)}{\widehat p_{z^*d^*}^{\text{dr}} - k\widehat p_{01}^{\text{dr}}}\frac{\mathbb{I}(D=d',Z=z')}{\widehat p_{0d_0}^{\text{par}}(\bcx)\widehat \pi_{0}^{\text{par}}(\bcx)}\left\{\widehat{\rho}_{d_1d_0}^{(10)}(M,\bcx)\widehat \mu_{1d_1}^{\text{par}}(M,\bcx)-\widehat \eta_{10}^{\rho,\text{par}}(\bcx)\right\} \nonumber \\
& + \frac{\widehat p_{z^*d^*}^{\text{par}}(\bcx)-k\widehat p_{01}^{\text{par}}(\bcx)}{\widehat p_{z^*d^*}^{\text{dr}} - k\widehat p_{01}^{\text{dr}}}\widehat \eta_{10}^{\rho,\text{par}}(\bcx) \Big\},
\end{align*}
where $\widehat \eta_{10}^{\rho,\text{par}}(\bcx)=\displaystyle\sum_{m=0}^{m_{\max}}\widehat{\rho}_{d_1d_0}^{(10)}(m,\bx)\widehat \mu_{1d_1}^{\text{par}}(m,\bx) \widehat r_{0d_0}^{\text{par}}(m,\bx)$ and $\widehat{\rho}_{d_1d_0}^{(10)}(m,\bx)$ is ${\rho}_{d_1d_0}^{(10)}(m,\bx)$ evaluated at $\widehat r_{zd}^{\text{par}}(m,\bx)$. The following proposition shows  that $\widehat\theta^{(10),\text{mr}}_{d_1d_0}(t)$ is a triply robust estimator under $\mathcal M_{\pi} \cap \mathcal M_{e} \cap \mathcal M_{m}$, $\mathcal M_{e} \cap \mathcal M_{m} \cap \mathcal M_{o}$ or $\mathcal M_{\pi} \cap \mathcal M_{m} \cap \mathcal M_{o}$. 
\begin{proposition}\label{proposition:mr_t}
   Suppose that Assumptions 1--4 and 6 hold. Then, under either $\mathcal M_{\pi} \cap \mathcal M_{e} \cap \mathcal M_{m}$, $\mathcal M_{e} \cap \mathcal M_{m} \cap \mathcal M_{o}$, or $\mathcal M_{\pi} \cap \mathcal M_{m} \cap \mathcal M_{o}$, $\widehat\theta^{(10),\text{mr}}_{d_1d_0}(t)$ is consistent and asymptotically normal for any $d_1d_0\in \mathcal U_{\text{a}}$ under standard monotonicity and $d_1d_0\in \mathcal U_{\text{b}}$ under strong monotonicity.
\end{proposition}
To conduct the sensitivity analysis, one can specify a parametric form of $t$ indexed by a finite-dimensional parameter $\bm\zeta$,  $t_{\bm\zeta} =  t(z,m,d_1d_0,\bx;\bm\zeta)$. Then, one can report $\widehat\theta^{(zz'),\text{mr}}_{d_1d_0}(t_{\bm\zeta})$  over a range of choices of $\bm\zeta$, which captures the sensitivity of the conclusion under departure from Assumption 5.

\subsection{Illustration of the sensitivity analysis framework based on the JOBS II study}

This section revisits the JOBS II study in Section 6.1 to assess the robustness of our conclusions to the violation of the proposed structural assumptions. The ignorability assumption of treatment assignment (Assumption 2) and strong monotonicity assumption (Assumption 3) are satisfied in JOBS II study by design, but the  principal ignorability (Assumption 4) and the  ignorability of the mediator (Assumption 5) are generally not empirically verifiable without additional data. Henceforth, we apply the proposed sensitivity analysis framework to assess robustness of the estimated principal natural mediation effects to the violation of Assumptions 4 and 5, separately. For illustration, we only assess the range of the estimated principal natural mediation effects among the compliers stratum. While examining the violation of one assumption, we assume all other assumptions hold.

\subsubsection{Sensitivity analysis for principal ignorability}

As we discussed in Section \ref{sec:a1}, under a one-sided noncompliance scenario (so strong monotonicity holds) with a binary mediator, the confounding functions $\kappa = \{\xi_M^{(0)}(1,\bx),\xi_Y^{(0)}(1,\bx)\}$ can be used to measure the extent to deviation of the principal ignorability assumption. Specifically, $\xi_M^{(0)}(1,\bx) = \frac{f_{M_0|U,\bcx}(1|10,\bx)}{f_{M_0|U,\bcx}(1|00,\bx)}$ measures the relative risk between compliers against the never-takers on the sense of mastery under the control condition and $\xi_Y^{(0)}(1,\bx) = \frac{E_{Y_{0m}|U,\bcx}[Y_{0m}|10,\bx]}{E_{Y_{0m}|U,\bcx}[Y_{0m}|00,\bx]}$ measures the ratio of the potential outcome mean (under the control condition) between compliers and never-takers. For simplicity (and this is often a practical strategy for sensitivity analysis without additional content knowledge), we assume the two confounding functions do not depend on the measured baseline covariates such that $\xi_M^{(0)}(1,\bx)=\lambda_M$ and $\xi_Y^{(0)}(1,\bx) = \lambda_Y$. Our specified parametric confounding function is thus $\kappa_{\bm\lambda} = \{\lambda_M,\lambda_Y\}$.  

\begin{figure}[h]
\begin{center}
\includegraphics[width=1.05\textwidth]{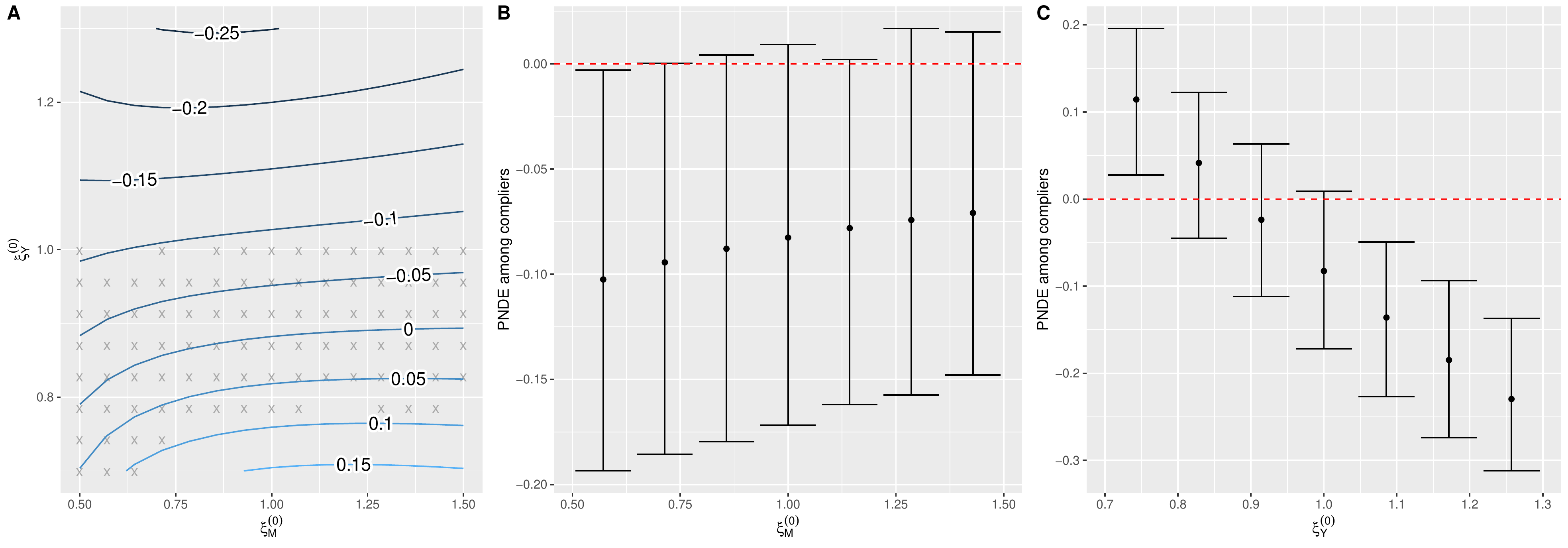}
\end{center}
\caption{Sensitivity analysis of $\text{PNDE}_{10}$ to violation of the principal ignorability assumption in JOBS II study. Panel A: Contour plot of  $\widehat{\text{PNDE}}_{10} = \widehat{\theta}_{10}^{(10),\text{mr}}(\xi_{\bm\lambda})-\widehat{\theta}_{10}^{(00),\text{mr}}(\xi_{\bm\lambda})$ for fixed values the sensitivity parameter $\{\lambda_M,\lambda_Y\} \in [0.5,1.5] \times [0.75,1.25]$, where pixels with the mark `$\times$' suggest that the corresponding 95\% interval estimate covers 0. Panel B: $\widehat{\text{PNDE}}_{10}$ with its 95\% interval estimate for $\lambda_M$ ranging from 0.5 to 1.5 while fixing $\lambda_Y=1$. Panel C: $\widehat{\text{PNDE}}_{10}$ with its 95\% interval estimate for $\lambda_Y$ ranging from 0.75 to 1.25 while fixing $\lambda_M=1$.}
\label{fig:sa_assumption4_1}
\end{figure}

Figure \ref{fig:sa_assumption4_1} presents the bias-corrected $\text{PNDE}_{10}$ estimate, $\widehat{\text{PNDE}}_{10}=\widehat \theta_{10}^{(10),\text{mr}}(\kappa_{\bm\lambda}) - \widehat \theta_{10}^{(00),\text{mr}}(\kappa_{\bm\lambda})$, with fixed values of $\{\lambda_M,\lambda_Y\}$ ranging within    $[0.5,1.5]\times[0.75,1.25]$. The results suggest that ${\text{PNDE}}_{10}$ is robust to violation of the principal ignorability on the mediator variable as  $\widehat{\text{PNDE}}_{10}$ has relatively small fluctuations with different values of $\lambda_M$. For example, $\widehat{\text{PNDE}}_{10}$ only increases from $-$0.102 (95\% CI: $[-0.193,-0.003]$) to $-$0.071 (95\% CI: $[-0.148,0.015]$) when varying $\lambda_M$ from 0.5 to 1.5 with $\lambda_Y$ fixed at 1 (Figure \ref{fig:sa_assumption4_1}, Panel B). In contrast, ${\text{PNDE}}_{10}$ is more sensitive to violation of the principal ignorability on the outcome variable, because $\widehat{\text{PNDE}}_{10}$ moved toward null when $\lambda_Y$ decreases from $1$ and the sign of $\widehat{\text{PNDE}}_{10}$ can even be reverted to positive when $\xi_Y \leq 0.87$. 

\begin{figure}[h]
\begin{center}
\includegraphics[width=0.5\textwidth]{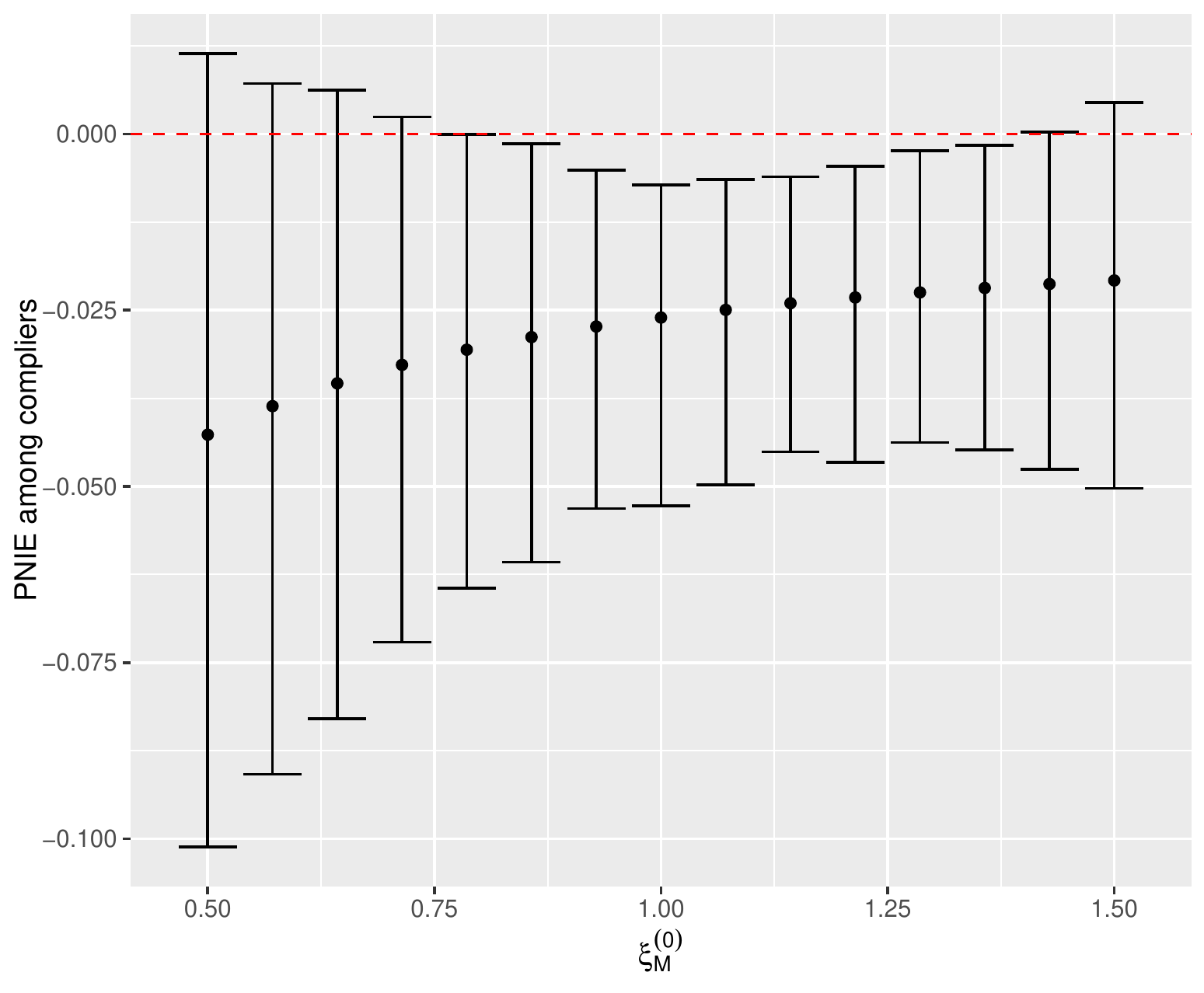}
\end{center}
\caption{Sensitivity analysis of $\text{PNIE}_{10}$ to violation of the principal ignorability assumption in JOBS II study. We report $\widehat{\text{PNIE}}_{10} = \widehat{\theta}_{10}^{(11),\text{mr}}(\xi_{\bm\lambda})-\widehat{\theta}_{10}^{(10),\text{mr}}(\xi_{\bm\lambda})$ and its 95\% bootstrap confidence interval with the sensitivity parameter $\lambda_M$ ranging from 0.5 to 1.5.}
\label{fig:sa_assumption4_2}
\end{figure}

Next, we assess robustness of our conclusions on $\text{PNIE}_{10}$ under departure from principal ignorability. In the one-sided noncompliance scenario, the validity of $\widehat{\text{PNIE}}_{10}$ only depends on the principal ignorability assumption on the mediator variable (as we clarify in Section \ref{sec:b9}, violation of principal ignorabilty on the outcome variable has no impact on $\widehat{\text{PNIE}}_{10}$). Therefore, we provide the bias-corrected $\text{PNIE}_{10}$ estimate, $\widehat{\text{PNIE}}_{10}=\widehat \theta_{10}^{(11),\text{mr}}(\kappa_{\bm\lambda}) - \widehat \theta_{10}^{(10),\text{mr}}(\kappa_{\bm\lambda})$, for $\lambda_M$ ranging from 0.5 to 1.5 in Figure \ref{fig:sa_assumption4_2}. The results suggest that estimates of $\text{PNIE}_{10}$ are robust against violations of principal ignorability, because $\widehat{\text{PNIE}}_{10}$ remains negative among all values of $\lambda_M$ considered. The estimated 95\% confidence intervals, however, straddle zero when $\lambda_M<0.75$ or $\lambda_M>1.35$. 

\subsubsection{Sensitivity analysis for  ignorability of the mediator}

We then investigate whether the conclusion about the principal natural mediation effects among the compliers will be subject to change if the ignorability of the mediator is violated (while assuming all remaining assumptions hold). As indicated in Section \ref{sec:a2}, the confounding function $t(z,1,d_1d_0,\bx)=\frac{\E_{Y_{1m}|Z,M,U,\bcx}[Y_{1m}|z,1,d_1d_0,\bx]}{\E_{Y_{1m}|Z,M,U,\bcx}[Y_{1m}|z,0,d_1d_0,\bx]}$ can be used to quantify the degree of violation of the ignorability of the mediator assumption. For simplicity, we assume $t(z,1,d_1d_0,\bx)$ is constant across all levels of $z$, $d_1d_0$, and $\bx$ and therefore focus on a one-dimensional sensitivity parameter $\zeta$ for $t(z,1,d_1d_0,\bx)$; in other words, the parametric confounding function is simply taken as $t_{\bm\zeta}:=t(z,1,d_1d_0,\bx;\zeta)=\zeta$.

\begin{figure}[h]
\begin{center}
\includegraphics[width=1\textwidth]{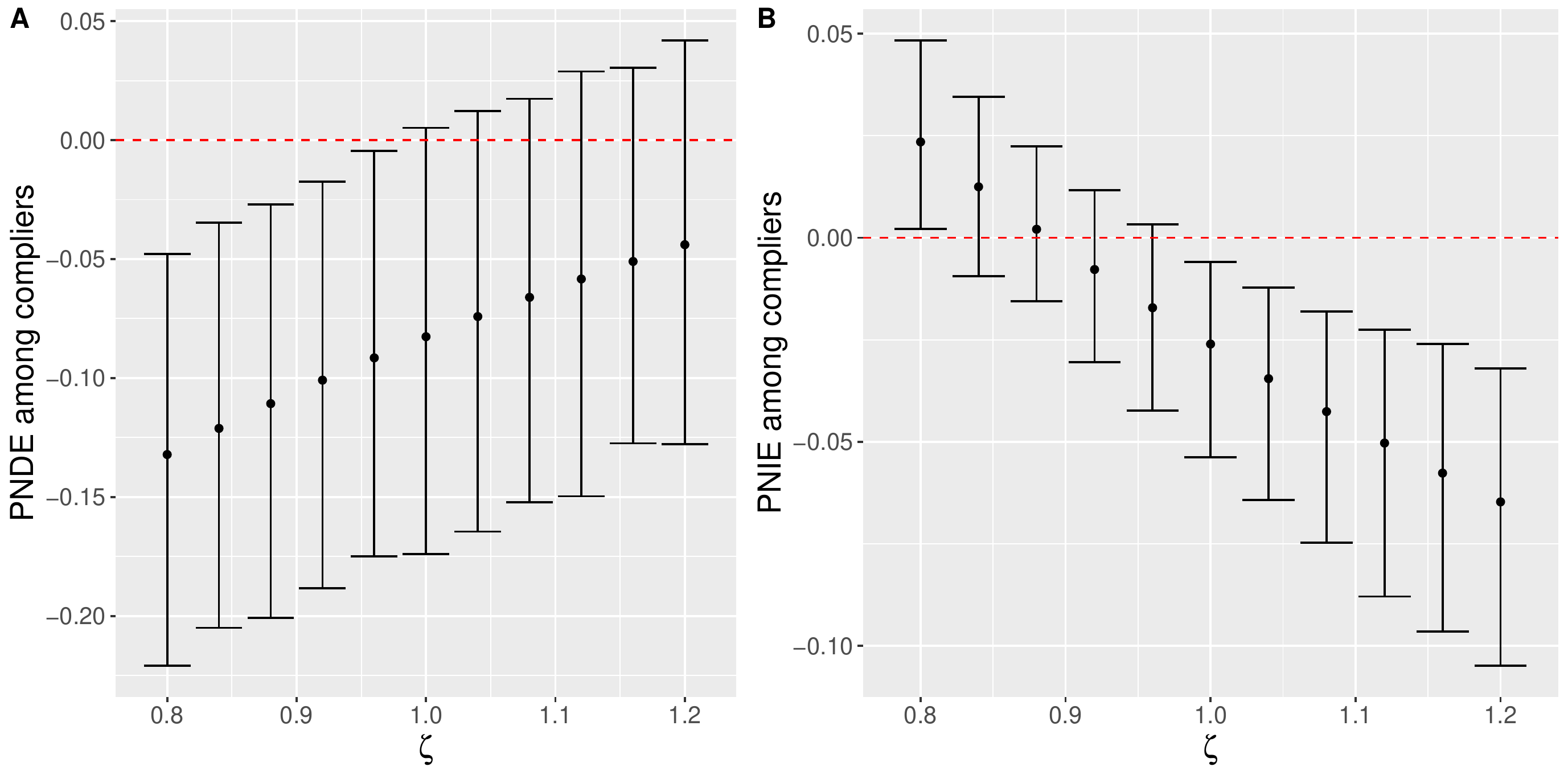}
\end{center}
\caption{Sensitivity analysis of principal natural mediation effects among complies to violation of the  ignorability of mediator assumption in JOBS II study. Panel A: $\widehat{\text{PNDE}}_{10} = \widehat{\theta}_{10}^{(10),\text{mr}}(t_{\bm\zeta})-\widehat{\theta}_{10}^{(00),\text{mr}}$ and its 95\% interval estimate with  $\zeta$ ranging from 0.8 to 1.2. Panel B: $\widehat{\text{PNIE}}_{10} = \widehat{\theta}_{10}^{(11),\text{mr}}-\widehat{\theta}_{10}^{(10),\text{mr}}(t_{\bm\zeta})$ and its 95\% interval estimate with  $\zeta$ ranging from 0.8 to 1.2. }
\label{fig:sa_assumption5}
\end{figure}

Figure \ref{fig:sa_assumption5} presents the bias-corrected estimates of $\text{PNDE}_{10}$, by $\widehat{\text{PNDE}}_{10} = \widehat{\theta}_{10}^{(10),\text{mr}}(t_{\bm\zeta})-\widehat{\theta}_{10}^{(00),\text{mr}}$, and the bias-corrected estimates of $\text{PNIE}_{10}$, by $\widehat{\text{PNIE}}_{10} = \widehat{\theta}_{10}^{(11),\text{mr}}-\widehat{\theta}_{10}^{(10),\text{mr}}(t_{\bm\zeta})$, with $\zeta$ varying from 0.8 to 1.2. We observe that $\widehat{\text{PNDE}}_{10}$ and $\widehat{\text{PNIE}}_{10}$ move towards null with a larger and smaller value of $\zeta$, respectively. Specifically, we observe that $\widehat{\text{PNDE}}_{10}$ remains negative under all assumed values of $\zeta$, but the point estimate increases from $-$0.132 (95\% CI: $[-0.221 -0.047]$) to $-$0.043 (95\% CI: $[-0.128,0.041]$) when $\zeta$ moves from 0.8 to 1.2. On the other hand, $\widehat{\text{PNIE}}_{10}$ decreases from 0.023 (95\% CI: $[0.002, 0.048]$) to $-$0.065 (95\% CI: $[-0.105, -0.031]$), 
when $\zeta$ increases from 0.8 and 1.2, suggesting that $\widehat{\text{PNIE}}_{10}$ is relatively more sensitive to violation of Assumption 5.


\section{Proofs and technical details}\label{sec:b}

\subsection{The nonparametric identification result (Theorem 1)}

\begin{lemma}\label{lemma:expectation}
Let $X$ and $V$ be two random variables with densities $f_X(x)$ and $f_V(v)$. Then, we have that $\E[h(X)|V=v]=\int_x\frac{f_{V|X}(v|x)}{f_V(v)}h(x) \differential\Prob_X(x)$.
\end{lemma}

\begin{proof}
The proof is straightforward and omitted here.
\end{proof}

\begin{lemma}\label{lemma:independent}
Let $X$, $V$, and $G$ be three random variables, then 
$$
X \indep \{V,G\} \Longleftrightarrow X\indep V | G \text{ and } X\indep G | V. 
$$
\end{lemma}

\begin{proof}
First suppose that $X \indep \{V,G\}$ holds, then we have that for any $x$, $v$, $g$,
$$
f_{X,V|G}(c,v|g) = \frac{f_{X,V,G}(x,v,g)}{f_{G}(g)} = \frac{f_{X}(x)f_{V,G}(v,g)}{f_{G}(g)} = f_X(x)f_{V|G}(v|g) = f_{X|G}(x|g)f_{V|G}(v|g),
$$
which implies that $X\indep V | G$. Using the same argument but switching the role of $G$ and $V$, we can show $X\indep G | V$ under  $X \indep \{V,G\}$. Next suppose $X\indep V | G \text{ and } X\indep G | V$, which imply that
\begin{align}
f_X(x) & = \int_g f_{X|G}(x|g) \differential\Prob_G(g) = \int_g f_{X|G,V}(x|g,v) \differential \Prob_G(g) = \int_g f_{X|V}(x|v) \differential\Prob_G(g) \nonumber \\
& = f_{X|V}(x|v),\label{eq:lemma2_1}
\end{align}
for any $x$ and $v$. Therefore, we can show that for any $x,v,g$:
\begin{align*}
f_{X,V,G}(x,v,g) & =  f_{X|V,G}(x|v,g) f_{V,G}(v,g) = f_{X|V}(x|v) f_{V,G}(v,g) = f_{X}(x) f_{V,G}(v,g),
\end{align*}
where the last equality follows from equation \eqref{eq:lemma2_1}. This concludes $X\indep \{V,G\}$.
\end{proof}

\begin{lemma}\label{lemma:pi_v2}
The principal ignorability assumption (Assumption 4) indicates that $\{M_{zd},Y_{z'dm'}\}\indep \{D_1,D_0\} |\bcx$ for any $z$, $z'$, $d$, and $m'$, which further implies
$$
M_{zd} \indep D_{1-z} | D_{z}, \bcx \text{ and } Y_{z'dm'} \indep D_{1-z'} | M_{zd}, D_{z'}, \bcx. 
$$ 
\end{lemma}

\begin{proof}
Observing $U=\{D_1,D_0\}$, we can see that Assumption 4 is equivalent to $\{M_{zd},Y_{z'dm'}\}\indep \{D_1,D_0\} |\bcx$, which implies
$$
M_{zd} \indep \{D_1,D_0\} | \bcx \text{ and } \{Y_{z'dm'},M_{zd}\} \indep \{D_1,D_0\} | \bcx.
$$
In addition, since $\{D_1,D_0\}$ is equivalent to $\{D_z,D_{1-z}\}$ or $\{D_{z'},D_{1-z'}\}$, one can verify that
$$
M_{zd} \indep \{D_z,D_{1-z}\} | \bcx \text{ and }\{Y_{z'dm'},M_{zd}\} \indep \{D_{z'},D_{1-z'}\} | \bcx
$$
hold. Therefore, $M_{zd} \indep D_{1-z} | D_{z}, \bcx$ follows from Lemma \ref{lemma:independent}, with $X=M_{zd}$, $V=D_{1-z}$, and $G=D_{z}$, conditional on $\bcx$. Similarly, one can show $\{Y_{z'dm'},M_{zd}\} \indep D_{1-z'} | D_{z'}, \bcx$ by applying Lemma \ref{lemma:independent}, with $X=\{Y_{z'dm'},M_{zd}\}$, $V=D_{1-z'}$, and $G=D_{z'}$, conditional on $\bcx$. Finally, $Y_{z'dm'} \indep D_{1-z'} | M_{zd}, D_{z'}, \bcx$ follows by applying Lemma \ref{lemma:independent} again, with $X=D_{1-z'}$, $V=Y_{z'dm'}$, and $G=M_{zd}$, conditional on $\bcx$.
\end{proof}\medskip

\begin{lemma}\label{lemma:forastiere2018}
Let $V$ and $G$ be two binary random variables satisfying $V\geq G$ and $X$ be any random variable, then we have
$$
X \indep \{V,G\} \Longleftrightarrow X\indep V  \text{ and } X\indep G. 
$$
\end{lemma}

\begin{proof}
This follows from Lemma S1 in \cite{forastiere2018principal}.
\end{proof}\medskip

\begin{lemma}\label{lemma:randomization2}
Under monotonicity (either Assumption 3a or 3b), Assumption 2 implies that
$
\{U,M_{z'd'},Y_{z^*d^*m^*}\} \indep Z | \bcx,
$ for any $z'$, $z^*$, $d'$, $d^*$, $m^*$.
\end{lemma}
\begin{proof}
Assumption 2 suggests that
\begin{equation*}
\{D_1, M_{z'd'},Y_{z^*d'm^*}\} \indep Z | \bcx \text{ and } \{D_0, M_{z'd'},Y_{z^*d'm^*}\} \indep Z | \bcx,
\end{equation*}
for any $z'$, $z^*$, $d'$, and $m^*$. Therefore, 
\begin{equation}\label{eq:lemma5_1}
D_1 \indep Z | M_{z'd'},Y_{z^*d'm^*}, \bcx \text{ and } D_0 \indep Z | \bcx, M_{z'd'},Y_{z^*d'm^*}
\end{equation}
follow from Lemma \ref{lemma:independent}. Moreover, \eqref{eq:lemma5_1} further implies
\begin{equation}\label{eq:lemma5_2}
\{D_1,D_0\} \indep Z | M_{z'd'},Y_{z^*d'm^*}, \bcx \Longleftrightarrow U \indep Z | M_{z'd'},Y_{z^*d'm^*}, \bcx,
\end{equation}
by applying Lemma \ref{lemma:forastiere2018}, with $V=D_1$, $G=D_0$, and $X=Z$, conditional on $\{M_{z'd'},Y_{z^*d'm^*}, \bcx\}$. Therefore, we have that
\begin{align}
& f(Z,U,M_{z'd'},Y_{z^*d'm^*}|\bcx) \nonumber \\
= & f(Z,U|M_{z'd'},Y_{z^*d'm^*},\bcx)f(M_{z'd'},Y_{z^*d'm^*}|\bcx)  \nonumber \\
= &f(Z|M_{z'd'},Y_{z^*d'm^*},\bcx)f(U|M_{z'd'},Y_{z^*d'm^*},\bcx)f(M_{z'd'},Y_{z^*d'm^*}|\bcx) \quad(\text{by \eqref{eq:lemma5_2}}) \nonumber \\
= &f(Z|\bcx)f(U|M_{z'd'},Y_{z^*d'm^*},\bcx)f(M_{z'd'},Y_{z^*d'm^*}|\bcx) \quad(\text{by Assumption 2}) \nonumber \\
= & f(Z|\bcx)f(U,M_{z'd'},Y_{z^*d'm^*}|\bcx). \nonumber
\end{align}
This equation then shows that $
\{U,M_{z'd'},Y_{z^*d'm^*}\} \indep Z | \bcx
$ for any $z'$, $z^*$, $d'$, and $m^*$. 
\end{proof}\medskip

\begin{lemma}\label{lemma:randomization3}
Under Assumptions 2--4, Assumption 5 is equivalent to
$M_{zd} \indep Y_{z'd'm'} | \bcx$ $\forall$ $z$, $z'$, $d$, $d'$, $m'$.
\end{lemma}
\begin{proof}
Observe that
\begin{align*}
\text{Assumption 5} & \Longleftrightarrow f(M_{zd},Y_{z'd'm'}|Z,U,\bcx) = f(M_{zd}|Z,U,\bcx)f(Y_{z'd'm'}|Z,U,\bcx) \\
& \Longleftrightarrow f(M_{zd},Y_{z'd'm'}|U,\bcx) = f(M_{zd}|U,\bcx)f(Y_{z'd'm'}|U,\bcx) \\
& \Longleftrightarrow f(M_{zd},Y_{z'd'm'}|\bcx) = f(M_{zd}|\bcx)f(Y_{z'd'm'}|\bcx) \\
& \Longleftrightarrow M_{zd}\indep Y_{z'd'm'}|\bcx,
\end{align*}
where the first to the second row follows from Lemma \ref{lemma:randomization2} (as a consequence of Assumptions 2--3), and the second to the third row follows from Assumption 4. This completes the proof.
\end{proof}\medskip

\noindent \textbf{\textit{Proof of Theorem 1.}} Define $d_z=\mathbb{I}(z=1)d_1+\mathbb{I}(z=0)d_0$ such that $d_z=d_1$ and $d_0$ if $z$ in $\theta_{d_1d_0}^{(zz')}$ is 1 and 0, respectively. Similarly, define $d_{z'}=\mathbb{I}(z'=1)d_1+\mathbb{I}(z'=0)d_0$, $d_{1-z}=\mathbb{I}(z=0)d_1+\mathbb{I}(z=1)d_0$, and $d_{1-z'}=\mathbb{I}(z'=0)d_1+\mathbb{I}(z'=1)d_0$. By the definition of $\theta_{d_1d_0}^{(zz')}$, we have that
\begin{align*}
\theta_{d_1d_0}^{(zz')} & = \E[Y_{zM_{z'}}|U=d_1d_0] \\
& = \E\left[\E[Y_{zM_{z'}}|U=d_1d_0,\bcx]\Big|U=d_1d_0\right] \quad \text{(by law of iterated expectations)} \\
& = \E\left[\E[Y_{zM_{z'}}|Z=z,U=d_1d_0,\bcx]\Big|U=d_1d_0\right] \quad \text{(by Lemma \ref{lemma:randomization2})}\\
& = \E\left[\int_m \E[Y_{zm}|Z=z,M_{z'}=m,U=d_1d_0,\bcx]\differential\Prob_{M_{z'}|Z,U,\bcx}(m|z,d_1d_0,\bcx) \Big|U=d_1d_0\right] \\
& = \E\left[\int_m \E[Y_{zm}|Z=z,M_{z}=m,U=d_1d_0,\bcx]\differential\Prob_{M_{z'}|Z,U,\bcx}(m|z,d_1d_0,\bcx) \Big|U=d_1d_0\right] \\
& \quad\quad\quad \text{(by Assumption 5)} \\
& = \E\left[\int_m \E[Y_{zm}|M_{z}=m,U=d_1d_0,\bcx]\differential\Prob_{M_{z'}|U,\bcx}(m|d_1d_0,\bcx) \Big|U=d_1d_0\right] \quad \text{(by Lemma \ref{lemma:randomization2})} \\
& = \E\left[\int_m \E[Y_{zD_zm}|M_{zD_z}=m,U=d_1d_0,\bcx]\differential\Prob_{M_{z'D_{z'}}|U,\bcx}(m|d_1d_0,\bcx) \Big|U=d_1d_0\right] \\
& \quad\quad\quad \text{(by composition of potential values)} \\
& = \E\left[\int_m \E[Y_{zD_zm}|D_z\!=\!d_z,D_{1\!-\!z}\!=\!d_{1-z},M_{zD_z}\!=\!m,\bcx]\differential\Prob_{M_{z'D_{z'}}|D_{z'},D_{1-z'},\bcx}(m|d_{z'},d_{1-z'},\bcx) \Big|U\!=\!d_1d_0\right] \\
& = \E\left[\int_m \E[Y_{zD_zm}|D_z\!=\!d_z,M_{zD_z}\!=\!m,\bcx]\differential\Prob_{M_{z'D_{z'}}|D_{z'},\bcx}(m|d_{z'},\bcx) \Big|U\!=\!d_1d_0\right] \quad\text{(by Lemma \ref{lemma:pi_v2})}\\
& = \E\left[\int_m \E[Y_{zD_zm}|Z=z,D_z\!=\!d_z,M_{zD_z}\!=\!m,\bcx]\differential\Prob_{M_{z'D_{z'}}|Z,D_{z'},\bcx}(m|z',d_{z'},\bcx) \Big|U\!=\!d_1d_0\right]  \\
&  = \E\left[\int_m \E[Y|Z=z,D=d_z,M=m,\bcx]\differential\Prob_{M|Z,D,\bcx}(m|z',d_{z'},\bcx) \Big|U=d_1d_0\right] \quad \text{(by Assumption 1)} \\
&  = \int_{\bx} \frac{f_{U|\bcx}(d_1d_0|\bx)}{f_{U}(d_1d_0)}\int_m \E_{Y|Z,D,M,\bcx}[Y|z,d_z,m,\bcx]\differential\Prob_{M|Z,D,\bcx}(m|z',d_{z'},\bcx) \differential\Prob_{\bcx}(\bx) \quad \text{(by Lemma \ref{lemma:expectation})}\\
&  = \int_{\bx} \frac{e_{d_1d_0}(\bx)}{e_{d_1d_0}}\int_m \E_{Y|Z,D,M,\bcx}[Y|z,d_z,m,\bcx]\differential\Prob_{M|Z,D,\bcx}(m|z',d_{z'},\bcx) \differential\Prob_{\bcx}(\bx),
\end{align*}
where $e_{d_1d_0}(\bx)=p_{z^*d^*}(\bx)-kp_{01}(\bx)$ and $e_{d_1d_0}=p_{z^*d^*}-kp_{01}$ are identified in equation (3) of the main manuscript under the monotonicity assumption (either Assumption 3a or 3b). This completes the proof.  \hfill$\square$

\subsection{Connections to existing literature (Remarks 2 and 3)}

We compare the identification assumptions used in the current article to the identification assumptions in \cite{zhou2022semiparametric} and \cite{tchetgen2014identification}. Specifically, \cite{zhou2022semiparametric} considers the identification of path-specific effects in the presence of multiple causally-ordered mediators and \cite{tchetgen2014identification} considers the identification of mediation effects in the presence of an exposure-induced confounder. 

We focus on a comparable scenario with two intermediate variables, a binary variable $D$ and a binary/continuous variable $M$, both of which sit in the causal pathway between the treatment assignment $Z$ and an outcome $Y$, and we further assume the monotonicity assumption of $Z$ on $D$ (either Assumption 3a or Assumption 3b) holds. These two intermediate variables, $(D,M)$, have different names in these three papers; they are referred to as the post-treatment event and the mediator in the  current paper, as the first mediator and the second mediator in \cite{zhou2022semiparametric}, and as the treatment-induced confounder and the mediator in \cite{tchetgen2014identification}. All three papers consider the consistency assumption (Assumption 1) and slightly different versions of the positivity assumption. To offer a common ground for the comparison of key identification assumptions, throughout the comparison, we always assume the consistency (Assumption 1) and the positivity (Assumption 6) hold. Besides the consistency and positivity assumptions, \cite{tchetgen2014identification} consider the monotonicity assumption (Assumption 3a) and the following NPSEM-IE holds:
\begin{assumption}[NPSEM-IE in \citealp{tchetgen2014identification}]\label{assum:vanderweele}
Suppose the following nonparametric structural equation models with independent errors hold:
\begin{compactitem}
\item[(\romannumeral1)] $\bcx=g_\bcx(\epsilon_\bcx)$,
\item[(\romannumeral2)] $Z=g_Z(\bcx,\epsilon_Z)$,
\item[(\romannumeral3)] $D=g_D(Z,\bcx,\epsilon_D)$,
\item[(\romannumeral4)] $M=g_M(Z,D,\bcx,\epsilon_M)$,
\item[(\romannumeral5)] $Y=g_Y(Z,D,M,\bcx,\epsilon_Y)$,
\end{compactitem}
where $\{g_\bcx,g_Z,g_D,g_M,g_Y\}$ are nonparametric functions and the errors $\{\epsilon_\bcx,\epsilon_Z,\epsilon_D,\epsilon_M,\epsilon_Y\}$ are mutually independent.
\end{assumption}

Besides the consistency and positivity assumptions, \cite{zhou2022semiparametric} consider the following generalized sequential ignorability assumptions:
\begin{assumption} [Assumption 2 in \citealp{zhou2022semiparametric}]\label{assum:zhou}
Suppose the following set of ignorability assumptions hold:
\begin{compactitem}
\item[(\romannumeral1)] $\{D_z,M_{z'd'},Y_{z^*d'm^*}\} \indep Z|\bcx$ for any $z$, $z'$, $d'$, $z^*$, $m^*$,
\item[(\romannumeral2)] $\{M_{zd},Y_{z'dm'}\} \indep D_{z^*}|Z,\bcx$ for any $z$, $z'$, $z^*$, $d$, $m'$,
\item[(\romannumeral3)] $M_{zd} \indep Y_{z'd'm'}|Z,D,\bcx$ for any $z$, $z'$, $d$, $d'$, $m'$.
\end{compactitem}
\end{assumption}

Besides the consistency and positivity assumptions, the current work considers Assumptions 2, 4, 5. To facilitate exposition, we restate these three assumptions:

\begin{assumption}[Assumptions 2, 4 and 5 in current work]\label{assum:s3}
Suppose the following ignorability assumptions hold:
\begin{compactitem}
\item[(\romannumeral1)] $\{D_z,M_{z'd'},Y_{z^*d'm^*}\} \indep Z|\bcx$ for any $z$, $z'$, $z^*$, $d'$, and $m^*$,
\item[(\romannumeral2)] $\{M_{zd},Y_{z'dm'}\} \indep U|\bcx$ for any $z$, $z'$, $d$, and $m'$,
\item[(\romannumeral3)] $M_{zd} \indep Y_{z'd'm'} | Z, U, \bcx$ for any $z$, $z'$, $d$, $d'$, $m'$.
\end{compactitem}
\end{assumption}

The following two lemmas are useful to prove Remarks 2 and 3 in the paper.

\begin{lemma}\label{lemma:zhou_vanderweele}
Suppose that Assumptions 1 and 3 hold. Then, if Assumption \ref{assum:vanderweele} holds, Assumption \ref{assum:zhou} also holds, but not vice versa. 
\end{lemma}

\begin{proof}
First suppose that Assumption \ref{assum:vanderweele} holds. According to Assumption \ref{assum:vanderweele} and by the consistency (Assumption 1) and composition of potential values, we have
\begin{align}
Z & = g_Z(\bcx,\epsilon_Z) \nonumber \\
D_z & = g_D(z,\bcx,\epsilon_D) \label{eq:lemma4_1}\\
M_{z'd'} & = g_M(z',d',\bcx,\epsilon_M) \label{eq:lemma4_2} \\
Y_{z^*d'm^*} & =g_Y(z^*,d',m^*,\bcx,\epsilon_Y) \label{eq:lemma4_3}
\end{align}
for any $z$, $z'$, $z^*$, $d'$, and $m^*$, which indicates that $\{D_z,M_{z'd'},Y_{z^*d'm^*}\} \indep Z|\bcx$ because $\{\epsilon_Z,\epsilon_D,\epsilon_M,\epsilon_Y\}$ are mutually independent. Therefore, Assumption \ref{assum:zhou}(\romannumeral1) holds. Moreover, equations \eqref{eq:lemma4_1}, \eqref{eq:lemma4_2}, and \eqref{eq:lemma4_3} suggest that
\begin{equation}\label{eq:lemma4_4}
\{M_{zd},Y_{z'dm'}\} \indep D_{z^*}|\bcx,
\end{equation}
for any $z$, $z'$, $z^*$, $d$, $m'$. This implies
$$
f(M_{zd},Y_{z'dm'},D_{z^*}|\bcx) = f(M_{zd},Y_{z'dm'}|\bcx)f(D_{z^*}|\bcx),
$$
which, together with Assumption \ref{assum:zhou}(\romannumeral1), implies that
$$
f(M_{zd},Y_{z'dm'},D_{z^*}|Z,\bcx) = f(M_{zd},Y_{z'dm'}|Z,\bcx)f(D_{z^*}|Z,\bcx).
$$
Therefore, Assumption \ref{assum:zhou}(\romannumeral2) hold. Similarly, equations \eqref{eq:lemma4_2} and \eqref{eq:lemma4_3} suggest that $M_{zd} \indep Y_{z'd'm'}|\bcx$ for any $z$, $z'$, $d$, $d'$, $m'$, which indicates
$$
f(M_{zd},Y_{z'd'm'}|\bcx) = f(M_{zd}|\bcx)f(Y_{z'd'm'}|\bcx).
$$
This, coupled with \eqref{eq:lemma4_4}, suggests that
$$
f(M_{zd},Y_{z'd'm'}|D_{z^*},\bcx) = f(M_{zd}|D_{z^*},\bcx)f(Y_{z'd'm'}|D_{z^*},\bcx),
$$
for any $z^*$, which further implies
$$
f(M_{zd},Y_{z'd'm'}|Z=z^*,D_{z^*},\bcx) = f(M_{zd}|Z=z^*,D_{z^*},\bcx)f(M_{zd}|Z=z^*,D_{z^*},\bcx)
$$
as a consequence of Assumption \ref{assum:zhou}(\romannumeral1). Because $D_{z^*}=D$ if $Z=z^*$, we conclude that Assumption \ref{assum:zhou}(\romannumeral3) holds. Now we complete the proof that Assumption \ref{assum:zhou} also holds if Assumption \ref{assum:vanderweele} is valid. However, Assumption \ref{assum:vanderweele} may not hold under Assumption \ref{assum:zhou}; that is, Assumption \ref{assum:vanderweele} is stronger than Assumption \ref{assum:zhou}.  For example, Assumption \ref{assum:zhou} does not require $\{M_{zd},M_{z'd'}\}\indep Z|\bcx$ for $zd\neq z'd'$, but Assumption \ref{assum:vanderweele} implicitly require this by the following set of nonparametric structural equations:
\begin{align*}
Z & = g_Z(\bcx,\epsilon_Z), \\
M_{zd} & = g_M(z,d,\bcx,\epsilon_M),  \\
M_{z'd'} & = g_M(z',d',\bcx,\epsilon_M).
\end{align*}

\end{proof}

\begin{lemma}\label{lemma:zhou_current}
Suppose that Assumptions 1 and 3 hold. Then, Assumption \ref{assum:zhou} is equivalent to Assumption \ref{assum:s3}. 
\end{lemma}

\begin{proof}
Assumption \ref{assum:zhou}(\romannumeral1) is same to Assumption \ref{assum:s3}(\romannumeral1) by direct comparison. Next, we show Assumption \ref{assum:zhou}(\romannumeral2) is equivalent to Assumption \ref{assum:s3}(\romannumeral2) under Assumption \ref{assum:zhou}(\romannumeral1) (or equivalently Assumption \ref{assum:s3}(\romannumeral1)). Specifically, under Assumption \ref{assum:zhou}(\romannumeral1), Assumption \ref{assum:zhou}(\romannumeral2) suggests that
\begin{align*}
f(M_{zd},Y_{z'dm'},D_{z^*}|\bcx) & =f(M_{zd},Y_{z'dm'},D_{z^*}|Z,\bcx) \\
&=  f(M_{zd},Y_{z'dm'}|Z,\bcx)f(D_{z^*}|Z,\bcx) \\
& = f(M_{zd},Y_{z'dm'}|\bcx)f(D_{z^*}|\bcx).
\end{align*}
That is, $
\{M_{zd},Y_{z'dm'}\} \indep D_{z^*} | \bcx
$ for any $z$, $z'$, $z^*$, $d$, and $m'$, which further implies that
$$
\{M_{zd},Y_{z'dm'}\} \indep D_{1} | \bcx \text{ and } \{M_{zd},Y_{z'dm'}\} \indep D_{0} | \bcx.
$$
Applying Lemma \ref{lemma:forastiere2018} to the previous equation and noting that $D_1\geq D_0$ by monotonicity, we obtain
$
\{M_{zd},Y_{z'dm'}\} \indep U | \bcx
$; i.e., Assumption \ref{assum:s3}(\romannumeral2)  holds under Assumption \ref{assum:zhou}(\romannumeral1)--(\romannumeral2). On the other hand, suppose that Assumption \ref{assum:s3}(\romannumeral2) hold, then we can obtain that $
\{M_{zd},Y_{z'dm'}\} \indep D_{1} | \bcx \text{ and } \{M_{zd},Y_{z'dm'}\} \indep D_{0} | \bcx$, by Lemma \ref{lemma:forastiere2018}. This suggests that
$$
\{M_{zd},Y_{z'dm'}\} \indep D_{z^*} | \bcx,
$$
for any $z$, $z'$, $z^*$, $d$, and $m'$. Then applying Assumption \ref{assum:zhou}(\romannumeral1), we have
\begin{align*}
f(M_{zd},Y_{z'dm'},D_{z^*}|Z,\bcx) = &  f(M_{zd},Y_{z'dm'},D_{z^*}|\bcx) = f(M_{zd},Y_{z'dm'}|\bcx)f(D_{z^*}|\bcx),\\
= &  f(M_{zd},Y_{z'dm'}|Z,\bcx)f(D_{z^*}|Z,\bcx)
\end{align*}
thus Assumption \ref{assum:zhou}(\romannumeral2) also holds under Assumption \ref{assum:s3}(\romannumeral1)--(\romannumeral2). Therefore, we have verified that \ref{assum:zhou}(\romannumeral1)--(\romannumeral2) are equivalent to Assumption \ref{assum:s3}(\romannumeral1)--(\romannumeral2). 

Next, we show that Assumption \ref{assum:zhou}(\romannumeral3) is equivalent to Assumption \ref{assum:s3}(\romannumeral3) under Assumption \ref{assum:zhou}(\romannumeral1)--(\romannumeral2) (or equivalently, Assumption \ref{assum:s3}(\romannumeral1)--(\romannumeral2)). When the monotonicity assumption (Assumption 3) holds, the following statements are equivalent under Assumption \ref{assum:zhou}(\romannumeral1)--(\romannumeral2) and Assumption \ref{assum:s3}(\romannumeral1)--(\romannumeral2):
\begin{align*}
& \text{Assumption \ref{assum:s3}(\romannumeral3)} \\
\Longleftrightarrow & f(M_{zd},Y_{z'd'm'}|Z,U,\bcx) = f(Y_{z'd'm'}|Z,U,\bcx)f(M_{zd}|Z,U,\bcx) \\
& \quad \text{(by Lemma S5)} \\
\Longleftrightarrow & f(M_{zd},Y_{z'd'm'}|U,\bcx) = f(Y_{z'd'm'}|U,\bcx)f(M_{zd}|U,\bcx)  \\
& \quad \text{(by Assumption \ref{assum:s3}(\romannumeral2))} \\
\Longleftrightarrow & f(M_{zd},Y_{z'd'm'}|\bcx) = f(Y_{z'd'm'}|\bcx)f(M_{zd}|\bcx)  \\
& \quad \text{(by Assumption \ref{assum:s3}(\romannumeral1) or Assumption \ref{assum:zhou}(\romannumeral1))} \\ 
\Longleftrightarrow & f(M_{zd},Y_{z'd'm'}|Z,\bcx) = f(Y_{z'd'm'}|Z,\bcx)f(M_{zd}|Z,\bcx) \\
& \quad \text{(by Assumption \ref{assum:zhou}(\romannumeral2))} \\
\Longleftrightarrow & f(M_{zd},Y_{z'd'm'}|Z,D_{z^*},\bcx) = f(Y_{z'd'm'}|Z,D_{z^*},\bcx)f(M_{zd}|Z,D_{z^*},\bcx) \\
\Longleftrightarrow & f(M_{zd},Y_{z'd'm'}|Z=z^*,D_{z^*},\bcx) = f(Y_{z'd'm'}|Z=z^*,D_{z^*},\bcx)f(M_{zd}|Z=z^*,D_{z^*},\bcx) \\
\Longleftrightarrow & f(M_{zd},Y_{z'd'm'}|Z,D,\bcx) = f(Y_{z'd'm'}|Z,D,\bcx)f(M_{zd}|Z,D,\bcx)  \\
\Longleftrightarrow & \text{Assumption \ref{assum:zhou}(\romannumeral3)}.
\end{align*}
Then we conclude the proof.
\end{proof}\medskip

\noindent \textbf{\textit{Proof of Remark 2 and Remark 3.}} Remark 3 follows from Lemma \ref{lemma:zhou_current}. Also, noting that Lemma \ref{lemma:zhou_current} suggests that Assumptions 2, 4, and 5 are equivalent to Lemma \ref{lemma:zhou_current} when the consistency (Assumption 1) and monotonicity (Assumption 3) hold. Remark 2 then directly follows from Lemma \ref{lemma:zhou_vanderweele}. \hfill $\square$

\subsection{Moment type estimators (Theorem 2 and Proposition 1)}

\noindent \textbf{\textit{Proof of Theorem 2.}} One can easily verify that $\theta_{d_1d_0}^{(zz')}=\theta_{d_1d_0}^{(zz'),\text{d}}$ by direct comparison. Below we show $\theta_{d_1d_0}^{(zz'),\text{c}}=\theta_{d_1d_0}^{(zz'),\text{d}}$:
\begin{align*}
\theta_{d_1d_0}^{(zz'),\text{c}} = & \E\left[\frac{p_{z^*d^*}(\bcx)-kp_{01}(\bcx)}{p_{z^*d^*}-kp_{01}}\frac{\mathbb{I}(D=d_{z'},Z=z')}{p_{z'd_{z'}}(\bcx)\pi_{z'}(\bcx)}\mu_{zd_z}(M,\bcx)\right] \\
= &  \int_{\bx}\int_m\left\{\frac{p_{z^*d^*}(\bx)-kp_{01}(\bx)}{p_{z^*d^*}-kp_{01}}\frac{1}{p_{z'd_{z'}}(\bx)\pi_{z'}(\bx)}\mu_{zd_z}(m,\bx)  \right\}f_{D|Z,\bcx}(d_{z'}|z',\bx)f_{Z|\bcx}(z'|\bx)\\
& \quad \differential \Prob_{M|Z,D,\bcx}(m|z',d_{z'},\bx) \differential \Prob_{\bcx}(\bx) \\
= &  \int_{\bx}\int_m\left\{\frac{p_{z^*d^*}(\bx)-kp_{01}(\bx)}{p_{z^*d^*}-kp_{01}}\mu_{zd_z}(m,\bx) \differential \Prob_{M|Z,D,\bcx}(m|z',d_{z'},\bx) \differential \Prob_{\bcx}(\bx)\right\} \\
= & \int_{\bx}\frac{p_{z^*d^*}(\bx)-kp_{01}(\bx)}{p_{z^*d^*}-kp_{01}}\int_m\mu_{zd_z}(m,\bx) r_{zd_{z'}}(m,\bx)\differential m \differential \Prob_{\bcx}(\bx) \\
= & \int_{\bx}\frac{p_{z^*d^*}(\bx)-kp_{01}(\bx)}{p_{z^*d^*}-kp_{01}}\eta_{zz'}(\bx) \differential \Prob_{\bcx}(\bx) = \theta_{d_1d_0}^{(zz'),\text{d}}.
\end{align*}
Next, we show $\theta_{d_1d_0}^{(zz'),\text{b}}=\theta_{d_1d_0}^{(zz'),\text{d}}$:
\begin{align*}
\theta_{d_1d_0}^{(zz'),\text{b}} = & \E\left[\left\{\frac{\mathbb{I}(Z=z^*,D=d^*)}{\pi_{z^*}(\bcx)} - k \frac{(1-Z)D}{\pi_0(\bcx)}\right\}\frac{\eta_{zz'}(\bcx)}{p_{z^*d^*}-kp_{01}}\right] \\
= & \int_{\bx} \left\{\frac{1}{\pi_{z^*}(\bx)}\frac{\eta_{zz'}(\bx)}{p_{z^*d^*}-kp_{01}}\right\} f_{D|Z,\bcx}(d^*|z^*,\bx) f_{Z|\bcx}(z^*|\bx) \differential \Prob_\bcx(\bx)\\
& - \int_{\bx} \left\{\frac{k}{\pi_{0}(\bx)}\frac{\eta_{zz'}(\bx)}{p_{z^*d^*}-kp_{01}}\right\} f_{D|Z,\bcx}(1|0,\bx) f_{Z|\bcx}(0|\bx) \differential \Prob_\bcx(\bx)\\
= & \int_{\bx} \left\{f_{D|Z,\bcx}(d^*|z^*,\bx)-kf_{D|Z,\bcx}(1|0,\bx)\right\} \frac{\eta_{zz'}(\bx)}{p_{z^*d^*}-kp_{01}} \differential \Prob_\bcx(\bx) \\
= & \int_{\bx}\frac{p_{z^*d^*}(\bx)-kp_{01}(\bx)}{p_{z^*d^*}-kp_{01}}\eta_{zz'}(\bx) \differential \Prob_{\bcx}(\bx) = \theta_{d_1d_0}^{(zz'),\text{d}}.
\end{align*}
Finally, we show $\theta_{d_1d_0}^{(zz'),\text{a}}=\theta_{d_1d_0}^{(zz'),\text{d}}$: 
\begin{align*}
\theta_{d_1d_0}^{(zz'),\text{a}} = & \E\left[\frac{p_{z^*d^*}(\bcx)-kp_{01}(\bcx)}{p_{z^*d^*}-kp_{01}}\frac{\mathbb{I}(D=d_z,Z=z)}{p_{zd_z}(\bcx)\pi_{z}(\bcx)}\frac{r_{z'd_{z'}}(M,\bcx)}{r_{zd_z}(M,\bcx)}Y\right] \\
= &  \int_{\bx}\int_m \int_y \left\{\frac{p_{z^*d^*}(\bx)-kp_{01}(\bx)}{p_{z^*d^*}-kp_{01}}\frac{1}{p_{zd_z}(\bx)\pi_{z}(\bx)}\frac{r_{z'd_{z'}}(m,\bx)}{r_{zd_z}(m,\bx)}y\right\}f_{D|Z,\bcx}(d_z|z,\bx) f_{Z|\bcx}(z|\bx) \\
& \quad \differential \Prob_{Y|Z,D,M\bcx}(y|z,d_{z},m,\bx) \differential \Prob_{M|Z,D,\bcx}(m|z',d_{z'},\bx) \differential \Prob_{\bcx}(\bx) \\
= & \int_{\bx}\int_m \int_y\frac{p_{z^*d^*}(\bx)-kp_{01}(\bx)}{p_{z^*d^*}-kp_{01}}\frac{r_{z'd_{z'}}(m,\bx)}{r_{zd_z}(m,\bx)}y\differential \Prob_{Y|Z,D,M\bcx}(y|z,d_{z},m,\bx) \\
& \quad\quad \differential \Prob_{M|Z,D,\bcx}(m|z',d_{z'},\bx) \differential \Prob_{\bcx}(\bx) \\
= & \int_{\bx}\frac{p_{z^*d^*}(\bx)-kp_{01}(\bx)}{p_{z^*d^*}-kp_{01}}\int_m \frac{r_{z'd_{z'}}(m,\bx)}{r_{zd_z}(m,\bx)}\int_y y\differential \Prob_{Y|Z,D,M\bcx}(y|z,d_{z},m,\bx) \\
& \quad\quad \differential \Prob_{M|Z,D,\bcx}(m|z',d_{z'},\bx) \differential \Prob_{\bcx}(\bx) \\
= & \int_{\bx}\frac{p_{z^*d^*}(\bx)-kp_{01}(\bx)}{p_{z^*d^*}-kp_{01}} \int_m \frac{r_{z'd_{z'}}(m,\bx)}{r_{zd_z}(m,\bx)}\mu_{zd_z}(m,\bx) \differential \Prob_{M|Z,D,\bcx}(m|z',d_{z'},\bx) \differential \Prob_{\bcx}(\bx) \\
= & \int_{\bx}\frac{p_{z^*d^*}(\bx)-kp_{01}(\bx)}{p_{z^*d^*}-kp_{01}} \int_m r_{z'd_{z'}}(m,\bx) \mu_{zd_z}(m,\bx) \differential m \differential \Prob_{\bcx}(\bx) \\
= & \int_{\bx}\frac{p_{z^*d^*}(\bx)-kp_{01}(\bx)}{p_{z^*d^*}-kp_{01}}\eta_{zz'}(\bx) \differential \Prob_{\bcx}(\bx) = \theta_{d_1d_0}^{(zz'),\text{d}}.
\end{align*}
This concludes that $\theta_{d_1d_0}^{(zz')}=\theta^{(zz'),\textrm{a}}_{d_1d_0}=\theta^{(zz'),\textrm{b}}_{d_1d_0}=\theta^{(zz'),\textrm{c}}_{d_1d_0}=\theta^{(zz'),\textrm{d}}_{d_1d_0}$. \hfill $\square$\medskip

Next, we proceed with the proof of Proposition 1.\medskip 

\noindent \textbf{\textit{Proof of Proposition 1.}} Here we only prove the consistency and asymptotic normality of $\widehat\theta_{d_1d_0}^{(zz'),\text{d}}$ under $\mathcal M_{e} \cap \mathcal M_{m} \cap \mathcal M_{o}$, and the proof can be easily extended to the other three moment-type estimators, $\widehat\theta_{d_1d_0}^{(zz'),\text{a}}$, $\widehat\theta_{d_1d_0}^{(zz'),\text{b}}$, and $\widehat\theta_{d_1d_0}^{(zz'),\text{c}}$. 

Let $\bm\tau$ be all of the parameters in the parametric working models of $h_{nuisance}^{\text{par}}$, and let $\bm\tau^*$ be the probability limit of $\widehat{\bm\tau}$. Let $\widetilde h_{nuisance} =\{\widetilde \pi_z(\bx), \widetilde p_{zd}(\bx), \widetilde r_{zd}(m,\bx), \widetilde \mu_{zd}(m,\bx) \}$ be the value of $h_{nuisance}^{\text{par}}$ when it is evaluated at $\bm\tau^*$; $\widetilde h_{nuisance}$ is taken as the probability limit of  $\widehat h_{nuisance}^{\text{par}}$. Under $\mathcal M_{e} \cap \mathcal M_{m} \cap \mathcal M_{o}$, we have that $\widetilde p_{zd}(\bx)= p_{zd}(\bx)$, $\widetilde r_{zd}(m,\bx)=r_{zd}(m,\bx)$, $\widetilde \mu_{zd}(m,\bx)=\mu_{zd}(m,\bx)$, but we allow $\widetilde \pi_z(\bx)\neq\pi_z(\bx)$ due to possible misspecification of $\mathcal M_{\pi}$. Let $\widetilde p_{zd}=\E\left[\frac{\mathbb{I}(Z=z)(\mathbb{I}(D=d)-\widetilde p_{zd}(\bcx))}{\widetilde \pi_{z}(\bcx)}+\widetilde p_{zd}(\bcx)\right]$ be the probability limit of $p_{zd}^{\text{dr}}$. According to \cite{jiang2022multiply}, $\widetilde p_{zd} = p_{zd}$ under $\mathcal M_{\pi} \cup \mathcal M_{e}$, a condition that is nested within $\mathcal M_{e} \cap \mathcal M_{m} \cap \mathcal M_{o}$.

Next, we prove the consistency and asymptotic normality of $\widehat \theta^{(zz'),\text{d}}_{d_1d_0}$. Notice that $\widehat \theta^{(zz'),\text{d}}_{d_1d_0}$ can be viewed as the solution of the following estimating equation
$$
\Prob_n\left[S_{\text{d}}\left(\bco;\theta_{d_1d_0}^{(zz')},\widehat{\bm\tau}\right)\right] = \Prob_n\left[\mathcal S_1(\bm O; \widehat{\bm \tau})-\theta_{d_1d_0}^{(zz')} \mathcal S_0(\bm O; \widehat{\bm \tau}) \right] = 0, 
$$
where
$$
\mathcal S_{1}(\bm O; \bm \tau) = \frac{p_{z^*d^*}^{\text{par}}(\bcx)-kp_{01}^{\text{par}}(\bcx)}{p_{z^*d^*}^{\text{dr}}-kp_{01}^{\text{dr}}}\int_m \mu_{zd_z}^{\text{par}}(m,\bcx)r_{z'd_{z'}}^{\text{par}}(m,\bcx) dm  
$$ 
and 
$$
\mathcal S_{0}(\bm O; \bm \tau)=\Prob_n\left[\frac{\mathbb{I}(Z=z)\left\{\mathbb{I}(D=d)- p_{zd}^{\text{par}}(\bcx)\right\}}{{\pi}_z^{\text{par}}(\bcx)}+ p_{zd}^{\text{par}}(\bcx)\right]. 
$$
In addition, assume that the following regularity conditions hold:
\begin{itemize}
    \item[1.] Assume that $\sqrt{n}(\widehat{\bm\tau}-\bm \tau^*) = \sqrt{n}\Prob_n\left[\text{IF}_{\bm\tau}(\bm O;\bm \tau^*)\right]+o_p(1)$, where $\text{IF}_{\bm\tau}(\bm O;\bm \tau^*)$ is the influence function of $\widehat{\bm\tau}$ and $o_p(1)$ is a remainder term that converges in probability to 0. Also, assume that $\Prob_n\left[\left\{\text{IF}_{\bm\tau}(\bm O;\bm \tau^*)\right\}^{\otimes 2}\right]$ converges to a positive definite matrix.
    \item[2.] Let $\bm\Xi$ be a bounded convex neighborhood of $\bm\tau^*$. Assume that the class of functions $\Big\{
    \mathcal S_{1}(\bm O;\bm \tau),\frac{\partial }{\partial \bm\tau} \mathcal S_{1}(\bm O;\bm \tau), \left\{\mathcal S_{1}(\bm O;\bm \tau)\right\}^2,
     \mathcal  S_{0}(\bm O;\bm \tau),\frac{\partial }{\partial \bm\tau} \mathcal  S_{0}(\bm O;\bm \tau), \left\{\mathcal  S_{0}(\bm O;\bm \tau)\right\}^2,
     \mathcal  S_{0}(\bm O;\bm \tau),\\ \text{IF}_{\bm\tau}(\bm O;\bm \tau), \left\{\text{IF}_{\bm\tau}(\bm O;\bm \tau)\right\}^{\otimes 2}\Big\}$ is a Glivenko-Cantelli class in $\bm\Xi$.
    \item[3.] Assume that $\Prob_n[\mathcal S_0(\bm O; \bm\tau^*)]$ converges to a positive value. In addition, we assume that both $\Prob_n[\left\{\mathcal S_{1}(\bm O;\bm \tau^*)\right\}^2]$ and $\Prob_n[\left\{\mathcal S_{0}(\bm O;\bm \tau^*)\right\}^2]$ converge to a positive value.  
\end{itemize}
To prove asymptotic normality, we use a Taylor series, along with the above conditions, to deduce that
\begin{align*}
0 = \Prob_n\left[S_{\text{d}}\left(\bco;\widehat \theta_{d_1d_0}^{(zz'),\text{d}},\widehat{\bm\tau}\right)\right] = & \Prob_n\left[\mathcal S_1(\bm O; \widehat{\bm \tau})-\widehat \theta_{d_1d_0}^{(zz'),\text{d}} \mathcal S_0(\bm O; \widehat{\bm \tau})\right] \\
= & \Prob_n\left[\mathcal S_1(\bm O; \bm \tau^*)-\theta_{d_1d_0}^{(zz'),\text{d}} \mathcal S_0(\bm O; \bm \tau^*)\right] \\
& - \Prob_n\left[\mathcal S_0(\bm O; \bm \tau^*)\right]\left(\widehat\theta_{d_1d_0}^{(zz'),\text{d}}-\theta_{d_1d_0}^{(zz'),\text{d}}\right)\\
& + \Prob_n\left[\frac{\partial }{\partial \bm\tau}\mathcal S_1(\bm O; \bm \tau^*)-\theta_{d_1d_0}^{(zz'),\text{d}} \frac{\partial }{\partial \bm\tau} \mathcal S_0(\bm O; \bm \tau^*)\right](\widehat{\bm\tau}-\bm\tau^*) + o_p(n^{-1/2}),
\end{align*}
which suggests that
\begin{align*}
& \sqrt{n}\left(\widehat\theta_{d_1d_0}^{(zz'),\text{d}}-\theta_{d_1d_0}^{(zz'),\text{d}}\right) \\
= & \left\{\E[\mathcal S_0(\bm O;\bm\tau^*)]\right\}^{-1} \Prob_n\left\{\mathcal S_1(\bm O; \bm \tau^*)-\theta_{d_1d_0}^{(zz'),\text{d}} \mathcal S_0(\bm O; \bm \tau^*)+R(\theta_{d_1d_0}^{(zz'),\text{d}},\bm\tau^*)\text{IF}_{\bm\tau}(\bm O;\bm \tau^*) \right\} + o_p(1),
\end{align*}
where $R(\theta_{d_1d_0}^{(zz'),\text{d}},\bm\tau^*)=\E\left[\frac{\partial }{\partial \bm\tau}\mathcal S_1(\bm O; \bm \tau^*)-\theta_{d_1d_0}^{(zz'),\text{d}} \frac{\partial }{\partial \bm\tau} \mathcal S_0(\bm O; \bm \tau^*)\right]$. Then, by applying the central limit theorem and noticing that 
$\theta_{d_1d_0}^{(zz'),\text{d}}=\theta_{d_1d_0}^{(zz')}$ under $\mathcal M_{e} \cap \mathcal M_{m} \cap \mathcal M_{o}$, we can show that $\sqrt{n}\left(\widehat\theta_{d_1d_0}^{(zz'),\text{d}}-\theta_{d_1d_0}^{(zz')}\right)$ converges to a zero-mean normal distribution with variance
$$
V_{\text{d}} = \left\{\E[\mathcal S_0(\bm O;\bm\tau^*)]\right\}^{-2}\E\left[\left\{\mathcal S_1(\bm O; \bm \tau^*)-\theta_{d_1d_0}^{(zz')} \mathcal S_0(\bm O; \bm \tau^*)+R(\theta_{d_1d_0}^{(zz')},\bm\tau^*)\text{IF}_{\bm\tau}(\bm O;\bm \tau^*)\right\}^2\right].
$$

\subsection{The efficient influence function (Theorem 3)}
\label{asec:b4}

We derive the efficient influence function (EIF) of $\theta_{d_1d_0}^{(zz')}$ under the nonparamatic model over the observed data $\bm O= \{\bcx, Z,D,M,Y\}$. Define  $H_{d_1d_0}^{(zz')} = \E\left[(p_{z^*d^*}(\bcx)-kp_{01}(\bcx))\eta_{zz'}(\bcx)\right]$, where $k=|d_1-d_0|$ and $z^*d^*=11,10,01$ if $d_1d_0=10,00,11$, respectively. Then, $\theta_{d_1d_0}^{(zz')} = H_{d_1d_0}^{(zz')}/e_{d_1d_0}$, where $e_{d_1d_0} = p_{z^*d^*}-kp_{01}=\E[p_{z^*d^*}(\bcx)-kp_{01}(\bcx)]$. The following  lemma demonstrates the EIFs of $H_{d_1d_0}^{(zz')}$ and $e_{d_1d_0}$, separately. 

\begin{lemma}\label{lemma:eif_H}
For any $z,z'\in\{0,1\}$, $d_1d_0\in \mathcal U_{\text{a}}$ under standard monotonicity, and $d_1d_0\in \mathcal U_{\text{b}}$ under strong monotonicity, the EIF of  $H_{d_1d_0}^{(zz')}$ over $\mathcal M_{np}$ is $\mathcal D_{d_1d_0}^{(zz'),H}(\bm O) = \psi_{d_1d_0}^{(zz')}(\bco) - H_{d_1d_0}^{(zz')}$, where
\begin{align*}
\psi_{d_1d_0}^{(zz')}(\bco) = & \left(\frac{\mathbb{I}(Z=z^*)\left\{\mathbb{I}(D=d^*)-p_{z^*d^*}(\bcx)\right\}}{\pi_{z^*}(\bcx)} - k\frac{(1-Z)\left\{D-p_{01}(\bcx)\right\}}{\pi_{0}(\bcx)}\right)\eta_{zz'}(\bcx) \\
& + \left\{p_{z^*d^*}(\bcx)-kp_{01}(\bcx)\right\}\frac{\mathbb{I}(D=d_z,Z=z)}{p_{zd_z}(\bcx)\pi_{z}(\bcx)}\frac{r_{z'd_{z'}}(M,\bcx)}{r_{zd_z}(M,\bcx)}\left\{Y-\mu_{zd_z}(M,\bcx)\right\} \\
& + \left\{p_{z^*d^*}(\bcx)-kp_{01}(\bcx)\right\}\frac{\mathbb{I}(D=d_{z'},Z=z')}{p_{z'd_{z'}}(\bcx)\pi_{z'}(\bcx)}\left\{\mu_{zd_z}(M,\bcx)-\eta_{zz'}(\bcx)\right\} \\
& + \left\{p_{z^*d^*}(\bcx)-kp_{01}(\bcx)\right\}\eta_{zz'}(\bcx),
\end{align*}
and $d_z$, $d_{z'}$, and $\eta_{zz'}(\bcx)$ are defined in Theorem 1. The EIF of $e_{d_1d_0}$ over $\mathcal M_{np}$ is $\mathcal D_{d_1d_0}^{e}(\bm O) = \delta_{d_1d_0}(\bm O)- e_{d_1d_0}$, where
$$
\delta_{d_1d_0}(\bm O) = \frac{\mathbb{I}(Z=z^*)\left\{\mathbb{I}(D=d^*)-p_{z^*d^*}(\bcx)\right\}}{\pi_{z^*}(\bcx)}- k\frac{(1-Z)\left\{D-p_{01}(\bcx)\right\}}{\pi_{0}(\bcx)} + p_{z^*d^*}(\bcx) - k p_{01}(\bcx).
$$
\end{lemma}

\begin{proof}
To simplify notation, we abbreviate $f_{Y|Z,D,M,\bcx}(y|z,d,m,\bx)$, $f_{M|Z,D,\bcx}(m|z,d,\bx)$, $f_{D|Z,\bcx}(d|z,\bx)$, $f_{Z|\bcx}(z|\bx)$, and $f_{\bcx}(\bx)$ as $f(y|z,d,m,\bx)$, $f(m|z,d,\bx)$, $f(d|z,\bx)$, $f(z|\bx)$, and $f(\bx)$, respectively. We let $f_{\bm O}(\bm o)$ be the joint density of the observed data $\bm O$, which is abbreviated as $f(\bm o)$ hereafter. Notice that $f(\bm o)$ can be factorized as
$$
f(\bm o) = f(y|z,d,m,\bx)f(m|z,d,\bx) f(d|z,\bx) f(z|\bx) f(\bx). 
$$
We consider a parametric submodel $f_t(\bm o)$ for $f(\bm o)$, which depends on a one-dimensional parameter $t$. We assume that $f_t(\bm o)$ contains the true model $f(\bm o)$ at $t=0$; i.e., $f_{t=0}(\bm o)=f(\bm o)$. Let $S_t(y,m,d,z,\bx)=S_t(\bm o)$ be the score function of this parametric
submodel, which is defined as
$$
S_t(\bm o)  = \triangledown_{t} \log f_t(\bm o)
$$
where $\triangledown_{t}\log f_t(\cdot) = \frac{\partial \log f_t(\cdot)}{\partial t}$. We can decompose the score function as a summation of the following 5 parts:
\begin{align*}
S_t(\bm o)  
=   S_t(y|z,d,m,\bx)+S_t(m|z,d,\bx)+ S_t(d|z,\bx) + S_t(z|\bx)+S_t(\bx), 
\end{align*}
where $S_t(y|z,d,m,\bx)=\triangledown_{t} \log f_t(y|z,d,m,\bx)$, and $S_t(m|z,d,\bx)$, $S_t(d|z,\bx)$, $S_t(z|\bx)$, and $S_t(\bx)$ are similarly defined. According to the semiparametric efficiency theory (\citealp{bickel1993efficient}), the EIF of $H_{d_1d_0}^{(zz')}$, denoted by $\mathcal D_{d_1d_0}^{(zz'),H}(\bm O)$, must satisfy the following equation: 
$$
\E\left[\mathcal D_{d_1d_0}^{(zz'),H}(\bm O)S_{t=0}(\bm O)\right] = \triangledown_{t=0} H_{d_1d_0}^{(zz')}(t),
$$
where 
$$
H_{d_1d_0}^{(zz')}(t) \!=\! \iiint_{\bx,m,y} \!\!\!\!\!\left\{f_t(d^*|z^*,\bx) \!-\! kf_t(D\!=\!0|Z\!=\!1,\bx) \right\}  y f_t(y|z,d_z,m,\bx)  f_t(m|z',d_{z'},\bx)  f_t(\bx)\differential y\differential m\differential \bx
$$
is $H_{d_1d_0}^{(zz')}$ evaluated under the parametric submodel $f_t(\bm o)$. 

Below we derive $\mathcal D_{d_1d_0}^{(zz'),H}(\bm O)$ by solving $\triangledown_{t=0} H_{d_1d_0}^{(zz')}(t)$ directly. Specifically, we can show\begingroup\makeatletter\def\f@size{10}\check@mathfonts
\begin{align}
& \triangledown_{t=0}H_{d_1d_0}^{(zz')}(t) \nonumber  \\
= & \int_{\bx} \!\!\left\{\triangledown_{t=0}f_t(d^*|z^*,\bx) \!-\! k\triangledown_{t=0}f_t(D\!=\!0|Z\!=\!1,\bx) \right\}\iint_{m,y}  y f(y|z,d_z,m,\bx)  f(m|z',d_{z'},\bx)  f(\bx)\differential y\differential m\differential \bx \label{eq:eif1}\\
&+ \int_{\bx} \!\!\left\{f(d^*|z^*,\bx) \!-\! kf(D\!=\!0|Z\!=\!1,\bx) \right\}\iint_{m,y}  y \triangledown_{t=0}f_t(y|z,d_z,m,\bx)  f(m|z',d_{z'},\bx)  f(\bx)\differential y\differential m\differential \bx \label{eq:eif2} \\
&+ \int_{\bx} \!\!\left\{f(d^*|z^*,\bx) \!-\! kf(D\!=\!0|Z\!=\!1,\bx) \right\}\iint_{m,y}  y f(y|z,d_z,m,\bx)  \triangledown_{t=0}f_t(m|z',d_{z'},\bx)  f(\bx)\differential y\differential m\differential \bx \label{eq:eif3} \\
&+ \int_{\bx} \!\!\left\{f(d^*|z^*,\bx) \!-\! kf(D\!=\!0|Z\!=\!1,\bx) \right\}\iint_{m,y}  y f(y|z,d_z,m,\bx)  f(m|z',d_{z'},\bx)  \triangledown_{t=0}f_t(\bx)\differential y\differential m\differential \bx, \label{eq:eif4}
\end{align}\endgroup
where\begingroup\makeatletter\def\f@size{10}\check@mathfonts
\begin{align*}
 & \eqref{eq:eif1} \\
= &\iint_{\bx,m}\left\{\triangledown_{t=0}f_t(d^*|z^*,\bx) \!-\! k\triangledown_{t=0}f_t(D\!=\!0|Z\!=\!1,\bx) \right\}\bbE[Y|z,d_z,m,\bx]f(m|z',d_{z'},\bx) \differential m f(\bx) \differential\bx \\
= &\int_{\bx}\triangledown_{t=0}f_t(d^*|z^*,\bx) \eta_{zz'}(\bx)f(\bx) \differential\bx - k\int_{\bx}\triangledown_{t=0}f_t(D\!=\!0|Z\!=\!1,\bx) \eta_{zz'}(\bx)f(\bx) \differential\bx \\
= &\int_{\bx}f(d^*|z^*,\bx)\left\{\E_{Y,M|Z,D,\bcx}[S_{t=0}(Y,M,d^*,z^*,\bx)|z^*,d^*,\bx] -\E_{Y,M,D|Z,\bcx}[S_{t=0}(Y,M,D,z^*,\bx)|z^*,\bx] \right\} \eta_{zz'}(\bx)f(\bx) \differential\bx \\
 & - k\int_{\bx}f(D=0|Z=1,\bx)\left\{\E_{Y,M|D,Z,\bcx}[S_{t=0}(Y,M,0,1,\bx)|1,0,\bx] -\E_{Y,M,D|Z,\bcx}[S_{t=0}(Y,M,D,1,\bx)|1,\bx] \right\} \eta_{zz'}(\bx)f(\bx) \differential\bx \\
= & \iiint_{\bx,m,y}f(d^*|z^*,\bx)\eta_{zz'}(\bx)S_{t=0}(y,m,d^*,z^*,\bx) f(y|z^*,d^*,m,\bx)f(m|z^*,d^*,\bx)f(\bx) \differential y \differential m \differential\bx\\
&  - \iiiint_{\bx,d,m,y}f(d^*|z^*,\bx)\eta_{zz'}(\bx)S_{t=0}(y,m,d,z^*,\bx) f(y|z^*,d,m,\bx)f(m|z^*,d,\bx)f(d|z^*,\bx)f(\bx) \differential y \differential m \differential d \differential\bx\\
& - k\iiint_{\bx,m,y}f(D=0|Z=1,\bx)\eta_{zz'}(\bx)S_{t=0}(y,m,0,1,\bx) f(y|1,0,m,\bx)f(m|1,0,\bx)f(\bx) \differential y \differential m \differential\bx\\
&  + k\iiiint_{\bx,d,m,y}f(D=0|Z=1,\bx)\eta_{zz'}(\bx)S_{t=0}(y,m,d,1,\bx) f(y|1,d,m,\bx)f(m|1,d,\bx)f(d|1,\bx)f(\bx) \differential y \differential m \differential d \differential\bx\\
= & \E\left[\frac{\mathbb{I}(D=d^*,Z=z^*)}{p_{z^*d^*}(\bcx)\pi_z^*(\bcx)}p_{z^*d^*}(\bcx)\eta_{zz'}(\bcx)S_{t=0}(\bm O)\right] - \E\left[\frac{\mathbb{I}(Z=z^*)}{\pi_{z^*}(\bcx)}p_{z^*d^*}(\bcx)\eta_{zz'}(\bcx)S_{t=0}(\bm O)\right] \\
 & - k\E\left[\frac{\mathbb{I}(D=0,Z=1)}{p_{01}(\bcx)\pi_0(\bcx)}p_{01}(\bcx)\eta_{zz'}(\bcx)S_{t=0}(\bm O)\right] + k\E\left[\frac{\mathbb{I}(Z=0)}{\pi_{0}(\bcx)}p_{01}(\bcx)\eta_{zz'}(\bcx)S_{t=0}(\bm O)\right] \\
= &\bbE\left[\left(\frac{\mathbb{I}(Z=z^*)\left\{\mathbb{I}(D=d^*)-p_{z^*d^*}(\bcx)\right\}}{\pi_{z^*}(\bcx)} - k\frac{(1-Z)\left\{D-p_{01}(\bcx)\right\}}{\pi_{0}(\bcx)}\right)\eta_{zz'}(\bcx)S_{t=0}(\bco)\right]
\end{align*}\endgroup
and\begingroup\makeatletter\def\f@size{10}\check@mathfonts
\begin{align*}
\eqref{eq:eif2} 
= &\int_{\bx} \left\{p_{z^*d^*}(\bx)-kp_{01}(\bx)\right\}\iint_{m,y}  y \triangledown_{t=0}f_t(y|z,d_z,m,\bx)  f(m|z',d_{z'},\bx)  f(\bx)\differential y\differential m\differential \bx \\
= & \iiint_{\bx,m,y} \left\{p_{z^*d^*}(\bx)\!-\!kp_{01}(\bx)\right\}yf(y|z,d_z,m,\bx)\left\{S_{t=0}(y,m,d_z,z,\bx)-\E_{Y|M,D,Z,\bcx}[S(Y,m,d_{z},z,\bx)|m,d_z,z,\bx]\right\}\\
& \quad f(m|z',d_{z'},\bx) f(\bx) \differential y\differential m\differential \bx \\
= & \iiint_{\bx,m,y} \left\{p_{z^*d^*}(\bx)\!-\!kp_{01}(\bx)\right\}yf(y|z,d_z,m,\bx)S_{t=0}(y,m,d_z,z,\bx)f(m|z',d_{z'},\bx) f(\bx) \differential y\differential m\differential \bx \\
& - \iiint_{\bx,m,y} \left\{p_{z^*d^*}(\bx)\!-\!kp_{01}(\bx)\right\}\E_{Y|Z,D,M,\bcx}[Y|z,d_z,m,\bx]S_{t=0}(y,m,d_z,z,\bx)f(y|z,d_z,m,\bx)\\
& \quad\quad f(m|z',d_{z'},\bx) f(\bx) \differential y\differential m\differential \bx\\
= & \iiint_{\bx,m,y} \!\!\!\!\left\{p_{z^*d^*}(\bx)\!-\!kp_{01}(\bx)\right\}y\frac{f(m|z',d_{z'},\bx)}{f(m|z,d_{z},\bx)}S_{t=0}(y,m,d_z,z,\bx) f(y|z,d_z,m,\bx) f(m|z,d_{z},\bx)f(\bx) \differential y\differential m\differential \bx \\
 & - \iiint_{\bx,m,y} \!\!\!\!\left\{p_{z^*d^*}(\bx)\!-\!kp_{01}(\bx)\right\}\mu_{zd_z}(m,\bx)\frac{f(m|z',d_{z'},\bx)}{f(m|z,d_{z},\bx)}S_{t=0}(y,m,d_z,z,\bx) f(y|z,d_z,m,\bx) \\
 & \quad\quad f(m|z,d_{z},\bx)f(\bx) \differential y\differential m\differential \bx \\
= & \E\left[\left\{p_{z^*d^*}(\bcx)-kp_{01}(\bcx)\right\}\frac{\mathbb{I}(D=d_z,Z=z)}{p_{zd_z}(\bcx)\pi_{z}(\bcx)}\frac{r_{z'd_{z'}}(M,\bcx)}{r_{zd_z}(M,\bcx)}\left\{Y-\mu_{zd_z}(M,\bcx)\right\}S_{t=0}(\bco)\right]
\end{align*}\endgroup
and \begingroup\makeatletter\def\f@size{10}\check@mathfonts
\begin{align*}
\eqref{eq:eif3} 
= &\iint_{\bx,m} \left\{p_{z^*d^*}(\bx)-kp_{01}(\bx)\right\}\mu_{zd_z}(m,\bx)\triangledown_{t=0}f_t(m|z',d_{z'},\bx) f(\bx) \differential m  \differential \bx \\
= &\iint_{\bx,m} \left\{p_{z^*d^*}(\bx)-kp_{01}(\bx)\right\}\mu_{zd_z}(m,\bx)f(m|z',d_{z'},\bx)\Big\{\E_{Y|Z,D,M,\bcx}[S_{t=0}(Y,m,d_{z'},z',\bx)|z',d_{z'},m,\bx] \\
& \quad\quad -\E_{Y,M|Z,D,\bcx}[S_{t=0}(Y,M,d_{z'},z',\bx)|z',d_{z'},\bx]\Big\} f(\bx) \differential m  \differential \bx \\
= & \iiint_{\bx,m,y} \!\!\!\left\{p_{z^*d^*}(\bx)-kp_{01}(\bx)\right\}\mu_{zd_z}(m,\bx) S_{t=0}(y,m,d_{z'},z',\bx)f(y|z',d_{z'},m,\bx)f(m|z',d_{z'},\bx)f(\bx) \differential y \differential m  \differential\bx \\
& - \iiint_{\bx,m,y} \!\!\!\left\{p_{z^*d^*}(\bx)-kp_{01}(\bx)\right\} \eta_{zz'}(\bx)  S_{t=0}(y,m,d_{z'},z',\bx)f(y|z',d_{z'},m,\bx)f(m|z',d_{z'},\bx)f(\bx) \differential y \differential m  \differential\bx \\
= & \E\left[\left\{p_{z^*d^*}(\bcx)-kp_{01}(\bcx)\right\}\frac{\mathbb{I}(D=d_{z'},Z=z')}{p_{z'd_{z'}}(\bcx)\pi_{z'}(\bcx)}\left\{\mu_{zd_z}(M,\bcx)-\eta_{zz'}(\bcx)\right\}S_{t=0}(\bco)\right]
\end{align*}\endgroup
and \begingroup\makeatletter\def\f@size{10}\check@mathfonts
\begin{align*}
\eqref{eq:eif4} 
= &\int_{\bx} \left\{p_{z^*d^*}(\bx)-kp_{01}(\bx)\right\}\eta_{zz'}(\bx)\triangledown_{t=0}f_t(\bx) \differential\bx \\
= &\int_{\bx} \left\{p_{z^*d^*}(\bx)-kp_{01}(\bx)\right\}\eta_{zz'}(\bx) f(\bx)\left\{\E_{Y,M,D,Z|\bcx}[S_{t=0}(Y,M,D,Z,\bx)|\bx]-\E[S_{t=0}(Y,M,S,Z,\bcx)]\right\} \differential\bx \\
= & \E\left[\left\{p_{z^*d^*}(\bcx)-kp_{01}(\bcx)\right\}\eta_{zz'}(\bcx)S_{t=0}(\bco)\right]-H_{d_1d_0}^{(zz')}\E\left[S_{t=0}(\bco)\right] \\
= & \E\left[\left(\left\{p_{z^*d^*}(\bcx)-kp_{01}(\bcx)\right\}\eta_{zz'}(\bcx)-H_{d_1d_0}^{(zz')}\right)S_{t=0}(\bco)\right].
\end{align*}\endgroup
Therefore,
\begin{align*}
\triangledown_{t=0} H_{d_1d_0}^{(zz')}(t) & = \eqref{eq:eif1} + \eqref{eq:eif2} + \eqref{eq:eif3} + \eqref{eq:eif4} \\
& = \E\left[\left\{\psi_{d_1d_0}^{(zz')}(\bco) - H_{d_1d_0}^{(zz')}\right\}S_{t=0}(\bm O)\right].
\end{align*}
Now we conclude that the EIF of $H_{d_1d_0}^{(zz')}$ is $\mathcal D_{d_1d_0}^{(zz'),H}(\bm O) = \psi_{d_1d_0}^{(zz')}(\bco) - H_{d_1d_0}^{(zz')}$.

Next, we drive the EIF of $e_{d_1d_0}$, denoted by $\mathcal D_{d_1d_0}^{e}(\bm O)$. By the semiparametric efficiency theory (\citealp{bickel1993efficient}), $\mathcal D_{d_1d_0}^{e}(\bm O)$ must satisfy the following equation
$$
\E\left[\mathcal D_{d_1d_0}^{e}(\bm O)S_{t=0}(\bm O)\right] = \triangledown_{t=0} e_{d_1d_0}(t),
$$
where $e_{d_1d_0}(t) = \int_{\bx} \left\{f_t(d^*|z^*,\bx)-kf_t(D=0|Z=1,\bx)\right\}f_t(\bx) \differential \bx$ is $e_{d_1d_0}$ evaluated under the parametric submodel. We can show
\begin{align}
\triangledown_{t=0} e_{d_1d_0}(t) = &  \int_{\bx} \left\{\triangledown_{t=0}f_t(d^*|z^*,\bx)-k\triangledown_{t=0}f_t(D=0|Z=1,\bx)\right\}f(\bx) \differential \bx \label{eq:q_eif1}\\
& + \int_{\bx} \left\{f(d^*|z^*,\bx)-kf(D=0|Z=1,\bx)\right\}\triangledown_{t=0}f_t(\bx) \differential \bx, \label{eq:q_eif2} 
\end{align}
where
\begingroup\makeatletter\def\f@size{10}\check@mathfonts
\begin{align*}
\!\!\! \eqref{eq:q_eif1} 
= &\int_{\bx}\triangledown_{t=0}f_t(d^*|z^*,\bx) f(\bx) \differential\bx - k\int_{\bx}\triangledown_{t=0}f_t(D\!=\!0|Z\!=\!1,\bx) f(\bx) \differential\bx \\
= &\int_{\bx}f(d^*|z^*,\bx)\left\{\E_{Y,M|Z,D,\bcx}[S_{t=0}(Y,M,d^*,z^*,\bx)|z^*,d^*,\bx] -\E_{Y,M,D|Z,\bcx}[S_{t=0}(Y,M,D,z^*,\bx)|z^*,\bx] \right\} f(\bx) \differential\bx \\
 & - k\int_{\bx}f(D=0|Z=1,\bx)\left\{\E_{Y,M|D,Z,\bcx}[S_{t=0}(Y,M,0,1,\bx)|1,0,\bx] -\E_{Y,M,D|Z,\bcx}[S_{t=0}(Y,M,D,1,\bx)|1,\bx] \right\} f(\bx) \differential\bx \\
= & \iiint_{\bx,m,y}f(d^*|z^*,\bx)S_{t=0}(y,m,d^*,z^*,\bx) f(y|z^*,d^*,m,\bx)f(m|z^*,d^*,\bx)f(\bx) \differential y \differential m \differential\bx\\
&  - \iiiint_{\bx,d,m,y}f(d^*|z^*,\bx)S_{t=0}(y,m,d,z^*,\bx) f(y|z^*,d,m,\bx)f(m|z^*,d,\bx)f(d|z^*,\bx)f(\bx) \differential y \differential m \differential d \differential\bx\\
& - k\iiint_{\bx,m,y}f(D=0|Z=1,\bx)S_{t=0}(y,m,0,1,\bx) f(y|1,0,m,\bx)f(m|1,0,\bx)f(\bx) \differential y \differential m \differential\bx\\
&  + k\iiiint_{\bx,d,m,y}f(D=0|Z=1,\bx)S_{t=0}(y,m,d,1,\bx) f(y|1,d,m,\bx)f(m|1,d,\bx)f(d|1,\bx)f(\bx) \differential y \differential m \differential d \differential\bx\\
= & \E\left[\frac{\mathbb{I}(D=d^*,Z=z^*)}{p_{z^*d^*}(\bcx)\pi_z^*(\bcx)}p_{z^*d^*}(\bcx)S_{t=0}(\bm O)\right] - \E\left[\frac{\mathbb{I}(Z=z^*)}{\pi_{z^*}(\bcx)}p_{z^*d^*}(\bcx)S_{t=0}(\bm O)\right] \\
 & - k\E\left[\frac{\mathbb{I}(D=0,Z=1)}{p_{01}(\bcx)\pi_0(\bcx)}p_{01}(\bcx)S_{t=0}(\bm O)\right] + k\E\left[\frac{\mathbb{I}(Z=0)}{\pi_{0}(\bcx)}p_{01}(\bcx)S_{t=0}(\bm O)\right] \\
= &\bbE\left[\left(\frac{\mathbb{I}(Z=z^*)\left\{\mathbb{I}(D=d^*)-p_{z^*d^*}(\bcx)\right\}}{\pi_{z^*}(\bcx)} - k\frac{(1-Z)\left\{D-p_{01}(\bcx)\right\}}{\pi_{0}(\bcx)}\right)S_{t=0}(\bco)\right]
\end{align*}\endgroup
and 
\begingroup\makeatletter\def\f@size{10}\check@mathfonts
\begin{align*}
\eqref{eq:q_eif2} 
= &\int_{\bx} \left\{p_{z^*d^*}(\bx)-kp_{01}(\bx)\right\}\triangledown_{t=0}f_t(\bx) \differential\bx \\
= &\int_{\bx} \left\{p_{z^*d^*}(\bx)-kp_{01}(\bx)\right\} f(\bx)\left\{\E_{Y,M,D,Z|\bcx}[S_{t=0}(Y,M,D,Z,\bx)|\bx]-\E[S_{t=0}(Y,M,S,Z,\bcx)]\right\} \differential\bx \\
= & \E\left[\left\{p_{z^*d^*}(\bcx)-kp_{01}(\bcx)\right\}S_{t=0}(\bco)\right]-e_{d_1d_0}\E\left[S_{t=0}(\bco)\right] \\
= & \E\left[\left(\left\{p_{z^*d^*}(\bcx)-kp_{01}(\bcx)\right\}-e_{d_1d_0}\right)S_{t=0}(\bco)\right].
\end{align*}\endgroup
Henceforth, we have that
\begin{align*}
\triangledown_{t=0} e_{d_1d_0}(t) = \eqref{eq:q_eif1} + \eqref{eq:q_eif2} = \E\left[\left\{\delta_{d_1d_0}^{(zz')}(\bco) - e_{d_1d_0}\right\}S_{t=0}(\bm O)\right].
\end{align*}
Therefore, the EIF of $e_{d_1d_0}$ is $\mathcal D_{d_1d_0}^{e}(\bm O) = \delta_{d_1d_0}(\bco) - e_{d_1d_0}$.
\end{proof}\medskip

\begin{lemma}\label{lemma:eif_2}
Assume that $\alpha_1$ and $\alpha_2$ are two causal estimands and their  EIFs based on the nonparametric model $\mathcal M_{np}$ in the observed data $\bm O$ are $\mathcal D_1(\bm O)$ and $\mathcal D_2(\bm O)$, respectively. Then, the EIF of $\alpha_3 = \alpha_1+\alpha_2$ is
$$
\mathcal D_3(\bm O) = \mathcal D_1(\bm O) + \mathcal D_2(\bm O).
$$
Moreover, if $\alpha_2\neq 0$, the EIF of $\alpha_4 = \alpha_1/\alpha_2$ is
$$
\mathcal D_4(\bm O) = \frac{1}{\alpha_2}\left\{\mathcal D_1(\bm O) - \alpha_4 \mathcal D_2(\bm O)\right\}.
$$
\end{lemma}

\begin{proof}
We shall follow the notations in the proof of Lemma \ref{lemma:eif_H}. Define $\alpha_1(t)$ and $\alpha_2(t)$  as the nonparametric identification formulas of the causal estimands $\alpha_1$ and $\alpha_2$, which are evaluated under the parametric submodel $f_{t}(\bm o)$. By the semiparametric efficiency theory, we have that
$$
\E\left[\mathcal D_1(\bm O)S_{t=0}(\bm O)\right] = \triangledown_{t=0} \alpha_1(t) \text{ and } \E\left[\mathcal D_2(\bm O)S_{t=0}(\bm O)\right] = \triangledown_{t=0} \alpha_2(t).
$$
Also, since $\alpha_3=\alpha_1+\alpha_2$, a valid nonparametric identification formula of $\alpha_3$ under the parametric submodel is $\alpha_3(t)=\alpha_1(t)+\alpha_2(t)$. Then, noting
\begin{align*}
\triangledown_{t=0} \alpha_3(t) & = \triangledown_{t=0} \alpha_1(t) + \triangledown_{t=0} \alpha_2(t) \\
& = \E\left[\left\{\mathcal D_1(\bm O)+\mathcal D_2(\bm O)\right\}S_{t=0}(\bm O)\right] \\
& = \E\left[\mathcal D_3(\bm O)S_{t=0}(\bm O)\right],
\end{align*}
we conclude that $\mathcal D_3(\bm O)$ is the EIF of  $\alpha_3$. Similarly, because $\alpha_4=\alpha_1/\alpha_2$, a valid nonparametric identification formula of $\alpha_4$ under the parametric submodel is $\alpha_4(t)=\alpha_1(t)/\alpha_2(t)$. Then, we have that
\begin{align*}
\triangledown_{t=0} \alpha_4(t) & = \triangledown_{t=0} \left\{\frac{\alpha_1(t)}{\alpha_2(t)}\right\}  = \frac{\alpha_2 \triangledown_{t=0}\alpha_1(t) - \alpha_1 \triangledown_{t=0}\alpha_2(t)}{\alpha_2^2} \\
& = \frac{1}{\alpha_2}\E\left[\mathcal D_1(\bm O)S_{t=0}(\bm O)\right] - \frac{\alpha_1}{\alpha_2}\E\left[\mathcal D_1(\bm O)S_{t=0}(\bm O)\right] \\
& = \E\left[\frac{\mathcal D_1(\bm O) - \alpha_4 \mathcal D_2(\bm O)}{\alpha_2}S_{t=0}(\bm O)\right],
\end{align*}
thus $\mathcal D_4(\bm O)$ is the EIF of  $\alpha_4$.
\end{proof}\medskip

\noindent \textbf{\textit{Proof of Theorem 3.}} Notice that $\theta_{d_1d_0}^{(zz')} = \frac{\E\left[(p_{z^*d^*}(\bcx)-kp_{01}(\bcx))\eta_{zz'}(\bcx)\right]}{p_{z^*d^*}-kp_{01}}$ is a ratio parameter, and the EIFs of its nominator and denominator, $H_{d_1d_0}^{(zz')}=\E\left[(p_{z^*d^*}(\bcx)-kp_{01}(\bcx))\eta_{zz'}(\bcx)\right]$ and $e_{d_1d_0}=p_{z^*d^*}-kp_{01}$, have been derived in Lemma \ref{lemma:eif_H}. Therefore, one can verify that the EIF shown in Theorem 3 holds by applying Lemma \ref{lemma:eif_2}. \hfill $\square$

\subsection{The multiply robust estimator (Theorem 4)}\label{asec:b5}

\noindent \textbf{\textit{Proof of Theorem 4.}} Let $\bm\tau$ be all of the parameters in the parametric working models of $h_{nuisance}^{\text{par}}$, and let $\bm\tau^*$ be the probability limit of $\widehat{\bm\tau}$, where some components of $\bm\tau^*$ may not equal to either true value due to misspecification. Let $\widetilde h_{nuisance} =\{\widetilde \pi_z(\bx), \widetilde p_{zd}(\bx), \widetilde r_{zd}(m,\bx), \\ \widetilde \mu_{zd}(m,\bx) \}$ be the value of $h_{nuisance}^{\text{par}}$ when it is evaluated at $\bm\tau^*$, which is the probability limit of  $\widehat h_{nuisance}^{\text{par}}$. Notice that $\widetilde \pi_z(\bx)=\pi_z(\bx)$, $\widetilde p_{zd}(\bx)= p_{zd}(\bx)$, $\widetilde r_{zd}(m,\bx)=r_{zd}(m,\bx)$, $\widetilde \mu_{zd}(m,\bx)=\mu_{zd}(m,\bx)$, under $\mathcal M_\pi$, $\mathcal M_e$, $\mathcal M_m$, and $\mathcal M_o$, respectively, but the equalities generally do not hold when the corresponding working model is misspecified. Let $\widetilde p_{zd}=\E\left[\frac{\mathbb{I}(Z=z)(\mathbb{I}(D=d)-\widetilde p_{zd}(\bcx))}{\widetilde \pi_{z}(\bcx)}+\widetilde p_{zd}(\bcx)\right]$ be the probability limit of $p_{zd}^{\text{dr}}$. According to \cite{jiang2022multiply}, $\widetilde p_{zd} = p_{zd}$ under $\mathcal M_{\pi} \cup \mathcal M_{e}$. 
Therefore, $\widetilde p_{zd} = p_{zd}$ also holds under either $\mathcal M_{\pi} \cap \mathcal M_{e} \cap \mathcal M_{m}$, $\mathcal M_{\pi} \cap \mathcal M_{m} \cap \mathcal M_{o}$, $\mathcal  M_{\pi} \cap \mathcal M_{e} \cap \mathcal M_{o}$, or $\mathcal M_{e} \cap \mathcal M_{m} \cap \mathcal M_{o}$. The previous discussion suggests that the probability limit of $\widehat{\theta}_{d_1d_0}^{(zz'),\text{mr}}$ is
\begin{align}
\theta^{(zz'),\text{mr}}_{d_1d_0} = & \E\Big\{\left(\frac{\mathbb{I}(Z=z^*)\left\{\mathbb{I}(D=d^*)-\widetilde p_{z^*d^*}(\bcx)\right\}}{\widetilde\pi_{z^*}(\bcx)} - k\frac{(1-Z)\left\{D-\widetilde p_{01}(\bcx)\right\}}{\widetilde\pi_{0}(\bcx)}\right) \frac{\widetilde\eta_{zz'}(\bcx)}{\widetilde p_{z^*d^*} - k\widetilde p_{01}} \nonumber \\
& + \frac{\widetilde p_{z^*d^*}(\bcx)-k\widetilde p_{01}(\bcx)}{\widetilde p_{z^*d^*} - k\widetilde p_{01}}\frac{\mathbb{I}(D=d_z,Z=z)}{\widetilde p_{zd_z}(\bcx)\widetilde\pi_{z}(\bcx)}\frac{\widetilde r_{z'd_{z'}}(M,\bcx)}{\widetilde r_{zd_z}(M,\bcx)}\left\{Y-\widetilde \mu_{zd_z}(M,\bcx)\right\} \nonumber \\
& + \frac{\widetilde p_{z^*d^*}(\bcx)-k\widetilde p_{01}(\bcx)}{\widetilde p_{z^*d^*} - k\widetilde p_{01}}\frac{\mathbb{I}(D=d_{z'},Z=z')}{\widetilde p_{z'd_{z'}}(\bcx)\widetilde \pi_{z'}(\bcx)}\left\{\widetilde \mu_{zd_z}(M,\bcx)-\widetilde \eta_{zz'}(\bcx)\right\} \nonumber \\
& + \frac{\widetilde p_{z^*d^*}(\bcx)-k\widetilde p_{01}(\bcx)}{\widetilde p_{z^*d^*} - k\widetilde p_{01}}\widetilde \eta_{zz'}(\bcx) \Big\},\nonumber
\end{align}
where $\widetilde\eta_{zz'}(\bcx)=\int_m \widetilde{\mu}_{zd_z}(m,\bcx)\widetilde r_{z'd_{z'}}(m,\bcx) \differential m$. In what follows, we show that $\theta^{(zz'),\text{mr}}_{d_1d_0} = \theta_{d_1d_0}^{(zz')}$ under Scenario I ($\mathcal M_{\pi} \cap \mathcal M_{e} \cap \mathcal M_{m}$), II ($\mathcal M_{\pi} \cap \mathcal M_{m} \cap \mathcal M_{o}$), III ($\mathcal  M_{\pi} \cap \mathcal M_{e} \cap \mathcal M_{o}$), or IV ($\mathcal M_{e} \cap \mathcal M_{m} \cap \mathcal M_{o}$), which collectively verify the quadruple robustness of $\widehat{\theta}_{d_1d_0}^{(zz'),\text{mr}}$.\medskip

\noindent \textbf{Scenario I ($\mathcal M_{\pi} \cap \mathcal M_{e} \cap \mathcal M_{m}$):}\medskip

In Scenario I, $\widetilde \pi_z(\bx)=\pi_z(\bx)$, $\widetilde p_{zd}(\bx)= p_{zd}(\bx)$, $\widetilde r_{zd}(m,\bx)=r_{zd}(m,\bx)$, but generally $\widetilde \mu_{zd}(m,\bx)\neq \mu_{zd}(m,\bx)$. By the doubly robustness of $\widetilde p_{zd}$, we also have $\widetilde p_{zd} = p_{zd}$. Observing this, we can rewrite $\theta^{(zz'),\text{mr}}_{d_1d_0} = \sum_{j=1}^4 \Delta_j$, where\begingroup\makeatletter\def\f@size{10}\check@mathfonts
\begin{align*}
\Delta_1 & = \E\left[\left(\frac{\mathbb{I}(Z=z^*)\left\{\mathbb{I}(D=d^*)- p_{z^*d^*}(\bcx)\right\}}{\pi_{z^*}(\bcx)} - k\frac{(1-Z)\left\{D- p_{01}(\bcx)\right\}}{\pi_{0}(\bcx)}\right) \frac{\int_m \widetilde{\mu}_{zd_z}(m,\bcx) r_{z'd_{z'}}(m,\bcx) \differential m}{ p_{z^*d^*} - k p_{01}}\right], \\
\Delta_2 & = \E\left[\frac{ p_{z^*d^*}(\bcx)-k p_{01}(\bcx)}{ p_{z^*d^*} - k p_{01}}\frac{\mathbb{I}(D=d_z,Z=z)}{ p_{zd_z}(\bcx)\pi_{z}(\bcx)}\frac{ r_{z'd_{z'}}(M,\bcx)}{ r_{zd_z}(M,\bcx)}Y\right], \\
\Delta_3 & = \E\left[\frac{ p_{z^*d^*}(\bcx)-k p_{01}(\bcx)}{p_{z^*d^*} - k p_{01}}\left\{\frac{\mathbb{I}(D=d_{z'},Z=z')}{p_{z'd_{z'}}(\bcx) \pi_{z'}(\bcx)}-\frac{\mathbb{I}(D=d_z,Z=z)}{ p_{zd_z}(\bcx)\pi_{z}(\bcx)}\frac{ r_{z'd_{z'}}(M,\bcx)}{ r_{zd_z}(M,\bcx)}\right\}\widetilde \mu_{zd_z}(M,\bcx)\right],\\
\Delta_4 & = \E\left[\frac{ p_{z^*d^*}(\bcx)-k p_{01}(\bcx)}{ p_{z^*d^*} - k p_{01}} \left\{1-\frac{\mathbb{I}(D=d_{z'},Z=z')}{ p_{z'd_{z'}}(\bcx) \pi_{z'}(\bcx)}\right\}\int_m \widetilde{\mu}_{zd_z}(m,\bcx) r_{z'd_{z'}}(m,\bcx) \differential m\right].
\end{align*}\endgroup
It is obvious that $\Delta_2 = \theta_{d_1d_0}^{(zz'),\text{a}}$. Moreover, \begingroup\makeatletter\def\f@size{10}\check@mathfonts
\begin{align*}
\Delta_1 = &  \E\left[\frac{\mathbb{I}(Z=z^*)\left\{\mathbb{I}(D=d^*)- p_{z^*d^*}(\bcx)\right\}}{\pi_{z^*}(\bcx)}\frac{\int_m \widetilde{\mu}_{zd_z}(m,\bcx) r_{z'd_{z'}}(m,\bcx) \differential m}{ p_{z^*d^*} - k p_{01}}\right] \\
& - k\E\left[\frac{(1-Z)\left\{D- p_{01}(\bcx)\right\}}{\pi_{0}(\bcx)}\frac{\int_m \widetilde{\mu}_{zd_z}(m,\bcx) r_{z'd_{z'}}(m,\bcx) \differential m}{ p_{z^*d^*} - k p_{01}}\right] \\
= & \E\left[\frac{\mathbb{I}(Z=z^*)}{\pi_{z^*}(\bcx)}\frac{\int_m \widetilde{\mu}_{zd_z}(m,\bcx) r_{z'd_{z'}}(m,\bcx) \differential m}{ p_{z^*d^*} - k p_{01}}\underbrace{\left\{\E_{D|Z,\bcx}[\mathbb{I}(D=d^*)|z^*,\bcx]- p_{z^*d^*}(\bcx)\right\}}_{=0}\right] \\
& - k\E\left[\frac{(1-Z)}{\pi_{0}(\bcx)}\frac{\int_m \widetilde{\mu}_{zd_z}(m,\bcx) r_{z'd_{z'}}(m,\bcx) \differential m}{ p_{z^*d^*} - k p_{01}}\underbrace{\left\{\E_{D|Z,\bcx}[D|0,\bcx]- p_{01}(\bcx)\right\}}_{=0}\right] \\
 = & 0 -k \times 0 = 0,\\
 \Delta_3 = & \E\left[\frac{ p_{z^*d^*}(\bcx)-k p_{01}(\bcx)}{p_{z^*d^*} - k p_{01}}\frac{\mathbb{I}(D=d_{z'},Z=z')}{p_{z'd_{z'}}(\bcx) \pi_{z'}(\bcx)}\widetilde \mu_{zd_z}(M,\bcx)\right] \\
 & - \E\left[\frac{ p_{z^*d^*}(\bcx)-k p_{01}(\bcx)}{p_{z^*d^*} - k p_{01}}\frac{\mathbb{I}(D=d_z,Z=z)}{ p_{zd_z}(\bcx)\pi_{z}(\bcx)}\frac{ r_{z'd_{z'}}(M,\bcx)}{ r_{zd_z}(M,\bcx)}\widetilde \mu_{zd_z}(M,\bcx)\right]\\
= & \E\left[\frac{ p_{z^*d^*}(\bcx)-k p_{01}(\bcx)}{p_{z^*d^*} - k p_{01}}\frac{\mathbb{I}(D=d_{z'},Z=z')}{p_{z'd_{z'}}(\bcx) \pi_{z'}(\bcx)}\E_{M|Z,D,\bcx}[\widetilde \mu_{zd_z}(M,\bcx)|z',d_{z'},\bcx]\right] \\
& - \E\left[\frac{ p_{z^*d^*}(\bcx)-k p_{01}(\bcx)}{p_{z^*d^*} - k p_{01}}\frac{\mathbb{I}(D=d_z,Z=z)}{ p_{zd_z}(\bcx)\pi_{z}(\bcx)}\E_{M|Z,D,\bcx}\left[\frac{ r_{z'd_{z'}}(M,\bcx)}{ r_{zd_z}(M,\bcx)}\widetilde \mu_{zd_z}(M,\bcx)\Big| z,d_z,\bcx\right]\right] \\
= & \E\left[\frac{ p_{z^*d^*}(\bcx)-k p_{01}(\bcx)}{p_{z^*d^*} - k p_{01}}\E_{M|Z,D,\bcx}[\widetilde \mu_{zd_z}(M,\bcx)|z',d_{z'},\bcx]\right] \\
& - \E\left[\frac{ p_{z^*d^*}(\bcx)-k p_{01}(\bcx)}{p_{z^*d^*} - k p_{01}}\E_{M|Z,D,\bcx}\left[\frac{ r_{z'd_{z'}}(M,\bcx)}{ r_{zd_z}(M,\bcx)}\widetilde \mu_{zd_z}(M,\bcx)\Big| z,d_z,\bcx\right]\right] \\
= & \E\left[\frac{ p_{z^*d^*}(\bcx)-k p_{01}(\bcx)}{p_{z^*d^*} - k p_{01}}\E_{M|Z,D,\bcx}[\widetilde \mu_{zd_z}(M,\bcx)|z',d_{z'},\bcx]\right] \\
& - \E\left[\frac{ p_{z^*d^*}(\bcx)-k p_{01}(\bcx)}{p_{z^*d^*} - k p_{01}}\E_{M|Z,D,\bcx}\left[\widetilde \mu_{zd_z}(M,\bcx)\Big| z',d_{z'},\bcx\right]\right] \\
= & 0,\\
 \Delta_4 = &  \E\left[\frac{ p_{z^*d^*}(\bcx)-k p_{01}(\bcx)}{ p_{z^*d^*} - k p_{01}} \E_{M|Z,D,\bcx}[\widetilde{\mu}_{zd_z}(m,\bcx)|z',d_{z'},\bcx]\left\{1-\frac{\mathbb{I}(D=d_{z'},Z=z')}{ p_{z'd_{z'}}(\bcx) \pi_{z'}(\bcx)}\right\}\right]\\
 = & \E\left[\frac{ p_{z^*d^*}(\bcx)-k p_{01}(\bcx)}{ p_{z^*d^*} - k p_{01}} \E_{M|Z,D,\bcx}[\widetilde{\mu}_{zd_z}(m,\bcx)|z',d_{z'},\bcx]\left\{1-1\right\}\right] \\
 = & 0, 
\end{align*}\endgroup
which suggests that $\theta^{(zz'),\text{mr}}_{d_1d_0} = \sum_{j=1}^4 \Delta_j = \theta^{(zz'),\text{a}}_{d_1d_0}=\theta^{(zz')}_{d_1d_0}$ under $\mathcal M_{\pi} \cap \mathcal M_{e} \cap \mathcal M_{m}$.\medskip

\noindent \textbf{Scenario II ($\mathcal M_{\pi} \cap \mathcal M_{m} \cap \mathcal M_{o}$):}\medskip

In Scenario II, $\widetilde \pi_z(\bx)=\pi_z(\bx)$, $\widetilde r_{zd}(m,\bx)=r_{zd}(m,\bx)$, $\widetilde \mu_{zd}(m,\bx)= \mu_{zd}(m,\bx)$, but generally $\widetilde p_{zd}(\bx)\neq p_{zd}(\bx)$.  Observing this, we can rewrite $\theta^{(zz'),\text{mr}}_{d_1d_0} = \sum_{j=1}^4 \Delta_j$, where\begingroup\makeatletter\def\f@size{10}\check@mathfonts
\begin{align*}
\Delta_1 & = \E\left[\left\{\frac{\mathbb{I}(Z=z^*,D=d^*)}{\pi_{z^*}(\bcx)} - k \frac{(1-Z)D}{\pi_0(\bcx)}\right\}\frac{\eta_{zz'}(\bcx)}{p_{z^*d^*}-kp_{01}}\right], \\
\Delta_2 & = \E\left[\frac{\widetilde p_{z^*d^*}(\bcx)-k\widetilde p_{01}(\bcx)}{ p_{z^*d^*} - k p_{01}}\frac{\mathbb{I}(D=d_z,Z=z)}{\widetilde p_{zd_z}(\bcx)\pi_{z}(\bcx)}\frac{ r_{z'd_{z'}}(M,\bcx)}{ r_{zd_z}(M,\bcx)}\left\{Y- \mu_{zd_z}(M,\bcx)\right\}\right], \\
\Delta_3 & = \E\left[\frac{\widetilde p_{z^*d^*}(\bcx)-k\widetilde p_{01}(\bcx)}{ p_{z^*d^*} - k p_{01}}\frac{\mathbb{I}(D=d_{z'},Z=z')}{\widetilde p_{z'd_{z'}}(\bcx)\pi_{z'}(\bcx)}\left\{ \mu_{zd_z}(M,\bcx)- \eta_{zz'}(\bcx)\right\}\right],\\
\Delta_4 & = \E\left[\left(\left\{1-\frac{\mathbb{I}(Z=z^*)}{\pi_{z^*}(\bcx)}\right\}\widetilde p_{z^*d^*}(\bcx)-k\left\{1-\frac{1-Z}{\pi_{0}(\bcx)}\right\}\widetilde p_{01}(\bcx)\right) \frac{\eta_{zz'}(\bcx)}{p_{z^*d^*} - k p_{01}}\right].
\end{align*}\endgroup
One can verify $\Delta_1 = \theta_{d_1d_0}^{(zz'),\text{b}}$, \begingroup\makeatletter\def\f@size{10}\check@mathfonts
\begin{align*}
\Delta_2 = &  \E\left[\frac{\widetilde p_{z^*d^*}(\bcx)-k\widetilde p_{01}(\bcx)}{ p_{z^*d^*} - k p_{01}}\frac{\mathbb{I}(D=d_z,Z=z)}{\widetilde p_{zd_z}(\bcx)\pi_{z}(\bcx)}\frac{ r_{z'd_{z'}}(M,\bcx)}{ r_{zd_z}(M,\bcx)}\left\{Y- \mu_{zd_z}(M,\bcx)\right\}\right] \\
= & \E\left[\frac{\widetilde p_{z^*d^*}(\bcx)-k\widetilde p_{01}(\bcx)}{ p_{z^*d^*} - k p_{01}}\frac{\mathbb{I}(D=d_z,Z=z)}{\widetilde p_{zd_z}(\bcx)\pi_{z}(\bcx)}\frac{ r_{z'd_{z'}}(M,\bcx)}{ r_{zd_z}(M,\bcx)}\underbrace{\left\{\E_{Y|Z,D,M,\bcx}[Y|z,d_z,M,\bcx]- \mu_{zd_z}(M,\bcx)\right\}}_{=0}\right] \\
= & 0 \\
 \Delta_3 = & \E\left[\frac{\widetilde p_{z^*d^*}(\bcx)-k\widetilde p_{01}(\bcx)}{ p_{z^*d^*} - k p_{01}}\frac{\mathbb{I}(D=d_{z'},Z=z')}{\widetilde p_{z'd_{z'}}(\bcx)\pi_{z'}(\bcx)}\left\{ \E_{M|Z,D,\bcx}[\mu_{zd_z}(M,\bcx)|z',d_{z'},\bcx]- \eta_{zz'}(\bcx)\right\}\right] \\
 = & \E\left[\frac{\widetilde p_{z^*d^*}(\bcx)-k\widetilde p_{01}(\bcx)}{ p_{z^*d^*} - k p_{01}}\frac{\mathbb{I}(D=d_{z'},Z=z')}{\widetilde p_{z'd_{z'}}(\bcx)\pi_{z'}(\bcx)}\left\{ \eta_{zz'}(\bcx)- \eta_{zz'}(\bcx)\right\}\right] \\
 = & 0 \\
 \Delta_4 = & \E\left[\left(\left\{1-\frac{\mathbb{I}(Z=z^*)}{\pi_{z^*}(\bcx)}\right\}\widetilde p_{z^*d^*}(\bcx)-k\left\{1-\frac{1-Z}{\pi_{0}(\bcx)}\right\}\widetilde p_{01}(\bcx)\right) \frac{\eta_{zz'}(\bcx)}{p_{z^*d^*} - k p_{01}}\right] \\
 = & \E\left[\left(\left\{1-\frac{\E_{Z|\bcx}\left[\mathbb{I}(Z=z^*)|\bcx\right]}{\pi_{z^*}(\bcx)}\right\}\widetilde p_{z^*d^*}(\bcx)-k\left\{1-\frac{\E_{Z|\bcx}[1-Z|\bcx]}{\pi_{0}(\bcx)}\right\}\widetilde p_{01}(\bcx)\right) \frac{\eta_{zz'}(\bcx)}{p_{z^*d^*} - k p_{01}}\right] \\
  = & 0.
\end{align*}\endgroup
Therefore, we have obtained $\theta^{(zz'),\text{mr}}_{d_1d_0} = \sum_{j=1}^4 \Delta_j = \theta^{(zz'),\text{b}}_{d_1d_0}=\theta^{(zz')}_{d_1d_0}$ under $\mathcal M_{\pi} \cap \mathcal M_{m} \cap \mathcal M_{o}$.\medskip

\noindent \textbf{Scenario III ($\mathcal  M_{\pi} \cap \mathcal M_{e} \cap \mathcal M_{o}$):}\medskip

In Scenario III, $\widetilde \pi_z(\bx)=\pi_z(\bx)$, $\widetilde p_{zd}(\bx)= p_{zd}(\bx)$, $\widetilde \mu_{zd}(m,\bx)= \mu_{zd}(m,\bx)$, but generally $\widetilde r_{zd}(m,\bx)\neq r_{zd}(m,\bx)$. Observing this, we can rewrite $\theta^{(zz'),\text{mr}}_{d_1d_0} = \sum_{j=1}^4 \Delta_j$, where\begingroup\makeatletter\def\f@size{10}\check@mathfonts
\begin{align*}
\Delta_1 & = \E\left[\left(\frac{\mathbb{I}(Z=z^*)\left\{\mathbb{I}(D=d^*)- p_{z^*d^*}(\bcx)\right\}}{\pi_{z^*}(\bcx)} - k\frac{(1-Z)\left\{D- p_{01}(\bcx)\right\}}{\pi_{0}(\bcx)}\right) \frac{\int_m {\mu}_{zd_z}(m,\bcx) \widetilde r_{z'd_{z'}}(m,\bcx) \differential m}{ p_{z^*d^*} - k p_{01}}\right], \\
\Delta_2 & = \E\left[\frac{ p_{z^*d^*}(\bcx)-k p_{01}(\bcx)}{ p_{z^*d^*} - k p_{01}}\frac{\mathbb{I}(D=d_z,Z=z)}{ p_{zd_z}(\bcx)\pi_{z}(\bcx)}\frac{ \widetilde r_{z'd_{z'}}(M,\bcx)}{ \widetilde r_{zd_z}(M,\bcx)}\left\{Y- \mu_{zd_z}(M,\bcx)\right\}\right], \\
\Delta_3 & = \E\left[\frac{ p_{z^*d^*}(\bcx)-k p_{01}(\bcx)}{ p_{z^*d^*} - k p_{01}}\frac{\mathbb{I}(D=d_{z'},Z=z')}{ p_{z'd_{z'}}(\bcx)\pi_{z'}(\bcx)}\mu_{zd_z}(M,\bcx)\right],\\
\Delta_4 & = \E\left[\frac{ p_{z^*d^*}(\bcx)-k p_{01}(\bcx)}{ p_{z^*d^*} - k p_{01}} \left\{1-\frac{\mathbb{I}(D=d_{z'},Z=z')}{ p_{z'd_{z'}}(\bcx) \pi_{z'}(\bcx)}\right\}\int_m {\mu}_{zd_z}(m,\bcx) \widetilde r_{z'd_{z'}}(m,\bcx) \differential m\right].
\end{align*}\endgroup
Noting that $\Delta_3=\theta_{d_1d_0}^{(zz'),\text{c}}$,\begingroup\makeatletter\def\f@size{10}\check@mathfonts
\begin{align*}
\Delta_1 = &  \E\left[\frac{\mathbb{I}(Z=z^*)\left\{\mathbb{I}(D=d^*)- p_{z^*d^*}(\bcx)\right\}}{\pi_{z^*}(\bcx)}\frac{\int_m {\mu}_{zd_z}(m,\bcx) \widetilde r_{z'd_{z'}}(m,\bcx) \differential m}{ p_{z^*d^*} - k p_{01}}\right] \\
& - k\E\left[\frac{(1-Z)\left\{D- p_{01}(\bcx)\right\}}{\pi_{0}(\bcx)}\frac{\int_m {\mu}_{zd_z}(m,\bcx) \widetilde r_{z'd_{z'}}(m,\bcx) \differential m}{ p_{z^*d^*} - k p_{01}}\right] \\
= & \E\left[\frac{\mathbb{I}(Z=z^*)}{\pi_{z^*}(\bcx)}\frac{\int_m {\mu}_{zd_z}(m,\bcx) \widetilde r_{z'd_{z'}}(m,\bcx) \differential m}{ p_{z^*d^*} - k p_{01}}\underbrace{\left\{\E_{D|Z,\bcx}[\mathbb{I}(D=d^*)|z^*,\bcx]- p_{z^*d^*}(\bcx)\right\}}_{=0}\right] \\
& - k\E\left[\frac{(1-Z)}{\pi_{0}(\bcx)}\frac{\int_m {\mu}_{zd_z}(m,\bcx) \widetilde r_{z'd_{z'}}(m,\bcx) \differential m}{ p_{z^*d^*} - k p_{01}}\underbrace{\left\{\E_{D|Z,\bcx}[D|0,\bcx]- p_{01}(\bcx)\right\}}_{=0}\right] \\
 = & 0 -k \times 0 = 0,\\
 \Delta_2 = & \E\left[\frac{ p_{z^*d^*}(\bcx)-k p_{01}(\bcx)}{ p_{z^*d^*} - k p_{01}}\frac{\mathbb{I}(D=d_z,Z=z)}{ p_{zd_z}(\bcx)\pi_{z}(\bcx)}\frac{ \widetilde r_{z'd_{z'}}(M,\bcx)}{ \widetilde r_{zd_z}(M,\bcx)}\left\{Y- \mu_{zd_z}(M,\bcx)\right\}\right] \\
 = & \E\left[\frac{ p_{z^*d^*}(\bcx)-k p_{01}(\bcx)}{ p_{z^*d^*} - k p_{01}}\frac{\mathbb{I}(D=d_z,Z=z)}{ p_{zd_z}(\bcx)\pi_{z}(\bcx)}\frac{ \widetilde r_{z'd_{z'}}(M,\bcx)}{ \widetilde r_{zd_z}(M,\bcx)}\underbrace{\left\{\E_{Y|Z,D,M,\bcx}[Y|z,d_z,M,\bcx]- \mu_{zd_z}(M,\bcx)\right\}}_{=0}\right] \\
 = & 0,\\
 \Delta_4 = &  \E\left[\frac{ p_{z^*d^*}(\bcx)-k p_{01}(\bcx)}{ p_{z^*d^*} - k p_{01}} \int_m {\mu}_{zd_z}(m,\bcx) \widetilde r_{z'd_{z'}}(m,\bcx) \left\{1-\E_{Z,D|\bcx}\left[\frac{\mathbb{I}(D=d_{z'},Z=z')}{ p_{z'd_{z'}}(\bcx) \pi_{z'}(\bcx)}|\bcx\right]\right\} \differential m\right]\\
 = & \E\left[\frac{ p_{z^*d^*}(\bcx)-k p_{01}(\bcx)}{ p_{z^*d^*} - k p_{01}} \int_m {\mu}_{zd_z}(m,\bcx) \widetilde r_{z'd_{z'}}(m,\bcx) \left\{1-1\right\} \differential m\right]\\
 = & 0 ,
\end{align*}\endgroup
we have obtained $\theta^{(zz'),\text{mr}}_{d_1d_0} = \sum_{j=1}^4 \Delta_j = \theta^{(zz'),\text{c}}_{d_1d_0}=\theta^{(zz')}_{d_1d_0}$ under $\mathcal  M_{\pi} \cap \mathcal M_{e} \cap \mathcal M_{o}$.\medskip

\noindent \textbf{Scenario IV ($\mathcal M_{e} \cap \mathcal M_{m} \cap \mathcal M_{o}$):}\medskip

In Scenario IV, $\widetilde p_{zd}(\bx)= p_{zd}(\bx)$, $\widetilde r_{zd}(m,\bx)=r_{zd}(m,\bx)$, $\widetilde \mu_{zd}(m,\bx)= \mu_{zd}(m,\bx)$, but generally $\widetilde \pi_z(\bx) \neq \pi_z(\bx)$. Therefore, we have $\theta^{(zz'),\text{mr}}_{d_1d_0} = \sum_{j=1}^4 \Delta_j$, where\begingroup\makeatletter\def\f@size{10}\check@mathfonts
\begin{align*}
\Delta_1 & = \E\left[\left(\frac{\mathbb{I}(Z=z^*)\left\{\mathbb{I}(D=d^*)- p_{z^*d^*}(\bcx)\right\}}{\widetilde \pi_{z^*}(\bcx)} - k\frac{(1-Z)\left\{D- p_{01}(\bcx)\right\}}{\widetilde \pi_{0}(\bcx)}\right) \frac{\eta_{zz'}(\bcx)}{ p_{z^*d^*} - k p_{01}}\right], \\
\Delta_2 & = \E\left[\frac{ p_{z^*d^*}(\bcx)-k p_{01}(\bcx)}{ p_{z^*d^*} - k p_{01}}\frac{\mathbb{I}(D=d_z,Z=z)}{ p_{zd_z}(\bcx)\widetilde\pi_{z}(\bcx)}\frac{ r_{z'd_{z'}}(M,\bcx)}{ r_{zd_z}(M,\bcx)}\left\{Y- \mu_{zd_z}(M,\bcx)\right\}\right], \\
\Delta_3 & = \E\left[\frac{ p_{z^*d^*}(\bcx)-kp_{01}(\bcx)}{ p_{z^*d^*} - k p_{01}}\frac{\mathbb{I}(D=d_{z'},Z=z')}{ p_{z'd_{z'}}(\bcx)\widetilde \pi_{z'}(\bcx)}\left\{ \mu_{zd_z}(M,\bcx)- \eta_{zz'}(\bcx)\right\}\right],\\
\Delta_4 & = \E\left[\frac{ p_{z^*d^*}(\bcx)-k p_{01}(\bcx)}{ p_{z^*d^*} - k p_{01}} \eta_{zz'}(\bcx)\right].
\end{align*}\endgroup
Noting that $\Delta_4=\theta_{d_1d_0}^{(zz'),\text{d}}$,\begingroup\makeatletter\def\f@size{10}\check@mathfonts
\begin{align*}
\Delta_1 = &  \E\left[\frac{\mathbb{I}(Z=z^*)\left\{\mathbb{I}(D=d^*)- p_{z^*d^*}(\bcx)\right\}}{\widetilde\pi_{z^*}(\bcx)}\frac{\eta_{zz'}(\bcx)}{ p_{z^*d^*} - k p_{01}}\right] - k\E\left[\frac{(1-Z)\left\{D- p_{01}(\bcx)\right\}}{\widetilde \pi_{0}(\bcx)}\frac{\eta_{zz'}(\bcx)}{ p_{z^*d^*} - k p_{01}}\right] \\
= & \E\left[\frac{\mathbb{I}(Z=z^*)}{\widetilde\pi_{z^*}(\bcx)}\frac{\eta_{zz'}(\bcx)}{ p_{z^*d^*} - k p_{01}}\underbrace{\left\{\E_{D|Z,\bcx}[\mathbb{I}(D=d^*)|z^*,\bcx]- p_{z^*d^*}(\bcx)\right\}}_{=0}\right] \\
& - k\E\left[\frac{(1-Z)}{\widetilde\pi_{0}(\bcx)}\frac{\eta_{zz'}(\bcx)}{ p_{z^*d^*} - k p_{01}}\underbrace{\left\{\E_{D|Z,\bcx}[D|0,\bcx]- p_{01}(\bcx)\right\}}_{=0}\right] \\
 = & 0 -k \times 0 = 0,\\
 \Delta_2 = & \E\left[\frac{ p_{z^*d^*}(\bcx)-k p_{01}(\bcx)}{ p_{z^*d^*} - k p_{01}}\frac{\mathbb{I}(D=d_z,Z=z)}{ p_{zd_z}(\bcx)\widetilde\pi_{z}(\bcx)}\frac{  r_{z'd_{z'}}(M,\bcx)}{  r_{zd_z}(M,\bcx)}\left\{Y- \mu_{zd_z}(M,\bcx)\right\}\right] \\
 = & \E\left[\frac{ p_{z^*d^*}(\bcx)-k p_{01}(\bcx)}{ p_{z^*d^*} - k p_{01}}\frac{\mathbb{I}(D=d_z,Z=z)}{ p_{zd_z}(\bcx)\widetilde\pi_{z}(\bcx)}\frac{  r_{z'd_{z'}}(M,\bcx)}{  r_{zd_z}(M,\bcx)}\underbrace{\left\{\E_{Y|Z,D,M,\bcx}[Y|z,d_z,M,\bcx]- \mu_{zd_z}(M,\bcx)\right\}}_{=0}\right] \\
 = & 0,\\
  \Delta_3 = & \E\left[\frac{ p_{z^*d^*}(\bcx)-k p_{01}(\bcx)}{ p_{z^*d^*} - k p_{01}}\frac{\mathbb{I}(D=d_{z'},Z=z')}{ p_{z'd_{z'}}(\bcx)\widetilde\pi_{z'}(\bcx)}\left\{ \E_{M|Z,D,\bcx}[\mu_{zd_z}(M,\bcx)|z',d_{z'},\bcx]- \eta_{zz'}(\bcx)\right\}\right] \\
 = & \E\left[\frac{ p_{z^*d^*}(\bcx)-k p_{01}(\bcx)}{ p_{z^*d^*} - k p_{01}}\frac{\mathbb{I}(D=d_{z'},Z=z')}{ p_{z'd_{z'}}(\bcx)\widetilde\pi_{z'}(\bcx)}\left\{ \eta_{zz'}(\bcx)- \eta_{zz'}(\bcx)\right\}\right] \\
 = & 0,
\end{align*}\endgroup
we have that $\theta^{(zz'),\text{mr}}_{d_1d_0} = \sum_{j=1}^4 \Delta_j = \theta^{(zz'),\text{d}}_{d_1d_0}=\theta^{(zz')}_{d_1d_0}$ under $\mathcal M_{e} \cap \mathcal M_{m} \cap \mathcal M_{o}$.

Up until this point, we have confirmed that the probability limit of $\widehat \theta^{(zz'),\text{mr}}_{d_1d_0}$, i.e., $ \theta^{(zz'),\text{mr}}_{d_1d_0}$, equals to the true value $ \theta^{(zz')}_{d_1d_0}$ under either $\mathcal M_{\pi} \cap \mathcal M_{e} \cap \mathcal M_{m}$, $\mathcal M_{\pi} \cap \mathcal M_{m} \cap \mathcal M_{o}$, $\mathcal  M_{\pi} \cap \mathcal M_{e} \cap \mathcal M_{o}$, or $\mathcal M_{e} \cap \mathcal M_{m} \cap \mathcal M_{o}$. Next, we prove the asymptotic normality of $\widehat \theta^{(zz'),\text{mr}}_{d_1d_0}$. Notice that $\widehat \theta^{(zz'),\text{mr}}_{d_1d_0}$ can be viewed as the solution of the following estimating equation
$$
\Prob_n\left[S_{\text{mr}}\left(\bco;\theta_{d_1d_0}^{(zz')},\widehat{\bm\tau}\right)\right] = \Prob_n\left[\mathcal S_1(\bm O; \widehat{\bm \tau})-\theta_{d_1d_0}^{(zz')} \mathcal S_0(\bm O; \widehat{\bm \tau}) \right] = 0, 
$$
where $\mathcal S_1(\bm O; {\bm \tau})$ and $\mathcal S_0(\bm O; {\bm \tau})$ are $\psi_{d_1d_0}^{(zz')}(\bm O)$ and $\delta_{d_1d_0}(\bm O)$ evaluated at $h_{nuisance}^{\text{par}}$. Assume that the following regularity conditions hold:
\begin{itemize}
    \item[1.] Assume that $\sqrt{n}(\widehat{\bm\tau}-\bm \tau^*) = \sqrt{n}\Prob_n\left[\text{IF}_{\bm\tau}(\bm O;\bm \tau^*)\right]+o_p(1)$, where $\text{IF}_{\bm\tau}(\bm O;\bm \tau^*)$ is the influence function of $\widehat{\bm\tau}$ and $o_p(1)$ is a remainder term that converges in probability to 0. Also, assume that $\Prob_n\left[\left\{\text{IF}_{\bm\tau}(\bm O;\bm \tau^*)\right\}^{\otimes 2}\right]$ converges to a positive definite matrix.
    \item[2.] Let $\bm\Xi$ be a bounded convex neighborhood of $\bm\tau^*$. Assume that the class of functions $\Big\{
    \mathcal S_{1}(\bm O;\bm \tau),\frac{\partial }{\partial \bm\tau} \mathcal S_{1}(\bm O;\bm \tau), \left\{\mathcal S_{1}(\bm O;\bm \tau)\right\}^2,
     \mathcal S_{0}(\bm O;\bm \tau),\frac{\partial }{\partial \bm\tau} \mathcal S_{0}(\bm O;\bm \tau), \left\{\mathcal S_{0}(\bm O;\bm \tau)\right\}^2,
      \text{IF}_{\bm\tau}(\bm O;\bm \tau), \\ \left\{\text{IF}_{\bm\tau}(\bm O;\bm \tau)\right\}^{\otimes 2}\Big\}$ is a Glivenko-Cantelli class in $\bm\Xi$.
    \item[3.]Assume that $\Prob_n[\mathcal S_0(\bm O; \bm\tau^*)]$ converges to a positive value. In addition, we assume that both $\Prob_n[\left\{\mathcal S_{1}(\bm O;\bm \tau^*)\right\}^2]$ and $\Prob_n[\left\{\mathcal S_{0}(\bm O;\bm \tau^*)\right\}^2]$ converge to a positive value.  
\end{itemize}
To prove asymptotic normality, we use a Taylor series, along with the above conditions, to deduce that
\begin{align*}
0 = \Prob_n\left[S_{\text{mr}}\left(\bco;\widehat \theta_{d_1d_0}^{(zz'),\text{mr}},\widehat{\bm\tau}\right)\right] = & \Prob_n\left[\mathcal S_1(\bm O; \widehat{\bm \tau})-\widehat \theta_{d_1d_0}^{(zz'),\text{mr}} \mathcal S_0(\bm O; \widehat{\bm \tau})\right] \\
= & \Prob_n\left[\mathcal S_1(\bm O; \bm \tau^*)-\theta_{d_1d_0}^{(zz'),\text{mr}} \mathcal S_0(\bm O; \bm \tau^*)\right] \\
& - \Prob_n\left[\mathcal S_0(\bm O; \bm \tau^*)\right]\left(\widehat\theta_{d_1d_0}^{(zz'),\text{mr}}-\theta_{d_1d_0}^{(zz'),\text{mr}}\right)\\
& + \Prob_n\left[\frac{\partial }{\partial \bm\tau}\mathcal S_1(\bm O; \bm \tau^*)-\theta_{d_1d_0}^{(zz'),\text{mr}} \frac{\partial }{\partial \bm\tau} \mathcal S_0(\bm O; \bm \tau^*)\right](\widehat{\bm\tau}-\bm\tau^*) + o_p(n^{-1/2}),
\end{align*}
which suggests that
\begin{align*}
& \sqrt{n}\left(\widehat\theta_{d_1d_0}^{(zz'),\text{mr}}-\theta_{d_1d_0}^{(zz'),\text{mr}}\right) \\
= & \left\{\E[\mathcal S_0(\bm O;\bm\tau^*)]\right\}^{-1} \Prob_n\left\{\mathcal S_1(\bm O; \bm \tau^*)-\theta_{d_1d_0}^{(zz'),\text{mr}} \mathcal S_0(\bm O; \bm \tau^*)+R(\theta_{d_1d_0}^{(zz'),\text{mr}},\bm\tau^*)\text{IF}_{\bm\tau}(\bm O;\bm \tau^*) \right\} + o_p(1),
\end{align*}
where $R(\theta_{d_1d_0}^{(zz'),\text{mr}},\bm\tau^*)=\E\left[\frac{\partial }{\partial \bm\tau}\mathcal S_1(\bm O; \bm \tau^*)-\theta_{d_1d_0}^{(zz'),\text{mr}} \frac{\partial }{\partial \bm\tau} \mathcal S_0(\bm O; \bm \tau^*)\right]$. Then, by applying the central limit theorem and noticing that 
$\theta_{d_1d_0}^{(zz'),\text{mr}}=\theta_{d_1d_0}^{(zz')}$ under either $\mathcal M_{\pi} \cap \mathcal M_{e} \cap \mathcal M_{m}$,  $\mathcal M_{\pi} \cap \mathcal M_{m} \cap \mathcal M_{o}$, $\mathcal  M_{\pi} \cap \mathcal M_{e} \cap \mathcal M_{o}$, or $\mathcal M_{e} \cap \mathcal M_{m} \cap \mathcal M_{o}$, we can show that $\sqrt{n}\left(\widehat\theta_{d_1d_0}^{(zz'),\text{mr}}-\theta_{d_1d_0}^{(zz')}\right)$ converges to a zero-mean normal distribution with variance
$$
V_{\text{mr}} = \left\{\E[\mathcal S_0(\bm O;\bm\tau^*)]\right\}^{-2}\E\left[\left\{\mathcal S_1(\bm O; \bm \tau^*)-\theta_{d_1d_0}^{(zz')} \mathcal S_0(\bm O; \bm \tau^*)+R(\theta_{d_1d_0}^{(zz')},\bm\tau^*)\text{IF}_{\bm\tau}(\bm O;\bm \tau^*)\right\}^2\right].
$$

Finally, when all parametric working models are correctly specified, i.e., under the intersection model $\mathcal M_{\pi} \cap \mathcal M_{e} \cap \mathcal M_{m} \cap \mathcal M_{o}$, then $V_{\text{mr}}=\E\left[\left\{\mathcal D_{d_1d_0}^{(zz')}(\bm O)\right\}^2\right]$ achieves the semiparametric efficiency bound. We can easily verify this by observing the following facts:
\begin{compactitem}
\item[1.] $\mathcal S_1(\bm O; \bm \tau^*) = \psi_{d_1d_0}^{(zz')}(\bm O)$ and $\mathcal S_0(\bm O; \bm \tau^*) = \delta_{d_1d_0}(\bm O)$ under $\mathcal M_{\pi} \cap \mathcal M_{e} \cap \mathcal M_{m} \cap \mathcal M_{o}$. 

\item[2.] Following point 1, $\E[\mathcal S_0(\bm O;\bm\tau^*)]=e_{d_1d_0}=p_{z^*d^*}-kp_{01}$ under $\mathcal M_{\pi} \cap \mathcal M_{e} \cap \mathcal M_{m} \cap \mathcal M_{o}$.

\item[3.] $R(\theta_{d_1d_0}^{(zz'),\text{mr}},\bm\tau^*)=0$, because the EIF is orthogonal to the likelihood score of the parametric working models when they are correctly specified.
\end{compactitem}
Under the above three points, we have
$$
V_{\text{mr}} = \E\left[\left\{\frac{\psi_{d_1d_0}^{(zz')}(\bm O)-\theta_{d_1d_0}^{(zz')}\delta_{d_1d_0}(\bm O)}{p_{z^*d^*}-kp_{01}}\right\}^2\right].
$$
This has completed the proof. \hfill $\square$

\subsection{The nonparametric efficient estimator (Theorem 5)}\label{sec:np-proof}

\noindent \textbf{\textit{Proof of Theorem 5.}} Proof of the multiply robostness of $\widehat\theta_{d_1d_0}^{(zz'),\text{np}}$ is same to that of $\widehat\theta_{d_1d_0}^{(zz'),\text{mr}}$ as shown in the proof of Theorem 4. Here, we only prove the asymptotic normality and local efficiency of $\widehat\theta_{d_1d_0}^{(zz'),\text{np}}$. 

To simplify notation but without loss of clarity, we abbreviate $\theta_{d_1d_0}^{(zz')}$ and its nonparametric estimator ($\widehat\theta_{d_1d_0}^{(zz'),\text{np}}$) as $\theta$ and $\widehat \theta$, respectively. Also, we abbreviate the two unknown functions in the efficient influence function, $\psi_{d_1d_0}^{(zz')}(\bm O)$ and $\delta_{d_1d_0}(\bm O)$, as $\psi(\bm O)$ and $\delta(\bm O)$, respectively. Based on the cross-splitting procedure, the nonparametric estimator $\widehat\theta$ can be decomposed in to the ratio of two terms $\Prob_n\left[\widehat\psi(\bm O)\right]$ and $\Prob_n\left[\widehat\delta(\bm O)\right]$, where 
\begin{align*}
\Prob_n\left[\widehat\psi(\bm O)\right] = \frac{1}{n}\sum_{v=1}^V n_v \Prob_{n_v}\left[\widehat\psi^v(\bm O)\right] \text{ and } \Prob_n\left[\widehat\delta(\bm O)\right]= \frac{1}{n}\sum_{v=1}^V n_v \Prob_{n_v}\left[\widehat\delta^v(\bm O)\right].
\end{align*}
Here, $n_v$ is the size of the $v$-th group $\mathcal O_v$, $\Prob_{n_v}[\cdot]$ is the empirical mean operator on $\mathcal O_v$, and $\{\widehat\psi^v(\bm O),\widehat\delta^v(\bm O)\}$ is $\{\psi(\bm O),\delta(\bm O)\}$ evaluated under $\widehat h_{nuisance}^{\text{np},v}$, which is the nonparametric estimator of the nuisance functions based on the leave-one-out sample $\mathcal O_{-v}$. We further have the following decomposition of $\Prob_{n_v}\left[\widehat\psi^v(\bm O)\right]$ and $\Prob_{n_v}\left[\widehat\delta^v(\bm O)\right]$:\begingroup\makeatletter\def\f@size{12}\check@mathfonts
\begin{align}
\Prob_{n_v}[\widehat\psi^v(\bm O)]  &= \Prob_{n_v} [\psi(\bm O)] + (\Prob_{n_v} - \E )[\widehat\psi^v(\bm O)-\psi(\bm O)] + \underbrace{\E[\widehat\psi^v(\bm O)-\psi(\bm O)]}_{=:R_1(\widehat\psi^v,\psi)}, \label{eq:t5_1}\\
\Prob_{n_v}[\widehat\delta^v(\bm O)]  &= \Prob_{n_v} [\delta(\bm O)] + (\Prob_{n_v} - \E )[\widehat\delta^v(\bm O)-\delta(\bm O)] +  \underbrace{\E[\widehat\delta^v(\bm O)-\delta(\bm O)]}_{=:R_2(\widehat\delta^v,\delta)}. \label{eq:t5_2}
\end{align}\endgroup
In what follows, we show that $\eqref{eq:t5_1} = \Prob_{n_v} [\psi(\bm O)]+o_p(n^{-1/2})$. Using a similar strategy, one can also deduce that $\eqref{eq:t5_2} = \Prob_{n_v} [\delta(\bm O)]+o_p(n^{-1/2})$. 

We can show the second term in \eqref{eq:t5_1} is $o_p(n^{-1/2})$ by cross-splitting. Specifically, by the Markov's inequality, the independence induced by cross-splitting, and the fact that
\begin{align*}
\text{Var}\left\{(\Prob_{n_v} - \E )[\widehat\psi^v(\bm O)-\psi(\bm O)]\Big|\mathcal O_{-v}\right\} & = \text{Var}\left\{\Prob_{n_v} [\widehat\psi^v(\bm O)-\psi(\bm O)]\Big|\mathcal O_{-v}\right\} \\
& = \frac{1}{n_v}\text{Var}\left\{\widehat\psi^v(\bm O)-\psi(\bm O)\Big|\mathcal O_{-v}\right\} \\
& = \frac{1}{n_v}\|\widehat\psi^v-\psi\|^2,
\end{align*}
we have that\begingroup\makeatletter\def\f@size{12}\check@mathfonts
\begin{align*}
\Prob\left\{\frac{\sqrt{n_v}\left|(\Prob_{n_v} - \E )[\widehat\psi^v(\bm O)-\psi(\bm O)]\right|}{\|\widehat\psi^v-\psi\|} \geq \epsilon \right\} & = \E\left[\Prob\left\{\frac{\sqrt{n_v}\left|(\Prob_{n_v} - \E )[\widehat\psi^v(\bm O)-\psi(\bm O)]\right|}{\|\widehat\psi^v-\psi\|} \geq \epsilon \Big| \mathcal O_{-v}\right\}\right] \\
& \leq \frac{1}{\epsilon^2} \E\left[ \text{Var}\left\{\frac{\sqrt{n_v}\left|(\Prob_{n_v} - \E )[\widehat\psi^v(\bm O)-\psi(\bm O)]\right|}{\|\widehat\psi^v-\psi\|}  \Big| \mathcal O_{-v}\right\}\right] \\
& = \epsilon^{-2},
\end{align*}\endgroup
for any $\epsilon>0$. Therefore, $(\Prob_{n_v} - \E )[\widehat\psi^v(\bm O)-\psi(\bm O)] = O_p(n_v^{-1/2}\|\widehat\psi^v-\psi\|) = o_p(n_v^{-1/2})$ because $\widehat\psi^v(\bm O)$ converges to $\psi(\bm O)$ in probability and therefore $\|\widehat\psi^v-\psi\|$ converges to 0. Since $V$ is a finite number and we partition the dataset as evenly as possible, we have that $n_v/n=O_p(1)$ and thus $(\Prob_{n_v} - \E )[\widehat\psi^v(\bm O)-\psi(\bm O)]  = o_p(n^{-1/2})$. 

Next, we show that $R_1(\widehat\psi^v,\psi)=o_p(n^{-1/2})$. Specifically, we can show that\begingroup\makeatletter\def\f@size{11}\check@mathfonts\begin{align*}
& R_1(\widehat\psi^v,\psi) = \E[\widehat\psi^v(\bm O)]-\E[\psi(\bm O)] \\
= & \E\Big\{\left(\frac{\mathbb{I}(Z=z^*)\left\{\mathbb{I}(D=d^*)-\widehat p_{z^*d^*}^{\text{np},v}(\bcx)\right\}}{\widehat\pi_{z^*}^{\text{np},v}(\bcx)} - k\frac{(1-Z)\left\{D-\widehat p_{01}^{\text{np},v}(\bcx)\right\}}{\widehat\pi_{0}^{\text{np},v}(\bcx)}\right) \widehat\eta_{zz'}^{\text{np},v}(\bcx) \nonumber \\
& + \left\{\widehat p_{z^*d^*}^{\text{np},v}(\bcx)-k\widehat p_{01}^{\text{np},v}(\bcx)\right\}\frac{\mathbb{I}(D=d_z,Z=z)}{\widehat p_{zd_z}^{\text{np},v}(\bcx)\widehat\pi_{z}^{\text{np},v}(\bcx)}\frac{\widehat r_{z'd_{z'}}^{\text{np},v}(M,\bcx)}{\widehat r_{zd_z}^{\text{np},v}(M,\bcx)}\left\{Y-\widehat \mu_{zd_z}^{\text{np},v}(M,\bcx)\right\} \nonumber \\
& + \left\{\widehat p_{z^*d^*}^{\text{np},v}(\bcx)-k\widehat p_{01}^{\text{np},v}(\bcx)\right\}\frac{\mathbb{I}(D=d_{z'},Z=z')}{\widehat p_{z'd_{z'}}^{\text{np},v}(\bcx)\widehat \pi_{z'}^{\text{np},v}(\bcx)}\left\{\widehat \mu_{zd_z}^{\text{np},v}(M,\bcx)-\widehat \eta_{zz'}^{\text{np},v}(\bcx)\right\} \nonumber \\
& + \left\{\widehat p_{z^*d^*}^{\text{np},v}(\bcx)-k\widehat p_{01}^{\text{np},v}(\bcx)\right\}\widehat \eta_{zz'}^{\text{np},v}(\bcx) \Big\} \\
& - \E\left[\left\{ p_{z^*d^*}(\bcx)-k p_{01}(\bcx)\right\}\int_m \mu_{zd_z}(m,\bcx) r_{z'd_{z'}}(m,\bcx)\differential m\right]\\
= & \E\left[\left(\frac{\pi_{z^*}(\bcx)\left\{p_{z^*d^*}(\bcx)-\widehat p_{z^*d^*}^{\text{np},v}(\bcx)\right\}}{\widehat\pi_{z^*}^{\text{np},v}(\bcx)} - k\frac{\pi_{0}(\bcx)\left\{p_{01}(\bcx)-\widehat p_{01}^{\text{np},v}(\bcx)\right\}}{\widehat\pi_{0}^{\text{np},v}(\bcx)}\right) \int_m\widehat \mu_{zd_z}^{\text{np},v}(m,\bcx)\widehat r_{z'd_{z'}}^{\text{np},v}(m,\bcx)\differential m\right] \\
& + \E\left[\left\{\widehat p_{z^*d^*}^{\text{np},v}(\bcx)-k\widehat p_{01}^{\text{np},v}(\bcx)\right\}\frac{p_{zd_z}(\bcx)\pi_{z}(\bcx)}{\widehat p_{zd_z}^{\text{np},v}(\bcx)\widehat\pi_{z}^{\text{np},v}(\bcx)}\int_m\frac{\widehat r_{z'd_{z'}}^{\text{np},v}(m,\bcx)}{\widehat r_{zd_z}^{\text{np},v}(m,\bcx)}\left\{\mu_{zd_z}(m,\bcx)-\widehat \mu_{zd_z}^{\text{np},v}(m,\bcx)\right\}r_{zd_z}(m,\bcx)\differential m \right]\\
& + \E\left[\left\{\widehat p_{z^*d^*}^{\text{np},v}(\bcx)-k\widehat p_{01}^{\text{np},v}(\bcx)\right\}\frac{p_{z'd_{z'}}(\bcx)\pi_{z'}(\bcx)}{\widehat p_{z'd_{z'}}^{\text{np},v}(\bcx)\widehat \pi_{z'}^{\text{np},v}(\bcx)}\int_m\widehat \mu_{zd_z}^{\text{np},v}(m,\bcx)\left\{r_{z'd_{z'}}(m,\bcx)-\widehat r_{z'd_{z'}}^{\text{np},v}(m,\bcx)\right\}\differential m\right] \\
& + \E\left[\left\{\widehat p_{z^*d^*}^{\text{np},v}(\bcx)-k\widehat p_{01}^{\text{np},v}(\bcx)\right\}\int_m\widehat \mu_{zd_z}^{\text{np},v}(m,\bcx)\widehat r_{z'd_{z'}}^{\text{np},v}(m,\bcx)\differential m\right] \\
& - \E\left[\left\{ p_{z^*d^*}(\bcx)-k p_{01}(\bcx)\right\}\int_m \mu_{zd_z}(m,\bcx) r_{z'd_{z'}}(m,\bcx)\differential m\right]
\end{align*}\endgroup
Define $\pi_{*}$, $\pi_{0}$, $\pi_{z}$, $\pi_{z'}$, $p_{*}$, $p_{0}$, $p_{z}$, $p_{z'}$, $r_{z}$, $r_{z'}$, and $\mu_{z}$ as the abbreviations of the unknown nuisance functions $\pi_{z^*}(\bx)$, $\pi_{0}(\bx)$, $\pi_{z}(\bx)$, $\pi_{z'}(\bx)$, $p_{z^*d^*}(\bx)$, $p_{01}(\bx)$, $p_{zd_z}(\bx)$, $p_{z'd_{z'}}(\bx)$, $r_{zd_z}(m,\bx)$, $r_{z'd_{z'}}(m,\bx)$, and $\mu_{zd_z}(m,\bx)$, respectively. Also, let $\widehat \pi_{*}^{v}$, $\widehat \pi_{0}^{v}$, $\widehat \pi_{z}^{v}$, $\widehat \pi_{z'}^{v}$, $\widehat p_{*}^{v}$, $\widehat p_{0}^{v}$, $\widehat p_{z}^{v}$, $\widehat p_{z'}^{v}$, $\widehat r_{z}^{v}$, $\widehat r_{z'}^{v}$, and $\widehat \mu_{z}^{v}$ be their corresponding estimators evaluated under $\widehat h_{nuisance}^{\text{np},v}$. Using these abbreviations, we can rewrite $R_1(\widehat\psi^v,\psi) = \Delta_1+\Delta_2+\Delta_3+\Delta_4 - \Delta_5$ with\begingroup\makeatletter\def\f@size{11}\check@mathfonts
\begin{align*}
\Delta_1 & = \E\left[\left(\frac{\pi_{*}(p_{*}-\widehat p_{*}^{v})}{\widehat\pi_{*}^{v}} - k\frac{\pi_{0}(p_{0}-\widehat p_{0}^{v})}{\widehat\pi_{0}^{v}}\right) \int_m \widehat \mu_{z}^{v}\widehat r_{z'}^{v}\differential m\right]\\
& = \E\left[\left(\frac{(\pi_{*}-\widehat\pi_{*}^{v})(p_{*}-\widehat p^{v})}{\widehat\pi_{*}^{v}} - k\frac{(\pi_{0}-\widehat\pi_{0}^{v})(p_{0}-\widehat p_{0}^{v})}{\widehat\pi_{0}^{v}}\right) \int_m \widehat \mu_{z}^{v}\widehat r_{z'}^{v}\differential m\right] + \E\left[\left\{(p_{*}-\widehat p_*^{v})-k(p_{0}-\widehat p_0^{v})\right\} \int_m \widehat \mu_{z}^{v}\widehat r_{z'}^{v}\differential m\right],\\
\Delta_2 & = \E\left[(\widehat p_*^v - k\widehat p_0^v)\frac{p_z\pi_z}{\widehat p_z^v\widehat \pi_z^v}\int_m \frac{\widehat r_{z'}^v}{\widehat r_{z}^v}(\mu_{z}-\widehat \mu_{z}^{v})r_{z'}\differential m\right]\\
& = \E\left[(\widehat p_*^v - k\widehat p_0^v)\frac{p_z\pi_z}{\widehat p_z^v\widehat \pi_z^v}\int_m \frac{(r_z-\widehat r_{z}^v)}{\widehat r_{z}^v}(\mu_{z}-\widehat \mu_{z}^{v})\widehat r_{z'}^v\differential m\right] +  \E\left[(\widehat p_*^v - k\widehat p_0^v)\frac{p_z\pi_z}{\widehat p_z^v\widehat \pi_z^v}\int_m (\mu_{z}-\widehat \mu_{z}^{v})\widehat r_{z'}^v\differential m\right] \\
& = \E\left[(\widehat p_*^v - k\widehat p_0^v)\frac{p_z\pi_z}{\widehat p_z^v\widehat \pi_z^v}\int_m \frac{(r_z-\widehat r_{z}^v)}{\widehat r_{z}^v}(\mu_{z}-\widehat \mu_{z}^{v})\widehat r_{z'}^v\differential m\right] \\
& \quad + \E\left[(\widehat p_*^v - k\widehat p_0^v)\frac{(p_z-\widehat p_z^v)\pi_z}{\widehat p_z^v\widehat \pi_z^v}\int_m (\mu_{z}-\widehat \mu_{z}^{v})\widehat r_{z'}^v\differential m\right] + \E\left[(\widehat p_*^v - k\widehat p_0^v)\frac{\pi_z}{\widehat \pi_z^v}\int_m (\mu_{z}-\widehat \mu_{z}^{v})\widehat r_{z'}^v\differential m\right] \\
& = \E\left[(\widehat p_*^v - k\widehat p_0^v)\frac{p_z\pi_z}{\widehat p_z^v\widehat \pi_z^v}\int_m \frac{(r_z-\widehat r_{z}^v)}{\widehat r_{z}^v}(\mu_{z}-\widehat \mu_{z}^{v})\widehat r_{z'}^v\differential m\right] + \E\left[(\widehat p_*^v - k\widehat p_0^v)\frac{(p_z-\widehat p_z^v)\pi_z}{\widehat p_z^v\widehat \pi_z^v}\int_m (\mu_{z}-\widehat \mu_{z}^{v})\widehat r_{z'}^v\differential m\right] \\
& \quad  + \E\left[(\widehat p_*^v - k\widehat p_0^v)\frac{\pi_z-\widehat \pi_z^v}{\widehat \pi_z^v}\int_m (\mu_{z}-\widehat \mu_{z}^{v})\widehat r_{z'}^v\differential m\right] + \E\left[(\widehat p_*^v - k\widehat p_0^v)\int_m (\mu_{z}-\widehat \mu_{z}^{v})\widehat r_{z'}^v\differential m\right] \\
& = \E\left[(\widehat p_*^v - k\widehat p_0^v)\frac{p_z\pi_z}{\widehat p_z^v\widehat \pi_z^v}\int_m \frac{(r_z-\widehat r_{z}^v)}{\widehat r_{z}^v}(\mu_{z}-\widehat \mu_{z}^{v})\widehat r_{z'}^v\differential m\right] + \E\left[(\widehat p_*^v - k\widehat p_0^v)\frac{(p_z-\widehat p_z^v)\pi_z}{\widehat p_z^v\widehat \pi_z^v}\int_m (\mu_{z}-\widehat \mu_{z}^{v})\widehat r_{z'}^v\differential m\right] \\
& \quad  + \E\left[(\widehat p_*^v - k\widehat p_0^v)\frac{\pi_z-\widehat \pi_z^v}{\widehat \pi_z^v}\int_m (\mu_{z}-\widehat \mu_{z}^{v})\widehat r_{z'}^v\differential m\right] + \E\left[\{(\widehat p_*^v-p_*) - k(\widehat p_0^v-p_0)\}\int_m (\mu_{z}-\widehat \mu_{z}^{v})\widehat r_{z'}^v\differential m\right], \\
& \quad + \E\left[(p_*- kp_0)\int_m (\mu_{z}-\widehat \mu_{z}^{v})\widehat r_{z'}^v\differential m\right],\\
\Delta_3 & = \E\left[(\widehat p_*^v - k\widehat p_0^v)\frac{p_{z'}\pi_{z'}}{\widehat p_{z'}^v\widehat \pi_{z'}^v}\int_m \widehat \mu_z^v (r_{z'}-\widehat r_{z'}^v) \differential m\right] \\
& = \E\left[(\widehat p_*^v - k\widehat p_0^v)\frac{(p_{z'}-\widehat p_{z'}^v )\pi_{z'}}{\widehat p_{z'}^v\widehat \pi_{z'}^v}\int_m \widehat \mu_z^v (r_{z'}-\widehat r_{z'}^v) \differential m\right] + \E\left[(\widehat p_*^v - k\widehat p_0^v)\frac{\pi_{z'}}{\widehat \pi_{z'}^v}\int_m \widehat \mu_z^v (r_{z'}-\widehat r_{z'}^v) \differential m\right] \\
& = \E\left[(\widehat p_*^v - k\widehat p_0^v)\frac{(p_{z'}-\widehat p_{z'}^v )\pi_{z'}}{\widehat p_{z'}^v\widehat \pi_{z'}^v}\int_m \widehat \mu_z^v (r_{z'}-\widehat r_{z'}^v) \differential m\right] \\
& \quad + \E\left[(\widehat p_*^v - k\widehat p_0^v)\frac{\pi_{z'} - \widehat \pi_{z'}^v}{\widehat \pi_{z'}^v}\int_m \widehat \mu_z^v (r_{z'}-\widehat r_{z'}^v) \differential m\right] +  \E\left[(\widehat p_*^v - k\widehat p_0^v)\int_m \widehat \mu_z^v (r_{z'}-\widehat r_{z'}^v) \differential m\right] \\
& = \E\left[(\widehat p_*^v - k\widehat p_0^v)\frac{(p_{z'}-\widehat p_{z'}^v )\pi_{z'}}{\widehat p_{z'}^v\widehat \pi_{z'}^v}\int_m \widehat \mu_z^v (r_{z'}-\widehat r_{z'}^v) \differential m\right] + \E\left[(\widehat p_*^v - k\widehat p_0^v)\frac{\pi_{z'} - \widehat \pi_{z'}^v}{\widehat \pi_{z'}^v}\int_m \widehat \mu_z^v (r_{z'}-\widehat r_{z'}^v) \differential m\right]\\
& \quad +  \E\left[\{(\widehat p_*^v-p_*) - k(\widehat p_0^v-p_0)\}\int_m \widehat \mu_z^v (r_{z'}-\widehat r_{z'}^v) \differential m\right]+\E\left[(p_* - kp_0)\int_m \widehat \mu_z^v (r_{z'}-\widehat r_{z'}^v) \differential m\right],\\
\Delta_4 & = \E\left[(\widehat p_*^v - k\widehat p_0^v)\int_m \widehat \mu_z^v \widehat r_{z'}^v \differential m\right] \\
& = \E\left[\{(\widehat p_*^v-p_*) - k(\widehat p_0^v-p_0)\}\int_m \widehat \mu_z^v \widehat r_{z'}^v \differential m\right]+\E\left[(p_* - kp_0)\int_m \widehat \mu_z^v \widehat r_{z'}^v \differential m\right],\\
\Delta_5 & = \E\left[(p_* - kp_0)\int_m \mu_z r_{z'} \differential m\right].
\end{align*}\endgroup
Therefore, we have that\begingroup\makeatletter\def\f@size{12}\check@mathfonts
\begin{align*}
R_1(\widehat\psi^v,\psi) = & \Delta_1+\Delta_2+\Delta_3+\Delta_4 - \Delta_5 \\
= &  \E\left[\left(\frac{(\pi_{*}-\widehat\pi_{*}^{v})(p_{*}-\widehat p^{v})}{\widehat\pi_{*}^{v}} - k\frac{(\pi_{0}-\widehat\pi_{0}^{v})(p_{0}-\widehat p_{0}^{v})}{\widehat\pi_{0}^{v}}\right) \int_m \widehat \mu_{z}^{v}\widehat r_{z'}^{v}\differential m\right] \\
& + \E\left[(\widehat p_*^v - k\widehat p_0^v)\frac{p_z\pi_z}{\widehat p_z^v\widehat \pi_z^v}\int_m \frac{(r_z-\widehat r_{z}^v)}{\widehat r_{z}^v}(\mu_{z}-\widehat \mu_{z}^{v})\widehat r_{z'}^v\differential m\right] \\
& + \E\left[(\widehat p_*^v - k\widehat p_0^v)\frac{(p_z-\widehat p_z^v)\pi_z}{\widehat p_z^v\widehat \pi_z^v}\int_m (\mu_{z}-\widehat \mu_{z}^{v})\widehat r_{z'}^v\differential m\right] \\
&   + \E\left[(\widehat p_*^v - k\widehat p_0^v)\frac{\pi_z-\widehat \pi_z^v}{\widehat \pi_z^v}\int_m (\mu_{z}-\widehat \mu_{z}^{v})\widehat r_{z'}^v\differential m\right] \\
& + \E\left[\{(\widehat p_*^v-p_*) - k(\widehat p_0^v-p_0)\}\int_m (\mu_{z}-\widehat \mu_{z}^{v})\widehat r_{z'}^v\differential m\right]\\
& + \E\left[(\widehat p_*^v - k\widehat p_0^v)\frac{(p_{z'}-\widehat p_{z'}^v )\pi_{z'}}{\widehat p_{z'}^v\widehat \pi_{z'}^v}\int_m \widehat \mu_z^v (r_{z'}-\widehat r_{z'}^v) \differential m\right] \\
& + \E\left[(\widehat p_*^v - k\widehat p_0^v)\frac{\pi_{z'} - \widehat \pi_{z'}^v}{\widehat \pi_{z'}^v}\int_m \widehat \mu_z^v (r_{z'}-\widehat r_{z'}^v) \differential m\right] \\
& + \E\left[\{(\widehat p_*^v-p_*) - k(\widehat p_0^v-p_0)\}\int_m \widehat \mu_z^v (r_{z'}-\widehat r_{z'}^v) \differential m\right] \\
& + \E\left[(p_*-kp_0)\int_m (\widehat\mu_z-\mu_z^v) (\widehat r_{z'}^v-r_{z'}) \differential m\right] \\
= &  \E\left[f_{m}^{-1}\left(\frac{(\pi_{*}-\widehat\pi_{*}^{v})(p_{*}-\widehat p_*^{v})}{\widehat\pi_{*}^{v}} - k\frac{(\pi_{0}-\widehat\pi_{0}^{v})(p_{0}-\widehat p_{0}^{v})}{\widehat\pi_{0}^{v}}\right)  \widehat \mu_{z}^{v}\widehat r_{z'}^{v}\right] \\
& + \E\left[f_{m}^{-1}(\widehat p_*^v - k\widehat p_0^v)\frac{p_z\pi_z}{\widehat p_z^v\widehat \pi_z^v} \frac{(r_z-\widehat r_{z}^v)}{\widehat r_{z}^v}(\mu_{z}-\widehat \mu_{z}^{v})\widehat r_{z'}^v\right] \\
& + \E\left[f_{m}^{-1}(\widehat p_*^v - k\widehat p_0^v)\frac{(p_z-\widehat p_z^v)\pi_z}{\widehat p_z^v\widehat \pi_z^v}(\mu_{z}-\widehat \mu_{z}^{v})\widehat r_{z'}^v\right] \\
&   + \E\left[f_{m}^{-1}(\widehat p_*^v - k\widehat p_0^v)\frac{\pi_z-\widehat \pi_z^v}{\widehat \pi_z^v}(\mu_{z}-\widehat \mu_{z}^{v})\widehat r_{z'}^v\right] \\
& + \E\left[f_{m}^{-1}\{(\widehat p_*^v-p_*) - k(\widehat p_0^v-p_0)\}(\mu_{z}-\widehat \mu_{z}^{v})\widehat r_{z'}^v\right]\\
& + \E\left[f_{m}^{-1}(\widehat p_*^v - k\widehat p_0^v)\frac{(p_{z'}-\widehat p_{z'}^v )\pi_{z'}}{\widehat p_{z'}^v\widehat \pi_{z'}^v}\widehat \mu_z^v (r_{z'}-\widehat r_{z'}^v) \right] \\
& + \E\left[f_{m}^{-1}(\widehat p_*^v - k\widehat p_0^v)\frac{\pi_{z'} - \widehat \pi_{z'}^v}{\widehat \pi_{z'}^v}\widehat \mu_z^v (r_{z'}-\widehat r_{z'}^v) \right] \\
& + \E\left[f_{m}^{-1}\{(\widehat p_*^v-p_*) - k(\widehat p_0^v-p_0)\} \widehat \mu_z^v (r_{z'}-\widehat r_{z'}^v) \right] \\
& + \E\left[f_{m}^{-1}(p_*-kp_0) (\widehat\mu_z^v-\mu_z) (\widehat r_{z'}^v-r_{z'}) \right],
\end{align*}\endgroup
where $f_m = f_{M|Z,D,\bcx}(M|Z,D,\bcx)$ and the last equality of the previous equation follows from the law of iterated expectation. Using the Cauchy-Schwartz inequality, we then have\begingroup\makeatletter\def\f@size{12}\check@mathfonts
\begin{align*}
|R_1(\widehat\psi^v,\psi)| \leq & \|f_m^{-1} \widehat \pi_*^{v^{-1}} \widehat \mu_{z}^{v}\widehat r_{z'}^{v}\|_{\infty} \|\widehat\pi_{*}^{v}-\pi_{*}\|\|\widehat p_{*}^{v}-p_{*}\| + \|f_m^{-1} \widehat \pi_0^{v^{-1}} \widehat \mu_{z}^{v}\widehat r_{z'}^{v}\|_{\infty} \|\widehat\pi_{0}^{v}-\pi_{0}\|\|\widehat p_0^{v}-p_{0}\|\\
& + \left\{\|f_{m}^{-1}\widehat p_*^vp_z\pi_z\widehat r_{z'}^v\widehat p_z^{v^{-1}}\widehat \pi_z^{v^{-1}}\widehat r_{z}^{v^{-1}}\|_{\infty}+\|f_{m}^{-1}\widehat p_0^vp_z\pi_z\widehat r_{z'}^v\widehat p_z^{v^{-1}}\widehat \pi_z^{v^{-1}}\widehat r_{z}^{v^{-1}}\|_{\infty}\right\}\|\widehat \mu_{z}^{v}-\mu_{z}\|\|\widehat r_{z}^v-r_z\| \\
& + \left\{\|f_{m}^{-1}\widehat p_*^v\pi_z\widehat r_{z'}^v\widehat p_z^{v^{-1}}\widehat \pi_z^{v^{-1}}\|_{\infty}+\|f_{m}^{-1}\widehat p_0^v\pi_z\widehat r_{z'}^v\widehat p_z^{v^{-1}}\widehat \pi_z^{v^{-1}}\|_{\infty}\right\}\|\widehat \mu_{z}^{v}-\mu_{z}\|\|\widehat p_z^v-p_z\|\\
& + \left\{\|f_{m}^{-1}\widehat p_*^v\widehat \pi_z^{v^{-1}}\widehat r_{z'}^v\|_{\infty}+\|f_{m}^{-1}\widehat p_*^v\widehat \pi_z^{v^{-1}}\widehat r_{z'}^v\|_{\infty}\right\}\|\widehat \mu_{z}^{v}-\mu_{z}\|\|\widehat \pi_z^v-\pi_z\| \\
& + \|f_m^{-1}\widehat r_{z'}^v\|_{\infty}\|\widehat p_*^v-p_*\|\|\widehat \mu_{z}^{v}-\mu_{z}\| + \|f_m^{-1}\widehat r_{z'}^v\|_{\infty}\|\widehat p_0^v-p_0\|\|\widehat \mu_{z}^{v}-\mu_{z}\| \\
& + \left\{\|f_{m}^{-1}\widehat p_*^v\pi_{z'}\widehat p_{z'}^{v^{-1}}\widehat \pi_{z'}^{v^{-1}}\mu_z^v\|_{\infty}+\|f_{m}^{-1}\widehat p_0^v\pi_{z'}\widehat p_{z'}^{v^{-1}}\widehat \pi_{z'}^{v^{-1}}\mu_z^v\|_{\infty}\right\}\|\widehat r_{z'}^v-r_{z'}\|\|\widehat p_{z'}^v-p_{z'}\| \\
& + \left\{\|f_m^{-1}\widehat p_*^v\widehat \pi_{z'}^{v^{-1}}\widehat \mu_z^v\|_{\infty}+\|f_m^{-1}\widehat p_*^v\widehat \pi_{z'}^{v^{-1}}\widehat \mu_z^v\|_{\infty}\right\}\|\widehat r_{z'}^v-r_{z'}\|\|\widehat \pi_{z'}^v-\pi_{z'}\| \\
& + \|f_m^{-1}\widehat \mu_z^v\|_{\infty}\|\widehat r_{z'}^v-r_{z'}\|\|\widehat p_{*}^v-p_{*}\| + \|f_m^{-1}\widehat \mu_z^v\|_{\infty}\|\widehat r_{z'}^v-r_{z'}\|\|\widehat p_{0}^v-p_{0}\| \\
& + \left\{\|f_m^{-1}p_*\|_{\infty} + \|f_m^{-1}p_0\|_{\infty}\right\}\|\widehat\mu_z^v-\mu_z\| \|\widehat r_{z'}^v-r_{z'}\|. 
\end{align*}\endgroup
Noting that it is assumed $\|\widehat l^{\text{np},v} - l\| \times \|\widehat g^{\text{np},v} - g\|=o_p(n^{-1/2})$ for any $l\neq g \in \{\pi_z(\bx),p_{zd}(\bx),r_{zd}(m,\bx),\\ \mu_{zd}(m,\bx)\}$, then $R_1(\widehat\psi^v,\psi)=o_p(n^{-1/2})$. Now, we have confirmed that
$$
\Prob_{n_v}[\widehat\psi^v(\bm O)] = \Prob_{n_v} [\psi(\bm O)] +o_p(n^{-1/2}),
$$
thus
\begin{align}
\Prob_n\left[\widehat\psi(\bm O)\right] & = \frac{1}{n}\sum_{v=1}^V n_v \Prob_{n_v}\left[\widehat\psi^v(\bm O)\right] \nonumber \\
& = \sum_{v=1}^V \left\{\frac{n_v}{n} \Prob_{n_v}\left[\psi(\bm O)\right]+o_p\left(\frac{n_v}{n^{3/2}}\right)\right\}\nonumber \\
& = \Prob_n\left[\psi(\bm O)\right] + o_p(n^{-1/2})\label{eq:np1}
\end{align}
Using similar arguments, we can show  
$
\Prob_{n_v}[\widehat\delta^v(\bm O)] = \Prob_{n_v} [\delta(\bm O)] +o_p(n^{-1/2})
$ and therefore
\begin{align}
\Prob_n\left[\widehat\delta(\bm O)\right] &  = \Prob_n\left[\delta(\bm O)\right] + o_p(n^{-1/2}).\label{eq:np2}
\end{align}
Notice that $\widehat\theta$ can be cast as the solution of the following equation
$$
\Prob_n\left[\widehat\psi(\bm O)-\widehat \theta \widehat\delta(\bm O)\right] = 0.
$$
This, along with \eqref{eq:np1} and \eqref{eq:np2} suggests that
\begin{align*}
& \Prob_n\left[\psi(\bm O)-\widehat \theta \delta(\bm O)\right] = o_p(n^{-1/2})\\
\Longleftrightarrow & \Prob_n\left[\psi(\bm O)-\theta \delta(\bm O)\right] - \Prob_n\left[\delta(\bm O)\right](\widehat\theta - \theta ) = o_p(n^{-1/2}).
\end{align*}
Moreover, since $\Prob_n\left[\delta(\bm O)\right] = \E\left[\delta(\bm O)\right]+O_p(n^{-1/2})$ and $\widehat\theta=\theta+o_p(1)$, it follows that $\Prob_n\left[\delta(\bm O)\right](\widehat\theta - \theta )=\E\left[\delta(\bm O)\right](\widehat\theta - \theta )+o_{p}(n^{-1/2})$. Therefore, we further obtain
$$
\Prob_n\left[\psi(\bm O)-\theta \delta(\bm O)\right] - \E\left[\delta(\bm O)\right](\widehat\theta - \theta ) = o_p(n^{-1/2}).
$$
After simple algebra and observing $\E\left[\delta(\bm O)\right]=e_{d_1d_0}=p_{z^*d^*}-kp_{01}$, we have
$$
\sqrt{n}(\widehat\theta - \theta ) = \sqrt{n}\Prob_n\left[\frac{\psi(\bm O)-\theta \delta(\bm O)}{p_{z^*d^*}-kp_{01}}\right] + o_p(1)
$$
which suggests that $\widehat\theta$ is asymptotically normal and its asymptotic variance achieves the efficiency lower bound discussed in Appendix \ref{asec:b5}. \hfill $\square$

\subsection{Estimation of natural mediation effects}

This section elaborates on the multiple robust estimator and nonparametric estimator for the mediation effects, $\text{PNIE}_{d_1d_0}$, $\text{PNDE}_{d_1d_0}$, $\text{ITT-NIE}$ and $\text{ITT-NDE}$. The following lemma provides the form of the EIF of the aforementioned mediation effects.
\begin{lemma}\label{lemma:eif_nme}
For any $d_1d_0\in \mathcal U_{\text{a}}$ or $d_1d_0\in \mathcal U_{\text{b}}$ under standard or strong monotonicity, the EIFs of $\text{PNIE}_{d_1d_0}$ and $\text{PNDE}_{d_1d_0}$ are
$$
\mathcal D_{d_1d_0}^{\text{PNIE}}(\bm O) = \frac{\psi_{d_1d_0}^{(11)}(\bm O)-\psi_{d_1d_0}^{(10)}(\bm O) - \text{PNIE}_{d_1d_0}\times \delta_{d_1d_0}(\bm O)}{p_{z^*d^*}-kp_{01}} 
$$
and 
$$
\mathcal D_{d_1d_0}^{\text{PNDE}}(\bm O) = \frac{\psi_{d_1d_0}^{(10)}(\bm O)-\psi_{d_1d_0}^{(00)}(\bm O) - \text{PNDE}_{d_1d_0}\times \delta_{d_1d_0}(\bm O)}{p_{z^*d^*}-kp_{01}} 
$$
respectively, where $\psi_{d_1d_0}^{(zz')}(\bm O)$ and $\delta_{d_1d_0}(\bm O)$ are given in Theorem 3, $k=|d_1-d_0|$, and $z^*d^*=$11, 10, 01 if $d_1d_0=$10, 00, and 11, respectively. In addition, the EIFs of $\text{ITT-NIE}$ and $\text{ITT-NDE}$ are 
$$
\mathcal D^{\text{ITT-NIE}}(\bm O) = \sum_{d_1d_0\in \mathcal U} \left\{\psi_{d_1d_0}^{(11)}(\bm O)-\psi_{d_1d_0}^{(10)}(\bm O)\right\} - \text{ITT-NIE}
$$
and 
$$
\mathcal D^{\text{ITT-NDE}}(\bm O) = \sum_{d_1d_0\in \mathcal U} \left\{\psi_{d_1d_0}^{(10)}(\bm O)-\psi_{d_1d_0}^{(00)}(\bm O)\right\} - \text{ITT-NDE}
$$
respectively, where $\mathcal U=\mathcal U_{\text{a}}$ or $\mathcal U=\mathcal U_{\text{b}}$ under the standard or strong monotonicity. 
\end{lemma}
\begin{proof}
Because $\text{PNIE}_{d_1d_0}$ is identified as the difference between $\theta_{d_1d_0}^{(11)}$ and $\theta_{d_1d_0}^{(10)}$, and the EIFs of $\theta_{d_1d_0}^{(11)}$ and $\theta_{d_1d_0}^{(10)}$ are $\mathcal D_{d_1d_0}^{(11)}(\bm O)$ and $\mathcal D_{d_1d_0}^{(10)}(\bm O)$ as derived in Theorem 3. According to Lemma \ref{lemma:eif_2}, we have that the EIF of $\text{PNIE}_{d_1d_0}$ is
$$
\mathcal D_{d_1d_0}^{\text{PNIE}}(\bm O) = \mathcal D_{d_1d_0}^{(11)}(\bm O)-\mathcal D_{d_1d_0}^{(10)}(\bm O)=\frac{\psi_{d_1d_0}^{(11)}(\bm O)-\psi_{d_1d_0}^{(10)}(\bm O) - \text{PNIE}_{d_1d_0}\times \delta_{d_1d_0}(\bm O)}{p_{z^*d^*}-kp_{01}}. 
$$
The EIF of $\text{PNDE}_{d_1d_0}$ can be similarly obtained. 

Also, $\text{ITT-NIE}$ is identified as 
$$
\text{ITT-NIE}=\sum_{d_1d_0 \in \mathcal U} e_{d_1d_0}\times \left(\theta_{d_1d_0}^{(11)}-\theta_{d_1d_0}^{(10)}\right) = \sum_{d_1d_0 \in \mathcal U}  \left(H_{d_1d_0}^{(11)}-H_{d_1d_0}^{(10)}\right),
$$
where $H_{d_1d_0}^{(zz')}=\E\left[(p_{z^*d^*}(\bcx)-kp_{01}(\bcx))\eta_{zz'}(\bcx)\right]$ is defined in Section \ref{asec:b4} and its EIF has been derived in Lemma \ref{lemma:eif_H}. Based on Lemma \ref{lemma:eif_2}, one can easily show
$$
\mathcal D^{\text{ITT-NIE}}(\bm O) = \sum_{d_1d_0 \in \mathcal U} \left\{\mathcal D_{d_1d_0}^{(11),H}(\bm O)-\mathcal D_{d_1d_0}^{(10),H}(\bm O)\right\} = \sum_{d_1d_0\in \mathcal U} \left\{\psi_{d_1d_0}^{(11)}(\bm O)-\psi_{d_1d_0}^{(10)}(\bm O)\right\} - \text{ITT-NIE}.
$$
We can calculate the EIF of \text{ITT-NDE} following the same strategy.
\end{proof}\medskip

The following proposition demonstrates the multiply robust estimator for the mediation effects is still quadruply robust and locally efficient.
\begin{proposition}\label{prop:mr_nme}
    Under either $\mathcal M_{\pi} \cap \mathcal M_{e} \cap \mathcal M_{m}$, $\mathcal M_{\pi} \cap \mathcal M_{m} \cap \mathcal M_{o}$, $\mathcal  M_{\pi} \cap \mathcal M_{e} \cap \mathcal M_{o}$ or $\mathcal M_{e} \cap \mathcal M_{m} \cap \mathcal M_{o}$, the multiply robust estimator $\widehat\tau^{\text{mr}}$ is consistent and asymptotically normal for all $\tau \in \{\text{PNIE}_{d_1d_0},\text{PNDE}_{d_1d_0},\text{ITT-NIE}, \text{ITT-NDE}\}$. Moreover, $\widehat\tau^{\text{mr}}$ is semiparametrically efficient under $\mathcal M_{\pi} \cap \mathcal M_{e} \cap \mathcal M_{m} \cap \mathcal M_{o}$.
\end{proposition}

\begin{proof}
The quadruple robustness and asymptotically normality of $\widehat{\text{PNIE}}_{d_1d_0}^{\text{mr}}$ and $\widehat{\text{PNDE}}_{d_1d_0}^{\text{mr}}$  follow directly from the properties of $\widehat\theta_{d_1d_0}^{(zz'),\text{mr}}$  in Theorem 4. Next, we prove that $\widehat{\text{PNIE}}_{d_1d_0}^{\text{mr}}$ is locally efficient under $\mathcal M_{\pi} \cap \mathcal M_{e} \cap \mathcal M_{m} \cap \mathcal M_{o}$ and similar results extend to $\widehat{\text{PNDE}}_{d_1d_0}^{\text{mr}}$. Based on the proof of Theorem 4 in Section \ref{asec:b5}, we know that the influence function of $\widehat\theta_{d_1d_0}^{(zz'),\text{mr}}$ under $\mathcal M_{\pi} \cap \mathcal M_{e} \cap \mathcal M_{m} \cap \mathcal M_{o}$ is
$$
\sqrt{n}(\widehat\theta_{d_1d_0}^{(zz'),\text{mr}}-\theta_{d_1d_0}^{(zz')}) = \sqrt{n}\Prob_n\left[\frac{\psi_{d_1d_0}^{(zz')}(\bm O)-\theta_{d_1d_0}^{(zz')}\delta_{d_1d_0}(\bm O)}{p_{z^*d^*}-kp_{01}}\right] + o_p(1).
$$
Then, it follows from $\widehat{\text{PNIE}}_{d_1d_0}^{\text{mr}}=\widehat\theta_{d_1d_0}^{(11),\text{mr}}-\widehat\theta_{d_1d_0}^{(10),\text{mr}}$ that
\begin{align*}
\sqrt{n}(\widehat{\text{PNIE}}_{d_1d_0}^{\text{mr}}-\text{PNIE}_{d_1d_0}) & = \sqrt{n}\Prob_n\left[\frac{\psi_{d_1d_0}^{(11)}(\bm O)-\psi_{d_1d_0}^{(10)}(\bm O) - \text{PNIE}_{d_1d_0}\times \delta_{d_1d_0}(\bm O)}{p_{z^*d^*}-kp_{01}}\right] + o_p(1) \\
& = \sqrt{n}\Prob_n\left[\mathcal D_{d_1d_0}^{\text{PNIE}}(\bm O)
\right] + o_p(1),
\end{align*}
where the second equality holds due to Lemma \ref{lemma:eif_nme}. This suggests that $\widehat{\text{PNIE}}_{d_1d_0}^{\text{mr}}$ is semiparametrically efficient when all working models are correctly specified. 

The multiply robust estimator of \text{ITT-NIE} is $\widehat{\text{ITT-NIE}}^{\text{mr}} = \sum_{d_1d_0\in\mathcal U} \widehat e_{d_1d_0}^{\text{dr}}\times (\widehat \theta_{d_1d_0}^{(11),\text{mr}}-\widehat \theta_{d_1d_0}^{(10),\text{mr}})$. Theorem 4 suggests that $\widehat \theta_{d_1d_0}^{(11),\text{mr}} \xrightarrow{p} \theta_{d_1d_0}^{(11)}$ and $\widehat \theta_{d_1d_0}^{(11),\text{mr}} \xrightarrow{p} \theta_{d_1d_0}^{(10)}$ under either $\mathcal M_{\pi} \cap \mathcal M_{e} \cap \mathcal M_{m}$, $\mathcal M_{\pi} \cap \mathcal M_{m} \cap \mathcal M_{o}$, $\mathcal  M_{\pi} \cap \mathcal M_{e} \cap \mathcal M_{o}$, or $\mathcal M_{e} \cap \mathcal M_{m} \cap \mathcal M_{o}$. Also, \cite{jiang2022multiply} suggests that the doubly robust estimator for the marginal principal score $\widehat e_{d_1d_0}^{\text{dr}}=\widehat p_{z^*d^*}^{\text{dr}}-k \widehat p_{01}^{\text{dr}}$ is consistent to $e_{d_1d_0}$ under $\mathcal M_{\pi} \cup \mathcal M_e$. This further implies that $
\widehat{\text{ITT-NIE}}^{\text{mr}} \xrightarrow{p} \sum_{d_1d_0\in\mathcal U} e_{d_1d_0} \times (\theta_{d_1d_0}^{(11)}-\theta_{d_1d_0}^{(10)}) = \text{ITT-NIE}$ under either $\mathcal M_{\pi} \cap \mathcal M_{e} \cap \mathcal M_{m}$, $\mathcal M_{\pi} \cap \mathcal M_{m} \cap \mathcal M_{o}$, $\mathcal  M_{\pi} \cap \mathcal M_{e} \cap \mathcal M_{o}$, or $\mathcal M_{e} \cap \mathcal M_{m} \cap \mathcal M_{o}$. To prove asymptotic normality, notice that $\widehat{\text{ITT-NIE}}^{\text{mr}}$
can be re-expressed as
\begin{align*}
\widehat{\text{ITT-NIE}}^{\text{mr}} & = \sum_{d_1d_0\in\mathcal U} \widehat e_{d_1d_0}^{\text{dr}}\times (\widehat \theta_{d_1d_0}^{(11),\text{mr}}-\widehat \theta_{d_1d_0}^{(10),\text{mr}}) \\
& = \sum_{d_1d_0\in\mathcal U} \mathbb{P}_n[\widehat \delta_{d_1d_0}^{\text{par}}(\bm O)]\times \left(\frac{\mathbb{P}_n[\widehat \psi_{d_1d_0}^{(11),\text{par}}(\bm O)]}{\mathbb{P}_n[\widehat \delta_{d_1d_0}^{\text{par}}(\bm O)]}-\frac{\mathbb{P}_n[\widehat \psi_{d_1d_0}^{(10),\text{par}}(\bm O)]}{\mathbb{P}_n[\widehat \delta_{d_1d_0}^{\text{par}}(\bm O)]}\right)\\
& =  \mathbb{P}_n\left[\sum_{d_1d_0\in\mathcal U}\widehat \psi_{d_1d_0}^{(11),\text{par}}(\bm O)-\widehat \psi_{d_1d_0}^{(10),\text{par}}(\bm O)\right].
\end{align*}
Define $S_{\text{mr}}(\bm O;\widehat{\bm\tau})=\{\sum_{d_1d_0\in\mathcal U}\widehat \psi_{d_1d_0}^{(11),\text{par}}(\bm O)-\widehat \psi_{d_1d_0}^{(10),\text{par}}(\bm O)\}$, where $\widehat{\bm\tau}$ is the estimator of the parameters in nuisance parametric working models. Then, under mild regularity conditions (similar to what we listed in the proof of Theorem 4), one can easily deduce that
\begin{align*}
& \sqrt{n}\left(\widehat{\text{ITT-NIE}}^{\text{mr}}-\text{ITT-NIE}\right) \\
= & \sqrt{n}\Prob_n\left\{S_{\text{mr}}(\bm O;{\bm\tau}^*) - \text{ITT-NIE} + \E\left[\frac{\partial}{\partial \bm\tau^*} S_{\text{mr}}(\bm O;{\bm\tau}^*)\right] \text{IF}_{\bm\tau}(\bm O;\bm \tau^*)\right\} + o_p(1),
\end{align*}
where $\bm\tau^*$ is the probability limit of $\widehat{\bm\tau}$ and $\text{IF}_{\bm\tau}(\bm O;\bm \tau^*)$ is the influence function of $\bm\tau$. This have confirmed that $\widehat{\text{ITT-NIE}}^{\text{mr}}$ is asmptotically normal. Under $\mathcal M_{\pi} \cap \mathcal M_{e} \cap \mathcal M_{m} \cap \mathcal M_{o}$, we can verify that $\widehat{\text{ITT-NIE}}^{\text{mr}}$ is semiparametrically efficient by observing that $S_{\text{mr}}(\bm O;{\bm\tau}^*)=\psi_{d_1d_0}^{(11)}(\bm O)-\psi_{d_1d_0}^{(10)}(\bm O)$ and $\E\left[\frac{\partial}{\partial \bm\tau^*} S_{\text{mr}}(\bm O;{\bm\tau}^*)\right]=0$  such that
\begin{align*}
\sqrt{n}\left(\widehat{\text{ITT-NIE}}^{\text{mr}}-\text{ITT-NIE}\right) 
& =  \sqrt{n}\Prob_n\left[\sum_{d_1d_0\in \mathcal U}\left\{\psi_{d_1d_0}^{(11)}(\bm O)-\psi_{d_1d_0}^{(10)}(\bm O)\right\}-\text{ITT-NIE}\right] + o_p(1)\\
& = \sqrt{n}\Prob_n\left[\mathcal D^{\text{ITT-NIE}}(\bm O)\right] + o_p(1),
\end{align*}
where the second equality from Lemma \ref{lemma:eif_nme}. This suggests that $\widehat{\text{ITT-NIE}}^{\text{mr}}$ is locally efficient. Using a similar strategy, one can prove that $\widehat{\text{ITT-NDE}}^{\text{mr}}$ is quadruply robust, asymptotically normal, and locally efficient.
\end{proof}\medskip

In parallel to results in Section \ref{sec:np-proof}, the following proposition demonstrates the properties of the nonparametric estimator of the mediation effects.
\begin{proposition}\label{prop:np_nme}
     For any $\tau \in \{\text{PNIE}_{d_1d_0},\text{PNDE}_{d_1d_0},\text{ITT-NIE}, \text{ITT-NDE}\}$, $\widehat{\tau}^{\text{np}}$ is consistent if any three of the four nuisance functions in $\widehat h_{nuisance}^{\text{np}}$ are consistently estimated in the $L_2(\Prob)$-norm.  Furthermore, if all elements in $\widehat h_{nuisance}^{\text{np}}$ are consistent in the $L_2(\Prob)$-norm and $\|\widehat l^{\text{np}} - l\| \times \|\widehat g^{\text{np}} - g\|=o_p(n^{-1/2})$ for any $l\neq g \in \{\pi_z(\bx),p_{zd}(\bx),r_{zd}(m,\bx),\mu_{zd}(m,\bx)\}$, then $\widehat{\tau}^{\text{np}}$ is asymptotically normal and semiparametrically efficient.
\end{proposition}

\begin{proof}
Following the proof of Proposition \ref{prop:mr_nme}, one can show that $\widehat\tau^{\text{np}}$ is consistent to $\tau$ for $\tau \in \{\text{PNIE}_{d_1d_0},\text{PNDE}_{d_1d_0},\text{ITT-NIE}, \text{ITT-NDE}\}$, if any three of the four functions in $\widehat h_{nuisance}^{\text{np}} = \{\widehat \pi_z^{\text{np}}(\bx),\widehat p_{zd}^{\text{np}}(\bx),\widehat r_{zd}^{\text{np}}(m,\bx),\widehat \mu_{zd}^{\text{np}}(m,\bx)\}$ are consistently estimated. Here, we only prove the asymptotic normality and local efficiency of $\widehat\tau^{\text{np}}$ when $\|\widehat l^{\text{np}} - l\| \times \|\widehat g^{\text{np}} - g\|=o_p(n^{-1/2})$ for any $l\neq g \in \{\pi_z(\bx),p_{zd}(\bx),r_{zd}(m,\bx),\mu_{zd}(m,\bx)\}$. 

We show in the proof of Theorem 5 that
$$
\sqrt{n}(\widehat\theta_{d_1d_0}^{(zz'),\text{np}} - \theta_{d_1d_0}^{(zz')} ) = \sqrt{n}\Prob_n\left[\frac{\psi_{d_1d_0}^{(zz')}(\bm O)-\theta_{d_1d_0}^{(zz')} \delta_{d_1d_0}(\bm O)}{p_{z^*d^*}-kp_{01}}\right] + o_p(1),
$$
when $\|\widehat l^{\text{np}} - l\| \times \|\widehat g^{\text{np}} - g\|=o_p(n^{-1/2})$ for any $l\neq g \in \{\pi_z(\bx),p_{zd}(\bx),r_{zd}(m,\bx),\mu_{zd}(m,\bx)\}$. Therefore,
\begin{align*}
\sqrt{n}(\widehat{\text{PNIE}}_{d_1d_0}^{\text{np}}-\text{PNIE}_{d_1d_0}) & = \sqrt{n}\Prob_n\left[\frac{\psi_{d_1d_0}^{(11)}(\bm O)-\psi_{d_1d_0}^{(10)}(\bm O) - \text{PNIE}_{d_1d_0}\times \delta_{d_1d_0}(\bm O)}{p_{z^*d^*}-kp_{01}}\right] + o_p(1) \\
& = \sqrt{n}\Prob_n\left[\mathcal D_{d_1d_0}^{\text{PNIE}}(\bm O)
\right] + o_p(1),\\
\sqrt{n}(\widehat{\text{PNDE}}_{d_1d_0}^{\text{np}}-\text{PNDE}_{d_1d_0}) & = \sqrt{n}\Prob_n\left[\frac{\psi_{d_1d_0}^{(10)}(\bm O)-\psi_{d_1d_0}^{(00)}(\bm O) - \text{PNDE}_{d_1d_0}\times \delta_{d_1d_0}(\bm O)}{p_{z^*d^*}-kp_{01}}\right] + o_p(1) \\
& = \sqrt{n}\Prob_n\left[\mathcal D_{d_1d_0}^{\text{PNDE}}(\bm O)
\right] + o_p(1).
\end{align*}
This implies that $\widehat{\text{PNIE}}_{d_1d_0}^{\text{np}}$ and $\widehat{\text{PNDE}}_{d_1d_0}^{\text{np}}$ are asymptotically normal and semiparametrically efficient under the required convergence rate conditions for the nuisance function estimates. Also, we show in the proof of Theorem 5 that $\Prob_n[\widehat \psi_{d_1d_0}^{(zz'),\text{np}}(\bm O)] = \Prob_n[\psi_{d_1d_0}^{(zz')}(\bm O)]+o_p(n^{-1/2})$, which suggests that\begingroup\makeatletter\def\f@size{10}\check@mathfonts
\begin{align*}
\widehat{\text{ITT-NIE}}^{\text{np}} & = \sum_{d_1d_0\in\mathcal U} \widehat e_{d_1d_0}^{\text{np}}\times (\widehat \theta_{d_1d_0}^{(11),\text{np}}-\widehat \theta_{d_1d_0}^{(10),\text{np}}) \\
& = \sum_{d_1d_0\in\mathcal U} \mathbb{P}_n[\widehat \delta_{d_1d_0}^{\text{np}}(\bm O)]\times \left(\frac{\mathbb{P}_n[\widehat \psi_{d_1d_0}^{(11),\text{np}}(\bm O)]}{\mathbb{P}_n[\widehat \delta_{d_1d_0}^{\text{np}}(\bm O)]}-\frac{\mathbb{P}_n[\widehat \psi_{d_1d_0}^{(10),\text{np}}(\bm O)]}{\mathbb{P}_n[\widehat \delta_{d_1d_0}^{\text{np}}(\bm O)]}\right)\\
& =  \mathbb{P}_n\left[\sum_{d_1d_0\in\mathcal U}\widehat \psi_{d_1d_0}^{(11),\text{np}}(\bm O)-\widehat \psi_{d_1d_0}^{(10),\text{np}}(\bm O)\right]\\
& =  \mathbb{P}_n\left[\sum_{d_1d_0\in\mathcal U}\psi_{d_1d_0}^{(11)}(\bm O)- \psi_{d_1d_0}^{(10)}(\bm O)\right] + o_p(n^{-1/2})
\end{align*}\endgroup
and thus $\sqrt{n}\left(\widehat{\text{ITT-NIE}}^{\text{np}}-\text{ITT-NIE}\right)=\sqrt{n}\Prob_n\left[\displaystyle\sum_{d_1d_0\in \mathcal U}\left\{\psi_{d_1d_0}^{(11)}(\bm O)-\psi_{d_1d_0}^{(10)}(\bm O)\right\}-\text{ITT-NIE}\right] + o_p(1)= \sqrt{n}\Prob_n\left[\mathcal D^{\text{ITT-NIE}}(\bm O)\right] + o_p(1)$. Similarly, we can show $\sqrt{n}\left(\widehat{\text{ITT-NDE}}^{\text{np}}-\text{ITT-NDE}\right)= \sqrt{n}\Prob_n\left[\mathcal D^{\text{ITT-NDE}}(\bm O)\right] + o_p(1)$. This implies that $\widehat{\text{ITT-NIE}}^{\text{np}}$ and $\widehat{\text{ITT-NDE}}^{\text{np}}$ are asymptotically normal and semiparametrically efficient under the required convergence rate conditions for the nonparametric nuisance function estimators. 
\end{proof}

\begin{remark}
(Variance estimation of the principal and ITT mediation effects) For the purpose of inference, nonparametric bootstrap can be used for the moment-type and multiply robust estimators. The asymptotic variance of the nonparametric efficient estimators can be obtained by using the empirical variance of the estimated EIF given in Lemma \ref{lemma:eif_nme}. For example, the asymptotic variance of $\widehat{\text{PNIE}}_{d_1d_0}^{\text{np}}$ can be estimated by 
$$
\widehat{\text{Var}}\left(\widehat{\text{PNIE}}_{d_1d_0}^{(zz')}\right) = \frac{1}{n}\mathbb{P}_n\left[\{\widehat{\mathcal D}_{d_1d_0}^{\text{PNIE,np}}(\bm O)\}^2\right],
$$
where $\widehat{\mathcal D}_{d_1d_0}^{\text{PNIE,np}}(\bm O)$ is $\mathcal D_{d_1d_0}^{\text{PNIE}}(\bm O)$ evaluated based on the nonparametric estimator of the nuisance functions, $\widehat h_{nuisance}^{\text{np}}$. The variance estimator of $\widehat{\text{PNDE}}_{d_1d_0}^{\text{np}}$, $\widehat{\text{ITT-NIE}}^{\text{np}}$, and $\widehat{\text{ITT-NDE}}^{\text{np}}$ can be similarly obtained. 
\end{remark}

\subsection{Sensitivity analysis for the principal ignorability assumption under standard monotonicity}\label{sec:b8}

This section provides the supporting information for the sensitivity analysis for the principal ignorability assumption, under standard monotonicity. We first present the explicit forms of the sensitivity weight $w_{d_1d_0}^{(zz')}(m,\bx)$ for all $zz'\in\{11,10,00\}$ and $d_1d_0\in\mathcal U_{\text{2-sided}}$: 
\begingroup\makeatletter\def\f@size{9}\check@mathfonts 
\begin{align*}
w_{10}^{(11)}(m,\bx) &=
\begin{cases}
    \frac{\xi_{M}^{(1)}(m,\bx)p_{11}(\bx)}{\xi_{M}^{(1)}(m,\bx)(p_{11}(\bx)-p_{01}(\bx))+p_{01}(\bx)}\frac{\xi_{M}^{(1)}(m,\bx)(p_{11}(\bx)-p_{01}(\bx))+p_{01}(\bx)}{p_{01}(\bx)/\xi_Y^{(1)}(m,\bx)+\xi_{M}^{(1)}(m,\bx)(p_{11}(\bx)-p_{01}(\bx))},& \text{if } m\geq 1,\\
    \left\{\frac{1}{r_{11}(0,\bx)}-\displaystyle\sum_{j=1}^{m_{\max}} \frac{\xi_{M}^{(1)}(j,\bx)p_{11}(\bx)r_{11}(j,\bx)/r_{11}(0,\bx)}{\xi_{M}^{(1)}(j,\bx)(p_{11}(\bx)-p_{01}(\bx))+p_{01}(\bx)}\right\}\frac{\xi_{M}^{(1)}(0,\bcx)(p_{11}(\bx)-p_{01}(\bx))+p_{01}(\bx)}{p_{01}(\bx)/\xi_Y^{(1)}(0,\bx)+\xi_{M}^{(1)}(0,\bx)(p_{11}(\bx)-p_{01}(\bx))},              & \text{if } m = 0.
\end{cases} \\
w_{10}^{(10)}(m,\bx) &=
\begin{cases}
    \frac{\xi_{M}^{(0)}(m,\bx)p_{00}(\bx)}{\xi_{M}^{(0)}(m,\bx)(p_{11}(\bx)-p_{01}(\bcx))+p_{10}(\bx)}\frac{\xi_{M}^{(1)}(m,\bcx)(p_{11}(\bx)-p_{01}(\bx))+p_{01}(\bx)}{p_{01}(\bx)/\xi_Y^{(1)}(m,\bx)+\xi_{M}^{(1)}(m,\bx)(p_{11}(\bx)-p_{01}(\bx))},& \text{if } m\geq 1,\\
    \left\{\frac{1}{r_{00}(0,\bx)}-\displaystyle\sum_{j=1}^{m_{\max}} \frac{\xi_{M}^{(0)}(j,\bx)p_{00}(\bx)r_{00}(j,\bx)/r_{00}(0,\bx)}{\xi_{M}^{(0)}(j,\bx)(p_{11}(\bx)-p_{01}(\bx))+p_{10}(\bx)}\right\}\frac{\xi_{M}^{(1)}(0,\bcx)(p_{11}(\bx)-p_{01}(\bx))+p_{01}(\bx)}{p_{01}(\bx)/\xi_Y^{(1)}(0,\bx)+\xi_{M}^{(1)}(0,\bx)(p_{11}(\bx)-p_{01}(\bx))},              & \text{if } m = 0.
\end{cases} \\
w_{10}^{(00)}(m,\bx) &=
\begin{cases}
    \frac{\xi_{M}^{(0)}(m,\bx)p_{00}(\bx)}{\xi_{M}^{(0)}(m,\bx)(p_{11}(\bx)-p_{01}(\bcx))+p_{10}(\bx)}\frac{\xi_{M}^{(0)}(m,\bx) \left(p_{11}(\bx)-p_{01}(\bx)\right)+p_{10}(\bx)}{p_{10}(\bx)/\xi_Y^{(0)}(m,\bx)+\xi_{M}^{(0)}(m,\bx) \left(p_{11}(\bx)-p_{01}(\bx)\right)},& \text{if } m\geq 1,\\
    \left\{\frac{1}{r_{00}(0,\bx)}-\displaystyle\sum_{j=1}^{m_{\max}} \frac{\xi_{M}^{(0)}(j,\bx)p_{00}(\bx)r_{00}(j,\bx)/r_{00}(0,\bx)}{\xi_{M}^{(0)}(j,\bx)(p_{11}(\bx)-p_{01}(\bx))+p_{10}(\bx)}\right\}\frac{\xi_{M}^{(0)}(0,\bx) \left(p_{11}(\bx)-p_{01}(\bx)\right)+p_{10}(\bx)}{p_{10}(\bx)/\xi_Y^{(0)}(0,\bx)+\xi_{M}^{(0)}(0,\bx) \left(p_{11}(\bx)-p_{01}(\bx)\right)},              & \text{if } m = 0.
\end{cases} \\
w_{00}^{(11)}(m,\bx) &= 1 \text{ for any $m$}.\\
w_{00}^{(10)}(m,\bx) &=
\begin{cases}
   \frac{p_{00}(\bx)}{\xi_{M}^{(0)}(m,\bx)(p_{11}(\bx)-p_{01}(\bx))+p_{10}(\bx)},& \text{if } m\geq 1,\\
    \frac{1}{r_{00}(0,\bx)}-\displaystyle\sum_{j=1}^{m_{\max}} \frac{p_{00}(\bx)r_{00}(j,\bx)/r_{00}(0,\bx)}{\xi_{M}^{(0)}(j,\bx)(p_{11}(\bx)-p_{01}(\bx))+p_{10}(\bx)},              & \text{if } m = 0.
\end{cases} \\
w_{00}^{(00)}(m,\bx) &=
\begin{cases}
   \frac{p_{00}(\bx)}{\xi_{M}^{(0)}(m,\bx)(p_{11}(\bx)-p_{01}(\bx))+p_{10}(\bx)}\frac{\xi_{M}^{(0)}(m,\bx) \left(p_{11}(\bx)-p_{01}(\bx)\right)+p_{10}(\bx)}{p_{10}(\bx)+\xi_Y^{(0)}(m,\bx)\xi_{M}^{(0)}(m,\bx) \left(p_{11}(\bx)-p_{01}(\bx)\right)},& \text{if } m\geq 1,\\
    \left\{\frac{1}{r_{00}(0,\bx)}-\displaystyle\sum_{j=1}^{m_{\max}} \frac{p_{00}(\bx)r_{00}(j,\bx)/r_{00}(0,\bx)}{\xi_{M}^{(0)}(j,\bx)(p_{11}(\bx)-p_{01}(\bx))+p_{10}(\bx)}\right\}\frac{\xi_{M}^{(0)}(0,\bx) \left(p_{11}(\bx)-p_{01}(\bx)\right)+p_{10}(\bx)}{p_{10}(\bx)+\xi_Y^{(0)}(0,\bx)\xi_{M}^{(0)}(0,\bx) \left(p_{11}(\bx)-p_{01}(\bx)\right)},              & \text{if } m = 0.
\end{cases} \\
w_{11}^{(11)}(m,\bx) &= \begin{cases}
   \frac{p_{11}(\bx)}{\xi_{M}^{(1)}(m,\bx)(p_{11}(\bx)-p_{01}(\bx))+p_{01}(\bx)}\frac{\xi_{M}^{(1)}(m,\bcx) \left(p_{11}(\bx)-p_{01}(\bx)\right)+p_{01}(\bx)}{p_{01}(\bx)+\xi_Y^{(1)}(m,\bx)\xi_{M}^{(1)}(m,\bx) \left(p_{11}(\bx)-p_{01}(\bx)\right)},& \text{if } m\geq 1,\\
    \left\{\frac{1}{r_{11}(0,\bx)}-\displaystyle\sum_{j=1}^{m_{\max}} \frac{p_{11}(\bx)r_{11}(j,\bx)/r_{11}(0,\bx)}{\xi_{M}^{(1)}(j,\bx)(p_{11}(\bx)-p_{01}(\bx))+p_{01}(\bx)}\right\}\frac{\xi_{M}^{(1)}(0,\bcx) \left(p_{11}(\bx)-p_{01}(\bx)\right)+p_{01}(\bx)}{p_{01}(\bx)+\xi_Y^{(1)}(0,\bx)\xi_{M}^{(1)}(0,\bx) \left(p_{11}(\bx)-p_{01}(\bx)\right)},              & \text{if } m = 0.
\end{cases}\\
w_{11}^{(10)}(m,\bx) &= \frac{\xi_{M}^{(1)}(m,\bcx) \left(p_{11}(\bx)-p_{01}(\bx)\right)+p_{01}(\bx)}{p_{01}(\bx)+\xi_Y^{(1)}(m,\bx)\xi_{M}^{(1)}(m,\bx) \left(p_{11}(\bx)-p_{01}(\bx)\right)} \text{ for any $m$.}\\
w_{11}^{(00)}(m,\bx) &= 1 \text{ for any $m$.} 
\end{align*}\endgroup

Next, we prove Propositions \ref{proposition:sen_pi} and \ref{proposition:mr_xi}, which include identification results and properties of the multiply robust estimator under violation of principal ignorability. We first provide two lemmas. 
\begin{lemma}\label{lemma:sa_1}
Under Assumptions 1, 2, 3a, and 6 and the proposed confounding function $\xi = \left\{\left(\xi_{M}^{(1)}(m,\bx), \xi_{M}^{(0)}(m,\bx)\right) \text{ for }m\geq 1\text{ and }\left(\xi_{Y}^{(1)}(m,\bx),\xi_{Y}^{(0)}(m,\bx)\right) \text{ for } m \geq 0\right\}$, we can nonparametrically identify the distribution of $M_z|\{U=d_1d_0,\bcx\}$ for all $z\in\{1,0\}$ and $d_1d_0\in \mathcal U_{2-sided}$. Specifically, we have that $f_{M_0|U,\bcx}(m|11,\bx)=r_{01}(m,\bx)$ and $f_{M_1|U,\bcx}(m|00,\bx)=r_{10}(m,\bx)$ for any $m=0,\dots,m_{\max}$ and 
\begin{align*}
f_{M_1|U,\bcx}(m|10,\bx) & = \frac{\xi_{M}^{(1)}(m,\bx)p_{11}(\bx)}{\xi_{M}^{(1)}(m,\bx)(p_{11}(\bx)-p_{01}(\bx))+p_{01}(\bx)}r_{11}(m,\bx), \\
f_{M_0|U,\bcx}(m|10,\bx) & = \frac{\xi_{M}^{(0)}(m,\bx)p_{00}(\bx)}{\xi_{M}^{(0)}(m,\bx)(p_{11}(\bx)-p_{01}(\bx))+p_{10}(\bx)}r_{00}(m,\bx), \\
f_{M_1|U,\bcx}(m|11,\bx) & = \frac{p_{11}(\bx)}{\xi_{M}^{(1)}(m,\bx)(p_{11}(\bx)-p_{01}(\bx))+p_{01}(\bx)}r_{11}(m,\bx), \\
f_{M_0|U,\bcx}(m|00,\bx) & = \frac{p_{00}(\bx)}{\xi_{M}^{(0)}(m,\bx)(p_{11}(\bx)-p_{01}(\bx))+p_{10}(\bx)}r_{00}(m,\bx), 
\end{align*}
for any $m=1,\dots,m_{\max}$. This suggests that, for $m=0$,
\begin{align*}
f_{M_1|U,\bcx}(0|10,\bx) & =1-\sum_{j=1}^{m_{\max}} \frac{\xi_{M}^{(1)}(j,\bx)p_{11}(\bx)}{\xi_{M}^{(1)}(j,\bx)(p_{11}(\bx)-p_{01}(\bx))+p_{01}(\bx)}r_{11}(j,\bx), \\
f_{M_0|U,\bcx}(0|10,\bx) & = 1-\sum_{j=1}^{m_{\max}} \frac{\xi_{M}^{(0)}(j,\bx)p_{00}(\bx)}{\xi_{M}^{(0)}(j,\bx)(p_{11}(\bx)-p_{01}(\bx))+p_{10}(\bx)}r_{00}(j,\bx), \\
f_{M_1|U,\bcx}(0|11,\bx) & = 1-\sum_{j=1}^{m_{\max}} \frac{p_{11}(\bx)}{\xi_{M}^{(1)}(j,\bx)(p_{11}(\bx)-p_{01}(\bx))+p_{01}(\bx)}r_{11}(j,\bx), \\
f_{M_0|U,\bcx}(0|00,\bx) & =1-\sum_{j=1}^{m_{\max}} \frac{p_{00}(\bx)}{\xi_{M}^{(0)}(j,\bx)(p_{11}(\bx)-p_{01}(\bx))+p_{10}(\bx)}r_{00}(j,\bx).
\end{align*}
\end{lemma}

\begin{proof}
We first show $f_{M_0|U,\bcx}(m|11,\bx)=r_{01}(m,\bx)$ for all $m\in\{0,\dots,m_{\max}\}$ and  similar results extend to $f_{M_1|U,\bcx}(m|00,\bx)=r_{10}(m,\bx)$. Specifically, for any $m\in\{0,\dots,m_{\max}\}$, we have that\begingroup\makeatletter\def\f@size{10}\check@mathfonts
\begin{align*}
f_{M_0|U,\bcx}(m|11,\bx) & = f_{M_0|Z,U,\bcx}(m|0,11,\bx) \\
& \quad \text{(by Lemma \ref{lemma:randomization2} and Lemma \ref{lemma:independent})} \\
& = f_{M_0|Z,D,U,\bcx}(m|0,1,11,\bx) \\
& \quad \text{(because $D$ must be 1 given $Z=0$ and $U=11$)} \\
& = f_{M_0|Z,D,\bcx}(m|0,1,\bx) \\
& \quad \text{(because the observed stratum with $D=0$ and $Z=1$ only contains the always-takers)} \\
& = r_{01}(m,\bx). 
\end{align*}\endgroup

Next, we derive the expression of $f_{M_1|U,\bcx}(m|10,\bx)$. Leveraging the same strategy, one can deduce the expressions of $f_{M_0|U,\bcx}(m|10,\bx)$, $f_{M_1|U,\bcx}(m|11,\bx)$, and $f_{M_0|U,\bcx}(m|00,\bx)$. For $m\geq 1$, we have that\begingroup\makeatletter\def\f@size{10}\check@mathfonts
\begin{align*}
r_{11}(m,\bx) = &  f_{M|Z,D,\bcx}(m|1,1,\bx) \\
= & f_{M_1|Z,D,U,\bcx}(m|1,1,10,\bx) f_{U|Z,D,\bcx}(10|1,1,\bx) + f_{M_1|Z,D,U,\bcx}(m|1,1,11,\bx) f_{U|Z,D,\bcx}(11|1,1,\bx) \\
& \quad\text{(followed by the law of total probability and $U=10$ or 11 in observed strata $(Z=1,D=1)$} \\
& \quad \quad \text{under standard monotonicity)}\\
= & f_{M_1|Z,U,\bcx}(m|1,10,\bx) f_{U|Z,D,\bcx}(10|1,1,\bx) + f_{M_1|Z,U,\bcx}(m|1,10,\bx) f_{U|Z,D,\bcx}(10|1,1,\bx)\\
& \quad \text{(because $D$ must be 1 given Z=1 and either $U=10$ or 11)} \\
= & f_{M_1|U,\bcx}(m|10,\bx) f_{U|Z,D,\bcx}(10|1,1,\bx) + f_{M_1|U,\bcx}(m|10,\bx) f_{U|Z,D,\bcx}(10|1,1,\bx)\\
& \quad \text{(by Lemma \ref{lemma:randomization2} and Lemma \ref{lemma:independent})} \\
= & f_{M_1|U,\bcx}(m|10,\bx) \frac{p_{11}(\bx)-p_{01}(\bx)}{p_{11}(\bx)} + f_{M_1|U,\bcx}(m|11,\bx) \frac{p_{01}(\bx)}{p_{11}(\bx)} \\
= & f_{M_1|U,\bcx}(m|10,\bx) \frac{p_{11}(\bx)-p_{01}(\bx)}{p_{11}(\bx)} + f_{M_1|U,\bx}(m|10,\bx) \frac{p_{01}(\bx)}{\xi_{M}^{(1)}(m,\bx)p_{11}(\bx)} \\
= & f_{M_1|U,\bcx}(m|10,\bx) \frac{\xi_{M}^{(1)}(m,\bx)\left(p_{11}(\bx)-p_{01}(\bx)\right)+p_{01}(\bx)}{\xi_{M}^{(1)}(m,\bx) p_{11}(\bx)},
\end{align*}\endgroup
which indicates 
$f_{M_1|U,\bcx}(m|10,\bx) = \frac{\xi_{M}^{(1)}(m,\bx) p_{11}(\bx)}{\xi_{M}^{(1)}(m,\bx)\left(p_{11}(\bx)-p_{01}(\bx)\right)+p_{01}(\bx)}r_{11}(m,\bx)$ for $m\in\{1,\dots,m_{\max}\}$. Because $\sum_{j=0}^{m_{\max}}f_{M_1|U,\bcx}(m|10,\bx) = 1$, we obtain 
$$
f_{M_1|U,\bcx}(0|10,\bx)  =1-\sum_{j=1}^{m_{\max}} \frac{\xi_{M}^{(1)}(j,\bx)p_{11}(\bx)}{\xi_{M}^{(1)}(j,\bx)(p_{11}(\bx)-p_{01}(\bx))+p_{01}(\bx)}r_{11}(j,\bx).
$$
\end{proof}

\begin{lemma}\label{lemma:sa_2}
Under Assumptions 1, 2, 3a, and 6 and the proposed confounding function $\xi = \left\{\left(\xi_{M}^{(1)}(m,\bx), \xi_{M}^{(0)}(m,\bx)\right) \text{ for }m\geq 1\text{ and }\left(\xi_{Y}^{(1)}(m,\bx),\xi_{Y}^{(0)}(m,\bx)\right) \text{ for } m \geq 0\right\}$, we can nonparametrically identify the $\E_{Y_{zm}|U,\bcx}(Y_{zm}|d_1d_0,\bx)$ for all $z\in\{1,0\}$, $m\in\{0,\dots,m_{\max}\}$, and $d_1d_0\in \mathcal U_{\text{2-sided}}$. Specifically, we have that 
\begin{align*}
\E_{Y_{1m}|U,\bcx}[Y_{1m}|10,\bx] & = \frac{\xi_{M}^{(1)}(m,\bx) \left(p_{11}(\bx)-p_{01}(\bx)\right)+p_{01}(\bx)}{p_{01}(\bx)/\xi_Y^{(1)}(m,\bx)+\xi_{M}^{(1)}(m,\bx) \left(p_{11}(\bx)-p_{01}(\bx)\right)} \mu_{11}(m,\bx), \\
\E_{Y_{0m}|U,\bcx}[Y_{0m}|10,\bx] & = \frac{\xi_{M}^{(0)}(m,\bx) \left(p_{11}(\bx)-p_{01}(\bx)\right)+p_{10}(\bx)}{p_{10}(\bx)/\xi_Y^{(0)}(m,\bx)+\xi_{M}^{(0)}(m,\bx) \left(p_{11}(\bx)-p_{01}(\bx)\right)} \mu_{00}(m,\bx),\\
\E_{Y_{1m}|U,\bcx}[Y_{1m}|11,\bx] & = \frac{\xi_{M}^{(1)}(m,\bcx) \left(p_{11}(\bx)-p_{01}(\bx)\right)+p_{01}(\bx)}{p_{01}(\bx)+\xi_Y^{(1)}(m,\bx)\xi_{M}^{(1)}(m,\bx) \left(p_{11}(\bx)-p_{01}(\bx)\right)} \mu_{11}(m,\bx),\\
\E_{Y_{0m}|U,\bcx}[Y_{0m}|11,\bx] & =\mu_{01}(m,\bx), \\
\E_{Y_{1m}|U,\bcx}[Y_{1m}|00,\bx] & =\mu_{10}(m,\bx), \\
\E_{Y_{0m}|U,\bcx}[Y_{0m}|00,\bx] & = \frac{\xi_{M}^{(0)}(m,\bx) \left(p_{11}(\bx)-p_{01}(\bx)\right)+p_{10}(\bx)}{p_{10}(\bx)+\xi_Y^{(0)}(m,\bx)\xi_{M}^{(0)}(m,\bx) \left(p_{11}(\bx)-p_{01}(\bx)\right)} \mu_{00}(m,\bx),
\end{align*}
for any $m=0,\dots,m_{\max}$, where $\xi_{M}^{(1)}(m,\bx)$ and $\xi_{M}^{(1)}(m,\bx)$ are given in $\xi$ for $m \geq 1$ and for $m=0$, 
\begin{align*}
\xi_{M}^{(1)}(0,\bx)  & =:\frac{f_{M_1|U,\bcx}(0|10,\bx)}{f_{M_1|U,\bcx}(0|11,\bx)} = \frac{1-\sum_{j=1}^{m_{\max}}\displaystyle\frac{\xi_{M}^{(1)}(j,\bx)p_{11}(\bx)}{\xi_{M}^{(1)}(j,\bx)(p_{11}(\bx)-p_{01}(\bx))+p_{01}(\bx)}r_{11}(j,\bx)}{1-\sum_{j=1}^{m_{\max}} \displaystyle\frac{p_{11}(\bx)}{\xi_{M}^{(1)}(j,\bx)(p_{11}(\bx)-p_{01}(\bx))+p_{01}(\bx)}r_{11}(j,\bx)}, \\
\xi_{M}^{(0)}(0,\bx) & =:\frac{f_{M_0|U,\bcx}(0|10,\bx)}{f_{M_0|U,\bcx}(0|00,\bx)} = \frac{1-\sum_{j=1}^{m_{\max}}\displaystyle\frac{\xi_{M}^{(0)}(j,\bx)p_{00}(\bx)}{\xi_{M}^{(0)}(j,\bx)(p_{11}(\bx)-p_{01}(\bx))+p_{10}(\bx)}r_{00}(j,\bx)}{1-\sum_{j=1}^{m_{\max}} \displaystyle\frac{p_{00}(\bx)}{\xi_{M}^{(0)}(j,\bx)(p_{11}(\bx)-p_{01}(\bx))+p_{10}(\bx)}r_{00}(j,\bx)}. 
\end{align*}
\end{lemma}

\begin{proof}
We first show $\E_{Y_{0m}|U,\bcx}(Y_{0m}|11,\bx) =\mu_{01}(m,\bx)$ and similar result extends to $\E_{Y_{1m}|U,\bcx}(Y_{1m}|00,\bx) =\mu_{10}(m,\bx)$. Specifically, one can verify\begingroup\makeatletter\def\f@size{10}\check@mathfonts
\begin{align*}
\E_{Y_{0m}|U,\bcx}(Y_{0m}|11,\bx) & = \E_{Y_{0m}|Z,U,\bcx}(Y_{0m}|0,11,\bx) \\
& \quad \text{(by Lemma \ref{lemma:randomization2} and Lemma \ref{lemma:independent})} \\
& = \E_{Y_{0m}|Z,M,U,\bcx}(Y_{0m}|0,m,11,\bx)\\
& \quad \text{(by Assumption 5)} \\
& = \E_{Y_{0m}|Z,D, M,U,\bcx}(Y_{0m}|0,1, m,11,\bx)\\
& \quad \text{(because $D$ must be 1 given $Z=0$ and $U=11$)} \\
& = \E_{Y_{0m}|Z,D, M,\bcx}(Y_{0m}|0,1, m,\bx) \\
& \quad \text{(because the observed stratum with $D=0$ and $Z=1$ only contains the always-takers)} \\
& = \mu_{01}(m,\bx). 
\end{align*}\endgroup

Next, we derive the expression of $\E_{Y_{1m}|U,\bcx}(Y_{1m}|10,\bx)$. Notice that the expressions of $\E_{Y_{0m}|U,\bcx}(Y_{0m}|10,\bx)$, $\E_{Y_{1m}|U,\bcx}(Y_{1m}|11,\bx)$, and $\E_{Y_{0m}|U,\bcx}(Y_{0m}|00,\bx)$ can be similarly obtained. Specifically, for any $m\geq 0$,\begingroup\makeatletter\def\f@size{10}\check@mathfonts
\begin{align*}
  & \mu_{11}(m,\bx) \\
= &  \E_{Y|Z,D,M,\bcx}[Y_{1m}|1,1,m,\bx] =  \E_{Y_{1m}|Z,D,M_1,\bcx}[Y_{1m}|1,1,m,\bx] \\
= & \E_{Y_{1m}|Z,D,M_1,U,\bcx}[Y_{1m}|1,1,m,10,\bx]f_{U|Z,D,M_1,\bcx}(10|1,1,m,\bx) \\
& + \E_{Y_{1m}|Z,D,M_1,U,\bcx}[Y_{1m}|1,1,m,11,\bx]f_{U|Z,D,M_1,\bcx}(11|1,1,m,\bx) \\
& \quad\text{(followed by the law of iterated expectation and $U=10$ or 11 in observed strata $(Z=1,D=1)$} \\
& \quad \quad \text{under standard monotonicity)}\\
= & \E_{Y_{1m}|Z,M_1,U,\bcx}[Y_{1m}|1,m,10,\bx]f_{U|Z,D,M_1,\bcx}(10|1,1,m,\bx) + \E_{Y_{1m}|Z,M_1,U,\bcx}[Y_{1m}|1,m,11,\bx]f_{U|Z,D,M_1,\bcx}(11|1,1,m,\bx)\\
& \quad \text{(because $D$ must be 1 given Z=1 and either $U=10$ or 11)} \\
= & \E_{Y_{1m}|Z,U,\bcx}[Y_{1m}|1,10,\bx]f_{U|Z,D,M_1,\bcx}(10|1,1,m,\bx) + \E_{Y_{1m}|Z,U,\bcx}[Y_{1m}|1,11,\bx]f_{U|Z,D,M_1,\bcx}(11|1,1,m,\bx)\\
& \quad \text{(followed by Assumption 5)} \\
= & \E_{Y_{1m}|U,\bcx}[Y_{1m}|10,\bx]f_{U|Z,D,M_1,\bcx}(10|1,1,m,\bx) + \E_{Y_{1m}|U,\bcx}[Y_{1m}|11,\bx]f_{U|Z,D,M_1,\bcx}(11|1,1,m,\bx),\\
& \quad \text{(by Lemma \ref{lemma:randomization2} and Lemma \ref{lemma:independent})} 
\end{align*}\endgroup
where\begingroup\makeatletter\def\f@size{10}\check@mathfonts
\begin{align*}
f_{U|Z,D,M_1,\bcx}(10|1,1,m,\bx)  
=& \frac{f_{M_1|Z,D,U,\bcx}(m|1,1,10,\bx)f_{U|Z,D,\bcx}(10|1,1,\bx)}{\displaystyle\sum_{u \in\{10,11\}} f_{M_1|Z,D,U,\bcx}(m|1,1,u,\bx)f_{U|Z,D,\bcx}(u|1,1,\bx)}  \\
= & \frac{f_{M_1|U,\bcx}(m|10,\bx)f_{U|Z,D,\bcx}(10|1,1,\bx)}{\displaystyle\sum_{u \in\{10,11\}} f_{M_1|U,\bcx}(m|u,\bx)f_{U|Z,D,\bcx}(u|1,1,\bx)}  \\
= & \frac{f_{M_1|U,\bcx}(m|10,\bx)\displaystyle\frac{p_{11}(\bx)-p_{01}(\bx)}{p_{11}(\bx)}}{f_{M_1|U,\bcx}(m|10,\bx)\displaystyle\frac{p_{11}(\bx)-p_{01}(\bx)}{p_{11}(\bx)}+f_{M_1|U,\bcx}(m|11,\bx)\displaystyle\frac{p_{01}(\bx)}{p_{11}(\bx)}} \\
= & \frac{\displaystyle\frac{p_{11}(\bx)-p_{01}(\bx)}{p_{11}(\bx)}}{\displaystyle\frac{p_{11}(\bx)-p_{01}(\bx)}{p_{11}(\bx)}+\displaystyle\frac{p_{01}(\bx)}{\xi_{M}^{(1)}(m,\bx)p_{11}(\bx)}} = \frac{\xi_{M}^{(1)}(m,\bx)\left(p_{11}(\bx)-p_{01}(\bx)\right)}{\xi_{M}^{(1)}(m,\bx) \left(p_{11}(\bx)-p_{01}(\bx)\right)+p_{01}(\bx)}.
\end{align*}\endgroup
and
\begin{align*}
f_{U|Z,D,M_1,\bcx}(11|1,1,m,\bx) & = 1-f_{U|Z,D,M_1,\bcx}(10|1,1,m,\bx) \\
& = \frac{p_{01}(\bx)}{\xi_{M}^{(1)}(m,\bx) \left(p_{11}(\bx)-p_{01}(\bx)\right)+p_{01}(\bx)}.
\end{align*}
This suggests that\begingroup\makeatletter\def\f@size{9.6}\check@mathfonts
\begin{align*}
& \mu_{11}(m,\bx)  \\
= &  \E_{Y_{1m}|U,\bcx}[Y_{1m}|10,\bx] \frac{\xi_{M}^{(1)}(m,\bx)\left(p_{11}(\bx)\!-\!p_{01}(\bx)\right)}{\xi_{M}^{(1)}(m,\bx) \left(p_{11}(\bx)\!-\!p_{01}(\bx)\right)+p_{01}(\bx)} \!+\! \E_{Y_{1m}|U,\bcx}[Y_{1m}|11,\bx]\frac{p_{01}(\bx)}{\xi_{M}^{(1)}(m,\bx) \left(p_{11}(\bx)-p_{01}(\bx)\right)+p_{01}(\bx)} \\
= & \E_{Y_{1m}|U,\bcx}[Y_{1m}|10,\bx] \frac{\xi_{M}^{(1)}(m,\bx)\left(p_{11}(\bx)\!-\!p_{01}(\bx)\right)}{\xi_{M}^{(1)}(m,\bx) \left(p_{11}(\bx)\!-\!p_{01}(\bx)\right)+p_{01}(\bx)} \!+\! \frac{\E_{Y_{1m}|U,\bcx}[Y_{1m}|11,\bx]}{\xi_Y^{(1)}(m,\bx)}\frac{p_{01}(\bx)}{\xi_{M}^{(1)}(m,\bx) \left(p_{11}(\bx)-p_{01}(\bx)\right)+p_{01}(\bx)} \\
= & \E_{Y_{1m}|U,\bcx}[Y_{1m}|10,\bx] \times \frac{p_{01}(\bx)/\xi_Y^{(1)}(m,\bx)+\xi_{M}^{(1)}(m,\bx) \left(p_{11}(\bx)-p_{01}(\bx)\right)}{\xi_{M}^{(1)}(m,\bx) \left(p_{11}(\bx)-p_{01}(\bx)\right)+p_{01}(\bx)}
\end{align*}\endgroup
and thus
$$
\E_{Y_{1m}|U,\bcx}[Y_{1m}|10,\bx] =  \frac{\xi_{M}^{(1)}(m,\bx) \left(p_{11}(\bx)-p_{01}(\bx)\right)+p_{01}(\bx)}{p_{01}(\bx)/\xi_Y^{(1)}(m,\bx)+\xi_{M}^{(1)}(m,\bx) \left(p_{11}(\bx)-p_{01}(\bx)\right)} \mu_{11}(m,\bx).
$$
This completes the proof.
\end{proof}

Next, we prove the nonparametric formulas $\theta_{d_1d_0}^{(zz')}$ in Proposition \ref{proposition:sen_pi}.\medskip

\noindent \textbf{\textit{Proof of Proposition \ref{proposition:sen_pi}.}} We will derive the nonparametric identification formula for $\theta_{10}^{(10)}$ and omit the similar proofs for all other $\theta_{d_1d_0}^{(zz')}$ since the steps are similar. By the definition of $\theta_{10}^{(10)}$, we have that
\begin{align*}
\theta_{10}^{(10)} & = \E[Y_{1M_{0}}|U=10] \\
& = \E\left[\E[Y_{1M_0}|U=10,\bcx]\Big|U=10\right] \quad \text{(by law of iterated expectations)} \\
& = \E\left[\sum_{m=0}^{m_{\max}} \E_{Y_{1m}|M_0,U,\bcx}[Y_{1m}|m,10,\bcx]f_{M_{0}|U,\bcx}(m|10,\bcx) \Big|U=10\right] \\
& = \E\left[\sum_{m=0}^{m_{\max}} \E_{Y_{1m}|Z,M_0,U,\bcx}[Y_{1m}|1,m,10,\bcx]f_{M_{0}|U,\bcx}(m|10,\bcx) \Big|U=10\right] \\
& \quad \quad \text{(by Lemma \ref{lemma:randomization2} coupled with Lemma \ref{lemma:independent})} \\
& = \E\left[\sum_{m=0}^{m_{\max}} \E_{Y_{1m}|Z,U,\bcx}[Y_{1m}|1,10,\bcx]f_{M_{0}|U,\bcx}(m|10,\bcx) \Big|U=10\right] \quad\quad \text{(by Assumption 5)} \\
& = \E\left[\sum_{m=0}^{m_{\max}} \E_{Y_{1m}|U,\bcx}[Y_{1m}|10,\bcx]f_{M_{0}|U,\bcx}(m|10,\bcx) \Big|U=10\right] \quad \text{(by Lemma \ref{lemma:randomization2})} \\
& = \E\left[\sum_{m=0}^{m_{\max}} w_{10}^{(10)}(m,\bcx) \mu_{11}(m,\bcx)r_{00}(m,\bcx)  \Big|U=10\right] \quad\quad\quad \text{(by Lemmas \ref{lemma:sa_1} and \ref{lemma:sa_2})} \\
&  = \int_{\bx} \frac{f_{U|\bcx}(10|\bx)}{f_{U}(10)} \left\{\sum_{m=0}^{m_{\max}} w_{10}^{(10)}(m,\bx) \mu_{11}(m,\bx)r_{00}(m,\bx) \right\}\differential\Prob_{\bcx}(\bx) \quad \text{(by Lemma \ref{lemma:expectation})}\\
&  = \int_{\bx} \frac{e_{10}(\bx)}{e_{10}} \left\{\sum_{m=0}^{m_{\max}} w_{10}^{(10)}(m,\bx) \mu_{11}(m,\bx)r_{00}(m,\bx) \right\}\differential\Prob_{\bcx}(\bx).
\end{align*}
This completes the proof.  \hfill$\square$

Finally, we prove properties of the multiply robust estimator $\widehat\theta_{d_1d_0}^{(zz')}(\xi)$ in Proposition \ref{proposition:mr_xi}.\medskip

\noindent \textbf{\textit{Proof of Proposition \ref{proposition:mr_xi}.}} Following the notation in the proof of Theorem 4, we let $\widetilde h_{nuisance} =\{\widetilde \pi_z(\bx), \widetilde p_{zd}(\bx), \widetilde r_{zd}(m,\bx),  \widetilde \mu_{zd}(m,\bx) \}$ be probability limit of $\widehat h_{nuisance}^{\text{par}}$, where $\widetilde \pi_z(\bx)=\pi_z(\bx)$, $\widetilde p_{zd}(\bx)= p_{zd}(\bx)$, $\widetilde r_{zd}(m,\bx)=r_{zd}(m,\bx)$, $\widetilde \mu_{zd}(m,\bx)=\mu_{zd}(m,\bx)$, under $\mathcal M_\pi$, $\mathcal M_e$, $\mathcal M_m$, and $\mathcal M_o$ respectively. Under the condition of either $\mathcal M_{\pi} \cap \mathcal M_{e} \cap \mathcal M_{m}$ or $ \mathcal M_{e} \cap \mathcal M_{m} \cap \mathcal M_{o}$, we always have $\widetilde p_{zd}(\bx)= p_{zd}(\bx)$ and $\widetilde r_{zd}(m,\bx)=r_{zd}(m,\bx)$. Also, because the sensitivity weight $w_{d_1d_0}^{(zz')}(m,\bx)$ only depends on the confounding function $\xi$ and the observed-data nuisance functions $\{p_{zd}(\bx),r_{zd}(m,\bx)\}$, we can confirm that the probability limit of $\widehat w_{d_1d_0}^{(zz'),\text{par}}(m,\bx)$ is equal to $w_{d_1d_0}^{(zz')}(m,\bx)$ under $\mathcal M_{\pi} \cap \mathcal M_{e} \cap \mathcal M_{m}$ or $ \mathcal M_{e} \cap \mathcal M_{m} \cap \mathcal M_{o}$. In addition, we can show that the probability limit of $\widehat p_{zd}^{\text{dr}}$, denoted by $\widetilde p_{zd}$, is equal to the true value $p_{zd}$ under $\mathcal M_{\pi} \cap \mathcal M_{e} \cap \mathcal M_{m}$ or $ \mathcal M_{e} \cap \mathcal M_{m} \cap \mathcal M_{o}$, because $\widehat p_{zd}^{\text{dr}}$ is a doubly robust estimator under $\mathcal M_{\pi} \cup \mathcal M_{e}$. The previous discussion suggests that the probability limit of $\widehat{\theta}_{d_1d_0}^{(zz'),\text{mr}}(\xi)$ is 
\begin{align}
\theta^{(zz'),\text{mr}}_{d_1d_0}(\xi) = & \E\Big\{\left(\frac{\mathbb{I}(Z=z^*)\left\{\mathbb{I}(D=d^*)- p_{z^*d^*}(\bcx)\right\}}{\widetilde\pi_{z^*}(\bcx)} - k\frac{(1-Z)\left\{D- p_{01}(\bcx)\right\}}{\widetilde\pi_{0}(\bcx)}\right) \frac{\widetilde\eta_{zz'}^{w}(\bcx)}{ p_{z^*d^*} - kp_{01}} \nonumber \\
& + \frac{p_{z^*d^*}(\bcx)-kp_{01}(\bcx)}{p_{z^*d^*} - kp_{01}}\frac{\mathbb{I}(D=d_z,Z=z)}{p_{zd_z}(\bcx)\widetilde\pi_{z}(\bcx)}\frac{ r_{z'd_{z'}}(M,\bcx)}{r_{zd_z}(M,\bcx)}w_{d_1d_0}^{(zz')}(M,\bcx)\left\{Y-\widetilde \mu_{zd_z}(M,\bcx)\right\} \nonumber \\
& + \frac{p_{z^*d^*}(\bcx)-kp_{01}(\bcx)}{ p_{z^*d^*} - kp_{01}}\frac{\mathbb{I}(D=d_{z'},Z=z')}{ p_{z'd_{z'}}(\bcx)\widetilde \pi_{z'}(\bcx)}\left\{w_{d_1d_0}^{(zz')}(M,\bcx)\widetilde \mu_{zd_z}(M,\bcx)-\widetilde \eta_{zz'}^w(\bcx)\right\} \nonumber \\
& + \frac{p_{z^*d^*}(\bcx)-kp_{01}(\bcx)}{p_{z^*d^*} - kp_{01}}\widetilde \eta_{zz'}^w(\bcx) \Big\}\nonumber
\end{align}
under $\mathcal M_{\pi} \cap \mathcal M_{e} \cap \mathcal M_{m}$ or $ \mathcal M_{e} \cap \mathcal M_{m} \cap \mathcal M_{o}$, where $\widetilde\eta_{zz'}^w(\bcx)=\sum_{m=0}^{m_{\max}} w_{d_1d_0}^{(zz')}(m,\bcx)\widetilde{\mu}_{zd_z}(m,\bcx)r_{z'd_{z'}}(m,\bcx)$. In what follows, we show that $\theta^{(zz'),\text{mr}}_{d_1d_0}(\xi) = \theta_{d_1d_0}^{(zz')}$ under Scenario I ($\mathcal M_{\pi} \cap \mathcal M_{e} \cap \mathcal M_{m}$) or Scenario II ($\mathcal M_{e} \cap \mathcal M_{m} \cap \mathcal M_{o}$), which concludes the double robustness of $\widehat{\theta}_{d_1d_0}^{(zz'),\text{mr}}(\xi)$.\medskip

\noindent \textbf{Scenario I ($\mathcal M_{\pi} \cap \mathcal M_{e} \cap \mathcal M_{m}$):.}\medskip

In Scenario I, $\widetilde \pi_z(\bx)=\pi_z(\bx)$ but generally $\widetilde \mu_{zd}(m,\bx)\neq \mu_{zd}(m,\bx)$. Therefore, we can rewrite $\theta^{(zz'),\text{mr}}_{d_1d_0}(\xi) = \sum_{j=1}^4 \Delta_j$, where\begingroup\makeatletter\def\f@size{9}\check@mathfonts
\begin{align*}
\Delta_1 & = \E\left[\left(\frac{\mathbb{I}(Z=z^*)\left\{\mathbb{I}(D=d^*)- p_{z^*d^*}(\bcx)\right\}}{\pi_{z^*}(\bcx)} - k\frac{(1-Z)\left\{D- p_{01}(\bcx)\right\}}{\pi_{0}(\bcx)}\right) \frac{\displaystyle\sum_{m=0}^{m_{\max}} w_{d_1d_0}^{(zz')}(m,\bcx)\widetilde{\mu}_{zd_z}(m,\bcx) r_{z'd_{z'}}(m,\bcx) }{ p_{z^*d^*} - k p_{01}}\right], \\
\Delta_2 & = \E\left[\frac{ p_{z^*d^*}(\bcx)-k p_{01}(\bcx)}{ p_{z^*d^*} - k p_{01}}\frac{\mathbb{I}(D=d_z,Z=z)}{ p_{zd_z}(\bcx)\pi_{z}(\bcx)}\frac{ r_{z'd_{z'}}(M,\bcx)}{ r_{zd_z}(M,\bcx)}w_{d_1d_0}^{(zz')}(M,\bcx)Y\right], \\
\Delta_3 & = \E\left[\frac{ p_{z^*d^*}(\bcx)-k p_{01}(\bcx)}{p_{z^*d^*} - k p_{01}}\left\{\frac{\mathbb{I}(D=d_{z'},Z=z')}{p_{z'd_{z'}}(\bcx) \pi_{z'}(\bcx)}-\frac{\mathbb{I}(D=d_z,Z=z)}{ p_{zd_z}(\bcx)\pi_{z}(\bcx)}\frac{ r_{z'd_{z'}}(M,\bcx)}{ r_{zd_z}(M,\bcx)}\right\}w_{d_1d_0}^{(zz')}(M,\bcx)\widetilde \mu_{zd_z}(M,\bcx)\right],\\
\Delta_4 & = \E\left[\frac{ p_{z^*d^*}(\bcx)-k p_{01}(\bcx)}{ p_{z^*d^*} - k p_{01}} \left\{1-\frac{\mathbb{I}(D=d_{z'},Z=z')}{ p_{z'd_{z'}}(\bcx) \pi_{z'}(\bcx)}\right\}\sum_{m=0}^{m_{\max}}w_{d_1d_0}^{(zz')}(m,\bcx)\widetilde{\mu}_{zd_z}(m,\bcx) r_{z'd_{z'}}(m,\bcx) \right].
\end{align*}\endgroup
One can show that $\Delta_1=\Delta_3=\Delta_4=0$ by using the law of iterated expectation and\begingroup\makeatletter\def\f@size{9}\check@mathfonts 
\begin{align*}
\Delta_2 = &  \int_{\bx}\frac{p_{z^*d^*}(\bx)-kp_{01}(\bx)}{p_{z^*d^*}-kp_{01}}\frac{1}{p_{zd_z}(\bx)\pi_{z}(\bx)}\sum_{m=0}^{m_{\max}}  \left\{\frac{r_{z'd_{z'}}(m,\bx)}{r_{zd_z}(m,\bx)}w_{d_1d_0}^{(zz')}(m,\bx)\int_y y\differential \Prob_{Y|Z,D,M\bcx}(y|z,d_{z},m,\bx)r_{zd_z}(m,\bx)\right\} \\
& \quad  f_{D|Z,\bcx}(d_z|z,\bx) f_{Z|\bcx}(z|\bx) \differential \Prob_{\bcx}(\bx) \\
= &  \int_{\bx}\frac{p_{z^*d^*}(\bx)-kp_{01}(\bx)}{p_{z^*d^*}-kp_{01}}\sum_{m=0}^{m_{\max}}  \left\{\frac{r_{z'd_{z'}}(m,\bx)}{r_{zd_z}(m,\bx)}w_{d_1d_0}^{(zz')}(m,\bx)\int_y y\differential \Prob_{Y|Z,D,M\bcx}(y|z,d_{z},m,\bx)r_{zd_z}(m,\bx)\right\}\differential \Prob_{\bcx}(\bx) \\
= &  \int_{\bx}\frac{p_{z^*d^*}(\bx)-kp_{01}(\bx)}{p_{z^*d^*}-kp_{01}}\sum_{m=0}^{m_{\max}}  \left\{\frac{r_{z'd_{z'}}(m,\bx)}{r_{zd_z}(m,\bx)}w_{d_1d_0}^{(zz')}(m,\bx)\mu_{zd_z}(m,\bx)r_{zd_z}(m,\bx)\right\}\differential \Prob_{\bcx}(\bx) \\
= &  \int_{\bx}\frac{p_{z^*d^*}(\bx)-kp_{01}(\bx)}{p_{z^*d^*}-kp_{01}}\sum_{m=0}^{m_{\max}}  \left\{w_{d_1d_0}^{(zz')}(m,\bx)\mu_{zd_z}(m,\bx)r_{z'd_{z'}}(m,\bx)\right\}\differential \Prob_{\bcx}(\bx) \\
= &  \int_{\bx}\frac{e_{d_1d_0}(\bx)}{e_{d_1d_0}}\sum_{m=0}^{m_{\max}}  \left\{w_{d_1d_0}^{(zz')}(m,\bx)\mu_{zd_z}(m,\bx)r_{z'd_{z'}}(m,\bx)\right\}\differential \Prob_{\bcx}(\bx) \\
= & \theta_{d_1d_0}^{(zz')},
\end{align*}\endgroup
which suggests that $\theta^{(zz'),\text{mr}}_{d_1d_0}(\xi) = \sum_{j=1}^4 \Delta_j=\theta^{(zz')}_{d_1d_0}$ under $\mathcal M_{\pi} \cap \mathcal M_{e} \cap \mathcal M_{m}$.

\noindent \textbf{Scenario II ($\mathcal M_{e} \cap \mathcal M_{m} \cap \mathcal M_{o}$):}\medskip

In Scenario II, $\widetilde \mu_{zd}(m,\bx)= \mu_{zd}(m,\bx)$ but generally $\widetilde \pi_z(\bx) \neq \pi_z(\bx)$. Therefore, we have $\theta^{(zz'),\text{mr}}_{d_1d_0}(\xi) = \sum_{j=1}^4 \Delta_j$, where\begingroup\makeatletter\def\f@size{9}\check@mathfonts
\begin{align*}
\Delta_1 & = \E\left[\left(\frac{\mathbb{I}(Z=z^*)\left\{\mathbb{I}(D=d^*)- p_{z^*d^*}(\bcx)\right\}}{\widetilde \pi_{z^*}(\bcx)} - k\frac{(1-Z)\left\{D- p_{01}(\bcx)\right\}}{\widetilde \pi_{0}(\bcx)}\right) \frac{\displaystyle\sum_{m=0}^{m_{\max}} w_{d_1d_0}^{(zz')}(m,\bcx){\mu}_{zd_z}(m,\bcx)r_{z'd_{z'}}(m,\bcx)}{ p_{z^*d^*} - k p_{01}}\right], \\
\Delta_2 & = \E\left[\frac{ p_{z^*d^*}(\bcx)-k p_{01}(\bcx)}{ p_{z^*d^*} - k p_{01}}\frac{\mathbb{I}(D=d_z,Z=z)}{ p_{zd_z}(\bcx)\widetilde\pi_{z}(\bcx)}\frac{ r_{z'd_{z'}}(M,\bcx)}{ r_{zd_z}(M,\bcx)}w_{d_1d_0}^{(zz')}(M,\bcx)\left\{Y- \mu_{zd_z}(M,\bcx)\right\}\right], \\
\Delta_3 & = \E\left[\frac{ p_{z^*d^*}(\bcx)-kp_{01}(\bcx)}{ p_{z^*d^*} - k p_{01}}\frac{\mathbb{I}(D=d_{z'},Z=z')}{ p_{z'd_{z'}}(\bcx)\widetilde \pi_{z'}(\bcx)}\left\{ w_{d_1d_0}^{(zz')}(M,\bcx) \mu_{zd_z}(M,\bcx)- \sum_{m=0}^{m_{\max}} w_{d_1d_0}^{(zz')}(m,\bcx){\mu}_{zd_z}(m,\bcx)r_{z'd_{z'}}(m,\bcx)\right\}\right],\\
\Delta_4 & = \E\left[\frac{ p_{z^*d^*}(\bcx)-k p_{01}(\bcx)}{ p_{z^*d^*} - k p_{01}} \sum_{m=0}^{m_{\max}} w_{d_1d_0}^{(zz')}(m,\bcx){\mu}_{zd_z}(m,\bcx)r_{z'd_{z'}}(m,\bcx)\right].
\end{align*}\endgroup
Noting that $\Delta_1=\Delta_2=\Delta_3=0$ by the law of iterated expectations and $\Delta_4 = \theta_{d_1d_0}^{(zz')}$ as shown in Proposition \ref{proposition:sen_pi}, we obtained that $\theta^{(zz'),\text{mr}}_{d_1d_0}(\xi) = \theta^{(zz')}_{d_1d_0}$ under $\mathcal M_{e} \cap \mathcal M_{m} \cap \mathcal M_{o}$. 

Up until this point, we have confirmed that $\widehat \theta^{(zz'),\text{mr}}_{d_1d_0}(\xi)$ is consistent to $\theta^{(zz')}_{d_1d_0}$ under $\mathcal M_{\pi} \cap \mathcal M_{e} \cap \mathcal M_{m}$ or $\mathcal M_{e} \cap \mathcal M_{m} \cap \mathcal M_{o}$. Then, under mild regularity conditions (similar to what we listed in the proof of Theorem 4), one can easily show that $\widehat \theta^{(zz'),\text{mr}}_{d_1d_0}(\xi)$ is also asymptotically normal such that $\sqrt{n}\left(\widehat \theta^{(zz'),\text{mr}}_{d_1d_0}(\xi)-\theta_{d_1d_0}^{(zz')}\right)$ converges to a zero-mean normal distribution with finite variance under either $\mathcal M_{\pi} \cap \mathcal M_{e} \cap \mathcal M_{m}$ or $\mathcal M_{e} \cap \mathcal M_{m} \cap \mathcal M_{o}$. The proof is omitted for brevity.\hfill $\square$

\subsection{Sensitivity analysis for the principal ignorability assumption under strong monotonicity}\label{sec:b9}

We can adapt $\widehat{\theta}_{d_1d_0}^{(zz'),\text{mr}}(\xi)$ to address strong monotonicity, following a similar procedure shown in Section \ref{sec:b8}. Under strong monotonicity, the proposed multiply robust estimator $\widehat{\theta}_{d_1d_0}^{(zz'),\text{mr}}(\kappa)$ takes the following explicit expression for any $d_1d_0 \in \mathcal U_{\text{b}}$: 
\begin{align}
\widehat\theta^{(zz'),\text{mr}}_{d_1d_0}(\kappa) = & \Prob_n\Big\{\left(\frac{\mathbb{I}(Z=z^*)\left\{\mathbb{I}(D=d^*)-\widehat p_{z^*d^*}^{\text{par}}(\bcx)\right\}}{\widehat\pi_{z^*}^{\text{par}}(\bcx)} - k\frac{(1-Z)\left\{D-\widehat p_{01}^{\text{par}}(\bcx)\right\}}{\widehat\pi_{0}^{\text{par}}(\bcx)}\right) \frac{\widehat\eta_{zz'}^{w,\text{par}}(\bcx)}{\widehat p_{z^*d^*}^{\text{dr}} - k\widehat p_{01}^{\text{dr}}} \nonumber \\
& + \frac{\widehat p_{z^*d^*}^{\text{par}}(\bcx)-k\widehat p_{01}^{\text{par}}(\bcx)}{\widehat p_{z^*d^*}^{\text{dr}} - k\widehat p_{01}^{\text{dr}}}\frac{\mathbb{I}(D=d_z,Z=z)}{\widehat p_{zd_z}^{\text{par}}(\bcx)\widehat\pi_{z}^{\text{par}}(\bcx)}\frac{\widehat r_{z'd_{z'}}^{\text{par}}(M,\bcx)}{\widehat r_{zd_z}^{\text{par}}(M,\bcx)}\widehat w_{d_1d_0}^{(zz')}(M,\bcx)\left\{Y-\widehat \mu_{zd_z}^{\text{par}}(M,\bcx)\right\} \nonumber \\
& + \frac{\widehat p_{z^*d^*}^{\text{par}}(\bcx)-k\widehat p_{01}^{\text{par}}(\bcx)}{\widehat p_{z^*d^*}^{\text{dr}} - k\widehat p_{01}^{\text{dr}}}\frac{\mathbb{I}(D=d_{z'},Z=z')}{\widehat p_{z'd_{z'}}^{\text{par}}(\bcx)\widehat \pi_{z'}^{\text{par}}(\bcx)}\left\{\widehat w_{d_1d_0}^{(zz')}(M,\bcx)\widehat \mu_{zd_z}^{\text{par}}(M,\bcx)-\widehat \eta_{zz'}^{w,\text{par}}(\bcx)\right\} \nonumber \\
& + \frac{\widehat p_{z^*d^*}^{\text{par}}(\bcx)-k\widehat p_{01}^{\text{par}}(\bcx)}{\widehat p_{z^*d^*}^{\text{dr}} - k\widehat p_{01}^{\text{dr}}}\widehat \eta_{zz'}^{w,\text{par}}(\bcx) \Big\},\label{e:theta_pi2}
\end{align}
where $\widehat \eta_{zz'}^{w,\text{par}}(\bx)=\displaystyle\sum_{m=0}^{m_{\max}}\widehat w_{d_1d_0}^{(zz')}(m,\bx)\widehat \mu_{zd_z}^{\text{par}}(m,\bx) \widehat r_{z'd_{z'}}^{\text{par}}(m,\bx)$ and 
$\widehat w_{d_1d_0}^{(zz')}(m,\bx)$ is the estimated sensitivity weight based on the parametric working models. Here, the sensitivity weight $w_{d_1d_0}^{(zz')}(m,\bx)$ takes a slightly different form under standard monotonicity. We provide the explicit expressions of $w_{d_1d_0}^{(zz')}(m,\bx)$ for $zz'\in\{11,10,00\}$ and $d_1d_0\in\{10,00\}$ below. 
\begingroup\makeatletter\def\f@size{8.5}\check@mathfonts 
\begin{align*}
w_{10}^{(11)}(m,\bx) &= 1 \text{ for any $m$.} \\
w_{10}^{(10)}(m,\bx) &=
\begin{cases}
    \frac{\xi_{M}^{(0)}(m,\bx)p_{00}(\bx)}{\xi_{M}^{(0)}(m,\bx)(p_{11}(\bx)-p_{01}(\bcx))+p_{10}(\bx)},& \text{if } m\geq 1,\\
    \frac{1}{r_{00}(0,\bx)}-\displaystyle\sum_{j=1}^{m_{\max}} \frac{\xi_{M}^{(0)}(j,\bx)p_{00}(\bx)r_{00}(j,\bx)/r_{00}(0,\bx)}{\xi_{M}^{(0)}(j,\bx)(p_{11}(\bx)-p_{01}(\bx))+p_{10}(\bx)},              & \text{if } m = 0.
\end{cases} \\
w_{10}^{(00)}(m,\bx) &=
\begin{cases}
    \frac{\xi_{M}^{(0)}(m,\bx)p_{00}(\bx)}{\xi_{M}^{(0)}(m,\bx)(p_{11}(\bx)-p_{01}(\bcx))+p_{10}(\bx)}\frac{\xi_{M}^{(0)}(m,\bx) \left(p_{11}(\bx)-p_{01}(\bx)\right)+p_{10}(\bx)}{p_{10}(\bx)/\xi_Y^{(0)}(m,\bx)+\xi_{M}^{(0)}(m,\bx) \left(p_{11}(\bx)-p_{01}(\bx)\right)},& \text{if } m\geq 1,\\
    \left\{\frac{1}{r_{00}(0,\bx)}-\displaystyle\sum_{j=1}^{m_{\max}} \frac{\xi_{M}^{(0)}(j,\bx)p_{00}(\bx)r_{00}(j,\bx)/r_{00}(0,\bx)}{\xi_{M}^{(0)}(j,\bx)(p_{11}(\bx)-p_{01}(\bx))+p_{10}(\bx)}\right\}\frac{\xi_{M}^{(0)}(0,\bx) \left(p_{11}(\bx)-p_{01}(\bx)\right)+p_{10}(\bx)}{p_{10}(\bx)/\xi_Y^{(0)}(0,\bx)+\xi_{M}^{(0)}(0,\bx) \left(p_{11}(\bx)-p_{01}(\bx)\right)},              & \text{if } m = 0.
\end{cases} \\
w_{00}^{(11)}(m,\bx) &= 1 \text{ for any $m$}.\\
w_{00}^{(10)}(m,\bx) &=
\begin{cases}
   \frac{p_{00}(\bx)}{\xi_{M}^{(0)}(m,\bx)(p_{11}(\bx)-p_{01}(\bx))+p_{10}(\bx)},& \text{if } m\geq 1,\\
    \frac{1}{r_{00}(0,\bx)}-\displaystyle\sum_{j=1}^{m_{\max}} \frac{p_{00}(\bx)r_{00}(j,\bx)/r_{00}(0,\bx)}{\xi_{M}^{(0)}(j,\bx)(p_{11}(\bx)-p_{01}(\bx))+p_{10}(\bx)},              & \text{if } m = 0.
\end{cases} \\
w_{00}^{(00)}(m,\bx) &=
\begin{cases}
   \frac{p_{00}(\bx)}{\xi_{M}^{(0)}(m,\bx)(p_{11}(\bx)-p_{01}(\bx))+p_{10}(\bx)}\frac{\xi_{M}^{(0)}(m,\bx) \left(p_{11}(\bx)-p_{01}(\bx)\right)+p_{10}(\bx)}{p_{10}(\bx)+\xi_Y^{(0)}(m,\bx)\xi_{M}^{(0)}(m,\bx) \left(p_{11}(\bx)-p_{01}(\bx)\right)},& \text{if } m\geq 1,\\
    \left\{\frac{1}{r_{00}(0,\bx)}-\displaystyle\sum_{j=1}^{m_{\max}} \frac{p_{00}(\bx)r_{00}(j,\bx)/r_{00}(0,\bx)}{\xi_{M}^{(0)}(j,\bx)(p_{11}(\bx)-p_{01}(\bx))+p_{10}(\bx)}\right\}\frac{\xi_{M}^{(0)}(0,\bx) \left(p_{11}(\bx)-p_{01}(\bx)\right)+p_{10}(\bx)}{p_{10}(\bx)+\xi_Y^{(0)}(0,\bx)\xi_{M}^{(0)}(0,\bx) \left(p_{11}(\bx)-p_{01}(\bx)\right)},              & \text{if } m = 0.
\end{cases} 
\end{align*}\endgroup
It is worth noting that the sensitivity weights $\{w_{d_1d_0}^{(10)}(m,\bx),w_{d_1d_0}^{(11)}(m,\bx)\}$ only depend on $\xi_{M}^{(0)}(m,\bx)$ but not $\xi_{Y}^{(0)}(m,\bx)$; this suggests that the estimated principal natural indirect effect, $\widehat{\theta}_{d_1d_0}^{(11),\text{mr}}(\kappa)-\widehat{\theta}_{d_1d_0}^{(10),\text{mr}}(\kappa)$, will not depend on the values of $\xi_{Y}^{(0)}(m,\bx)$. However, the estimated principal natural direct effect, $\widehat{\theta}_{d_1d_0}^{(10),\text{mr}}(\kappa)-\widehat{\theta}_{d_1d_0}^{(00),\text{mr}}(\kappa)$, is dependent on both $\xi_{Y}^{(0)}(m,\bx)$ and $\xi_{M}^{(0)}(m,\bx)$. 

We can show that $\widehat\theta^{(zz'),\text{mr}}_{d_1d_0}(\kappa)$ enjoys similar properties to $\widehat\theta^{(zz'),\text{mr}}_{d_1d_0}(\xi)$:
\begin{proposition}\label{proposition:mr_kappa}
   Suppose that Assumptions 1, 2, 3b, 5, and 6 hold. Under either $\mathcal M_{\pi} \cap \mathcal M_{e} \cap \mathcal M_{m}$ or $\mathcal M_{e} \cap \mathcal M_{m} \cap \mathcal M_{o}$, the estimator $\widehat\theta^{(zz'),\text{mr}}_{d_1d_0}(\kappa)$ is consistent and asymptotically normal for any $d_1d_0\in \mathcal U_{\text{b}}$.
\end{proposition}
Proof of Proposition \ref{proposition:mr_kappa} is similar to the proof of Proposition \ref{proposition:mr_xi} and has been omitted for brevity.

\subsection{Sensitivity analysis for the  ignorability of the mediator assumption}\label{sec:b10}

This section presents proofs for Propositions \ref{proposition:sen_m} and \ref{proposition:mr_t}, which refer to the identification results and properties of the multiply robust estimator under violation of Assumption 5. \medskip

\noindent \textbf{\textit{Proof of Proposition \ref{proposition:sen_m}.}}
For any $z\in\{0,1\}$ and $d_1d_0 \in \mathcal U$ with $\mathcal U=\mathcal U_{\text{a}}$ under standard monotonicity and $\mathcal U=\mathcal U_{\text{b}}$ under strong monotonicity, we can show that
\begin{align*}
  & \E_{Y_{1m}|Z,U,\bcx}[Y_{1m}|z,d_1d_0,\bx] \\
  = & \sum_{j=0}^{m_{\max}} \E_{Y_{1m}|Z,M,U,\bcx}[Y_{1m}|z,j,d_1d_0,\bx]f_{M|Z,U,\bcx}(j|z,d_1d_0,\bx) \\
  = & \E_{Y_{1m}|Z,M,U,\bcx}[Y_{1m}|z,m,d_1d_0,\bx] \sum_{j=0}^{m_{\max}} \frac{\E_{Y_{1m}|Z,M,U,\bcx}[Y_{1m}|z,j,d_1d_0,\bx]}{\E_{Y_{1m}|Z,M,U,\bcx}[Y_{1m}|z,m,d_1d_0,\bx]}f_{M|Z,U,\bcx}(j|z,d_1d_0,\bx) \\
  = & \E_{Y_{1m}|Z,M,U,\bcx}[Y_{1m}|z,m,d_1d_0,\bx] \sum_{j=0}^{m_{\max}} \frac{t(z,j,d_1d_0,\bx)}{t(z,m,d_1d_0,\bx)}f_{M|Z,U,\bcx}(j|z,d_1d_0,\bx)
\end{align*}
Therefore,
\begin{align*}
  & \frac{\E_{Y_{1m}|Z,M,U,\bcx}[Y_{1m}|0,m,d_1d_0,\bx]}{\E_{Y_{1m}|Z,M,U,\bcx}[Y_{1m}|1,m,d_1d_0,\bx]} \\
= &  \frac{\E_{Y_{1m}|Z,U,\bcx}[Y_{1m}|0,d_1d_0,\bx] \left\{\displaystyle\sum_{j=0}^{m_{\max}} \frac{t(0,j,d_1d_0,\bx)}{t(0,m,d_1d_0,\bx)}f_{M|Z,U,\bcx}(j|0,d_1d_0,\bx)\right\}^{-1} }{\E_{Y_{1m}|Z,U,\bcx}[Y_{1m}|1,d_1d_0,\bx]  \left\{\displaystyle\sum_{j=0}^{m_{\max}} \frac{t(1,j,d_1d_0,\bx)}{t(1,m,d_1d_0,\bx)}f_{M|Z,U,\bcx}(j|1,d_1d_0,\bx)\right\}^{-1} }\\
= & \left\{\displaystyle\sum_{j=0}^{m_{\max}} \frac{t(1,j,d_1d_0,\bx)}{t(1,m,d_1d_0,\bx)}f_{M|Z,U,\bcx}(j|1,d_1d_0,\bx)\right\} \Big/ \left\{\displaystyle\sum_{j=0}^{m_{\max}} \frac{t(0,j,d_1d_0,\bx)}{t(0,m,d_1d_0,\bx)}f_{M|Z,U,\bcx}(j|0,d_1d_0,\bx)\right\},
\end{align*}
where the last equality is followed by  $\E_{Y_{1m}|Z,U,\bcx}[Y_{1m}|0,d_1d_0,\bx]=\E_{Y_{1m}|Z,U,\bcx}[Y_{1m}|1,d_1d_0,\bx]$ due to Lemma \ref{lemma:randomization2}.  We can then identify $\E_{Y_{1m}|M_0,U,\bcx}[Y_{1m}|m,d_1d_0,\bx]$ using the following equations:
\begin{align*}
& \E_{Y_{1m}|M_0,U,\bcx}[m,d_1d_0,\bx] \\
= & \E_{Y_{1m}|Z,M,U,\bcx}[0,m,d_1d_0,\bx] \\
= &  \E_{Y_{1m}|Z,M,U,\bcx}[1,m,d_1d_0,\bx] \times \frac{\sum_{j=0}^{m_{\max}} \displaystyle\frac{t(1,j,d_1d_0,\bx)}{t(1,m,d_1d_0,\bx)}f_{M|Z,U,\bcx}(j|1,d_1d_0,\bx)}{\sum_{j=0}^{m_{\max}} \displaystyle\frac{t(0,j,d_1d_0,\bx)}{t(0,m,d_1d_0,\bx)}f_{M|Z,U,\bcx}(j|0,d_1d_0,\bx)}\\
= &  \E_{Y_{1m}|M_1,U,\bcx}[m,d_1d_0,\bx] \times \frac{\sum_{j=0}^{m_{\max}} \displaystyle\frac{t(1,j,d_1d_0,\bx)}{t(1,m,d_1d_0,\bx)}f_{M_1|U,\bcx}(j|d_1d_0,\bx)}{\sum_{j=0}^{m_{\max}} \displaystyle\frac{t(0,j,d_1d_0,\bx)}{t(0,m,d_1d_0,\bx)}f_{M_0|U,\bcx}(j|d_1d_0,\bx)} \\
= &  \E_{Y_{1m}|D_1,D_0,M_1,\bcx}[d_1,d_0,m,\bx] \times \frac{\sum_{j=0}^{m_{\max}} \displaystyle\frac{t(1,j,d_1d_0,\bx)}{t(1,m,d_1d_0,\bx)}f_{M_1|D_1,D_0,\bcx}(j|d_1,d_0,\bx)}{\sum_{j=0}^{m_{\max}} \displaystyle\frac{t(0,j,d_1d_0,\bx)}{t(0,m,d_1d_0,\bx)}f_{M_0|D_1,D_0,\bcx}(j|d_1,d_0,\bx)} \\
= &  \E_{Y_{1m}|D_1,M_1,\bcx}[d_1,m,\bx] \times \frac{\sum_{j=0}^{m_{\max}} \displaystyle\frac{t(1,j,d_1d_0,\bx)}{t(1,m,d_1d_0,\bx)}f_{M_1|D_1,\bcx}(j|d_1,\bx)}{\sum_{j=0}^{m_{\max}} \displaystyle\frac{t(0,j,d_1d_0,\bx)}{t(0,m,d_1d_0,\bx)}f_{M_0|D_0,\bcx}(j|d_0,\bx)} \quad \quad \text{(by Lemma \ref{lemma:pi_v2})} \\
= & \E_{Y|Z,D,M,\bcx}[1,d_1,m,\bx] \times \frac{\sum_{j=0}^{m_{\max}} \displaystyle\frac{t(1,j,d_1d_0,\bx)}{t(1,m,d_1d_0,\bx)}f_{M|Z,D,\bcx}(j|1,d_1,\bx)}{\sum_{j=0}^{m_{\max}} \displaystyle\frac{t(0,j,d_1d_0,\bx)}{t(0,m,d_1d_0,\bx)}f_{M|Z,D,\bcx}(j|0,d_0,\bx)} \quad \quad (\text{by Assumption 2}) \\
= &  \mu_{1d_1}(m,\bx)\frac{\sum_{j=0}^{m_{\max}} \displaystyle\frac{t(1,j,d_1d_0,\bx)}{t(1,m,d_1d_0,\bx)}r_{1d_1}(j,\bx)}{\sum_{j=0}^{m_{\max}} \displaystyle\frac{t(0,j,d_1d_0,\bx)}{t(0,m,d_1d_0,\bx)}r_{0d_0}(j,\bx)}.
\end{align*}
Therefore, we can identify $\theta_{d_1d_0}^{(10)}$ as
\begin{align*}
\theta_{d_1d_0}^{(10)} & = \E[Y_{1M_{0}}|U=d_1d_0] = \E\left[\E[Y_{1M_{0}}|U=d_1d_0,\bcx]\Big|U=d_1d_0\right] \\
& = \E\left[\sum_{m=0}^{m_{\max}} \E_{Y_{1m}|M_0,U,\bcx}[Y_{1m}|m,d_1d_0,\bcx]f_{M_{0}|U,\bcx}(m|d_1d_0,\bcx) \Big|U=d_1d_0\right] \\
& = \E\left[\sum_{m=0}^{m_{\max}} \frac{\sum_{j=0}^{m_{\max}} \displaystyle\frac{t(1,j,d_1d_0,\bcx)}{t(1,m,d_1d_0,\bcx)}r_{1d_1}(j,\bcx)}{\sum_{j=0}^{m_{\max}} \displaystyle\frac{t(0,j,d_1d_0,\bcx)}{t(0,m,d_1d_0,\bcx)}r_{0d_0}(j,\bcx)}\mu_{1d_1}(m,\bcx)f_{M_{0}|U,\bcx}(m|d_1d_0,\bcx) \Big|U=d_1d_0\right] \\
& = \E\left[\sum_{m=0}^{m_{\max}} \rho_{d_1d_0}^{(10)}(m,\bcx)\mu_{1d_1}(m,\bcx)f_{M_{0}|U,\bcx}(m|d_1d_0,\bcx) \Big|U=d_1d_0\right] \\
& = \E\left[\sum_{m=0}^{m_{\max}} \rho_{d_1d_0}^{(10)}(m,\bcx)\mu_{1d_1}(m,\bcx)f_{M_{0}|D_0,\bcx}(m|0,d_0,\bcx) \Big|U=d_1d_0\right] \quad \text{(by Lemma \ref{lemma:pi_v2})}\\
& = \E\left[\sum_{m=0}^{m_{\max}} \rho_{d_1d_0}^{(10)}(m,\bcx)\mu_{1d_1}(m,\bcx)f_{M_{0}|Z,D_0,\bcx}(m|0,d_0,\bcx) \Big|U=d_1d_0\right] \quad \text{(by Assumption 2)} \\
& = \E\left[\sum_{m=0}^{m_{\max}} \rho_{d_1d_0}^{(10)}(m,\bcx)\mu_{1d_1}(m,\bcx)r_{0d_0}(m,\bcx) \Big|U=d_1d_0\right] \\
&  = \int_{\bx} \frac{f_{U|\bcx}(d_1d_0|\bx)}{f_{U}(d_1d_0)} \left\{\sum_{m=0}^{m_{\max}} \rho_{d_1d_0}^{(10)}(m,\bx)\mu_{1d_1}(m,\bx)r_{0d_0}(m,\bx) \right\}\differential\Prob_{\bcx}(\bx) \quad \text{(by Lemma \ref{lemma:expectation})}\\
&  = \int_{\bx} \frac{e_{d_1d_0}(\bx)}{e_{d_1d_0}} \left\{\sum_{m=0}^{m_{\max}} \rho_{d_1d_0}^{(10)}(m,\bx)\mu_{1d_1}(m,\bx)r_{0d_0}(m,\bx) \right\}\differential\Prob_{\bcx}(\bx).
\end{align*}
This completes the proof. \hfill $\square$\medskip

\noindent \textbf{\textit{Proof of Proposition \ref{proposition:mr_t}.}} Following notation in the proofs of Theorem 4 and Proposition \ref{proposition:mr_xi}, we let $\widetilde h_{nuisance} =\{\widetilde \pi_z(\bx), \widetilde p_{zd}(\bx), \widetilde r_{zd}(m,\bx),  \widetilde \mu_{zd}(m,\bx) \}$ be the probability limit of $\widehat h_{nuisance}^{\text{par}}$. Because we assume the condition under either $\mathcal M_{\pi} \cap \mathcal M_{e} \cap \mathcal M_{m}$, $ \mathcal M_{e} \cap \mathcal M_{m} \cap \mathcal M_{o}$, or $ \mathcal M_{\pi} \cap \mathcal M_{m} \cap \mathcal M_{o}$, we always have $\widetilde r_{zd}(m,\bx)=r_{zd}(m,\bx)$, which suggests that the probability limit of $\widehat{\rho}_{d_1d_0}^{(zz'),\text{par}}(m,\bx)$ always equals to $\rho_{d_1d_0}^{(zz')}(m,\bx)$ because $\widehat{\rho}_{d_1d_0}^{(zz'),\text{par}}(m,\bx)$ is only a function of the sensitivity weight $t$ and the mediator model $\mathcal M_m$. Also, we can show that the probability limit of $\widehat p_{zd}^{\text{dr}}$, denoted by $\widetilde p_{zd}$, always equals to the true value $p_{zd}$, because $\widehat p_{zd}^{\text{dr}}$ is doubly robust under $\mathcal M_{\pi} \cup \mathcal M_{e}$. The previous discussion suggests that the probability limit of $\widehat{\theta}_{d_1d_0}^{(zz'),\text{mr}}(t)$ is 
\begin{align}
\theta^{(10),\text{mr}}_{d_1d_0}(t) = & \E\Big\{\left(\frac{\mathbb{I}(Z=z^*)\left\{\mathbb{I}(D=d^*)- \widetilde p_{z^*d^*}(\bcx)\right\}}{\widetilde\pi_{z^*}(\bcx)} - k\frac{(1-Z)\left\{D- \widetilde p_{01}(\bcx)\right\}}{\widetilde\pi_{0}(\bcx)}\right) \frac{\widetilde\eta_{10}^{\rho}(\bcx)}{ p_{z^*d^*} - kp_{01}} \nonumber \\
& + \frac{\widetilde p_{z^*d^*}(\bcx)-k\widetilde p_{01}(\bcx)}{p_{z^*d^*} - kp_{01}}\frac{\mathbb{I}(D=d_1,Z=1)}{\widetilde p_{1d_1}(\bcx)\widetilde\pi_{1}(\bcx)}\frac{ r_{0d_{0}}(M,\bcx)}{r_{1d_1}(M,\bcx)}\rho_{d_1d_0}^{(10)}(M,\bcx)\left\{Y-\widetilde \mu_{zd_z}(M,\bcx)\right\} \nonumber \\
& + \frac{\widetilde p_{z^*d^*}(\bcx)-k\widetilde p_{01}(\bcx)}{ p_{z^*d^*} - kp_{01}}\frac{\mathbb{I}(D=d_{0},Z=0)}{ \widetilde p_{0d_{0}}(\bcx)\widetilde \pi_{0}(\bcx)}\left\{\rho_{d_1d_0}^{(10)}(M,\bcx)\widetilde \mu_{1d_1}(M,\bcx)-\widetilde \eta_{10}^{\rho}(\bcx)\right\} \nonumber \\
& + \frac{\widetilde p_{z^*d^*}(\bcx)-k\widetilde p_{01}(\bcx)}{p_{z^*d^*} - kp_{01}}\widetilde \eta_{10}^{\rho}(\bcx) \Big\}\nonumber
\end{align}
under $\mathcal M_{\pi} \cap \mathcal M_{e} \cap \mathcal M_{m}$ $\mathcal M_{e} \cap \mathcal M_{m} \cap \mathcal M_{o}$, or $ \mathcal M_{\pi} \cap \mathcal M_{m} \cap \mathcal M_{o}$, where 
$$
\widetilde\eta_{10}^{\rho}(\bx)=\sum_{m=0}^{m_{\max}} \rho_{d_1d_0}^{(10)}(m,\bx)\widetilde{\mu}_{1d_1}(m,\bx)r_{0d_{0}}(m,\bx).
$$ 
In what follows, we show that $\theta^{(10),\text{mr}}_{d_1d_0}(t) = \theta_{d_1d_0}^{(10)}$ under Scenario I ($\mathcal M_{\pi} \cap \mathcal M_{e} \cap \mathcal M_{m}$),  Scenario II ($\mathcal M_{\pi} \cap \mathcal M_{m} \cap \mathcal M_{o}$), or Scenario III ($\mathcal M_{e} \cap \mathcal M_{m} \cap \mathcal M_{o}$), which concludes the triple robustness of $\widehat{\theta}_{d_1d_0}^{(zz'),\text{mr}}(t)$.\medskip

\noindent \textbf{Scenario I ($\mathcal M_{\pi} \cap \mathcal M_{e} \cap \mathcal M_{m}$):}\medskip

In Scenario I, $\widetilde \pi_z(\bx)=\pi_z(\bx)$ and $\widetilde p_{zd}(\bx)=p_{zd}(\bx)$ but generally $\widetilde \mu_{zd}(m,\bx)\neq \mu_{zd}(m,\bx)$. Therefore, we can rewrite $\theta^{(10),\text{mr}}_{d_1d_0}(t) = \sum_{j=1}^4 \Delta_j$, where\begingroup\makeatletter\def\f@size{9}\check@mathfonts
\begin{align*}
\Delta_1 & = \E\left[\left(\frac{\mathbb{I}(Z=z^*)\left\{\mathbb{I}(D=d^*)- p_{z^*d^*}(\bcx)\right\}}{\pi_{z^*}(\bcx)} - k\frac{(1-Z)\left\{D- p_{01}(\bcx)\right\}}{\pi_{0}(\bcx)}\right) \frac{\displaystyle\sum_{m=0}^{m_{\max}} \rho_{d_1d_0}^{(10)}(m,\bcx)\widetilde{\mu}_{1d_1}(m,\bcx) r_{0d_{0}}(m,\bcx) }{ p_{z^*d^*} - k p_{01}}\right], \\
\Delta_2 & = \E\left[\frac{ p_{z^*d^*}(\bcx)-k p_{01}(\bcx)}{ p_{z^*d^*} - k p_{01}}\frac{\mathbb{I}(D=d_1,Z=1)}{ p_{1d_1}(\bcx)\pi_{1}(\bcx)}\frac{ r_{0d_{0}}(M,\bcx)}{ r_{1d_1}(M,\bcx)}\rho_{d_1d_0}^{(10)}(M,\bcx)Y\right], \\
\Delta_3 & = \E\left[\frac{ p_{z^*d^*}(\bcx)-k p_{01}(\bcx)}{p_{z^*d^*} - k p_{01}}\left\{\frac{\mathbb{I}(D=d_{0},Z=0)}{p_{0d_{0}}(\bcx) \pi_{0}(\bcx)}-\frac{\mathbb{I}(D=d_1,Z=1)}{ p_{1d_1}(\bcx)\pi_{1}(\bcx)}\frac{ r_{0d_{0}}(M,\bcx)}{ r_{1d_1}(M,\bcx)}\right\}\rho_{d_1d_0}^{(10)}(M,\bcx)\widetilde \mu_{1d_1}(M,\bcx)\right],\\
\Delta_4 & = \E\left[\frac{ p_{z^*d^*}(\bcx)-k p_{01}(\bcx)}{ p_{z^*d^*} - k p_{01}} \left\{1-\frac{\mathbb{I}(D=d_{0},Z=0)}{ p_{0d_{0}}(\bcx) \pi_{0}(\bcx)}\right\}\sum_{m=0}^{m_{\max}}\rho_{d_1d_0}^{(10)}(m,\bcx)\widetilde{\mu}_{1d_1}(m,\bcx) r_{0d_{0}}(m,\bcx) \right].
\end{align*}\endgroup
One can show that $\Delta_1=\Delta_3=\Delta_4=0$ by using the law of iterated expectation and\begingroup\makeatletter\def\f@size{9}\check@mathfonts 
\begin{align*}
\Delta_2 = &  \int_{\bx}\frac{p_{z^*d^*}(\bx)-kp_{01}(\bx)}{p_{z^*d^*}-kp_{01}}\frac{1}{p_{1d_1}(\bx)\pi_{1}(\bx)}\sum_{m=0}^{m_{\max}}  \left\{\frac{r_{0d_{0}}(m,\bx)}{r_{1d_1}(m,\bx)}\rho_{d_1d_0}^{(10)}(m,\bx)\int_y y\differential \Prob_{Y|Z,D,M\bcx}(y|1,d_{1},m,\bx)r_{1d_1}(m,\bx)\right\} \\
& \quad  f_{D|Z,\bcx}(d_1|1,\bx) f_{Z|\bcx}(1|\bx) \differential \Prob_{\bcx}(\bx) \\
= &  \int_{\bx}\frac{p_{z^*d^*}(\bx)-kp_{01}(\bx)}{p_{z^*d^*}-kp_{01}}\sum_{m=0}^{m_{\max}}  \left\{\frac{r_{0d_{0}}(m,\bx)}{r_{1d_1}(m,\bx)}\rho_{d_1d_0}^{(10)}(m,\bx)\int_y y\differential \Prob_{Y|Z,D,M\bcx}(y|1,d_{1},m,\bx)r_{1d_1}(m,\bx)\right\}\differential \Prob_{\bcx}(\bx) \\
= &  \int_{\bx}\frac{p_{z^*d^*}(\bx)-kp_{01}(\bx)}{p_{z^*d^*}-kp_{01}}\sum_{m=0}^{m_{\max}}  \left\{\frac{r_{0d_{0}}(m,\bx)}{r_{1d_1}(m,\bx)}\rho_{d_1d_0}^{(10)}(m,\bx)\mu_{1d_1}(m,\bx)r_{1d_1}(m,\bx)\right\}\differential \Prob_{\bcx}(\bx) \\
= &  \int_{\bx}\frac{p_{z^*d^*}(\bx)-kp_{01}(\bx)}{p_{z^*d^*}-kp_{01}}\sum_{m=0}^{m_{\max}}  \left\{\rho_{d_1d_0}^{(10)}(m,\bx)\mu_{1d_1}(m,\bx)r_{0d_{0}}(m,\bx)\right\}\differential \Prob_{\bcx}(\bx) \\
= & \theta_{d_1d_0}^{(zz')},
\end{align*}\endgroup
where the last equality follows from Proposition \ref{proposition:sen_m}. This  suggests that $\theta^{(10),\text{mr}}_{d_1d_0}(t) = \theta^{(10)}_{d_1d_0}$ under $\mathcal M_{\pi} \cap \mathcal M_{e} \cap \mathcal M_{m}$.

\noindent \textbf{Scenario II ($\mathcal M_{\pi} \cap \mathcal M_{m} \cap \mathcal M_{o}$):}\medskip

In Scenario II, $\widetilde \pi_z(\bx)=\pi_z(\bx)$ and $\widetilde \mu_{zd}(m,\bx)= \mu_{zd}(m,\bx)$ but generally $\widetilde p_{zd}(\bx)\neq p_{zd}(\bx)$.  Observing this, we can rewrite $\theta^{(zz'),\text{mr}}_{d_1d_0} = \sum_{j=1}^4 \Delta_j$, where\begingroup\makeatletter\def\f@size{10}\check@mathfonts
\begin{align*}
\Delta_1 & = \E\left[\left\{\frac{\mathbb{I}(Z=z^*,D=d^*)}{\pi_{z^*}(\bcx)} - k \frac{(1-Z)D}{\pi_0(\bcx)}\right\}\frac{\eta_{10}^{\rho}(\bcx)}{p_{z^*d^*}-kp_{01}}\right], \\
\Delta_2 & = \E\left[\frac{\widetilde p_{z^*d^*}(\bcx)-k\widetilde p_{01}(\bcx)}{ p_{z^*d^*} - k p_{01}}\frac{\mathbb{I}(D=d_1,Z=1)}{\widetilde p_{1d_1}(\bcx)\pi_{1}(\bcx)}\frac{ r_{0d_{0}}(M,\bcx)}{ r_{1d_1}(M,\bcx)}\rho_{d_1d_0}^{(10)}(M,\bcx)\left\{Y- \mu_{1d_1}(M,\bcx)\right\}\right], \\
\Delta_3 & = \E\left[\frac{\widetilde p_{z^*d^*}(\bcx)-k\widetilde p_{01}(\bcx)}{ p_{z^*d^*} - k p_{01}}\frac{\mathbb{I}(D=d_{0},Z=0)}{\widetilde p_{0d_{0}}(\bcx)\pi_{0}(\bcx)}\left\{ \rho_{d_1d_0}^{(10)}(M,\bcx) \mu_{1d_1}(M,\bcx)- \eta_{10}^{\rho}(\bcx)\right\}\right],\\
\Delta_4 & = \E\left[\left(\left\{1-\frac{\mathbb{I}(Z=z^*)}{\pi_{z^*}(\bcx)}\right\}\widetilde p_{z^*d^*}(\bcx)-k\left\{1-\frac{1-Z}{\pi_{0}(\bcx)}\right\}\widetilde p_{01}(\bcx)\right) \frac{\eta_{10}^{\rho}(\bcx)}{p_{z^*d^*} - k p_{01}}\right],
\end{align*}\endgroup
where $\eta_{10}^{\rho}(\bcx)=\displaystyle\sum_{m=0}^{m_{\max}}\rho_{d_1d_0}(m,\bx)\mu_{1d_1}(m,\bx) r_{0d_0}(m,\bx)$. One can verify $\Delta_2=\Delta_3=\Delta_4 = 0$ by using the law of iterated expectation and
\begin{align*}
\Delta_1 
= & \int_{\bx} \left\{\frac{1}{\pi_{z^*}(\bx)}\frac{\eta_{10}^{\rho}(\bx)}{p_{z^*d^*}-kp_{01}}\right\} f_{D|Z,\bcx}(d^*|z^*,\bx) f_{Z|\bcx}(z^*|\bx) \differential \Prob_\bcx(\bx)\\
& - \int_{\bx} \left\{\frac{k}{\pi_{0}(\bx)}\frac{\eta_{10}^{\rho}(\bx)}{p_{z^*d^*}-kp_{01}}\right\} f_{D|Z,\bcx}(1|0,\bx) f_{Z|\bcx}(0|\bx) \differential \Prob_\bcx(\bx)\\
= & \int_{\bx} \left\{f_{D|Z,\bcx}(d^*|z^*,\bx)-kf_{D|Z,\bcx}(1|0,\bx)\right\} \frac{\eta_{10}^{\rho}(\bx)}{p_{z^*d^*}-kp_{01}} \differential \Prob_\bcx(\bx) \\
= & \int_{\bx}\frac{p_{z^*d^*}(\bx)-kp_{01}(\bx)}{p_{z^*d^*}-kp_{01}}\eta_{10}^{\rho}(\bx) \differential \Prob_{\bcx}(\bx) \\
= & \int_{\bx}\frac{p_{z^*d^*}(\bx)-kp_{01}(\bx)}{p_{z^*d^*}-kp_{01}}\sum_{m=0}^{m_{\max}}  \left\{\rho_{d_1d_0}^{(10)}(m,\bx)\mu_{1d_1}(m,\bx)r_{0d_{0}}(m,\bx)\right\}\differential \Prob_{\bcx}(\bx)\\
= & \theta_{d_1d_0}^{(10)}.
\end{align*}
Therefore, we have obtained $\theta^{(10),\text{mr}}_{d_1d_0}(t) = \theta^{(10)}_{d_1d_0}$ under $\mathcal M_{\pi} \cap \mathcal M_{m} \cap \mathcal M_{o}$.\medskip

\noindent \textbf{Scenario III ($\mathcal M_{e} \cap \mathcal M_{m} \cap \mathcal M_{o}$):}\medskip

In Scenario III, $\widetilde \mu_{zd}(m,\bx)= \mu_{zd}(m,\bx)$ and $\widetilde p_{zd}(\bx)= p_{zd}(\bx)$ but generally $\widetilde \pi_z(\bx) \neq \pi_z(\bx)$. Therefore, we have $\theta^{(10),\text{mr}}_{d_1d_0}(\xi) = \sum_{j=1}^4 \Delta_j$, where\begingroup\makeatletter\def\f@size{9}\check@mathfonts
\begin{align*}
\Delta_1 & = \E\left[\left(\frac{\mathbb{I}(Z=z^*)\left\{\mathbb{I}(D=d^*)- p_{z^*d^*}(\bcx)\right\}}{\widetilde \pi_{z^*}(\bcx)} - k\frac{(1-Z)\left\{D- p_{01}(\bcx)\right\}}{\widetilde \pi_{0}(\bcx)}\right) \frac{\eta_{10}^{\rho}(\bcx)}{ p_{z^*d^*} - k p_{01}}\right], \\
\Delta_2 & = \E\left[\frac{ p_{z^*d^*}(\bcx)-k p_{01}(\bcx)}{ p_{z^*d^*} - k p_{01}}\frac{\mathbb{I}(D=d_1,Z=1)}{ p_{1d_1}(\bcx)\widetilde\pi_{1}(\bcx)}\frac{ r_{0d_{0}}(M,\bcx)}{ r_{1d_1}(M,\bcx)}\rho_{d_1d_0}^{(10)}(M,\bcx)\left\{Y- \mu_{1d_1}(M,\bcx)\right\}\right], \\
\Delta_3 & = \E\left[\frac{ p_{z^*d^*}(\bcx)-kp_{01}(\bcx)}{ p_{z^*d^*} - k p_{01}}\frac{\mathbb{I}(D=d_{0},Z=0)}{ p_{0d_{0}}(\bcx)\widetilde \pi_{0}(\bcx)}\left\{ \rho_{d_1d_0}^{(10)}(M,\bcx) \mu_{1d_1}(M,\bcx)- \eta_{10}^{\rho}(\bcx)\right\}\right],\\
\Delta_4 & = \E\left[\frac{ p_{z^*d^*}(\bcx)-k p_{01}(\bcx)}{ p_{z^*d^*} - k p_{01}} \eta_{10}^{\rho}(\bcx)\right],
\end{align*}\endgroup
where $\eta_{10}^{\rho}(\bcx)=\displaystyle\sum_{m=0}^{m_{\max}}\rho_{d_1d_0}(m,\bx)\mu_{1d_1}(m,\bx) r_{0d_0}(m,\bx)$. Noting that $\Delta_1=\Delta_2=\Delta_3=0$ by the law of iterated expectations and $\Delta_4 = \theta_{d_1d_0}^{(10)}$ as shown in Proposition \ref{proposition:sen_m}, we obtained that $\theta^{(10),\text{mr}}_{d_1d_0}(t) = \theta^{(10)}_{d_1d_0}$ under $\mathcal M_{e} \cap \mathcal M_{m} \cap \mathcal M_{o}$. 

Now we have proved that $\widehat \theta^{(10),\text{mr}}_{d_1d_0}(\xi)$ is consistent to $\theta^{(10)}_{d_1d_0}$ under $\mathcal M_{\pi} \cap \mathcal M_{e} \cap \mathcal M_{m}$, $\mathcal M_{\pi} \cap \mathcal M_{m} \cap \mathcal M_{o}$, or $\mathcal M_{e} \cap \mathcal M_{m} \cap \mathcal M_{o}$. We can also show that $\widehat \theta^{(10),\text{mr}}_{d_1d_0}(t)$ is asymptotically normal under certain regularity conditions similar to what we provide in the proof of Theorem 4; the proofs are omitted for brevity. \hfill $\square$


\section{Figures and Tables}\label{sec:c}

\begin{figure}[h]
\begin{center}
\includegraphics[width=0.81\textwidth]{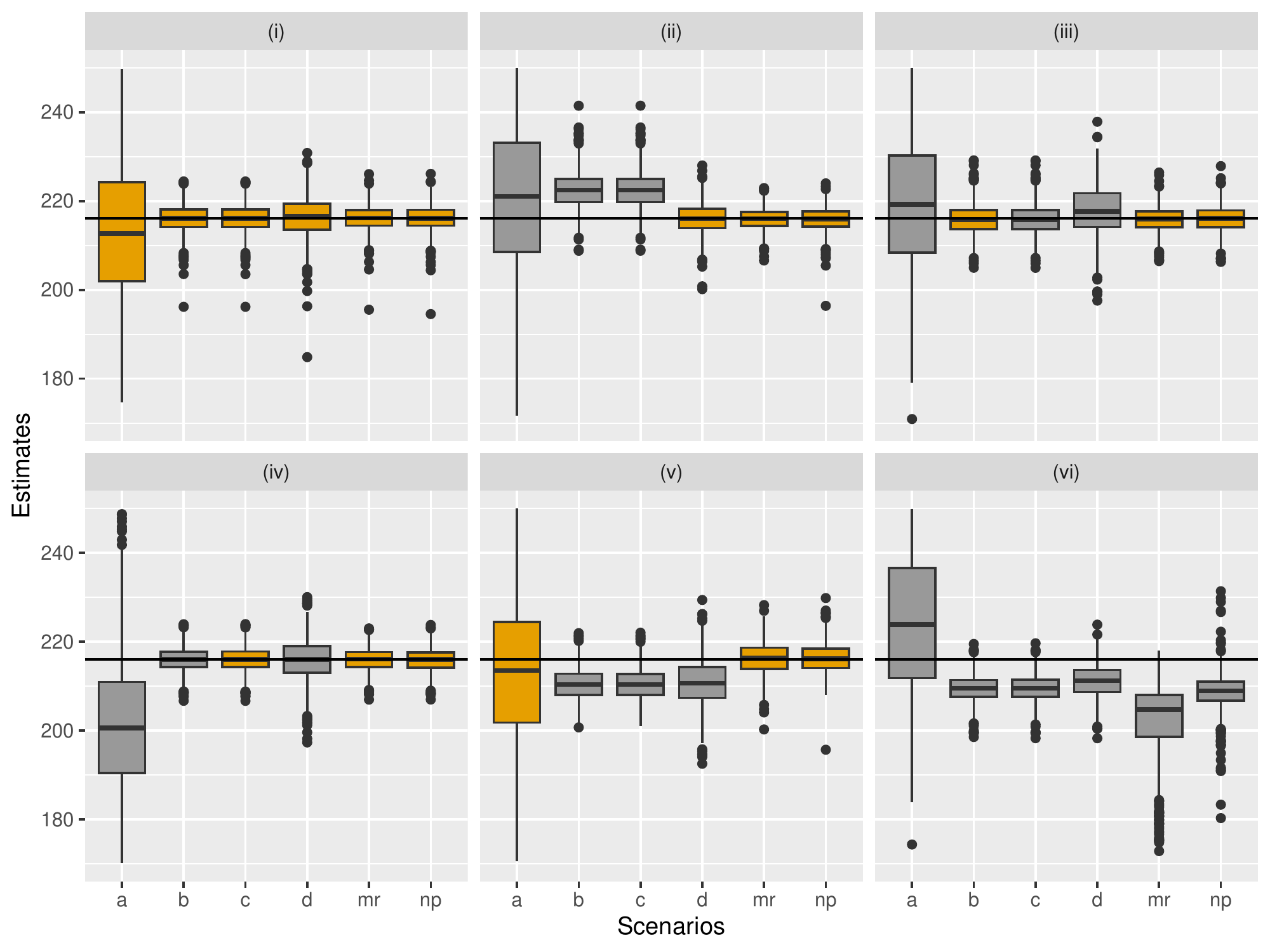}
\end{center}
\caption{Simulation results for estimators of $\theta_{11}^{(10)}$ among 6 different scenarios with sample size $n=1,000$. Scenarios (\romannumeral1)--(\romannumeral6) are described in Section 5. The horizontal line in each panel is the true value of $\theta_{11}^{(10)}$. The yellow highlighted boxplots indicate that the corresponding estimators are expected to be consistent by large-sample theory. }
\label{fig:sim1}
\end{figure}

\begin{figure}[h]
\begin{center}
\includegraphics[width=0.82\textwidth]{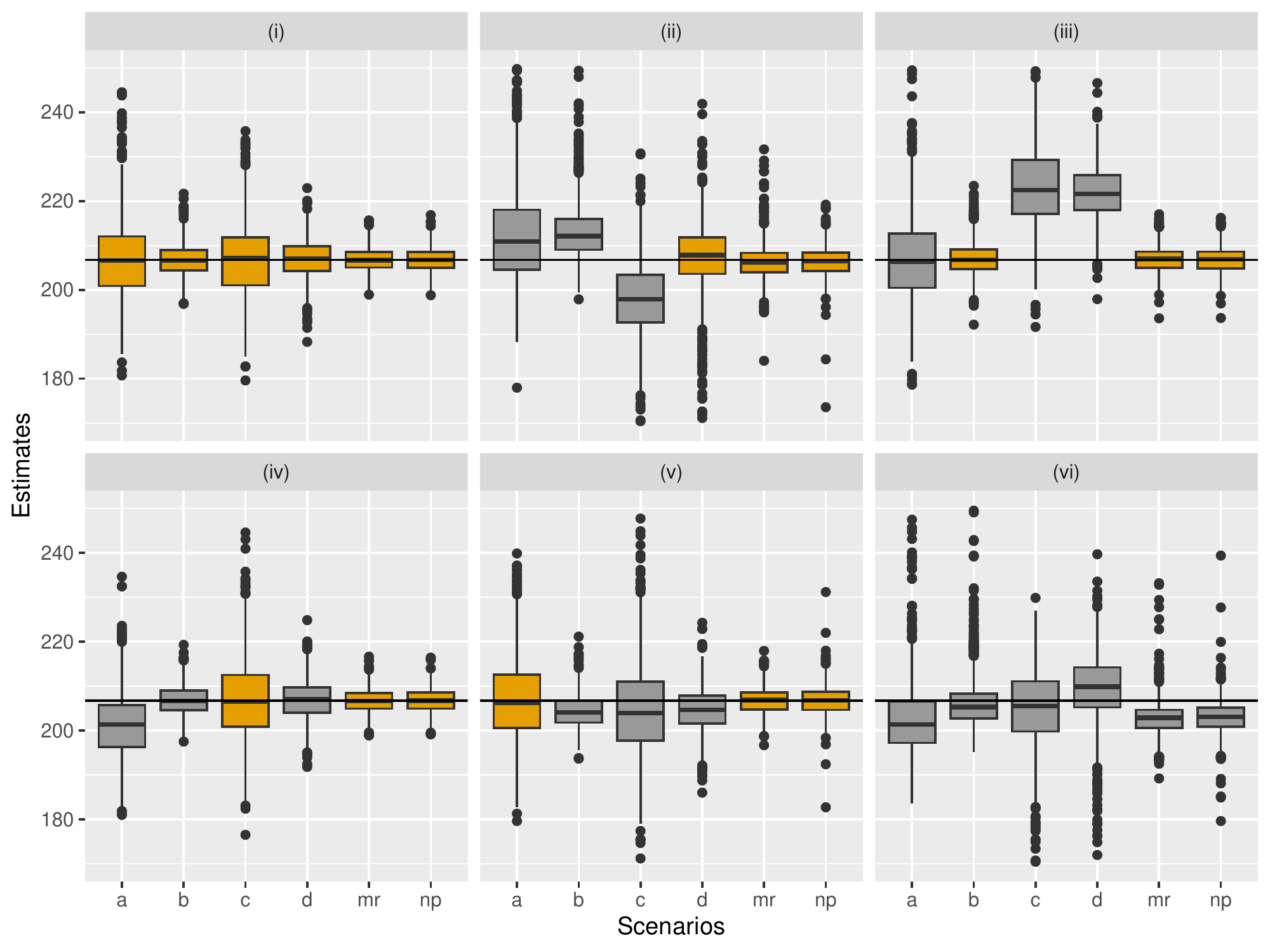}
\end{center}
\caption{Simulation results for estimators of $\theta_{00}^{(10)}$ among 6 different scenarios with sample size $n=1,000$. Scenarios (\romannumeral1)--(\romannumeral6) are described in Section 5. The horizontal line in each panel is the true value of $\theta_{00}^{(10)}$. The yellow highlighted boxplots indicate that the corresponding estimators are expected to be consistent by large-sample theory.}
\label{fig:sim2}
\end{figure} 

\begin{figure}[h]
\begin{center}
\includegraphics[width=1\textwidth]{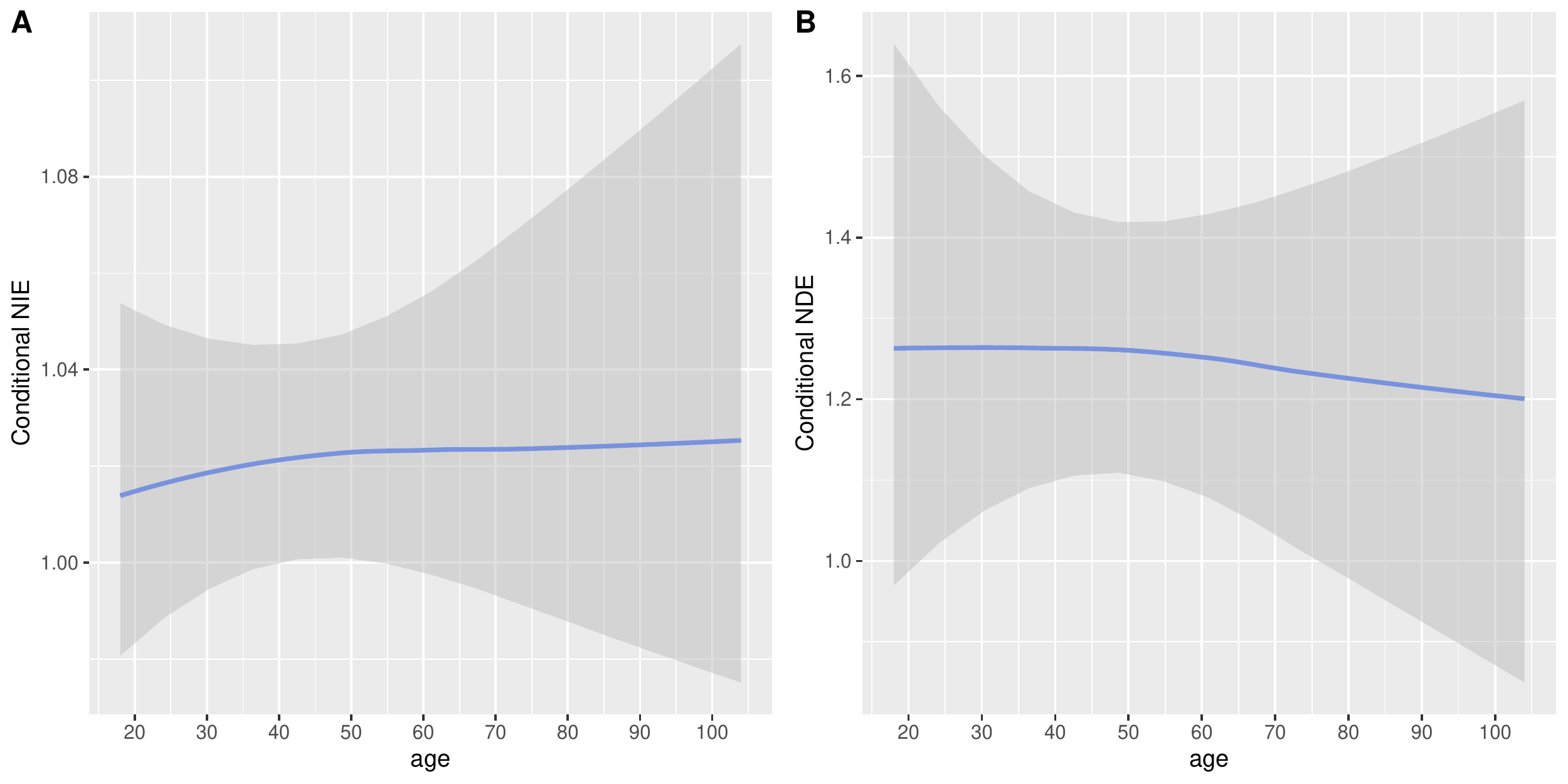}
\end{center}
\caption{Causal moderated mediation analysis conditional on participants' age, WHO-LARES study. The conditional natural (in)direct effects are defined on the risk ratio scale, which are calculated using the the R package \texttt{moderate.mediation} based on the methodology in \cite{qin2024causal}. }
\label{fig:causal_moderated_mediation_LARES_age}
\end{figure} 

\clearpage

\begin{table}[t]
\centering
\caption{Analysis of the ITT natural mediation effects in the JOBS II study.}\label{tab:itt_jobs2}
\scalebox{0.9}{
\begin{threeparttable}
\begin{tabular}{cccc}
\hline
Method & ITT-NIE & ITT-NDE & ITT                         \\
\hline
a & $-$0.017 ($-$0.036, $-$0.005) & $-$0.075 ($-$0.156, $-$0.001) & $-$0.092 ($-$0.173, $-$0.020) \\ 
  b & $-$0.014 ($-$0.031, $-$0.003) & $-$0.073 ($-$0.148, $-$0.001) & $-$0.087 ($-$0.165, $-$0.017) \\ 
  c & $-$0.021 ($-$0.045, $-$0.005) & $-$0.072 ($-$0.148, $-$0.002) & $-$0.092 ($-$0.174, $-$0.021) \\ 
  d & $-$0.014 ($-$0.031, $-$0.003) & $-$0.072 ($-$0.147, 0.000) & $-$0.086 ($-$0.165, $-$0.016) \\ 
  mr & $-$0.015 ($-$0.032, $-$0.003) & $-$0.072 ($-$0.151, $-$0.001) & $-$0.088 ($-$0.167, $-$0.016) \\ 
np        &  $-$0.017 ($-$0.031, $-$0.004)   
                               & $-$0.067 ($-$0.146, 0.013)           & $-$0.084 ($-$0.162, $-$0.005)  \\ 
\hline
\end{tabular}
\end{threeparttable}}
\end{table}

\begin{table}[!htbp]
\centering
\caption{Analysis of the principal natural mediation effects in the JOBS II study. 
}\label{tab:pce_jobs2}
\scalebox{0.86}{
\begin{threeparttable}
\begin{tabular}{cccc}
\hline
\multirow{2}{*}{Method} & \multirow{2}{*}{Estimand} & \multicolumn{1}{c}{Compliers}                & \multicolumn{1}{c}{Never takers}          \\
 & & $(\widehat e_{10}^{\text{np}}=0.55)$               & $(\widehat e_{00}^{\text{np}}=0.45)$          \\
\hline
\multirow{3}{*}{a}        & PNIE     & $-$0.030 ($-$0.062, $-$0.009)   & 0.000 ($-$0.007, 0.005) \\
                           & PNDE     & $-$0.083 ($-$0.176, 0.002)   & $-$0.063 ($-$0.174, 0.029) \\
                           & PCE      & $-$0.113 ($-$0.206, $-$0.027)  & $-$0.063 ($-$0.172, 0.027)  \\
\hline
\multirow{3}{*}{b}        & PNIE     & $-$0.024 ($-$0.052, $-$0.006)   & 0.000 ($-$0.006, 0.005) \\
                           & PNDE     & $-$0.082 ($-$0.167, 0.003)   & $-$0.062 ($-$0.166, 0.028) \\
                           & PCE      & $-$0.106 ($-$0.186, $-$0.026)  & $-$0.062 ($-$0.165, 0.027)  \\
\hline
\multirow{3}{*}{c}        & PNIE     & $-$0.030 ($-$0.071, 0.001)   & $-$0.007 ($-$0.032, 0.015) \\
                           & PNDE     & $-$0.083 ($-$0.171, 0.003)   & $-$0.056 ($-$0.160, 0.032) \\
                           & PCE      & $-$0.113 ($-$0.207, $-$0.029)  & $-$0.063 ($-$0.171, 0.029)  \\
\hline
\multirow{3}{*}{d}        & PNIE     & $-$0.024 ($-$0.052, $-$0.006)   & 0.000 ($-$0.006, 0.005) \\
                           & PNDE     & $-$0.081 ($-$0.166, 0.003)   & $-$0.060 ($-$0.166, 0.029) \\
                           & PCE      & $-$0.105 ($-$0.187, $-$0.025)  & $-$0.060 ($-$0.167, 0.029)  \\
\hline
\multirow{3}{*}{mr}        & PNIE     & $-$0.026 ($-$0.055, 0.006)   & 0.000 ($-$0.006, 0.006) \\
                           & PNDE     & $-$0.083 ($-$0.170, 0.002)   & $-$0.058 ($-$0.160, 0.031) \\
                           & PCE      & $-$0.109 ($-$0.191, $-$0.027)  & $-$0.058 ($-$0.160, 0.030)  \\ 
\hline
\multirow{3}{*}{np}        & PNIE     & $-$0.029 ($-$0.052, $-$0.006)  & $-$0.001 ($-$0.004, 0.003) \\
                           & PNDE     & $-$0.066 ($-$0.156, 0.023)  & $-$0.066 ($-$0.163, 0.032) \\
                           & PCE      & $-$0.096 ($-$0.182, $-$0.009) & $-$0.067 ($-$0.163, 0.032)  \\ 
\hline
\multirow{3}{*}{Rudolph et al.$^\mathsection$ }      & PNIE     & $-$0.030 ($-$0.053, $-$0.006)  & -- \\
                           & PNDE     & $-$0.115 ($-$0.251, 0.022)  & -- \\
                           & PCE      & $-$0.145 ($-$0.280, $-$0.008) & -- \\ 
                           
\hline
\end{tabular}
\begin{tablenotes}
      \footnotesize
      \item[$\mathsection$] `Rudolph et al.' is the nonparametric efficient estimator given by \cite{rudolph2024using}; see Remark 1 for more details of this approach. 
    \end{tablenotes}
\end{threeparttable}}
\end{table}

\begin{table}[h]
    \centering
    \caption{Mean and standard deviation of baseline characteristics among the compliers and never-takers, JOBS II study$^\ddagger$.}\label{tab:basline_ps}
    \scalebox{1}[1]{
    \begin{threeparttable}
    \begin{tabular}{lccc}
    \hline
        Variable & Compliers & Never-takers & ASD$^\mathsection$ \\ 
        \hline
        Proportion & 55\% & 45\% & \\
\textbf{Gender} (male) & 0.53 (0.50) & 0.62 (0.48) & 0.19 \\  
  \textbf{Age} (years) & 39.14 (10.37) & 35.51 (9.40) & 0.37 \\  
  \textbf{White race} & 0.84 (0.37) & 0.77 (0.42) & 0.15 \\ 
  \textbf{Depression} & 1.89 (0.55) & 1.87 (0.55) & 0.03 \\  
  \textbf{Economic hardship} & 3.04 (0.93) & 3.11 (0.94) & 0.07 \\  
  \textbf{Motivation} & 5.28 (0.63) & 5.05 (0.64) & 0.35 \\ 
  \textbf{Marriage} (baseline: never married) & \\
  ~~Married & 0.49 (0.50) & 0.43 (0.50) & 0.11 \\ 
  ~~Separated & 0.03 (0.17) & 0.03 (0.18) & 0.03 \\ 
  ~~Divorced & 0.18 (0.38) & 0.19 (0.4) & 0.04 \\ 
  ~~Widowed & 0.02 (0.14) & 0.02 (0.15) & 0.02 \\ 
  \textbf{Education} (baseline: less than high school) & \\
  ~~High school & 0.32 (0.47) & 0.33 (0.47) & 0.02 \\ 
  ~~Post-secondary non-tertiary education & 0.32 (0.47) & 0.40 (0.49) & 0.17 \\ 
  ~~Bachelor's degree & 0.19 (0.39) & 0.10 (0.31) & 0.23 \\ 
  ~~Higher than a Bachelor's degree & 0.13 (0.33) & 0.09 (0.28) & 0.13 \\ 
  \textbf{Assertiveness} & 3.39 (0.87) & 3.56 (0.83) & 0.21 \\ 
        \hline
    \end{tabular}
    \begin{tablenotes}
    \item[$\ddagger$] Calculation of stratum-specific mean and standard deviation of the baseline characteristics follows the method in \cite{cheng2023multiply}, which are weighted average of the mean and standard deviation of baseline characteristics based on the principal scores (estimated according to the nonparametric efficient estimator).
        \item[$\mathsection$] ASD is the absolute standardized difference across the two principal strata. Given a specific baseline covariate, its ASD is calculated as ${|\bar x_{10} - \bar x_{00}|}/{\sqrt{0.5(s_{10}^2+s_{00}^2)}}$, where $\bar x_{d_1d_0}$ and $s_{d_1d_0}$ are the estimated mean and standard deviation of this covariate in the stratum $U=d_1d_0$.
    \end{tablenotes}
    \end{threeparttable}}
\end{table}

\begin{table}[h]
    \centering
    \caption{Mean and standard deviation of baseline characteristics across the doomed, harmed, and immune strata, WHO-LARES study$^\ddagger$.}\label{tab:basline_ps_2}
    \scalebox{0.9}[0.9]{
    \begin{threeparttable}
    \begin{tabular}{lcccc}
    \hline
        Variable & Doomed & Harmed & Immune & Max ASD$^\mathsection$ \\ 
        \hline
        Proportion & 51\% & 8\% & 41\% & \\
        \textbf{Personal characteristics} & & & \\
~~Female & 0.63 (0.47) & 0.56 (0.52) & 0.48 (0.48) & 0.31 \\ 
  ~~Age (years) & 46.53 (17.89) & 46.80 (17.99) & 46.76 (17.3) & 0.02 \\ 
  ~~Married & 0.59 (0.49) & 0.63 (0.48) & 0.64 (0.48) & 0.09 \\ 
  ~~Post-secondary education & 0.27 (0.45) & 0.27 (0.45) & 0.29 (0.45) & 0.04 \\ 
  ~~Employed & 0.58 (0.49) & 0.58 (0.49) & 0.62 (0.49) & 0.09 \\ 
  ~~Non-smoking & 0.50 (0.50) & 0.49 (0.50) & 0.47 (0.5) & 0.06 \\ 
  \textbf{Dwelling condition} \\
  ~~Owning the house & 0.74 (0.44) & 0.72 (0.45) & 0.76 (0.43) & 0.08 \\ 
  ~~House size greater than $50\text{m}^2$ & 0.86 (0.35) & 0.84 (0.37) & 0.87 (0.34) & 0.08 \\ 
  ~~Crowding ($\geq 1$ resident per room) & 0.64 (0.48) & 0.64 (0.48) & 0.64 (0.48) & 0.01 \\ 
  \multicolumn{3}{l}{\textbf{Self-evaluation on dwelling condition}} \\
~~Satisfied with the heating system & 0.85 (0.35) & 0.85 (0.35) & 0.89 (0.32) & 0.10 \\ 
  ~~Satisfied with natural light & 0.75 (0.43) & 0.73 (0.44) & 0.77 (0.42) & 0.10 \\ 
        \hline
    \end{tabular}
    \begin{tablenotes}
    \item[$\ddagger$] Calculation of stratum-specific mean and standard deviation of the baseline characteristics follows the method in \cite{cheng2023multiply}, which are weighted average of the mean and standard deviation of baseline characteristics based on the principal scores (estimated according to the nonparametric efficient estimator).
        \item[$\mathsection$] Max ASD is the maximum pairwise absolute standardized difference across the three principal strata. Given a specific baseline covariate, its Max ASD is calculated as the maximum of ${|\bar x_{d_1d_0} - \bar x_{d_1'd_0'}|}/{\sqrt{0.5(s_{d_1d_0}^2+s_{d_1'd_0'}^2)}}$ for all $d_1d_0 \neq d_1'd_0'\in \{11,10,00\}$, where $\bar x_{d_1d_0}$ and $s_{d_1d_0}$ are the estimated mean and standard deviation of this covariate in the stratum $U=d_1d_0$.
    \end{tablenotes}
    \end{threeparttable}}
\end{table}

\begin{table}[htbp]
\centering
\caption{Analysis of the ITT natural mediation effects on a risk ratio scale, WHO-LARES study. 
}\label{tab:jobs2-supp}
\begin{threeparttable}
\begin{tabular}{cccc}
\hline
\multicolumn{1}{l}{Method} & ITT-NIE$^{\text{RR}}$ & ITT-NDE$^{\text{RR}}$ & ITT$^{\text{RR}}$                         \\
\hline
a & 1.024 (1.003, 1.045) & 1.242 (1.108, 1.385) & 1.271 (1.127, 1.418) \\ 
  b & 1.021 (1.002, 1.039) & 1.239 (1.105, 1.387) & 1.266 (1.121, 1.407) \\ 
  c & 1.029 (1.005, 1.055) & 1.232 (1.098, 1.386) & 1.268 (1.127, 1.414) \\ 
  d & 1.021 (1.002, 1.040) & 1.230 (1.095, 1.381) & 1.256 (1.110, 1.418) \\ 
mr        & 1.021 (1.003, 1.043)    & 1.248 (1.114, 1.405)    & 1.274 (1.137, 1.438)  \\ 
np        &  1.031 (1.010, 1.053)   
                               & 1.219 (1.078, 1.361)           & 1.257 (1.114, 1.400)  \\ 
\hline
\end{tabular}
\end{threeparttable}
\end{table}

\begin{table}[htbp]
\centering
\caption{Analysis of the principal natural mediation effects on a risk ratio scale, WHO-LARES study. 
}\label{tab:pce_who_lares}
\begin{threeparttable}
\begin{tabular}{ccccc}
\hline
\multirow{2}{*}{Method} & \multirow{2}{*}{Estimand} & \multicolumn{1}{c}{Doomed}                & \multicolumn{1}{c}{Harmed} &    \multicolumn{1}{c}{Immune}      \\
 & & $(\widehat e_{11}^{\text{np}}=0.51)$               & $(\widehat e_{10}^{\text{np}}=0.08)$  &   $(\widehat e_{00}^{\text{np}}=0.41)$     \\
\hline
\multirow{3}{*}{a}        & PNIE$^{\text{RR}}$     & 1.019 (0.992, 1.041) & 1.028 (0.990, 1.061) & 1.030 (0.995, 1.073) \\
                           & PNDE$^{\text{RR}}$     & 1.252 (1.066, 1.421) & 2.112 (1.780, 2.478) & 1.087 (0.896, 1.306) \\ 
                           & PCE$^{\text{RR}}$      & 1.276 (1.087, 1.444) & 2.171 (1.862, 2.508) & 1.120 (0.919, 1.323) \\ 
\hline
\multirow{3}{*}{b}        & PNIE$^{\text{RR}}$     & 1.021 (0.999, 1.043) & 1.030 (0.998, 1.065) & 1.018 (0.987, 1.053) \\ 
                           & PNDE$^{\text{RR}}$     &  1.226 (1.043, 1.408) & 2.141 (1.659, 2.775) & 1.096 (0.908, 1.310) \\ 
                           & PCE$^{\text{RR}}$      & 1.252 (1.073, 1.438) & 2.205 (1.705, 2.833) & 1.115 (0.931, 1.311) \\ 
\hline
\multirow{3}{*}{c}        & PNIE$^{\text{RR}}$     & 1.040 (0.998, 1.081) & 1.026 (0.984, 1.061) & 1.010 (0.943, 1.065) \\ 
                           & PNDE$^{\text{RR}}$    & 1.226 (1.044, 1.407) & 2.113 (1.816, 2.458) & 1.099 (0.904, 1.322) \\ 
                           & PCE$^{\text{RR}}$      & 1.275 (1.086, 1.452) & 2.168 (1.844, 2.495) & 1.111 (0.913, 1.308) \\ 
\hline
\multirow{3}{*}{d}        & PNIE$^{\text{RR}}$     & 1.021 (0.999, 1.043) & 1.033 (0.998, 1.070) & 1.017 (0.987, 1.052) \\ 
                           & PNDE$^{\text{RR}}$     & 1.227 (1.040, 1.403) & 2.093 (1.795, 2.441) & 1.096 (0.903, 1.322) \\ 
                           & PCE$^{\text{RR}}$      & 1.253 (1.070, 1.433) & 2.162 (1.852, 2.490) & 1.115 (0.918, 1.319) \\ 
\hline
\multirow{3}{*}{mr}        & PNIE$^{\text{RR}}$     & 1.017 (0.999, 1.034) & 1.029 (1.001, 1.059) & 1.025 (0.983, 1.074) \\ 
                           & PNDE$^{\text{RR}}$     & 1.223 (1.044, 1.403) & 2.296 (1.841, 2.982) & 1.102 (0.911, 1.328) \\ 
                           & PCE$^{\text{RR}}$      & 1.244 (1.070, 1.433) & 2.363 (1.897, 3.045) & 1.129 (0.945, 1.336) \\
\hline
\multirow{3}{*}{np}        & PNIE$^{\text{RR}}$     & 1.025 (1.001, 1.050) & 1.046 (1.013, 1.079) & 1.035 (0.995, 1.075) \\ 
                           & PNDE$^{\text{RR}}$     & 1.181 (1.014, 1.348) & 2.142 (1.724, 2.560) & 1.111 (0.881, 1.340) \\ 
                           & PCE$^{\text{RR}}$      & 1.212 (1.044, 1.379) & 2.241 (1.817, 2.666) & 1.150 (0.916, 1.385) \\
\hline
\end{tabular}
\end{threeparttable}
\end{table}

\begin{table}[h]
    \centering
    \caption{Causal moderated mediation analysis for each discrete baseline characteristic based on the R package \texttt{moderate.mediation} \citep{qin2024causal}, WHO-LARES study. The natural indirect effect and natural direct effect for the entire population, as output from the \texttt{moderate.mediation} package, are 1.021 (1.002, 1.043) and 1.251 (1.114, 1.394), respectively. All mediation effects are defined on the risk ratio scale.}\label{tab:moderated_analysis}
    \scalebox{0.9}[0.9]{
    \begin{threeparttable}
    \begin{tabular}{lcccc}
    \hline
        Subpopulation & Conditional NIE & Conditional NDE \\ 
        \hline
    \textbf{Gender} \\ 
~~Male & 1.008 (0.983, 1.038) & 1.233 (1.011, 1.500)\\
~~Female & 1.029 (1.005, 1.058) & 1.261 (1.099, 1.440)\\
  \textbf{Current marital status} \\
  ~~Unmarried & 1.028 (0.997, 1.065) & 1.246 (1.043, 1.488)\\
  ~~Married & 1.014 (0.992, 1.037) & 1.256 (1.081, 1.447) \\
  \textbf{Education} \\
  ~~Secondary school or less & 1.017 (0.999, 1.037) & 1.305 (1.161, 1.476) \\
  ~~Post-secondary education & 1.039 (0.979, 1.119) & 0.972 (0.678, 1.336) \\
  \textbf{Employment status} \\
  ~~Unemployed & 1.022 (0.996, 1.054) & 1.305 (1.127, 1.507) \\
  ~~Employed & 1.020 (0.995, 1.046) &  1.198 (1.001, 1.408) \\
   \textbf{Smoking status} \\   
   ~~Smoking & 1.040 (1.007, 1.078) & 1.204 (1.023, 1.412) \\
   ~~Non-smoking & 1.003 (0.981, 1.025) &  1.300 (1.105, 1.508) \\
   \textbf{Owning the house} \\  
   ~~No & 1.025 (0.977, 1.079) & 1.110 (0.868, 1.425)\\
   ~~Yes & 1.019 (1.001, 1.038) & 1.300 (1.145, 1.466) \\
   \textbf{House size} \\
   ~~$\leq 50\text{m}^2$ & 1.053 (1.005, 1.128) & 1.256 (0.978, 1.594) \\
   ~~$> 50\text{m}^2$ & 1.012 (0.993, 1.030) & 1.251 (1.103, 1.413) \\
   \textbf{Crowding} \\
   ~~$< 1$ resident per room & 1.031 (1.001, 1.072) & 1.038 (0.869, 1.239) \\
   ~~$\geq 1$ resident per room & 1.014 (0.989, 1.038) & 1.434 (1.235, 1.649) \\
   \textbf{Heating system} \\
   ~~Unsatisfied & 1.023 (1.002, 1.047) & 1.244 (1.095, 1.420) \\
   ~~Satisfied & 1.008 (0.978, 1.042) & 1.279 (1.000, 1.582) \\
   \textbf{Natural light} \\
   ~~Unsatisfied & 1.032 (0.995, 1.076) & 1.248 (1.032, 1.501) \\
   ~~Satisfied & 1.013 (0.994, 1.034) & 1.252 (1.087, 1.431) \\
   \hline
    \end{tabular}
    \end{threeparttable}}
\end{table}

\clearpage

\end{document}